\pdfoutput=1
\documentclass[
   draft=false,
   fontsize=11pt,
   headings=big,
   ngerman,           
   paper=a4,
   headlines=1.1,     
   pagesize,         
   twocolumn=false,       
   twoside,          
   titlepage,        
   raggedbottom,	
   headsepline=yes,      
   chapterprefix=no,  
   listof=totoc,      
   index=totoc,        
   bibliography=totoc,         
   bibliography=totocnumbered 	, 
   listof=numbered, 
   toc=graduated,        
   listof=graduated,      
   listof=flat,       
   numbers=noenddot, 
   ,sort&compress
]{scrbook}
\def\SJSAuthor{Simon J.\ Schilling}
\def\SJSTitle{Contribution to\\Temporal Fault Tree Analysis\\without Modularization and Transformation into the State Space}
\def\SJSSubTitle{Original german title:\\Beitrag zur dynamischen Fehlerbaumanalyse ohne Modulbildung\\und zustandsbasierte Erweiterungen}

\def\printSprueche{SpruecheAn}

\areaset[16mm]{158.4mm}{240.94mm}       

\usepackage{scrpage2}

\ofoot[\pagemark]{} 
\ohead[]{\pagemark}
\ihead[]{\headmark} 
\usepackage{ragged2e}

\usepackage{upgreek} 
\usepackage{textcomp}

\usepackage{timing}
\renewcommand{\timescalefactor}{1.25}

\usepackage{pgf}
\usepackage{tikz}

\usetikzlibrary{external}
\tikzsetexternalprefix{tikz_autocreate/}

\usetikzlibrary{arrows,automata,shapes,trees,shapes.gates.logic.US,patterns,shapes.gates.logic.IEC,plotmarks}

\usepackage{lastpage}

\usepackage{bbm}
\usepackage{savesym}
\usepackage{amsmath}
\usepackage{amssymb}
\usepackage{stmaryrd}
\usepackage{marvosym}
\usepackage{manfnt}

\usepackage{morefloats}

\usepackage{icomma}

\usepackage{flafter}

\usepackage{pifont}

\usepackage[normalem]{ulem}

\usepackage{makeidx}             
\usepackage{graphicx}            
\usepackage{multicol}            

\usepackage{float}
\usepackage{rotating}

\usepackage{cancel}

\usepackage{color}

\immediate\write18{bash ./text/vc -m}
\gdef\GITAbrHash{9089092}%
\gdef\GITAuthorName{Simon J. Schilling}%
\gdef\VCRevision{\GITAbrHash}%
\gdef\VCAuthor{\GITAuthorName}%
\gdef\VCDateISO{2015-05-17}%
\usepackage[T1]{fontenc}
\usepackage[utf8x]{inputenc}

\usepackage[squaren,Gray,textstyle]{SIunits}

\usepackage[
]{acronym}

\usepackage[numbers]{natbib}
\usepackage{hypernat} 

\usepackage{datetime}

\allowdisplaybreaks

\usepackage[%
  bookmarks=true,        
  breaklinks=true,       
  bookmarksnumbered=true,
  plainpages=false,      
  pdfpagelabels,         
  colorlinks,
  linkcolor= black,      
  anchorcolor= black,    
  citecolor = black,     
  filecolor= black,      
  menucolor= black,      
  urlcolor= black,       
  setpagesize=false      
]{hyperref}

\hypersetup{%
pdftitle =    \SJSTitle \SJSSubTitle,
pdfsubject =  {GIT-Revision: \VCRevision},
pdfauthor =   \VCAuthor,
pdfkeywords = {},
pdfcreator =  {LaTeX and other Free Software}
}

\usepackage{booktabs}
\usepackage{longtable}

\usepackage{centernot}

\usepackage[activate]{pdfcprot}
\def\clap#1{\hbox to 0pt{\hss#1\hss}}

\def\mathrlap{\mathpalette\mathrlapinternal}
\def\mathclap{\mathpalette\mathclapinternal}

\def\mathrlapinternal#1#2{%
\rlap{$\mathsurround=0pt#1{#2}$}}
\def\mathclapinternal#1#2{%
\clap{$\mathsurround=0pt#1{#2}$}}

\newcommand{\D}{\mathrm{d}}
\DeclareMathOperator{\E}{e}
\newcommand{\ISO}{ISO\,26262}
\newcommand{\IEC}{IEC\,61508}
\newcommand{\sjshour}{\text{h}}
\newcommand{\sjsyear}{\text{a}}

\DeclareMathOperator{\pand}{\smash{\overset{\shortrightarrow}{\vphantom{\text{=}}\smash{\wedge}}}}
\DeclareMathOperator{\sand}{\smash{\overset{=}{\vphantom{\text{=}}\smash{\wedge}}}}
\DeclareMathOperator{\ist}{{}={}}
\DeclareMathOperator{\booland}{\wedge{}}
\DeclareMathOperator{\boolor}{\vee{}}
\DeclareMathOperator{\boolnot}{\neg{}}
\DeclareMathOperator{\nist}{{}\neq{}}
\newcommand{\False}{\mathit{False}}
\newcommand{\True}{\mathit{True}}
\newcommand{\ES}{\mathit{ES}}
\newcommand{\eES}{\mathit{eES}}
\newcommand{\MS}{\mathit{MS}}
\newcommand{\MCSS}{\mathit{MCSS}}
\newcommand{\formatthetokens}[1]{\text{\sffamily #1}}
\newcommand{\AEtoken}{\formatthetokens{ae}}
\newcommand{\NAEtoken}{\formatthetokens{nae}}
\newcommand{\CEtoken}{\formatthetokens{ce}}
\newcommand{\NCEtoken}{\formatthetokens{nce}}
\newcommand{\EStoken}{\formatthetokens{es}}
\newcommand{\NEStoken}{\formatthetokens{nes}}
\newcommand{\TTtoken}{\formatthetokens{tdnf}}
\newcommand{\eCEtoken}{\formatthetokens{ece}}
\newcommand{\eEStoken}{\formatthetokens{ees}}
\newcommand{\NeEStoken}{\formatthetokens{nees}}
\newcommand{\eTTtoken}{\formatthetokens{etdnf}}

\newcommand{\TERMtoken}{\formatthetokens{tt}}
\newcommand{\minneg}[1]{\boolnot{}#1}

\newlength\mylenisMinimal    
\DeclareMathOperator{\isMinimal}{\text{$ \begin{smallmatrix} \smash{\supseteq}\vphantom{\supset} \\ \subset \end{smallmatrix}$} \hspace{-0.5\mylenisMinimal} \mathclap{\diagup} \hspace{0.5\mylenisMinimal} }
\newlength{\gnat}
\DeclareMathOperator{\corresponds}{\overset{\wedge}{=}}
\DeclareMathOperator{\EW}{E}

\newcommand{\oInfinitOperat}{\mathrm{o}(\Delta t)}

\makeatletter
 \def\@makessubsectionhead#1{%
 \vspace*{50\p@}
 {\parindent \z@ \raggedright
 \normalfont
 \interlinepenalty\@M
 \Huge \bfseries #1\par\nobreak
 \vskip 40\p@ 
 }}
 \makeatother

\makeindex

\begin{document}


\frontmatter
\date{{\tiny~}}\titlehead{{\tiny~}}

%
%
%
%
\title{
  \begin{huge}
  	\SJSTitle
  \end{huge}
}

\author{}
%
\date{
	\begin{Large}
	{\sffamily
	    ~\vspace{-2cm}\\
		\rule{\textwidth}{0.5pt}\\
		~\vspace{0.8cm}\\
		Translation into English\\
		of the doctoral thesis of\\
		Dr.\ Ing.\ Simon J.\ Schilling\\
		at the\\
		\textbf{Bergische Universität Wuppertal}.\\
		\vspace{0.5em}
		Date of examination:\\
		21. December 2009\\
		\vspace{0.5em}
		Reviewer/Supervisor:\\
		Univ.-Prof.\ Dr.-Ing.\ A.\ Meyna\\
		Univ.-Prof.\ Dr.\ rer.nat.\ P.\ C. Müller\\
		\vspace{0.5em}
		\rule{\textwidth}{0.5pt}\\
		~\vspace{0.5em}\\
		{\small{}
			The german original can be downloaded from
			\\
			\href{http://nbn-resolving.de/urn/resolver.pl?urn=urn:nbn:de:hbz:468-20100070}{http://nbn-resolving.de/urn/resolver.pl?urn=urn:nbn:de:hbz:468-20100070}
			\\
			Translated version of \today.
			\\
			This work is licensed under a Creative Commons Attribution-ShareAlike 4.0 International License.
			\\
			See inside for more details.
			\\
			~
		}
	}
	\end{Large}
}


\maketitle 
%
%
%
%
%
%
%
%

\thispagestyle{empty}
\chapter*{~}
  ~\vspace{4cm}

\begin{center}
  \begin{large}
    For\\
    Albert\ and\ Alexandra\ and\ Liselotte
\end{large}

\end{center}

%
%
%
%
\thispagestyle{empty}
\chapter*{Preface to the Translation}
This translation into English was done in order to present my work to a broader audience.
I aimed at staying as close to the german original as possible.
This is especially relevant for the state of the art chapter which was not updated.
Thus, newer work, as well as additional work by authors that were already referenced in the original, was not taken into account.

The german original is an official doctoral (i.e.\ Ph.D.) thesis and was published and is hosted as PDF by the university itself.
I chose to publish this translation -- including the complete latex sources -- under a CreativeCommons license and host it at github because I was looking for a simple, stable and open -- as in open source -- solution for the benefit of potential readers.
As English is not my first language, I surely made some mistakes and would greatly appreciate any comments and suggestions for improvements.

~\\
\noindent
Munich, May 2015
\hfill
Simon Schilling

\vfill{}

\vfill{}

\section*{License}
``Contribution to Temporal Fault Tree Analysis without Modularization and Transformation into the State Space'' by {\SJSAuthor} is licensed under the Creative Commons Attribution-ShareAlike 4.0 International License.
\\
To view a copy of this license, visit \href{http://creativecommons.org/licenses/by-sa/4.0/}{http://creativecommons.org/licenses/by-sa/4.0/}.
\\
It is based on the work at \href{http://nbn-resolving.de/urn/resolver.pl?urn=urn:nbn:de:hbz:468-20100070}{http://nbn-resolving.de/urn/resolver.pl?urn=urn:nbn:de:hbz:468-20100070}, which is the german original version of this thesis, and also by {\SJSAuthor}. Note, that the german original is not published under a Creative Commons License.
~\\
~\\
\includegraphics[width=50mm]{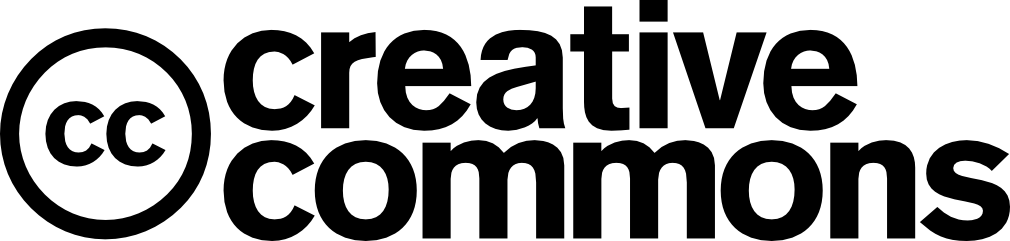}
\hfill
\includegraphics[width=50mm]{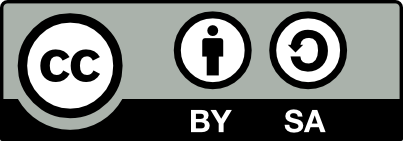}

\chapter*{Preface}
This work was accomplished during my time as scientific member of the Central Functional Safety Team at BMW Group in Munich, Germany.

I want to specifically thank Univ.-Prof.~Dr.-Ing.~Arno Meyna and Dipl.-Ing.~Christoph Jung.

I thank Professor Meyna, for his support during my external promotion at the department of safety engineering, safety theory and traffic engineering at the Bergische Universit\"{a}t Wuppertal.

I thank Mr.~Jung, who was head of the Central Functional Safety Team at BMW Group and convenor of ISO TC\,22 SC\,3 WG\,16 and as such one of the main creative heads behind and responsible for ISO~26262, for making this work possible and I thank him for repeatedly trusting and supporting me throughout the last years.

I thank Prof.~Dr.~rer.~nat.~P.~C.~M\"{u}ller for writing the second assessment on this work and being part of the graduation comittee.
I thank Prof.~Dr.-Ing.~Dipl.-Wirtsch.-Ing.~B.~H.~M\"{u}ller for chairing the graduation comittee.
I thank Prof.~Dr.-Ing.~U.~Barth for being part of the graduation comittee.

I thank my collegues at BMW for their support and interest.

I especially thank Dr.-Ing.~Martin Woltereck, who brought me to the field of functional safety and to fault tree analysis.

~\\
\noindent
Munich, December 2009
\hfill
Simon Schilling

\thispagestyle{empty}
\addchap*{Abstract}
\minisec{Background}
\emph{Fault tree analysis} (FTA) is a well established method for qualitative as well as probabilistic reliability and safety analysis.
Fault trees are Boolean models and thus do not support modelling of dynamic effects like sequence dependencies between fault events.
In order to overcome this limitations, \emph{dynamic fault tree} methods were defined previously.
Most of these are based on complete or partial transformation of the fault tree model into state-space-models like Markov chains or Petri nets.
These state-space-models generally suffer from exponential state explosion which imposes the necessity to define  small ``dynamic'' modules which need to be independet from the rest of the model.
Moreover, these state-space-models lack some of the FTA's benefits like logical simplification of complex system functions or a real cutset analysis.
Because of these deficiencies, a method is needed that allows consideration of sequence dependencies without transformations into state-space.
This work describes such a new approach.
\minisec{Concept}
The new \emph{temporal fault tree analysis} (TFTA) described in this work extends the Boolean FTA in order to take sequence dependencies into account.
The TFTA is based on a new \emph{temporal logic} which adds a \emph{concept of time} to the Boolean logic and algebra.
This allows modelling of temporal relationships between events using Boolean operators (AND ``{\small$\booland$}'', OR ``{\small$\boolor$}'', NOT ``{\small$\boolnot$}'') and two new temporal operators (PAND ``{\small$\pand$}'' and SAND ``{\small$\sand$}'').
With a set of \emph{temporal logic rules}, a given \emph{temporal term} may be simplified to its \emph{temporal disjunctive normal form} (TDNF) which is similar to the Boolean DNF but includes event sequencies.
In TDNF the top event's temporal system function may be reduced to a list of \emph{minimal cutset sequences} (MCSS). These allow qualitative analyses similar to Boolean cutset analysis in normal FTA.
Furthermore the TFTA may also be used for probabilistic analyses. Probabilities and rates of MCSS may be calculated without using state-space models. Again the procedure is similar to the normal FTA: top event failure probabilities and rates are derived from the failure probabilities and rates of the basic events including sequence dependencies.
\minisec{Realisation}
Starting with the Boolean FTA this work describes a new notation and new rules for a temporal logic. This temporal logic aims at transforming temporal terms into a TDNF, which then may be transformed further into a form where all terms are mutually exclusive. This form is well suited for quantification, too. Several examples are provided which explain each step in detail. Furthermore, there are two probabilistic approximation methods described, which allow a significant reduction of the calculatory effort.
\minisec{Results}
One significant aspect of the new TFTA described in this work is the possibility to take sequence dependencies into account for qualitative and probabilistic analyses without state-space transformations. Among others, this allows for modelling of event sequencies at all levels within a fault tree, a real qualitative analysis similar to the FTA's cutset analysis, and quantification of sequence dependencies within the same model.
\clearpage
\thispagestyle{empty}
\minisec{General Remark and Disclaimer}
All safety and reliability analyses in this work are presented solely for the purpose of demonstrating new analysis methods and are to be seen as simplifications and examples only.
While they use, among others, technical functions and data similar to those of real systems,
they must not be taken as evidence for the safety or reliability of
existing or planned ``real life'' systems, functions, or components.
%
%
%
%
\tableofcontents                  
%
%
%
%
\mainmatter
%
%
%
%
\clearpage
\ifx \printSprueche\undefined
\else
  \renewcommand*{\dictumwidth}{.40\textwidth}
  \setchapterpreamble[ur]{%
    \dictum[Mueller]{System safety is organized common sense.}
    \vspace{3cm}
  }
\fi

\chapter{Introduction}\label{_chap_1308002}

%
%
%
%
%
%
%
\section{Motivation}\label{chap080401-070}
The fault tree analysis (FTA) is one of the most important methods of modelling and analyzing the realibility and safety of systems qualitatively as well as probabilistically.
In the automative domain there is a trend towards more safety critical electronics \cite{Florecke2004}, and thus functional safety is increasingly important \cite{Meyer2003Methoden}.
Therefore the domain specific functional safety standard {\ISO} \cite{ISO2009} is currently being derived from the more generic {\IEC} \cite{2002IECa}.

In the automotive domain the {FTA} is used during development for several reasons:  the allocation of safety requirements, as well as the confirmation and verification of requirements (e.g. failure rates as required by {\ISO}), and the comparison of safety architectures.

Today, the {FTA} is generally considered as state of the art, e.g.\ \cite{Veseley1981NUREG-0492,2004IEC,19811990DIN,Schneeweiss1999Die,Verband2003Sicherung}. 
Nevertheless certain problems remain, and there is an ongoing scientific interest for the {FTA} method.

This thesis results from years of practise experience during my time at the functional safety department of a german automotive manufacturer.
Contrary to expectations, the conventional -- i.e.\ static -- {FTA} is still having  difficulties at providing realistic and not too conservative results when applied to modern electric/electronic ({EE}) systems.

The operational behaviour and failures of such systems are highly dynamic in a sense that 
subsystems, functions and components (or their failures) depend on each other (structural dependencies) or depend on their relative timing (temporal dependencies) \cite{Schilling2006Bedeutung}.

The fault tree methode on the other hand is limited to binary parameters as it is based on Boolean (failure-)logic.
As a consequence, temporal dependencies and dependencies between failure rates of fault tree basic events must be omitted.
Both limitations may usually be circumvented, or at least mitigated, by taking specific assumtions and approximations into accout.
But both problems can not be completly solved from within the conventional {FTA}.

Furthermore, when using fault trees one has to keep in mind that conservative approximations (less modelling effort) usually conflict with the wish to avoid an unnecessarily expensive system design.
Unprecise (approximated) fault tree models must not lead to overly complex and overly expensive technical solutions in the system under consideration.

This problem and conflict is well known \cite{Heidtmann1992,Manian1998,Hauschild2006}.
In general, there is always the possibility to analyze the system using other methods that can take dynamic effects into account, like e.g.\ state based methods.
On the other hand there is a reason for the FTA's success as one of the most widely used methods for analyzing the reliability and safety of complex systems \cite{Meyna2003Taschenbuch}:
in comparison to other methods fault trees are easy to use, to read, to understand, and they are scalable.
This is, because a system's fault tree is similarly structured as the system architecture.
Especially state based methods (e.g.\ markov diagrams) lack this feature.

For years there have been several approaches to combine state based methods with the conventional {FTA}.
These aim at combining the benefits of both methods while circumventing their disadvantages.
Usually the user shall stay within the more intuitive fault tree, while modelling the system under consideration; then, the system's dynamic effects and dependencies are hidden from the user by state based models that do the calculations in the background automatically.

Such hybrid techniques are often called \emph{dynamic {FTA}}; but they also have some specific disadvantages.
Mostly they use fault trees as a tool for easy visualization or relatively simple creation of models; but they do not also fully use the fault tree for the analysis and calculation, and thus they do without some of the FTA's biggest benfits.  

These problems, as well as pure scientific curiosity, lead to intense research on a more efficient way to handle dynamic effects and dependencies from within fault trees.

This thesis presents the results of this research.

%
%
%
%
%
\section{Structure of this Thesis}\label{chap080401-071}
This thesis deals with dynamic effects in safety and reliability analyses, and specifically with the modelling of failure sequences in fault trees.
It is structured as follows.

Chapter \ref{chap080401-001} presents the state of the art as relevant for this thesis; specific focus goes to the conventional Boolean {FTA} (chapter \ref{chap080401-010}), as well as to dynamic extensions of the {FTA} (chapter \ref{chap080401-011}); the latter includes methods where the fault tree model is transformed into a state based model, as well as methods using temporal logics.

This survey points to several shortcomings of the current state of the art; specifically these result from changing the modelling and analysis and calculation's focus and are listed in chapter \ref{chap80401-002} which also derives criteria and requirements for improvements.

Chapter \ref{_chap_080401-004} describes the proposed new approach for including 
failure event sequences into the fault tree without changing to the state space.
This new \emph{temporal fault tree analysis} ({TFTA}) relies on an temporal extension to the conventional Boolean algebra and logic; this \emph{temporal logic} has its own notation  (chapter \ref{chap080401-030}) and its own laws of transformation (chapter \ref{chap080401-031}).
Chapter \ref{chap080809-001} then shows how to transform temporal terms into disjunct minimal failure event sequences.
There is also an extended form of the {TFTA} which is presented in chapter \ref{080817-010}; it allows for reduced calculatory effort when solving more complex temporal failure functions.

Chapter \ref{chap080401-033} discusses the quantification of temporal terms, which in turn allows probabilistic evaluation of temporal fault trees.

Chapter \ref{chap080104_005} compares the new {TFTA} approach with
a) conventional Boolean {FTA},
b) the dynamic fault tree approach ({DFT}) as a typical dynamic extension of the Boolean {FTA}, and
c) markov diagrams.

Chapter \ref{chap080401_006} applies the {TFTA} to a more complex and complete example in order to demonstrate its practical use. A typical automotive ECU architecture is analyzed: beginning with its system analysis, followed by creation of a corresponding temporal fault tree, and finally the qualitativ as well as probabilistic fault tree transformation and analysis.

This thesis closes with a summary and outlook in chapter \ref{chap080401-008}.
\clearpage
\clearpage
\ifx \printSprueche\undefined
\else
\renewcommand*{\dictumwidth}{.3333\textwidth}

\setchapterpreamble[ur]{%
\dictum[An FAA administrator]{Of course, it is safe, we certified it.}\vspace{3cm}
}
\fi

\chapter{State of the Art: Static and Dynamic Fault Tree Analysis (FTA)} \label{chap080401-001}
This chapter provides an overview over the state of the art as relevant for the {TFTA} method.
\begin{itemize}
	\item Chapter \ref{chap080409-001} describes the field of safety related fault tree analysis in general.
	\item The conventional and solely static {FTA} is among the most common methods for systematic top down failure analysis of complex systems, see chapter \ref{chap080401-010}.
	\item As shown in chapter \ref{chap080401-011}, today there are several extensions to the conventional {FTA}; they take dynamic failure behaviour into account and try to mitigate the FTA's shortcomings in this field. Chapter \ref{chap080413-003} presents state based methods, and methods using temporal (failure) logics are discussed in chapter \ref{chap080413-006}.
	\item Chapter \ref{chap080630-001} summarizes the state of the art, which leads to the main problem description of this thesis in the following chapter \ref{chap80401-002}.                                                                  
\end{itemize}

%
%
%
%
%
%
%
\section{Background}\label{chap080409-001}

\subsection{Reliability and Safety Analyses}\label{chap080412-001}
The \emph{reliability} of a system or a component (in general: an entity) is defined as its ``capability [\ldots] to meet expected performance criteria, given by its intended use, during a defined time period \cite{DIN1990}.
An entity that has failed can no longer provide its functionality; therefore, conventional \emph{reliability analysis} reflects upon entities' failure behaviour.

Such an analysis usually covers the following steps \cite{Bitter1986Technische}:
it supports develoment of new systems by comparing different -- existing or proposed -- system designs among each other, as well as comparing them to objective requirements (i.e.\ \emph{reliability prediction}, \emph{reliability comparison}, \emph{reliability pursuit}, \emph{identification of weak spots}).
Additionally, it allows \emph{reliability verification} of existing systems and concepts.
The same methods and analytical approaches are usually used for all these purposes. 

In comparison to reliability analysis, the \emph{safety analysis} is focused on only those system and component failures that lead to loss of ``safety'', where safety is defined as ``freedom from unacceptable risks'' \cite{2002IECa}.
From a safety perspective, an entity's relevant reliability is therefore its capability -- or, in case of a more probabilistic view, its probability -- to not induce \emph{dangerous effects} (i.e.\ damage) during a defined time period and under given circumstances.
Thus, reliability, from a safety perspective, takes failures consequences into account, too.

Safety analyses therefore need to define which risks and which damages are relevant.
In the context of conventional safety of technical systems these typically are the danger for life and limbs, or injuries and death of persons \cite{2002IECa}.
In general, the same analysis methods are used in other contexts, too; e.g.\ in the context of security of technical systems \cite{Isograph2005,Sulfredge2002}.
This thesis only addresses the safety context\footnote{Author's remark: in german there is only one term ``Sicherheit'' for both of the english ``safety'' and ``security''; therefore, a further distinction and limitation of this thesis' scope follows at this place, but is omitted in the english translation.}.
\subsection{Static and Dynamic Analyses}\label{chap080412-002}
\subsubsection{Dynamic System Behaviour}\label{chap080624-001}
A system behaves dynamically if \cite{Siu1994} the system response to a initial disturbance develops over time, while the system's components interact among each other, as well as with their surrounding.
In comparison, conventional fault tree analysis looks at unwanted events (i.e.\ system failures) as static, determined, and time invariant consequence to certain component failures \cite{Siu1994}.

In a world full of dynamic influences and interactions basically all technical systems also behave dynamically.
Statistical methods and models for reliability and safety analysis of systems therefore necessarily only approximate a system's real dynamic behaviour.

This simplification is the main reason why handling of statistical analysis like {FTA} or \emph{reliability block diagrams} ({RBD}) is relatively easy.
Actually, in many cases it is the assumption of static behaviour that makes an analysis feasible at all.
In practise the relevant question is which static approximations allow ``good enough'' representation of the actual dynamic failure behaviour.

It has been demonstrated that conventional {FTA} is very well suited for logical and probabilistical analyses of systems, if their failure behaviour is -- at least in the first approximation -- free of time dependencies or dynamic interactions between its components.

On the other hand, and since the very beginning of systematic failure behaviour analysis after the mid-20th century, researchers and users are complaining about static analysis being too imprecise \cite{NRC1975}.
Therefore, scientists are researching how static analysis methods like {FTA} may be extended by the most important dynamic effects -- but without excessively increasing modelling and calculatory effort. 
\subsubsection{Methods of Modelling}\label{chap080623-001}
In sight of \cite{Aldemir1994} and \cite{Woltereck2000} three types of \emph{dynamic realiability and safety analyses} (\acsu{ZSA}) may be distinguished by their different modelling approaches.
These are
\begin{itemize}
	\item state transition models, especially makov models, e.g.\ \cite{Fahrmeir1982},
	\item direct simulation of systems, especially using {MoCaS}, e.g.\ \cite{Hauschild2006,Hauschild2007a}, and
	\item extensions of static event sequence analysis and the {FTA} in order to also represent dynamic effects.
\end{itemize}
The following chapters cover those methods in more detail.

%
%
%
%
%
%
\section{Static {FTA} -- the Classical Approach}\label{chap080401-010}
The history of {FTA} can be traced to the mid-20th century and starts with the reliability analysis of the Minuteman missle \cite{Leveson2003White,Watson1961}.

The conventional fault tree \cite{19811990DIN,2004IEC,Schneeweiss1999Die} is a Boolean model, that systematically and methodically describes the interaction of failures within a system that lead to a system failure. 
It is a top down or deductive method.
Starting from an undesirable event or system state -- the so-called \emph{TOP} --, more detailled failure events are searched for iteratively, that cause this TOP.
Graphical representation of these failure events is done using a tree notation, the so-called \emph{fault tree}.
The components' failure events modelled in the fault tree are represented by events that can be in one of two states according to Boolean logic:
``intact/unfailed/failure has not occurred'' is represented by a Boolean $\False$ or $0$, and ``defect/failed/failure has occurred'' is represented by a Boolean $\True$ or $1$, respectively.

Evaluation of the fault tree is done qualitatively as well as probabilistically.
The system is comprised of clearly separable elements (components), each of which has its own reliability and safety characteristics, and that influence the system reliability and safety according to the components' logical interconnection. 
Using these connections, the fault tree model is then able to derive the system charateristics from its component characteristics.

With the simplifying laws of Boolean algebra the \emph{system function}/\emph{failure function}, i.e.\ the logical function of the TOP event, is transformed into a minimal disjunctive normal form.
Thereby determined \emph{minimal cutsets} of the fault tree may then be further used probabilistically together with the laws of probability calculus.
The probability or frequency of occurrence of the undesirable event or system state is -- for non-repairable systems -- the failure probability and the failure density or failure rate of the TOP event, respecively; for repairable systems, it is the unavailability and failure frequency of the TOP event, respecively, \cite{Amari2004}.

Furthermore, qualitative analysis of the system architecture is possible, too, because of the similarity of the fault tree model to the real system structure;
specifically, such qualitative analysis allows analysis of redundancy structures as well as sensitivity analysis \cite{Skorek2004Determination}, importance analysis \cite{Dutuit2001}, and confidence analysis \cite{Woltereck2004Reliability}.

Qualitative and probabilistic static {FTA} is state of the art in many domains like nuclear  \cite{Veseley1981NUREG-0492}, aerospace \cite{Veseley2002Fault}, and automotive industries \cite{Limbourg2007,Verband2003Sicherung}.
There is demand for further research on using {FTA} for analysis of software ``failures'' \cite{Weber2003Enhancing}, especially because of difficulties stemming from proper representation of dynamic effects, see below.

{FTA} is intuitive in its application -- in comparison to other methods like e.g.\ state based markov diagrams; thus, learning the {FTA} method is comparatively easy, and fault trees are easy to create, read, understand, rework, and edit, as well as to detail iteratively, and to use in modules.  

One main limitation of the {FTA} is that its event are (only) bivalent, i.e.\ $\True$ or $\False$;{}
another limitation is that the \emph{assumptions of monotony} or \emph{coherence} must be satisfied \cite{Heidtmann1997,Andrews2000To};
a third limitation is the implied independence of its basic events.
Furthermore, {FTA} has only very limited possibilities of representing dynamic failure and repair behaviour \cite{Manian1998}.
Reason for this is the underlying Boolean logic \cite{Gumm1981Boolesche}, that has no concept of time, and thus only covers structural aspects of failure combinations \cite{Heidtmann1997}.
No statement is made about the sequence in which events occur, as well as about other time dependencies, see chapter \ref{chap080413-001}.


%
%
%
%
%
%
\section{Dynamic {FTA}}\label{chap080401-011}
The expression \emph{dynamic {FTA}} is often used as a synonym for the \emph{dynamic fault tree} ({DFT}) approach according to Dugan \cite{Dugan1992}.
The {DFT} uses markov chains to extend the static {FTA} to model and to analyze \emph{sequence dependencies}.

The {DFT} approach therefore defines its ``dynamic'' with event sequences.
This thesis and the {TFTA} approach, as described in chapter \ref{_chap_080401-004}, are also based on this underlying interpretation of ``dynamic'', i.e.\ on the possibility of representing event sequences. 
%
\subsection{Defining Dynamic with Event Sequences}\label{chap080413-001}
Boolean logic with its AND, OR, and NOT operations is not capable of expressing temporal relationships.
For example, the failures of two components $A$ and $B$ in a system shall be considered.
The event ``$A$ AND $B$'' represents ``both components have failed''.
It does not, though, provide any information on the real points in time at which $A$ and $B$ occurred, and from that: the sequence, in which both events occur.
This Boolean view grasps only the static state that the two components are (or are not) failed.

In contrast to that, a dynamic view discriminates between different ways of reaching this event or state.
It extends the all-static analysis of only considering possible combinations of events \cite{Sullivan1999}.

For ``$A$ AND $B$'' there are three different such ways.
First, $A$ may fail before $B$, and then $B$ fails later, too.
Second, $B$ may fail before $A$, and then $A$ fails later, too.
Third, $A$ and $B$ may fail exactly simultaneously.

\begin{figure}
	\centering
        \input{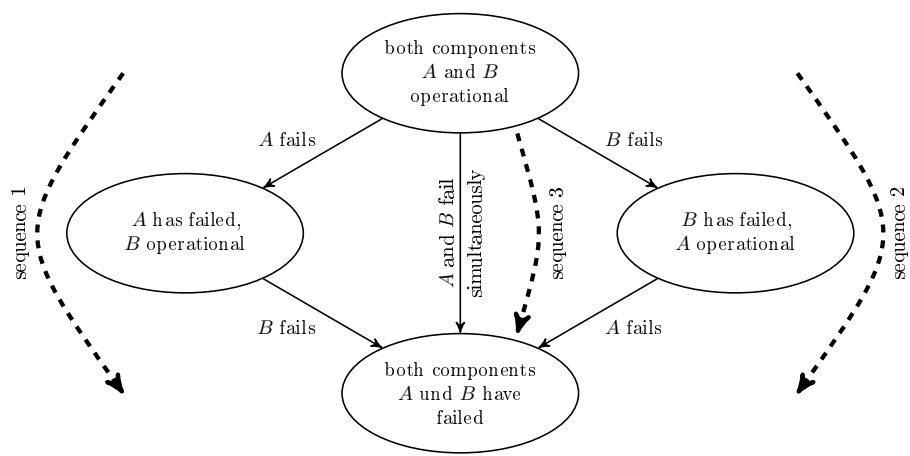}
	\caption{State-transition diagram of a simple, redundant, and non-reparable system, that consists of two components $A$ and $B$; this diagram shows the possible four states and five transitions.}
	\label{fig:bsp_zweiausfaelle_zustaende}
\end{figure}

Each of these ways leads to the -- from a Boolean point of view: identical -- state, that both components have failed.
This discrimination of possible ways to an event or state may be visualized using state-transition diagrams.
Figure \ref{fig:bsp_zweiausfaelle_zustaende} shown such a state-transition diagram, corresponding to the example above.
``Dynamic'' as discrimination of different ways to an event or state works with temporal expressions like ``before'', ``after'', ``first'', ``then'', ``simultaneous'', and so on.
Modelling such ``dynamics'' requires to differentiate the different points in time when events occur.
This capability requires that a \emph{concept of time} exists within the model \cite{Distefano2007}.
Conversely, differentiating points in time when events occur also allows to distinguish between different event sequences.
And with event sequences a multitude of dynamic effects can be described \cite{Dugan1999,Manian1998}.
\paragraph{Next Steps}~\\
The contribution to dynamic {FTA}, as presented in this thesis, also uses ``dynamic'' in the sense of representation of event sequences.
The next section \ref{chap080413-002} differentiates this meaning of ``dynamic'' from others that are also used in the context of {ZSA}, and specifically are used in the context of {FTA}.
Section \ref{chap080413-003} discusses typical implementations of this meaning of ``dynamic'', specifically implementations based on markov chains and petri nets.
Section \ref{chap080413-006} outlines a very different way of extending the {FTA} by event sequences, and for this purpose describes several approaches of extended (temporal) failure logics.
Chapter \ref{chap080630-001} summarizes this state of the art of dynamic {FTA}.
\subsection{Other Definitions of Dynamic}\label{chap080413-002}
Apart from the consideration of event sequences there are other temporal dependencies among (failure) events, and consequently other definitions of ``dynamic'' in the {ZSA} field, too, some of which are listed below.
One overview in \cite{Aldemir1994a} is not he most recent, but is still valid.

In \cite{Hirschberg1994} dynamic effects in analyses result either from time-dependent failure rates, or from time-dependent unavailabilities, or from reduction of uncertainty whether the reliability data used is correct, or from failure sequences.

Abstracting these categories, dynamic either results from variable reliability data, or from the failure events' sequence. 
Sometimes, \emph{phased mission} methods are seen as a third such category, see e.g.\ \cite{Distefano2007,Xing2002} or \cite{Garrett1995Dynamic}.
But these may as well be seen as belonging to either of the first two categories, or they may be interpreted as piecewise static analysis.

A further distinction into ``fast'' and ``slow'' dynamic temporal dependencies is given in \cite{Woltereck2000}.
Slow dynamic effects occur during normal operation, e.g.\ by aging, learning effecs, or changes in the system.
On the other hand, fast dynamic effects describe incidents, and thus dynamic {ZSA} focus on these.
In \cite{Woltereck2000} dynamic {ZSA} is based on {MoCaS}.

The referenced work comes mainly from the nuclear domain.
They emphasize explicit consideration of temporal dependencies as well as consideration of \emph{{HRA}} (\acsu{HRA}) \cite{Kolaczkowski2004Human} as another important contribution of dynamic {ZSA}. On the other hand, {HRA} is not as relevant in the automotive domain today; reasons for this are
\begin{enumerate}
  \item that safety critical systems are preferably designed as \emph{fail safe systems}, thus real \emph{fail operational systems} are rare \cite{Thane1996},
  \item the lack of human operators as part of the safety systems, which directly influence the system's behaviour during normal operation as well as during incidences, and
  \item the lack of inspection, maintenance, and repair crews, as they are known in plants or in the aerospace domain.
\end{enumerate}
It is expected that {HRA} will become more and more relevant for the functional safety of automotive systems, too, specifically because of the increase of high-voltage systems in electric and hybrid cars, and because of the increasing integration of active safety systems and driver assistance systems.

Moreover, there are special approaches to dynamic {ZSA} using {MoCaS} in the automotive domain, too.
For example, \cite{Hauschild2007a} considers the influence of dynamic system behaviour on the system's failure behaviour by taking time-dependent failure data into account.
As these approaches require comparably high effort, they are used only for special cases and are not (yet) widespread. 
\subsection{Dynamic FTA~-- Other Approaches}\label{chap080413-003}
From here on this thesis on dynamic {FTA} focusses on ``dynamic'' in the sense of representation of event sequences.

Known approaches to extending the {FTA} by dynamic effects typically are either simulations, or they automatically transform the fault tree model into a markov model, and then solve the resulting differential equation system.

\begin{figure}
	\centering
	\input{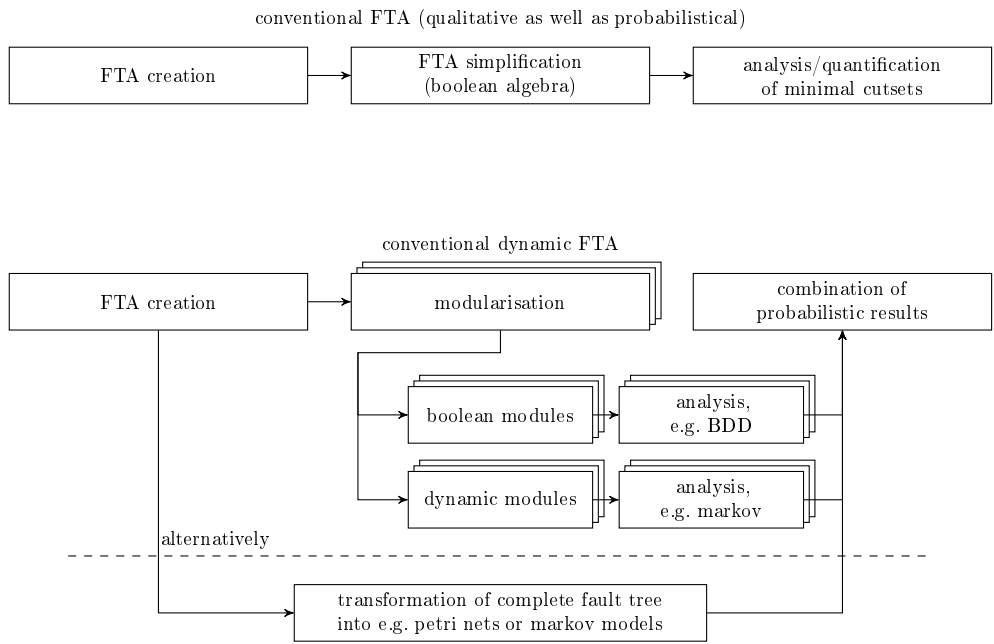}
	\caption{Main steps in conventional {FTA} (top) and - for comparison - main steps of a conventional dynamic fault tree extension using state-based modelling.}
	\label{fig:stand_der_technik_dfta_vorgehen}
\end{figure}

The well known {DFT} approach \cite{Dugan1992} is based on modularizing the fault tree into static and dynamic modules, that are then calculated using \emph{binary decision diagrams} (\acsu{BDD}) \cite{Coudert1994,Jung2004Development} and markov chains.
Static modules consist only of Boolean fault tree gates and events; dynamic modules also include dynamic fault tree gates.
The latter are used to represent effects like sequences, or cold, warm, and hot redundancies, or trigger events.
Figure \ref{fig:stand_der_technik_dfta_vorgehen} shown the main steps of this approach and compares them to the conventional static {FTA}.

The {DFT} method is included into numerous fault tree tools in differing completeness; e.g.\ in DIFTree \cite{Dugan1997} or Galileo \cite{Sullivan1999}, as well as in several commercial {FTA} tools like Isograph Faulttree+ \cite{Isograph2005FaultTree+}, ITEM Toolkit \cite{ITEM2007}, or RELAX Reliability Studio \cite{Relex2007}.
{DFT} are also mentioned in the recent edition of the \emph{Fault Tree Handbook} \cite{Veseley2002Fault}.

A similar approach is presented in \cite{Montani2006}, which uses \emph{dynamic bayesian networks} instead of creating and solving markov chains, a method for reducing calculatory costs.

Another alternative in \cite{Tang2003} solves {DFT} modules with modified {BDD}, which are called \emph{zero-suppressed binary decision diagrams}; this approach requires to manually include sequenc information into the relevant minimal cutsets, instead of using markov models.
This manual step limits the use of dynamic gates to relatively simple structures, though.
Another similar such method is discussed in \cite{Bozzano2003}.

The approach introduced in \cite{Bouissou2003,Bousissou2007} is based on \emph{Boolean logic driven markov processes} (BDMP) and, compared to the listed approaches from above, improves qualitative system analysis, and to some extent also allows taking repairable components into account.

A different approach to dynamic {FTA} based on petri nets, and without markov models, is chosen in \cite{Bobbio2003} and \cite{Schneeweiss1999a,Schneeweiss1999b}; a further possibility are \emph{state-event-fault-trees} given in \cite{Kaiser2004State-Event-Fault-Trees}. 

\minisec{Discussion}
All these approaches to dynamic {FTA} are based on transforming the original fault tree model into state-based models.
The latter are able to consider temporal dependencies and thus event sequences, too.
The different approaches differ in their choice of transformation method -- on the one hand, the complete fault tree is transformed; on the other hand, modularization and transformation only of those sub-trees that carry relevant dynamic data --, and they differ in their choice of state-based method.

But they have in common that, firstly, their calculatory cost grows exponentially with the size of their dynamic modules.
Newer methods in \cite{Gulati1997,Dutuit1996} reduce the time needed for the actual modularization, so that the calculatory effort grows only linearly with the number of modelled elements.
But the complexity for solving the markov chains is always $\text{O}\{K\cdot N^3\}$ \cite{Amari2003}.
$K$ is dependent on the number of computation-steps, and thus from the mission time and the calculations precission.
And $N$ is dependent on the number of states within the markov model; this number in in the range of $N\ist n^n$ for $n$ elements under consideration.
This \emph{state explosion} \cite{Malhotra1995} requires modularization with as small dynamic modules as possible.
On the other hand, these markov models and their resulting differential equation systems can, in many cases, only be solved approximately, even despite of modularization (see e.g.\ \cite{Amari2003}).

Secondly, modularization requires that the modules are independent from each other.
This limits the dynamic dependencies between the system's elements that can be considered in the model; or it implies increasing the size of the dynamic modules -- with the described negative impact on calculatory effort.

Thirdly, qualitative analyses are not possible, or possible only for very simple structures.
This is owed to the transformation into the state space which does not follow the real system architecture as closely as the Boolean system model.
One of the main benefits of the {FTA} is therefore missing in state based models: they can not ``automatically'' transform the modelled structure into a minimal form.
For example, the {DFT} provides -- depending on its specific implementation -- either ``normal'' Boolean minimal cutsets without any event sequence information, or provides minimal cutsets with ``meta-events'', that cover complete markov models without further breaking them apart.

\begin{figure}
  \centering
    \input{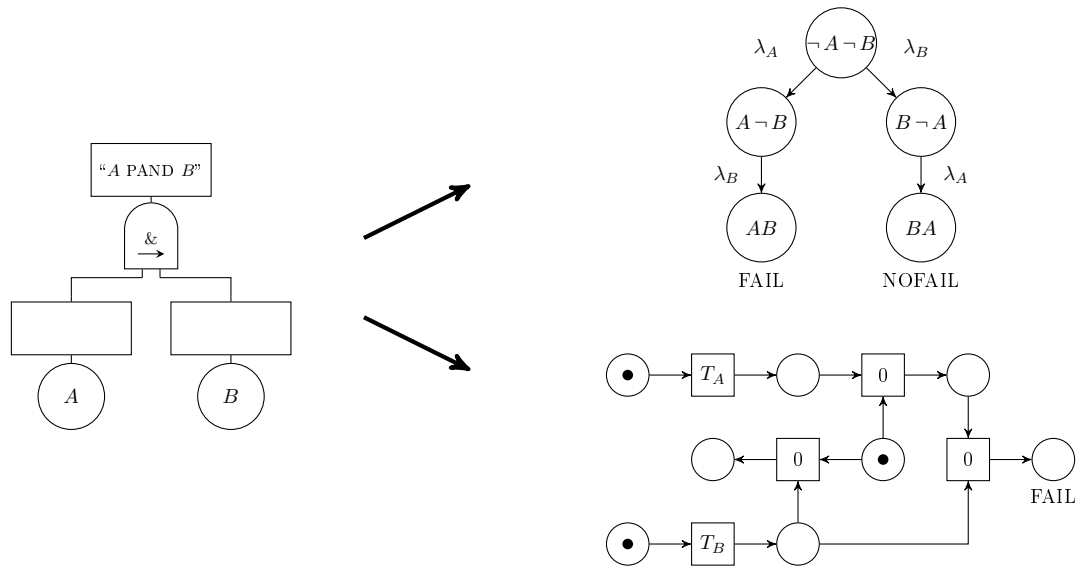}
  \caption{On the left side, a PAND gate with inputs $A$ and $B$, that need to occur in the sequence ``$A$ first, and then $B$'', in order for the gate event to occur.
  The symbol used is not the one used in \cite{Veseley1981NUREG-0492} but is taken from the {TFTA} approach in chapter \ref{_chap_080401-004}.
  The top right side shows a markov model from \cite{Manian1999}, which is equivalent to the PAND gate; the bottom right side shows a petri net from \cite{Schneeweiss1999a}, which is also equivalent to the PAND gate ($T_A$ and $T_B$ represent the time-to-failure of $A$ and $B$).}
  \label{fig:stand_der_technik_markov_petri}
\end{figure}

Fourthly, state based models lack the ``user-friendliness'' of Boolean methods, also resulting from the Boolean model's closeness to the real system architecture.
Instead, components and their dependencies are, for example, expressed by states and state-transitions (in markov models), or by places and transitions and marks (in petri nets).
Figure \ref{fig:stand_der_technik_markov_petri} shows an example.
One effect resulting from these differences is that state-based methods and models are less easy readable, less comprehensible, less easy in maintenance, and less scalable than the conventional {FTA} \cite{Scientific2000}.
\subsection{Dynamic FTA -- Based on a Temporal Failure Logic}\label{chap080413-006}
Another possibility to include temporal dependencies is to use a \emph{temporal logic} that extends the conventional Boolean logic.
A temporal logic describes not only structural combinations of different events -- that is the Boolean approach --, but also has a concept of time.
The latter is used to make statements on the points in time at which events occur, and to include such statements into the logic function.

Applied to the field of reliability and safety, there are several approaches to use temporal logic for fault trees.
One early approach of describing event sequences is found in \cite{Fussel1976}.
It concentrates on probabilistic modelling aspects for individual event sequences; this is an approach that has later been revived and refined, e.g.\ in \cite{Schneeweiss1999a} and \cite{Long2000}.
All these works do not expand onto a general temporal logic, which goes beyond taking individual event sequences into account.
Therefore, they require that the relevant minimal failure sequences, that lead to the TOP event, have been found with other methods.
This, of course, severely limits their application for complex projects. 

\begin{figure}
	\centering
		\input{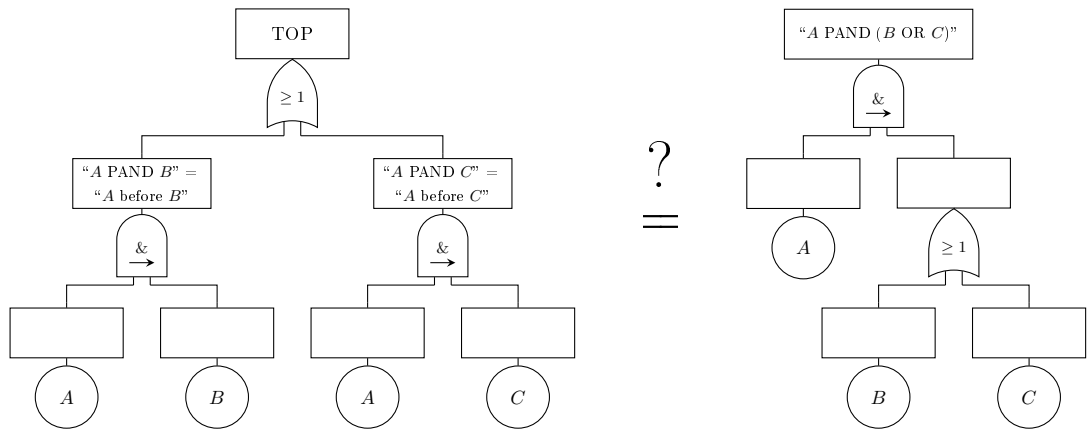}
	\caption{Questions on the state of the art of dynamic {FTA} using temporal logic.
	As the PAND gates introduced in \cite{Veseley1981NUREG-0492} lack a universal temporal logic, it is undefined, whether both shown fault trees are equivalent or not.
	The PAND gates' symbols used in this figure are not the ones from \cite{Veseley1981NUREG-0492}, but from the {TFTA} approach presented in chapter \ref{_chap_080401-004}.}
	\label{fig:stand_der_technik_temporale_logik_fragen}
\end{figure}

The first version of the \emph{fault tree handbook} \cite{Veseley1981NUREG-0492} was a de facto standard for fault tree analysis for a long time; it also describes a so-called \emph{priority AND} ({PAND}) gate.
This gate is used exclusively for qualitative modelling of event sequences; probabilistically it is treated as a conventional AND gate.
This approach again focusses on individual event sequences, and it does not provide a further and generic temporal logic.
For example, it is not discussed, whether -- and how -- the fault tree structure shown on the left side of figure \ref{fig:stand_der_technik_temporale_logik_fragen} may be simplified, and/or if it is equivalent to the structure shown on the right side of figure \ref{fig:stand_der_technik_temporale_logik_fragen}.
In the Boolean model with AND instead of PAND gates, both fault trees are equivalent, as the Boolean distributive law -- see \eqref{080129-005} on page \pageref{080129-005} -- yields $
(A\booland{}  B)\boolor{}   (A\booland{}  C)\ist A\booland{} (B\boolor{}   C)$

The interval-based temporal logic of the so-called \emph{AND-Then gates} in \cite{Wijayarathna2000} pursue a broader approach, as well as the work presented in \cite{Gorski1994,Gorski1997} and the so-called \emph{temporal fault trees}
in \cite{Palshikar2002}.
They all stem from the field of formal fault tree analysis, which is mainly motivated by adopting the conventional fault tree analysis method, so as to model software based systems and their ``failures''.
Failure analysis of software based systems is fundamentally different from the conventional and hardware orientated {ZSA}, especially because of their very different failure mechanisms.
An overview on the state of the art of {FTA} for software based systems is given in \cite{Thums2004}.
Because of the high dynamic of software based systems, the temporal logics presented in the works above are also complex and complicated;
furthermore, their application is quite different to conventional {FTA}, because of their very strict defininitions.

In earlier work, Heidtmann interpreted \emph{modal logic} \cite{Galton1987}, which originates in the field of theoretical philosophy, for reliability modelling, see \cite{Heidtmann1992} and \cite{Heidtmann1997}.
His temporal logic describes event sequences not directly, but asserts so-called \emph{anytime-} and \emph{always-relationships} between events.
Using these, many temporal dependencies and contexts may be portrayed, including event sequences.
Heidtmann discusses the qualitative as well as the probabilistic application of his temporal logic, and he is not limited to the fault tree method.
On the other hand and because of its power, his logic involves comparably complex models and calculations.

The dedicated aim of the \emph{Pandora} approach in \cite{Walker2006,Walker2007} is to provide a ``useable'' method that is similar to conventional {FTA}.
The term ``Pandora'' puns on the figure from greek legend, as well as it is a composite of ``Priority AND'' and the greek term  $\acute{\upomega}\uprho\upalpha$ (\emph{ora}), which means ``time'' \cite{Walker2006}.
Creation and analysis of Pandora fault trees is similar to conventional Boolean {FTA}.
By using additional temporal gates -- which are called PAND, SAND, and POR --, a temporal failure function of the TOP event is built.
This function is then transformed into a minimal form by applying temporal logic simplification laws that are sketched in \cite{Walker2007}.
Central to these laws is the concept of so-called ``doublets''.
A doublet describes the temporal relationship between exactly two events, and is itself treated like a basic event.
Temporal relationships are given only relatively to each other, i.e.\ the absolute points in time when events occur are not considered.
The minimal form is the equivalent to the minimal cutsets in conventional {FTA}; it allows a qualitative analysis of the failure behaviour including event sequence information.
The concept of doublets simplifies the analysis greatly; but it also limits the Pandora approach in terms of probabilistic analysis, specifically because it leaves unresolved (temporal) dependencies between doublets.
For example, in Pandora \cite{Walker2007} the expression ``$A$ occurs first, and then $B$ and $C$ occur'' is written as
\begin{align}
  A\pand{} (B\booland{}  C ) & \ist{}  \Bigl[ (A\pand{} C )\booland{}  (B\pand{} C ) \Bigr] \boolor{}  \Bigl[   (A\pand{} B )\booland{}  (B\sand{} C ) \Bigr] \boolor{}   \Bigl[  (A\pand{} B )\booland{}   (C\pand{} B )\Bigr] ~.
\end{align}
Instead of the original Pandora notation, the notation from chapter \ref{_chap_080401-004} is used here, in order to improve comparability of the results.
Each term in round brackets on the right side indicates one doublet.

These doublets allow qualitative analyse, but they can not be simply quantified, as shown by the following considerations.

A Boolean conjunction, e.g.\ $ (A\booland{}  C )\booland{}  (B\booland{}  C )$, must not, in general, be quantified by simple multiplication of the individual event probabilities; i.e.\
\begin{align}
	F_{ (A\booland{}  C )\booland{}  (B\booland{}  C )}&\neq
		 (F_A\cdot F_C )\cdot (F_B\cdot F_C )~,
\end{align}
if it is not given in a minimal form, already, or the individual events are not independent from each other.
If these conditions are satisfied, e.g.\ after transforming into
\begin{align}
	 (A\booland{}  C )\booland{}  (B\booland{}  C )&\ist A\booland{}  B\booland{}  C ~,
\end{align}
then a direct quantification is possible.
\begin{align}
	F_{ (A\booland{}  C )\booland{}  (B\booland{}  C )}&\ist
		F_{A\booland{}  B\booland{}  C}\ist
		F_A\cdot F_B\cdot F_C~.
\end{align}
In analogy, Pandora expressions, like the one shown above, must not be quantified directly.
For example, the ``joint'' event $C$ in both doublets, i.e.\ an unresolved dependency between both doublets, is the reason for
\begin{align}
    F_{ (A\pand{} C )\booland{}   (B\pand{} C )}&\neq F_{ (A\pand{} C )}\cdot F_{ (B\pand{} C )}~.
\end{align}
The {TFTA} approach presented in this work adopts some aspects of Pandora.
But the {TFTA} goes beyond Pandora by (among others)
\begin{itemize}
  \item providing a complete and systematic set of logic transformation laws of universal validity and applicability, where Pandora only sketches temporal logic rules in \cite{Walker2007}, and
  \item allowing probabilistic, as well as qualitative modelling and analysis, where Pandora stays qualitative, and
  \item not pursuing the concept of doublets, that is not well-suited for probabilistic analysis, and
  \item not using a POR operator.
\end{itemize}

The differences from that may be demonstrated by comparing the Pandora expression from above with an equivalent expression according to the {TFTA} approach.
Anticipating the chapters below, the latter is given as 
\begin{align} \begin{split}
  A\pand{}(B\booland{}   C)\ist &
  		\Bigl[ A \pand{} B \pand{} C \Bigr] \boolor{}
  		 \Bigl[ B \pand{} A \pand{} C  \Bigr]  \boolor{}
  		 \Bigl[ A \pand{} C \pand{} B  \Bigr]  \boolor{}
  		 \Bigl[ C \pand{} A \pand{} B  \Bigr]  \boolor{}   \\
  	& \boolor{}
  		 \Bigl[  (A \sand{} B ) \pand{} C  \Bigr]  \boolor{}
  		 \Bigl[ A \pand{}  (B \sand{} C )  \Bigr]  \boolor{}
  		 \Bigl[  (A \sand{} C ) \pand{} B  \Bigr] ~.
\end{split}
\end{align}
As shown in this thesis, these terms may be quantified directly -- and they may also be transformed into a more compact form in order to reduce the calculatory effort:
\begin{align}
  &A\pand{}(B\booland{}   C)\ist
       \Bigl[ (A \booland{} B) \pand{} C  \Bigr] \boolor{}
       \Bigl[ (A \booland{} C) \pand{} B  \Bigr] \boolor{}
       \Bigl[  A \pand{} ( B \sand{} C ) \Bigr] ~. 
\end{align}
The right side expressions are mutually exclusive (disjoint), thus
\begin{align}\begin{split}
   F_{A\pand{}(B\booland{}   C)}(t) \ist &
        F_{(A \booland{} B) \pand{} C}(t) +
        F_{(A \booland{} C) \pand{} B }(t) +
        F_{A \pand{} ( B \sand{} C ) }(t) \ist \\
    \ist &
        \int\limits_0^t \Bigl( F_A(\tau) F_B(\tau) f_C(\tau) + F_A(\tau) F_C(\tau) f_B(\tau)\Bigr) \cdot \D \tau
                ~.
\end{split}
\end{align}


%
%
%
%
%
%
\section{Summary}\label{chap080630-001}
Conventional Boolean {FTA} is state of the art for systematic, top-down, and qualitative as well as probabilistic analysis of the failure behaviour of complex systems in several industries and application fields (see chapters \ref{chap080409-001} and \ref{chap080401-010}).

The call for an improved consideration of time-dependencies lead to development of several extensions of the Boolean {FTA} in order to take into account dynamic effects and specifically \emph{sequence dependencies}, see chapter \ref{chap080413-001}.
There are two main strategies for such consideration of event sequences:
On the one hand the Boolean fault tree model is transformed into a state-based model, which allows the calculation of dynamic effects (see chapter \ref{chap080413-003}).
On the other hand, an extended and temporal logic is used instead of the Boolean (failure) logic, see chapter \ref{chap080413-006}.

In the past several proposals for each of the two strategies were presented.
Moreover, some of the state-based extensions are being used for solving real-world problems today.
But by switching into the state-space these approaches loose some of the main advantages of conventional {FTA}, specifically with respect to the necessary calculatory effort, its intuitive useability, and its ability to provide meaningful qualitative analyses. 

Very powerful but also very complex methods dominate the field of extensions by temporal logic; they stem mainly from research on applying the {FTA} on software.
Further research is needed for improved useability, in order to convey the conventional Boolean FTA's ``user-friendliness'' onto dynamic {FTA}.

\begin{figure}
  \centering
    \input{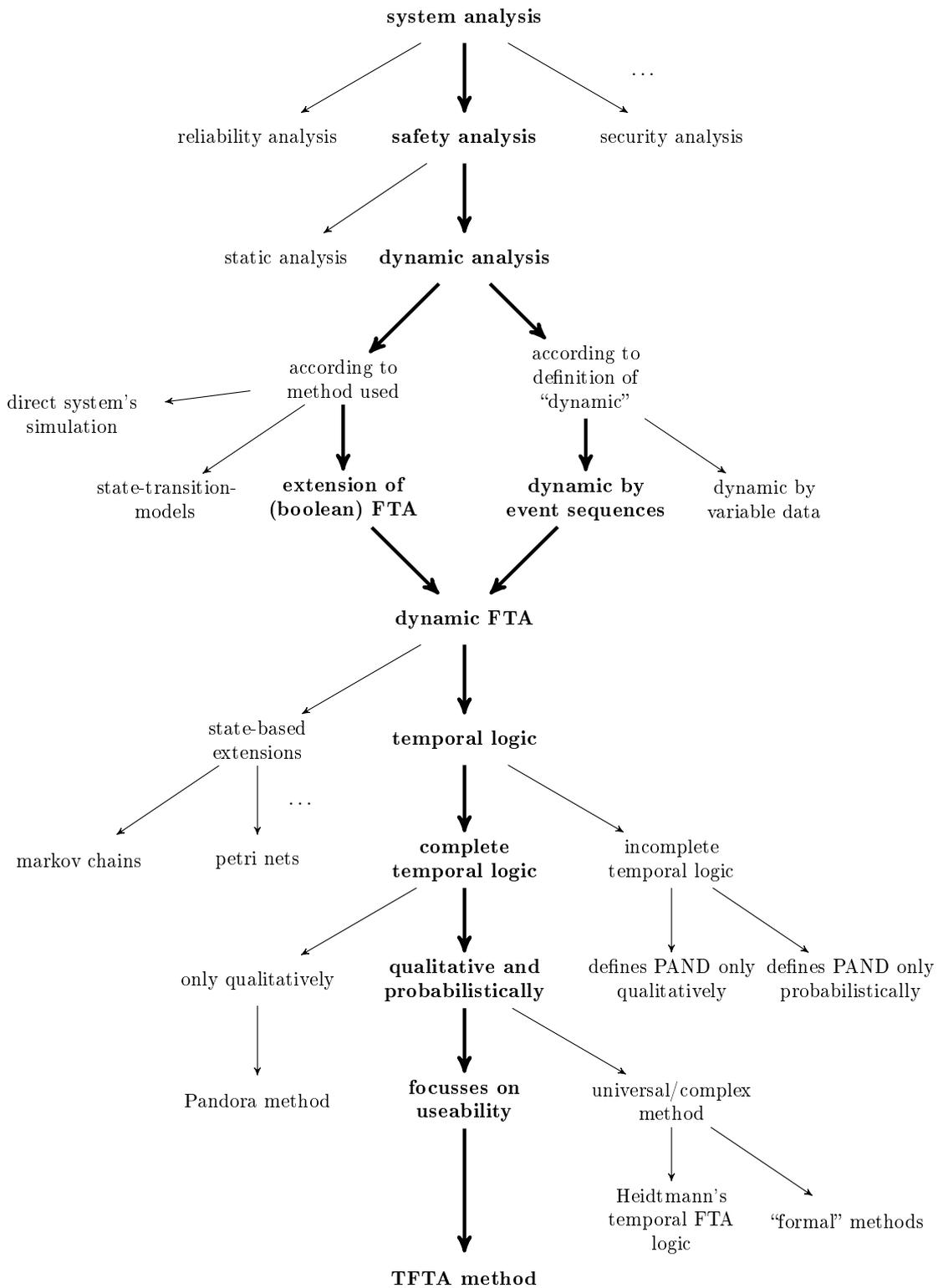}
  \caption{{TFTA} approach in comparison to the state of the art and other methods of considering ``dynamic effects'' within {ZSA}.}
  \label{fig:stand_der_technik_einordnung_tfta_ansatz}
\end{figure}

Figure \ref{fig:stand_der_technik_einordnung_tfta_ansatz} shows how the {TFTA} approach presented in this thesis fits into the state of the art, and it differentiates the {TFTA} from other methods.
%
\clearpage
\clearpage
\ifx \printSprueche\undefined
\else
\renewcommand*{\dictumwidth}{.36\textwidth}

\setchapterpreamble[ur]{
\dictum[Frédéric Chopin]{Simplicity is the final achievement.}
\vspace{3cm}
}
\fi

\chapter{Problem Definition: Event Sequences in FTA without Modularization}\label{chap80401-002}
%
%
%
%
%
%
%
%
\section{Demand for Improved Methods}\label{chap080401-020}
\subsection{Demand for Dynamic {FTA}}
One of the FTA's vital objectives is the probabilistic evidence that the failure rate and failure probability of a system are lower than given target values.
Practical experience shows that in many cases reaching these target values -- derived from e.g.\ safety standards like {\IEC} or {\ISO} -- is a close call. 
Modelling the same system with a dynamic {FTA} provides less conservative results than the conventional {FTA}; this, of course, helps to comply with probabilistic target values.
It is much more credible to improve one's system analysis by using such a dynamic and more detailed method than to reach compliance with one's objectives by improving (reducing) the failure data input to the basic events; the latter is often hardly justifiable.

For systems with higher safety levels the conventional qualitative single failure analysis using FMEA is not sufficient \cite{Tietjen2003FMEA,Verband2003Sicherung}
In such cases and for complex system architectures the qualitative {FTA} improves systematic understanding of multiple failure interaction.
For example, it is very efficient to improve the safety of programmable systems by making the conditions of switching elements dependend on sequential information.
Fail-activation is reduced as only certain sequences of trigger events are relevant.
In many cases such sequential conditions can be added into integrated circuits with only negliable costs.
When compared to the conventional {FTA}, an {FTA} that takes such sequences into account can then provide a much more meaningful view on the system under consideration.

Chapter \ref{chap080104_005} shows an example system where conventional Boolean fault tree modelling and analysis provides only unprecise results.
\subsection{Demand for Improved Dynamic FTA}
Dynamic extensions to {FTA}, as listed in chapter \ref{chap080401-011}, aim at the correct probabilistic calculation of fault trees; this is especially true for the state based methods like {DFT}.
Chapter \ref{090212003} shows an example where the {DFT} succeeds in this respect and thus proves to be a real improvement when compared to the conventional Boolean {FTA}.

Criticism of state based extensions comprises mainly from the following aspects:
\begin{itemize}
 \item state based extensions are limited in their use for qualitative analysis of sequence effects.
 This comes from the forced change between methods with Boolean fault tree logic on the one hand and a state based dynamic model on the other hand. 
 \item they are limited in case of interdependencies between dynamic and non-dynamic parts (modules) of the same fault tree.
 \item probabilistic calculation is rather costly and approximations are not easy to identify and use.
\end{itemize}
Practical experience shows that there is a certain correlation between the necessities of probabilistic and qualitative analyses of dynamic effects.
Therefore, from an effort point of view it is beneficial to cover both aspects with the same modelling method.
Methods are needed that allow both analyses with reasonable effort and idealy also allow a step wise workflow: first the results are only approximated, then the most important contributors are identified, and then only for those the more complex but exact calculations are done. 
\subsection{Remarks on Using Dynamic {FTA}}\label{080630-002}
In general, an analysis' effort and its benefit must not be disproportionate to each other even if there is a very understandable quest to model the reality (which is dynamic, see chapter \ref{chap080624-001}) as exact and detailled as possible.
Today there are several attempts to extend the Boolean {FTA} with dynamic effects and event sequenes; but many of those extensions are limited to simple and mostly academic examples.
This is especially true for approaches based on a temporal logic; their very high complexity conflicts with their practical useability.

Useability, (relative) ease of use, and scalability are three critical success factors of the conventional {FTA}; and they have added tremendously to the {FTA} being first choice for safety and reliability analyses in many domains. 

In order to transfer this success, the dynamic {FTA} needs to satisfy the following generic requirements:
\begin{itemize}
 \item real system effects must translate into the model's logic easily, 
 \item the actual implementation into a fault tree needs to be possible with reasonable effort, 
 \item qualitative as well as probabilistic calculations must be possible without changing the analysis method,
 \item computing time must be reasonable,
 \item the fault tree as well as its results must be easily readable and comprehensible,
 \item scalability and possibility to detail and extend parts of the fault tree.
\end{itemize}

%
%
%
%
%
%
\section{Concept}\label{chap080401-021}
\subsection{Requirements for {TFTA}}\label{090222001}
By taking useability and practical considerations into account the following is required from the new {TFTA} method:
\begin{enumerate}
  \item The temporal {TFTA} logic shall be able to model sequence dependencies between events.
  \item The temporal {TFTA} logic shall be a detailing (extension) of the Boolean logic.
  \item The {TFTA} shall be similar to the conventinal {FTA} regarding notation, abstract concept, workflow, work products.
  \item The qualitative {TFTA} shall provide minimal event sequences similar to the Boolean minimal cutsets.
  Each ``minimal cutset sequence'' shall consist of ``temporal conjunction terms'' similar to the Boolean AND term but including event sequence information.
  The TOP or system failure function shall then consist of such ``minimal cutset sequences'' given in ``temporal disjunctive normal form''.
  \item In order to allow for probabilistic analysis the ``minimal cutset sequences'' shall be disjoint (i.e.\ mutually exclusive); this allows for easy quantification by convolution of the failure densities/frequencies.
  \item In order to reduce calculation efforts the {TFTA} shall support step-wise modelling: a first step provides only approximations; more exact calculations follow only for the most important contributors. It shall be possible to calculate exact results if necessary.
\end{enumerate}

\paragraph{Assumptions on {TFTA}}~\\
The following discussions are based on two assumptions: 

\begin{enumerate}
 \item fault trees are monotone (sometime also called coherent) and
 \item all component failures are non repairable.
\end{enumerate}

\subsection{{TFTA} -- Step by Step}
\begin{figure}[ht]
  \centering
    {\small
  \input{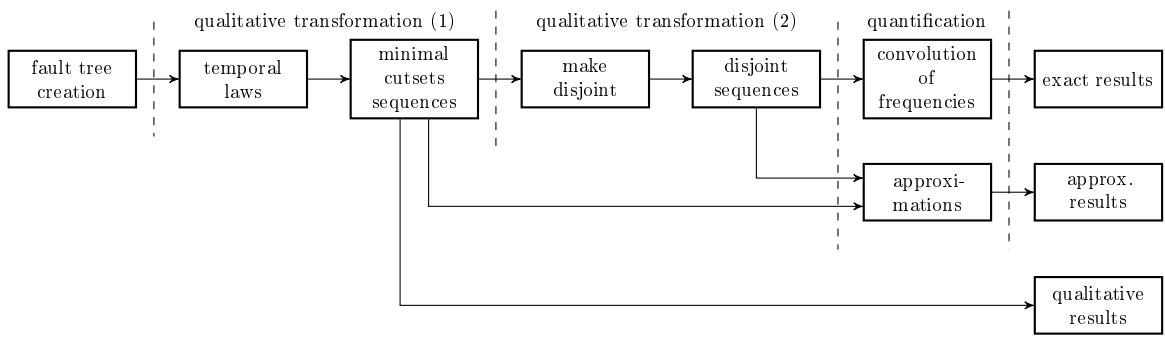}
}
  \caption{Step by step workflow of the {TFTA} with its two-step transformation of a temporal expression into a minimal and then mutually exclusive form, and probabilistic quantification.
  Approximations are possible, first, based on the mutually exclusive event sequences or, second and a little more unprecise, directly from the minimal event sequences.}
  \label{fig02}
\end{figure}

Figure \ref{fig02} shows the {TFTA} workflow with its multiple steps.
First, there is the two step qualitative transformation of the initial logic expression into a minimal and later disjunct form;
in a second step, this is then quantified probabilistically.
This workflow is very similar to the workflow of conventional {FTA}; there, too, minimal cutsets need not automatically be mutually exclusive.
The {TFTA} workflow is split into two steps because of the potentially very high effort necessary for transforming a minimal temporal expression into mutually exclusive terms. 

The structure of chapter \ref{_chap_080401-004} is influenced by this workflow steps, too; chapter \ref{chap080401-030} provides the notation of the temporal logic; chapter \ref{chap080401-031} provides the TFTA's (temporal) rules of transformation; chapter \ref{chap080809-001} describes the transformation into mutually exclusive sequences; and chapter \ref{chap080401-033} provides the probabilistic evaluation of temporal expressions.
\clearpage
\clearpage
\ifx \printSprueche\undefined
\else
\renewcommand*{\dictumwidth}{.35\textwidth}

\setchapterpreamble[ur]{%
\dictum[Douglas Adams]{Time is the worst place, so to speak, to get lost in.}
\vspace{3cm}
}
\fi

\chapter{Temporal Fault Tree Analysis (TFTA): A New Approach to Dynamic FTA}\label{_chap_080401-004}
This chapter describes the \emph{temporal fault tree analysis} (TFTA) which extends the Boolean {FTA} and allows analysis of event sequences. 
\begin{itemize}
\item Chapter \ref{chap080401-030} presents the notation of the new temporal TFTA logic. 
Specifically, there are two new temporal operators corresponding to two temporal fault tree gates.
\item At the heart of the new temporal logic there are several rules of transformation (``temporal logic laws'') described in chapter \ref{chap080401-031}.
They allow the transformation of a temporal expression into its temporal disjunctive normal form (TDNF).
\item Chapter \ref{chap080809-001} discusses minimal and disjoint temporal expressions.
\item There is an extended form of temporal expressions, as shown in chapter \ref{080817-010}, which reduces the effort necessary for describing and calculating complex temporal failure functions~-- especially if such failure functions only include few real temporal relationships between events. 
\end{itemize}
%
%
%
%
%
%
%
\section{TFTA Notation}\label{chap080401-030}
First of all, some remarks on the terms used:
In the fault tree method basic events represent atomic failure events of real life entities (i.e.\ systems, components, parts, functions).
Likewise, fault tree gates represent non-atomic ``higher level'' failure events.
The terminology is sometimes confused so that there is no discrimination between ``incidence of a real world failure event'' and ``fault tree event becomes $\True$'', where the latter represents the real life event in the fault tree model.
\subsection{Boolean Algebra and the {FTA} Failure Logic}
In the context of {FTA} events are failure events.
Contrary to uses of the Boolean algebra for reliability calculations, the {FTA} therefore uses a \emph{negated logic} \cite[chapter 14.4.2]{Meyna2003Taschenbuch}.
In the following text negating all events in their written form is ommited for reasons of better readability.
For all failure events  
\begin{align}
			X_i\ist&  \begin{cases}
							\True~\text{or}~1 & \quad \text{entity $i$ has failed}\\
							\False~\text{or}~0 & \quad \text{entity $i$ is operational}
						\end{cases}~.\label{080155-015}
\end{align}
For the {TFTA} approach most of the Boolean logic and its application on the fault tree stays the same:

The \emph{conjunction} using the \emph{AND operator} and
\begin{align}
			X_{\text{AND}}\ist A \booland B \label{080129-013}
\end{align}
is $\True$, if and only if both events $A $ and $B $ are $\True$.
In fault trees the conjunction is represented by AND gates.

The \emph{disjunction} using the \emph{OR operator} and
\begin{align}
			X_{\text{OR}}\ist A \boolor B \label{080129-014}
\end{align}
is $\True$, if either only event $A $ or only event $B $ is $\True$, or if both events are $\True$.
In fault trees the disjunction is represented by OR gates.

The \emph{negation} using the \emph{NOT operator} and
\begin{align}
			X_{\text{NOT}}\ist \boolnot A \label{080129-015}
\end{align}
is $\True$, if and only if event $A $ is $\False$.
The shorter $A\boolnot B$ is used below instead of $A\booland\boolnot B$.
In fault trees the negation is represented by NOT gates.
\subsection{Temporal Logic Operators}
\begin{figure}
	\centering
		\input{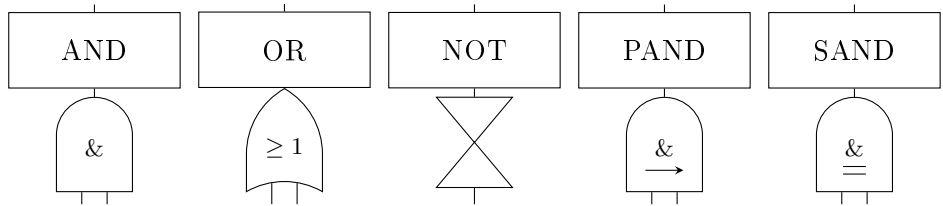}
	\caption{Fault tree gates of the TFTA: Boolean gates (left) and temporal gates (right)}
	\label{fig:tftagatter}
\end{figure}

The TFTA uses two temporal operators and their corresponding gates in addition to the Boolean operators and gates in order to describe temporal event relationships (see figure \ref{fig:tftagatter}).
\paragraph{PAND: The Sequence of Events}~\\
The \emph{PAND operation} (\emph{Priority AND}) using the \emph{PAND operator} and
\begin{align}
			X_{\text{PAND}} &\ist A \pand B \label{080129-010}
\end{align}
is $\True$, if and only if
		\begin{itemize}
			\item both events $A $ and $B $ are $\True$ and
			\item $A $ has become $\True$ before $B $ has become $\True$.
		\end{itemize}
Therefore, PAND describes a chronology of events becomming $\True$ after each other. 
In fault trees the PAND operation is represented by PAND gates.
%
\paragraph{SAND: Concurrence of Events}~\\
The \emph{SAND operation} (\emph{Simultaneous AND}) using the \emph{SAND operator} and
\begin{align}
			X_{\text{SAND}}\ist A \sand B \label{080129-011}
\end{align}
is $\True$, if and only if
		\begin{itemize}
			\item both events $A $ and $B $ are $\True$ and
			\item $A $ and $B $ have become $\True$ simultaneously.
		\end{itemize}
Therefore, SAND describes events becomming $\True$ exactly at the same time. 
In fault trees the SAND operation is represented by SAND gates.

\emph{Remark:} PAND as well as SAND uses time indications relatively, i.e.\ no statement is made on the absolut (real) time at which an event becomes $\True$.
\subsection{Boolean and Temporal Operations Visualized as Sets}
\begin{figure}
  \centering
   \input{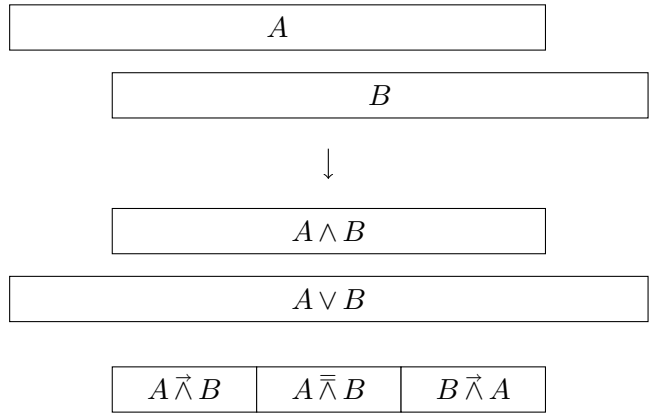}
  \caption{Temporal operations from top to bottom: events $A$ and $B$; their intersection (AND) and union (OR); the three subsets defined by distinction between the possible event sequences (PAND and SAND). }
  \label{fig:pandoramengen}
\end{figure}

Figure \ref{fig:pandoramengen} shows the different operators as sets and illustrates the relationshios among them.
First, there are two event $A$ and $B$ symbolized as sets.
If $A$ and $B$ are the operands to AND and OR operators (i.e.\ they are inputs to Boolean AND and OR gates in a fault tree), then two sets result:
$A\booland B=B\booland A$ (intersection) and $A\boolor B=B\boolor A$ (union).
If $A$ and $B$ are the operands to PAND and SAND operators (i.e.\ they are inputs to temporal PAND and SAND gates in a temporal fault tree), then three sets result:
$A\pand B$ and $B\pand A$ and $A\sand B=B\sand A$.
Note, that negated events and their corresponding ``sets'' are not shown here.
\begin{figure}
  \centering  
  \input{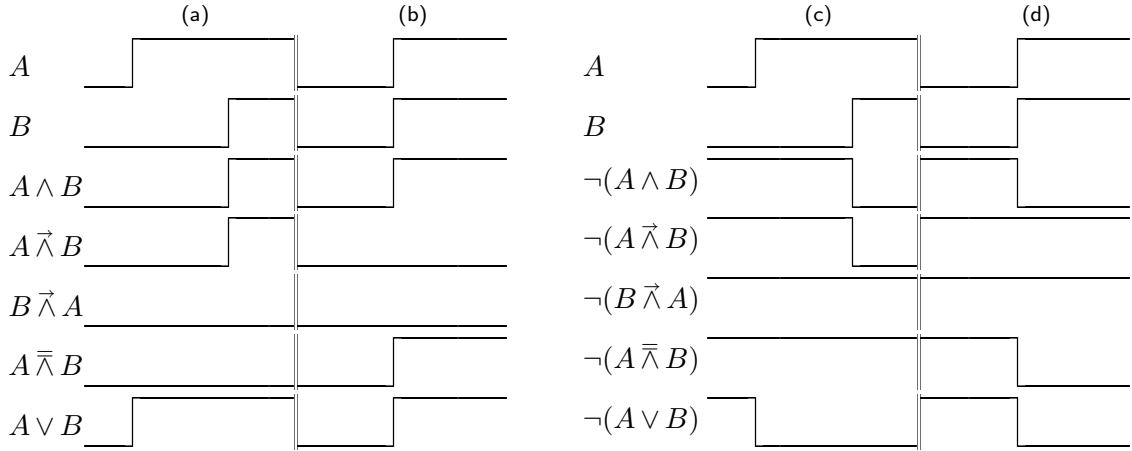}
  \caption{Temporal sequence of two events: In (a) and (c) event $A$ becomes $\True$ before $B$ (upper two rows); the following rows show which events formed by $A$ and $B$ become $\True$ at which time. In (b) and (d) events $A$ and $B$ become $\True$ simultaneously.}
  \label{fig:ZeitlicherAblaufZweierEreignisse}
\end{figure}
\begin{figure}
  \centering
 \input{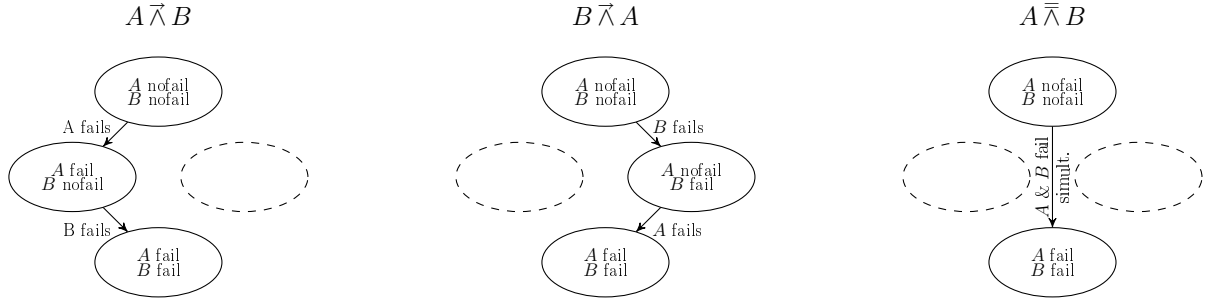}
  \caption{Illustration of the three possible sequences of state transitions (which are mutually exclusive) that lead to failure of both components of the example system in figure \ref{fig:bsp_zweiausfaelle_zustaende}.}
  \label{fig:pand_markov_01}
\end{figure}
The depiction in figure \ref{fig:pandoramengen}  allows a first qualitativ statement on the meaning of tempral operators/gates.

According to \eqref{080129-010} and \eqref{080129-011} PAND and SAND events are real subsets of the Boolean conjunction $A\booland{} B\ist{}B\booland{} A$ (``\ldots both events $A$ and $B$ are $\True$ \ldots'').
There are three possibile sequences how two events $A$ and $B$ can ``both be $\True$'' (see the law of completion in chapter \ref{chap080401-031}).

As sets this may be written as 
\begin{align}
  &A\pand B \subset{} A\booland{} B~, && A\sand B \subset{} A\booland{} B~, && B\pand A \subset{} A\booland{} B~, \label{090124001} \\
  &A\booland{} B \subset{} A~, && A\booland{} B \subset{} B~. \label{090124002}
\end{align}
Events $A\pand B$, $B\pand A$, and $A\sand B$ are pairwise mutually exclusive, i.e.\ there is no intersection between them (see chapter \ref{chap080809-001}):
\begin{align}
  &A\pand B \perp A\sand B ~, && A\pand B \perp B\pand A ~, && A\sand B \perp B\pand A \label{090124004} ~.
\end{align}

\subsection{Temporal Operations: Timing Behaviour}
Temporal sequence diagrams illustrate (temporal) relationships between events.
Figure \ref{fig:ZeitlicherAblaufZweierEreignisse} shows logic levels over time for Boolean and temporal operators used in the TFTA.
In general, events may become $\True$ in sequence or simultaneously (see sub-figures (a) and (c) and (b) and (d) respectively). 

The possibile failure sequences in a sytstem which result from those timings may be shown e.g.\ with state diagrams.
In a simple example system consisting of two redundant components (see state diagram in figure \ref{fig:bsp_zweiausfaelle_zustaende}), where events may become $\True$ after each other or simultaneously, there are the three possible state transition sequences which were already mentioned and which are shown in figure \ref{fig:pand_markov_01}.
These sequences correspond to the two PAND operations $A\pand B$ and $B\pand A$ on the one hand and the SAND operation $A\sand B$ on the other hand.

From page \pageref{090109010} on further examples of temporal sequence diagrams are compared with other methods of illustration.
%
\subsection{Syntax of Temporal Expressions} \label{090321009}
A logic expression with at least one temporal operator is called \emph{temporal-logic expression} or shorter: \emph{temporal expression}.

In conventional {FTA} a Boolean expression which is represented by the fault tree's TOP event is called \emph{Boolean failure function} and is symbolized by $\varphi$.{}
In the {TFTA} the TOP event represents a temporal expression which is called \emph{temporal failure function} and is symbolized by its own symbol $\varpi$ for better discrimination in the following text.

The next sections explain elements of a temporal logic grammar as used by {TFTA}.
This grammar is summarized in table \ref{090131002}.
The temporal logic's operators $\{\booland,\boolor,\pand,\sand,\boolnot\}$ are used as terminal symbols.
%
\begin{table}
\centering
{\small
  %
%
%
%
\begin{tabular}{llll} \toprule
 Token	&	Description	&	Format	&	Example	\\ 
 \midrule \addlinespace[7pt]
$\AEtoken$	&	atomic event 	&	-	&	$X$	\\
	&	$\quad$(basic event)	&		&	$Y$	\\
	&		&		&	$Z$	\\
$\NAEtoken$	&	negated atomic event	&	$\boolnot \AEtoken$	&	$\boolnot X$	\\
$\CEtoken$	&	core event	&	
  $\AEtoken$	&	see above\\
	&		&	$\CEtoken\sand{}\AEtoken$	&	$X\sand{}Y$	\\
	&		&		&	$X\sand{}Y\sand{}Z$	\\
\bottomrule\end{tabular}
}
\caption{The syntax of temporal expressions: the more complex tokens are based on the token of an atomic event (basic event) as an entity which is not further dividable; complex tokens are: core events, event sequences and temporal expressions in {TDNF}; they are composed in multiple ways. The examples given do not include all possible combinations. The lower part of the figure shows temporal expressions in a more generic form; those need to be transformed for further analysis. }
  \label{090131002}

\end{table}
\paragraph{Atomic Events/Basic Events}~\\
\emph{Atomic events} are the smallest event entities in temporal expressions, and are not further dividable.
Within the temporal fault tree they are represented by basic events which do not differ from those basic events used in conventional {FTA}.
Particularly, probabilistic (failure) data like failure rates may be assigned to them.

The formal grammar of the temporal logic uses the $\AEtoken$ token for atomic events.

\emph{Negated atomic events} with toke $\NAEtoken$ are -- as the name suggests -- the negation of atomic events:
\begin{align}
   \NAEtoken \quad \rightarrow \quad & \boolnot ~~ \AEtoken &&.&
\end{align}
Within the {TFTA} negated events have a special meaning, see chapter \ref{080220-002}.
\paragraph{General Temporal Expressions}~\\
In general, a temporal expression either consists of a basic event, or consists of two other temporal expressions, which are connected by a temporal (including Boolean) operator, or consist of a negation of another temporal expression.
Therefore
\begin{align}
  \TERMtoken \quad \rightarrow \quad & \AEtoken && |  &\\
       & \TERMtoken  ~~ \booland{}  ~~ \TERMtoken && |  & \nonumber\\
       & \TERMtoken  ~~ \boolor{}  ~~ \TERMtoken && |  & \nonumber\\
       & \TERMtoken  ~~ \pand{}  ~~ \TERMtoken && |  & \nonumber\\
       & \TERMtoken  ~~ \sand{}  ~~ \TERMtoken  && |  & \nonumber\\
       & \boolnot ~~ \TERMtoken   && . & \nonumber
\end{align}
Aside from the additional temporal operators this corresponds to the formal representation of Boolean expressions.

This general form is not suited for direct qualitative or probabilistic analysis.
From chapter \ref{chap080401-031} on transformation laws for temporal expressions are described that allow to transform any temporal expression into a TDNF -- which in turn allow further analysis.
The following sections explain the structure of this {TDNF}.
\subsubsection{Temporal Disjunctive Normal Form (TDNF, Sum of Products)}\label{090125001}
\paragraph{Core Events}~\\
In the temporal logic \emph{core events} describe that one or more events become $\True$ at a certain point in time.
Negated core events indicate that at a given time one or more events have not (yet) become $\True$.
Many equations in this thesis use $K$ for core events.

A core event event is represented by token $\CEtoken$ and consists of either one atomic event, or consists of a temporal expression (in braces), which itself consists of only SAND connected atomic events.
More formally,
\begin{align}
  \CEtoken \quad \rightarrow \quad 
       & \AEtoken && |  &\label{080828_003}\\
       & \CEtoken ~~  \sand{}  ~~ \AEtoken &&.& \nonumber
\end{align}
A \emph{negated core event} (token $\NCEtoken$) consists of either one negated atomic event, or consist of a temporal expression (in braces), which itself consists of only AND connected negated atomic events.
More formally,
\begin{align}
  \NCEtoken\quad \rightarrow \quad 
            & \NAEtoken && |  &\label{080828_004}\\
            & \NCEtoken  ~~ \booland{}  ~~ \NAEtoken && .  &\nonumber
\end{align}
\paragraph{Event Sequences}~\\
\emph{Event sequences} are the temporal logic's equivalent of Boolean cutsets.
They describe a temporal sequence of one or more core events.
In analogy to the Boolean minimal cutsets, minimal event sequences ({MCSS}, see chapter \ref{081018_001}) have a special significance in the temporal logic.

\emph{Event sequences with negated events} are important for transforming temporal expressions into disjoint, i.e.\ mutually exclusive, terms.
This is similar to the Boolean logic.
Many equations in this thesis use $\ES$ for event sequences.

Event sequences are represented by the token $\EStoken$ and either consist of exactly one core event, or consist of several PAND connected core events.
More formally
\begin{align}
  \EStoken \quad \rightarrow \quad & \CEtoken &&| & \\
            & \EStoken ~~  \pand{} ~~  \CEtoken && .  &\nonumber
\end{align}
Additionally, there are event sequences with negated events consisting of exactly one negated core event, which is AND connected with exactly one event sequence.
They are represented by the token $\NEStoken$.
Therefore
\begin{align}
  \NEStoken \quad \rightarrow \quad & \NCEtoken ~~\booland{} ~~ \EStoken && .  &
\end{align}
\paragraph{Temporal Expressions in {TDNF}}~\\
Event sequences, connected by OR operators, provide the \emph{temporal disjunctive normal form} (\acsu{TDNF}):
\begin{equation}
  \varpi \ist{}{} \bigvee\limits_{\mathclap{j=1}}^{\zeta} \ES_j \ist{}{} \ES_1 \vee{} \ES_2 \vee{} \ldots \vee{} \ES_{\zeta}~.
  \label{080828_001}
\end{equation}
The symbol $\zeta$ indicates the number of event sequences $\ES$ of $\varpi$, which themselves are not necessarily already in a minimal form.
More formally,
\begin{align}
  \TTtoken \quad \rightarrow \quad &  \EStoken &&| & \\
            & \NEStoken  && |  &\nonumber\\
             & \TTtoken ~~\boolor{}~~ \TTtoken  && .  &\nonumber
\end{align}
\subsubsection{Extended TDNF (Sum of Products)}\label{080817-002}
\paragraph{Temporal Expression in Extended {TDNF}}~\\
The extended {TDNF} of a temporal failure function $\varpi$ is given as $\zeta$ \emph{extended event sequences} $\eES_j$ which are connected by OR operators:
\begin{equation}
  \varpi \ist{}{} \bigvee\limits_{\mathclap{j=1}}^{\zeta} \eES_j \ist{}{} \eES_1 \vee{} \eES_2 \vee{} \ldots \vee{} \eES_{\zeta}~.
  \label{080828_005}
\end{equation}
This extended {TDNF} greatly simplifies the qualitative as well as probabilistic transformations and caluclations.
More formally,
\begin{align}
  \eTTtoken \quad \rightarrow \quad &  \eEStoken &&| & \\
            & \NeEStoken  && |  & \nonumber\\
            & \eTTtoken ~~\boolor{}~~ \TTtoken  && |  &\nonumber\\
            & \eTTtoken ~~\boolor{}~~ \eTTtoken  && .  &\nonumber
\end{align}
The extended {TDNF} consists of extended core events and extended event sequences with and without negated events.
\paragraph{Extended Core Events}~\\
An \emph{extended core event} is represented by the token $\eCEtoken$ and consists of two or more AND connected atomic events.
It is identical to the conventional conjunction of atomic events in Boolean algebra.
Therefore, 
\begin{align}
  \eCEtoken \quad \rightarrow \quad & \AEtoken ~~  \booland{}  ~~ \AEtoken  &&| & \\
       & \eCEtoken ~~  \booland{}  ~~ \AEtoken &&.& \nonumber
\end{align}
\paragraph{Extended Event Sequences}~\\
\emph{Extended event sequences} with token $\eEStoken$ either consist of exactly one extended core event or consist of only PAND connected extended core eventst or consist of a mixture of PAND connected normal and extended event sequences.
Thus,
\begin{align}
  \eEStoken \quad \rightarrow \quad & \eCEtoken &&| & \label{080828_006}\\
            & \eEStoken ~~  \pand{} ~~  \eCEtoken && |  & \nonumber \\
            & \eEStoken ~~  \pand{} ~~  \CEtoken && |  & \nonumber \\
            & \EStoken ~~  \pand{} ~~  \eCEtoken && .  &\nonumber
\end{align}
\emph{Extended event sequences with negated events} are defined as event sequences which consist of exactly one negated core event which is AND connected with exactly one extended event sequence; they are represented by the token $\NeEStoken$.{}
Formally,
\begin{align}
  \NeEStoken \quad \rightarrow \quad & \NCEtoken ~~ \booland{} ~~ \eEStoken && .  &
\end{align}
The following chapters at first don't touch the subject of the extended form of temporal expressions.
Chapter \ref{080817-010} then explains how the qualitative analysis is simplified by using extended event sequences.
Chapter \ref{080817-001} discusses the probabilistic quantification of extended event sequences.
\subsection{Events the are ``Part'' of an Expression}
For certain transformations of the temporal logic it is necessary to identify events that are ``part'' of a temporal expression, and accordingly, to know whether a given expression ``includes'' a certain event.
Specifically, it is necessary to know whether an event $X_{i}$ is part of an (extendend) core event or of an (extended) event sequence.

For a given event $X_i$ and a given expression $\varpi$,
\begin{align}
  \varpi \ist{} &
  \begin{cases}
      X_1 \booland{} X_2  \booland{} \ldots \booland{} X_n &,\\
      X_1 \pand{} X_2  \pand{} \ldots \pand{} X_n & , \\
      X_1 \sand{} X_2  \sand{} \ldots \sand{} X_n &
  \end{cases}
  && \text{and} &&
  i \in  \{1,2,\ldots,n\} ~ ,
\end{align}
$X_i$ is ``part'' of the expression $\varpi$; or in other words: expression $\varpi$ ``includes'' $X_i$.
We propose a new operator to denote this relationship:
\begin{align}
  &X_i \inplus \varpi \label{091115001} ~ .
\end{align}
For example,
\begin{align}
  &A \inplus A\booland{} B~, && A \inplus A\pand B~,  && B \inplus A\pand B \nonumber ~,
  &B \inplus A \pand{} B \pand{} C \nonumber ~.
\end{align}
\subsection{Visualization Using Sequential Failure Trees}\label{chap080726-021}
\emph{Sequential failure trees} illustrate possible failure sequences within a (non-repairable) system.
As such they help understanding the exact meaning and logical statement of temporal expressions, and they can also be used as a verification tool.
For instance, two different temporal expressions are logically identical if and only if they have identical sequential failure trees.

The explanations below for ``normal'' sequential failure trees (without simultaneous events, i.e.\ without SAND connected events) roughly follow the findings in \cite{Reinschke1988}.
Chapter \ref{090107002} then extends these ideas to general TFTA temporal expressions that may include SAND connections between events. 

\begin{figure}[H]
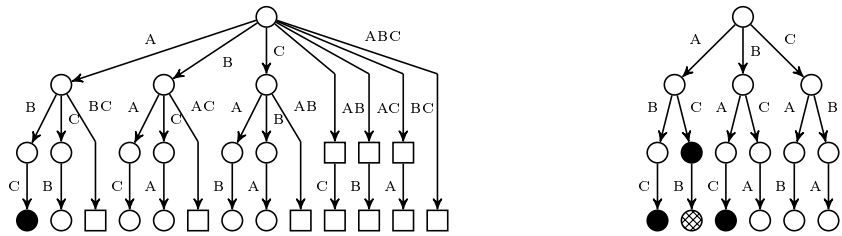

\centering
\hfill
\input{pics/minimaldisjunkt_sequ_ausfallbaum_gross1.tex}\hfill
\input{pics/minimaldisjunkt_sequ_ausfallbaum_gross2}\hfill{}
\caption{Sequential failure trees for the expressions $A\pand B \pand C$ (left side, including SAND connections) and $A\pand C$ (right side, without SAND connections, and shown as a ``simplified'' tree). 
Rectangular nodes include SAND connections; circular nodes do not include SAND connections.
Nodes that represent a system failure are filled in black; nodes that do not represent a system failure are filled in white; non-minimal failure nodes are crosshatched. 
}
\label{080910_020}
\end{figure}

Two examples:
Figure \ref{080910_020} shows sequential failure trees for the two temporal expressions $A\pand B \pand C$ (on the left) and $A\pand C$ (on the right), where both are used on a system with a total of three failure events $A$, $B$, and $C$.
%
%
\subsubsection{Normal Sequential Failure Trees (without SAND)}\label{090107001}
The sequential failure tree for a system comprised from $n$ elements (e.g.\ components) has $n+1$ levels with ${n \choose i}\cdot i!$ nodes on each level $i\in\{0, 1, \ldots, n\}$, see figure \ref{080910_020}.
Each node represents one specific system state $r$ and may be expressed as vector $\vec{K}_r \ist (X_1, X_2, \ldots, X_n)$; all elements that are not failed in this system state are written with $0$ ($False$), and all failed elements are written as $1, 2, \ldots, i$ according to the failure sequence that lead to this system state.

For example, the sequence $A\pand B\pand C$, i.e.\ "`$A$ before $B$ before $C$"', corresponds to vector $\vec{K} \ist (1, 2, 3)$.
the node on the top most level (level $0$) has the zero vector $\vec{K} \ist (0, 0, \ldots, 0)$.

A system's temporal failure function $\varpi$ may be expressed as function of vectors $\vec{K}_r$:
\begin{align}
  \varpi(\vec{K}_r) &=
    \begin{cases}
      1 & \text{, if the system is failed in state $r$.}\\
      0 & \text{, if the system is not failed in state $r$.}
    \end{cases} \label{090107004}
\end{align}
With the exception of the one node on level $0$, every node $\vec{K}$ has exactly one \emph{predecessor node} $\vec{K}'$.
With the exception of the nodes on the lowest level $n$, every node has at least one \emph{successor node} $\vec{K}''$.

Because of the definite sequence the following is always given:
\begin{align}
  \vec{K} & > \vec{K}' ~.
\end{align}
According to this ``vector inequation'', no element in $\vec{K}$ may be less than the corresponding element in $\vec{K}'$, and at least one element in $\vec{K}$ must be greater than the corresponding element in $\vec{K}'$.

Accordingly,
\begin{align}
  \vec{K} & < \vec{K}'' ~.
\end{align}
Taking the property of monotony into account, the follonwing statement holds for failure functions: 
\begin{align}
  \varpi(\vec{K}) & \geq \varpi(\vec{K}') ~.
\end{align}
Furthermore, the property of monotony yields that if $\varpi(\vec{K})= 0$ then the system function of a predecessor node $\vec{K}'$ of node $\vec{K}$ must also be $\varpi(\vec{K}')=0$.

A node $\vec{K}$ is a \emph{minimal failure node} if the failure sequence that is represented by $\vec{K}$ leads to a first-time failure of the system, i.e.\
\begin{align}
  \varpi(\vec{K}) & = 1 \qquad \text{and} \qquad \varpi(\vec{K}') = 0 ~. \label{090107005}
\end{align}
The succesor nodes of a minimal failure node are called \emph{non-minimal failure nodes}.
All succesor nodes of a non-minimal failure node are also non-minimal failure nodes.
And again, with the property of monotony the system function $ \varpi(\vec{K}'')$ of all successor nodes of a minimal (or non-minimal) node $\varpi(\vec{K})=1$ must also be $\varpi(\vec{K}'')=1$.

Sequential failure trees and the {TFTA} notation correspond to each other:
\emph{Nodes} (sequential failure tree) correspond to TFTA \emph{failure sequences};
\emph{minimal failure nodes} correspond to \emph{{MCSS}};
\emph{non-minimal failure nodes} correspond to \emph{non-minimal failure sequences}.

Providing all minimal failure nodes (or, respectively, all {MCSS}) completely describes the TOP event of a temporal fault tree and its failure function $\varpi$.

The left side of figure \ref{090107003} shows the simplified sequential failure tree (without SAND) of a system with three components $A$, $B$, and $C$ and the failure function $\varpi = ( C \pand B \pand A ) \boolor{} (B\pand C)$.

The sequential failure tree has $n+1=4$ levels.
Four of the $\sum\limits_{i=0}^{i=n=3}{n \choose i}\cdot i! = 16$ possible nodes (without SAND) are minimal failure nodes which correspond to the four {MCSS} $A\pand{} B\pand{} C$ and $\boolnot A\booland{}(B\pand{} C)$ and $B \pand{} A \pand{} C$ and $C\pand{}B\pand{}A$.
In addition, there is a non-minimal failure node, corresponding to the failure sequence $B\pand C \pand A$.

Nodes that do not represent a system failure state are filled white, minimal failure nodes are filled black, and non-minimal failure nodes are crosshatched.
\subsubsection{Sequential Failure Trees with Concurrent Events/SAND}\label{090107002}
The right side of figure \ref{090107003} shows the sequential failure tree of a system with failure function $\varpi = ( C \pand B \pand A ) \boolor{} (B\pand C)$; in this case SAND connections and corresponding nodes and transitions are also shown.
\begin{figure}[H]
  \centering
 \hfill
  \input{pics/sequ_ausfallbaum_BSP1} \hfill
  \input{pics/sequ_ausfallbaum_BSP2} \hfill\hfill
  \caption{Sequential failure tree without SAND (left side) and with SAND (right side) of a system with failure function $\varpi = ( C \pand B \pand A ) \boolor{} (B\pand C)$.}
  \label{090107003}

\end{figure}
For better discrimination failure nodes (system failure states) without SAND connection are depicted as circles and failure nodes with at least one SAND connection are depicted as rectangles.

Besides that, the notation, as introduced in chapter \ref{090107001}, stays the same.
For example, sequence $(A\sand{} B)\pand{} C$ corresponds to vector $\vec{K} \ist (1, 1, 2)$, and sequence $A\pand{}( B\sand{} C)$ corresponds to vector $\vec{K} \ist (1, 2, 2)$.
Equations \eqref{090107004} to \eqref{090107005} also hold for sequential failure trees with SAND connections.
\subsubsection{Using Sequential Failure Trees}\label{090107303}
Sequential failure trees allow an intuitive visualization of temporal expressions and thus ease their analysis:
\begin{itemize}
\item{}
They directly illustrate temporal expressions, comparable to logic tables as illustrations of Boolean expressions.
Moreover, different temporal expressions are equivalent, if they have identical sequential failure trees.
\item{}
They directly show if temporal expressions are minimal, or if they include each other, see chapter \ref{081018_001}.
Temporal expressions are minimal, if each of their sequential failure trees has at least one minimal failure node which is not a failure node in any of the other failure trees.
\item{}
They directly show if temporal expressions are mutually exclusive (disjoint), or if they have intersections, see chapter \ref{081018_002}. 
Temporal expressions are mutually exclusive, if their failure trees have no failure node in common.
 \end{itemize}
Two types of sequential failure trees are used below: the ``explicit form'' shown on the left side of figure \ref{090107304}, as well as a ``compact form'' shown on the right side of figure \ref{090107304}. 
\begin{figure}[H]
  \centering
  \input{pics/sequ_ausfallbaum_BSP3}
  \caption{Explicit and compact forms of the same sequential failure tree with failure function $\varpi = ( A\pand{}C ) \boolor (B\pand{} C)$.
  Both forms are used in this thesis.}
  \label{090107304}

\end{figure}
Based on some examples, creating and using sequential failure trees is demonstrated from page \pageref{090109010} on; there, sequential failure trees are compared to other visualization methods, too.
The appendix includes further explanantions on sequential failure trees, see page \pageref{091009020}.
\paragraph{Summary of Chapter \ref{chap080401-030}:}~\\
The TFTA's notation is based on the three Boolean operators AND, OR, and NOT, added by two new temporal operators PAND and SAND.
Temporal expressions may be reduced to their sum of products form (OR connected event sequences), which is called {TDNF} and consists of PAND connected core events; the TDNF corresponds to the Boolean disjunctive normal form.
The extended {TDNF} also allows AND connected core events, which reduces computing effort.
Sequential failure trees allow the visualization of temporal expressions and show if temporal expressions are minimal or mutually exclusive (disjoint).

%
%
%
%
%
\section{Laws of the TFTA Temporal Logic}\label{chap080401-031}\label{080129-002}\label{080723-010}
The temporal logic rules of the TFTA method are an extension to conventional Boolean logic and algebra.
These rules describe \emph{temporal relationships} between events, i.e.\ combinations and dependencies between events, while taking into account the individual points in time at which the events become $\True$, and taking into account possible sequences between events.
As it includes a concept of time, the temporal logic rules are more extensive and more complex than Boolean algebra.

There are two major differences between the application of the TFTA temporal logic and the Boolean logic: 
\begin{enumerate}
	\item Event sequences are expressed by the order in which events and operators are positioned in a temporal expression; therefore, the laws of commutation, laws of associativity, and distributive laws are not fully applicable.
	\item In temporal logic there are logical \emph{contradictions}, i.e.\ temporal relationships between events that are ``not possible''.
	Such contradictions always yield a logic $\False$. 
	For instance, an event can not become $\True$ after it has already become $\True$, and thus $X\pand{} X\ist \False$.
\end{enumerate}
\subsection{Boolean Algebra}\label{080129-001}
The conventional Boolean algebra describes \emph{Boolean relationships} between events, i.e.\ it makes statements on different events becoming $\True$; but it does not take into account the timing between those events.
Boolean logic basically consists of the rules listed below \cite{Meyna2003Taschenbuch,Schneeweiss1999Die}:\\
\newcommand{\DAbstandhalterSJS}{A \booland{} (B \boolor{} C ) \ist (A \booland{} B )\boolor{} (A \booland{} C )}
\textbf{laws of commutation}
\begin{align}
 & 
 A \booland{} B \ist B \booland{} A 
 \qquad \text{and} \qquad A \boolor{} B \ist B \boolor{} A ~.\label{080129-003}
\end{align}
\textbf{laws of associativity}
\begin{align}
 \begin{split}
 &A \booland{} (B \booland{} C ) \ist (A \booland{} B )\booland{} C \ist A \booland{} B \booland{} C \qquad \text{and} \\
 &A\boolor{}(B \boolor{} C )\ist(A\boolor{} B )\boolor{} C \ist A \boolor{} B \boolor{} C ~.
 \end{split}\label{080129-004}
\end{align}
\textbf{distributive laws}
\begin{align}
 &A \booland{} (B \boolor{} C ) \ist (A \booland{} B )\boolor{} (A \booland{} C )\qquad \text{and} \qquad
   A \boolor{} (B \booland{} C ) \ist (A \boolor{} B )\booland{} (A \boolor{} C ) ~.\label{080129-005}
\end{align}
\textbf{laws of idempotency}
\begin{align}
 &
 A \booland{} A \ist A 
    \qquad \text{and} \qquad A \boolor{} A \ist A ~.\label{080129-006}
\end{align}
\textbf{laws of absorption}
\begin{align}
 &
 A \booland{} (A \boolor{} B ) \ist A 
    \qquad \text{and} \qquad A \boolor{} (A \booland{} B ) \ist A ~.\label{080129-007}
\end{align}
\textbf{de Morgan's theorems}
\begin{align}
 &
 \boolnot{
 (A \booland{} B )} \ist
   \boolnot{A }\boolor\boolnot{B } 
    \qquad \text{and} \qquad \boolnot{(A \boolor{} B )} \ist
   \boolnot{A }\booland\boolnot{B } ~.\label{080129-008}
\end{align}
\textbf{operations with $\False$ and $\True$}
\begin{align}
 \begin{split}
 &\boolnot{\False} \ist \True ~,\\
 &
 A \booland{} \False \ist \False 
    \qquad \text{and} \qquad A \booland{} \True\ist A ~,\\
 &
 A \boolor{} \False \ist \mathrlap{A}\hphantom{\False} 
    \qquad \text{and} \qquad A \boolor{} \True\ist \True~.
 \end{split} \label{080129-009}
\end{align}
%
\subsection{Law of Completion}\label{080129-022}
The \emph{law of completion} in \eqref{071217-017} describes the main relationship between Boolean and temporal operators and fault tree gates, see figure \ref{fig:pandoramengen}:
\begin{align}
	A \booland{} B \ist& (A \pand{} B ) \boolor{} (A \sand{} B ) \boolor{} (B \pand{} A ) ~.
		\label{071217-017}
\end{align}
Terms on the right side of \eqref{071217-017} are mutually exclusive (disjoint).

The SAND connection between \emph{different} events expresses (structurally) dependend failures, which may be interpreted as \emph{common cause failures} (CCF).
It can be shown that the expectancy value of the failure probability/failure rate is zero for failure events which are connected by SANDs, if \emph{independent} failures are assumed.
For instance, $\EW[A\sand{} B] \ist{}{} 0$, see chapter \ref{080822-020} for details.
The SAND operator is also very important for transformations of temporal expressions and for qualitative analysis.
%
\subsection{Law of Contradiction}\label{080822-001}
In general, it is logically contradictory if the same event becomes $\True$ after itself.
This follows directly from the assumption of monotony combined with non-repairable components; see chapter \ref{090222001} for these two general assumptions of this thesis.

In the most simple case, 
\begin{align}
			A\pand{} A \ist& \False \label{071217-011}~.
\end{align}
More generally, an event sequence yields $\False$ if at least one event exists more than once in it; i.e.\
\begin{align}
			&X_1\pand{} X_2\pand{} \ldots\pand{} X_n \ist \False~,
				\label{080216-105}
\end{align}
if $\exists ~ X_i\ist X_j$ for $i, j\in\{1, 2, \ldots, n\}$ and $i\neq j$.
In a temporal fault tree a PAND gate therefore yields $\False$ if it has the same event as input more than once.

The law of contradiction applies to non-atomic core events analogously:
\begin{align}
	(A \sand{} B )\pand{} A \ist& (B \sand{} A )\pand{} A\ist \False
				\label{080206-319}~,\\
	A \pand{} (A \sand{} B )\ist& A \pand{} (B \sand{} A )\ist \False ~,
				\label{080206-322}
\end{align}
or, more generally,
\begin{align}
			&K_1\pand{} K_2\pand{} \ldots\pand{} K_n \ist \False~,
				\label{080216-005}
\end{align}
if there is at least one atomic event $X$ which is part of two or more core events $K$, i.e.\ if $\exists ~ (X\inplus K_i) \booland{} (X\inplus K_j)$ for $i, j\in\{1, 2, \ldots, n\}$ and $i\neq j$.

An example: $(A \sand{} B )\pand{} C\pand{} (A \sand{} D \sand{}E ) \ist \False$, as $(A \sand{} B )$ and $(A \sand{} D \sand{}E )$ both contain the same atomic event $A$.
%
\subsection{Temporal Law of Idempotency}\label{090107018}
A new \emph{temporal law of idempotency} may be derived from the laws of completion and the law of contradiction.
The temporal law of idempotency applies only to the SAND operator.
From \eqref{071217-017} and \eqref{071217-011} and the Boolean law of idempotency in \eqref{080129-006} follows that
\begin{align}
 &A \booland{} A \ist{} (A \pand{} A ) \boolor{} (A \sand{} A ) \boolor{} (A \pand{} A ) \ist{} \False \boolor{} (A \sand{} A ) \boolor{}\False \qquad \text{and}\nonumber\\
 &A \booland{} A \ist{} A \qquad \text{, and therefore} \nonumber\\
 &A \sand{} A \ist{} A \label{090107019} ~.
\end{align}
%
\subsection{Temporal Law of Commutativity}\label{080822_003}
A \emph{temporal law of commutativity} (or \emph{commutation}) applies only to the SAND operator, as
	\begin{align}
			A \sand{} B \ist& B \sand{} A ~,\label{071218-001}
		\intertext{but not for the PAND operator, as }
			A \pand{} B {}\neq{}& B \pand{} A~. 		\label{071218-002}
	\end{align}
%
\subsection{Temporal Law of Associativity}\label{080822_004}
The SAND operator also has the property of associativity; thus 
\begin{align}
		A \sand{}(B \sand{} C ) &\ist A \sand{} B \sand{} C \ist
		 (A \sand{} B) \sand{} C ~. \label{080206-040}
	\intertext{The PAND operator, on the other hand, is only left-associative, as in}
		(A \pand{} B )\pand{} C &\ist A \pand{} B \pand{} C
		 \neq A \pand{} (B \pand{} C ) ~. \label{080122-002}
\end{align}
\subsection{Further Temporal Logic Laws}\label{080822_805}
There are two more temporal laws with special significance:
\begin{align}
		&A \pand{}(B \pand{} C ) \ist (A \booland{} B )\pand{} C
				\qquad \text{and} \label{080122-001}\\
		&A \sand{}(B \pand{} C ) \ist B\pand{}(A\sand{} C)
				~. \label{080216-020}
\end{align}
\paragraph{Examples illustrating the laws of temporal TFTA logic}\label{090109010}~\\%
The correctness of these two laws is demonstrated using three different graphical methods:
\begin{itemize}
\item Table \ref{tab:WahrtabelleWeitereGrundlegendeRegeln} (page \pageref{tab:WahrtabelleWeitereGrundlegendeRegeln}) shows correctness of \eqref{080122-001} and \eqref{080216-020} using truth tables similar to the ones known from Boolean logic.
The main difference is, that in the temporal logic all possible event sequences have to be taken into account.
 \item Figure \ref{090109011} shows sequential failure trees for \eqref{080122-001} and \eqref{080216-020}, see page \pageref{090109011}, which are well suited to verify and visualize temporal expressions.
 \item Finally, figure \ref{fig:ablauf_grundlegende_ausdruecke1} shows the correctness of \eqref{080122-001} and \eqref{080216-020} using timing diagrams, see page \pageref{fig:ablauf_grundlegende_ausdruecke1}.
\end{itemize}
The number of entries, i.e.\ rows, in the truth table equals the number of nodes in the sequential failure tree.
Indeed, one can use sequential failure trees in order to simplify the process of creating the truth table.
Timing diagrams, on the other hand, are well suited for specific checks of more complex temporal expressions.
%
%
\begin{table}[h]
\small
 \centering
 \begin{tabular}{ccc} \toprule
 \vphantom{\Large Ö} & $A \pand{}(B \pand{} C )$ & $(A \booland{} B )\pand{} C $
 \\ \midrule 
\vphantom{\Large Ö} $\boolnot A\boolnot B\boolnot C   $ & $\False $ & $\False $\\
\vphantom{\Large Ö} $\boolnot B\boolnot C \booland{}A  $ & $\False $ & $\False $\\
\vphantom{\Large Ö} $\boolnot A\boolnot C \booland{}B  $ & $\False $ & $\False $\\
\vphantom{\Large Ö} $\boolnot A\boolnot B \booland{}C  $ & $\False $ & $\False $\\
\vphantom{\Large Ö} $\boolnot C\booland{}(A\pand{} B) $ & $\False $ & $\False $\\
\vphantom{\Large Ö} $\boolnot C\booland{}(B\pand{} A) $ & $\False $ & $\False $\\
\vphantom{\Large Ö} $\boolnot C\booland{}(A\sand{} B) $ & $\False $ & $\False $\\
\vphantom{\Large Ö} $\boolnot B\booland{}(A\pand{} C)  $ & $\False $ & $\False $\\
\vphantom{\Large Ö} $\boolnot B\booland{}(C\pand{} A)  $ & $\False $ & $\False $\\
\vphantom{\Large Ö} $\boolnot B\booland{}(A\sand{} C)  $ & $\False $ & $\False $\\
\vphantom{\Large Ö} $\boolnot A\booland{}(B\pand{} C)  $ & $\False $ & $\False $\\
\vphantom{\Large Ö} $\boolnot A\booland{}(C\pand{} B)  $ & $\False $ & $\False $\\
\vphantom{\Large Ö} $\boolnot A\booland{}(B\sand{} C)  $ & $\False $ & $\False $\\
\vphantom{\Large Ö} $A\pand{} B\pand{} C       $ & $\True $ & $\True $\\
\vphantom{\Large Ö} $B\pand{} A\pand{} C       $ & $\True $ & $\True $\\
\vphantom{\Large Ö} $A\pand{} C\pand{} B       $ & $\False $ & $\False $\\
\vphantom{\Large Ö} $C\pand{} A\pand{} B       $ & $\False $ & $\False $\\
\vphantom{\Large Ö} $B\pand{} C\pand{} A       $ & $\False $ & $\False $\\
\vphantom{\Large Ö} $C\pand{} B\pand{} A       $ & $\False $ & $\False $\\
\vphantom{\Large Ö} $A\pand{} (B\sand{} C)       $ & $\False $ & $\False $\\
\vphantom{\Large Ö} $B\pand{} (A\sand{} C)       $ & $\False $ & $\False $\\
\vphantom{\Large Ö} $C\pand{} (A\sand{} B)       $ & $\False $ & $\False $\\
\vphantom{\Large Ö} $(A\sand{} B)\pand{} C       $ & $\True $ & $\True $\\
\vphantom{\Large Ö} $(A\sand{} C)\pand{} B       $ & $\False $ & $\False $\\
\vphantom{\Large Ö} $(B\sand{} C)\pand{} A       $ & $\False $ & $\False $\\
\vphantom{\Large Ö} $A\sand{} B\sand{} C        $ & $\False $ & $\False $
\\ \bottomrule
\end{tabular}
 \begin{tabular}{ccc} \toprule
 \vphantom{\Large Ö} & $A \sand{}(B \pand{} C ) $& $B\pand{}(A\sand{} C) $
 \\ \midrule 
\vphantom{\Large Ö} $\boolnot A\boolnot B\boolnot C $ & $\False $ & $\False $\\
\vphantom{\Large Ö} $\boolnot B\boolnot C \booland{}A $ & $\False $ & $\False $\\
\vphantom{\Large Ö} $\boolnot A\boolnot C \booland{}B $ & $\False $ & $\False $\\
\vphantom{\Large Ö} $\boolnot A\boolnot B \booland{}C $ & $\False $ & $\False $\\
\vphantom{\Large Ö} $\boolnot C\booland{}(A\pand{} B) $ & $\False $ & $\False $\\
\vphantom{\Large Ö} $\boolnot C\booland{}(B\pand{} A) $ & $\False $ & $\False $\\
\vphantom{\Large Ö} $\boolnot C\booland{}(A\sand{} B) $ & $\False $ & $\False $\\
\vphantom{\Large Ö} $\boolnot B\booland{}(A\pand{} C) $ & $\False $ & $\False $\\
\vphantom{\Large Ö} $\boolnot B\booland{}(C\pand{} A) $ & $\False $ & $\False $\\
\vphantom{\Large Ö} $\boolnot B\booland{}(A\sand{} C) $ & $\False $ & $\False $\\
\vphantom{\Large Ö} $\boolnot A\booland{}(B\pand{} C) $ & $\False $ & $\False $\\
\vphantom{\Large Ö} $\boolnot A\booland{}(C\pand{} B) $ & $\False $ & $\False $\\
\vphantom{\Large Ö} $\boolnot A\booland{}(B\sand{} C) $ & $\False $ & $\False $\\
\vphantom{\Large Ö} $A\pand{} B\pand{} C$ & $\False $ & $\False $\\
\vphantom{\Large Ö} $B\pand{} A\pand{} C$ & $\False $ & $\False $\\
\vphantom{\Large Ö} $A\pand{} C\pand{} B$ & $\False $ & $\False $\\
\vphantom{\Large Ö} $C\pand{} A\pand{} B$ & $\False $ & $\False $\\
\vphantom{\Large Ö} $B\pand{} C\pand{} A$ & $\False $ & $\False $\\
\vphantom{\Large Ö} $C\pand{} B\pand{} A$ & $\False $ & $\False $\\
\vphantom{\Large Ö} $A\pand{} (B\sand{} C)$ & $\False $ & $\False $\\
\vphantom{\Large Ö} $B\pand{} (A\sand{} C)$ & $\True $ & $\True $\\
\vphantom{\Large Ö} $C\pand{} (A\sand{} B)$ & $\False $ & $\False $\\
\vphantom{\Large Ö} $(A\sand{} B)\pand{} C$ & $\False $ & $\False $\\
\vphantom{\Large Ö} $(A\sand{} C)\pand{} B$ & $\False $ & $\False $\\
\vphantom{\Large Ö} $(B\sand{} C)\pand{} A$ & $\False $ & $\False $\\
\vphantom{\Large Ö} $A\sand{} B\sand{} C$ & $\False $ & $\False $
 \\ \bottomrule
\end{tabular}
 \caption{Truth table which demonstrates that \eqref{080122-001} (left side) and \eqref{080216-020} (right side) are correct.
 Including SAND connections, there are $26$ possible sequences.
 Logical equivalence of both expressions is shown as in both cases all possible sequences yield identical results. }
 \label{tab:WahrtabelleWeitereGrundlegendeRegeln}
\end{table} %
%
%
%
%
%
%
%
%
%
%
%
%
\begin{figure}
 \centering
 \input{pics/sequ_ausfallbaum_VERGLEICH1}
 \caption{Sequential failure trees for \eqref{080122-001} and \eqref{080216-020}, which show their correctness.
 From left to right and from top to bottom:
 $A$, $B$, $C$, $B\pand{}C$, $A\booland{}B$, $A\sand{}C$, $A\pand{}(B\pand{}C)\ist{}(A\booland{}B)\pand{}C$, $A\sand{}(B\pand{}C)=B\pand{}(A\sand{}C)$.}
 \label{090109011}
\end{figure}
\begin{figure}
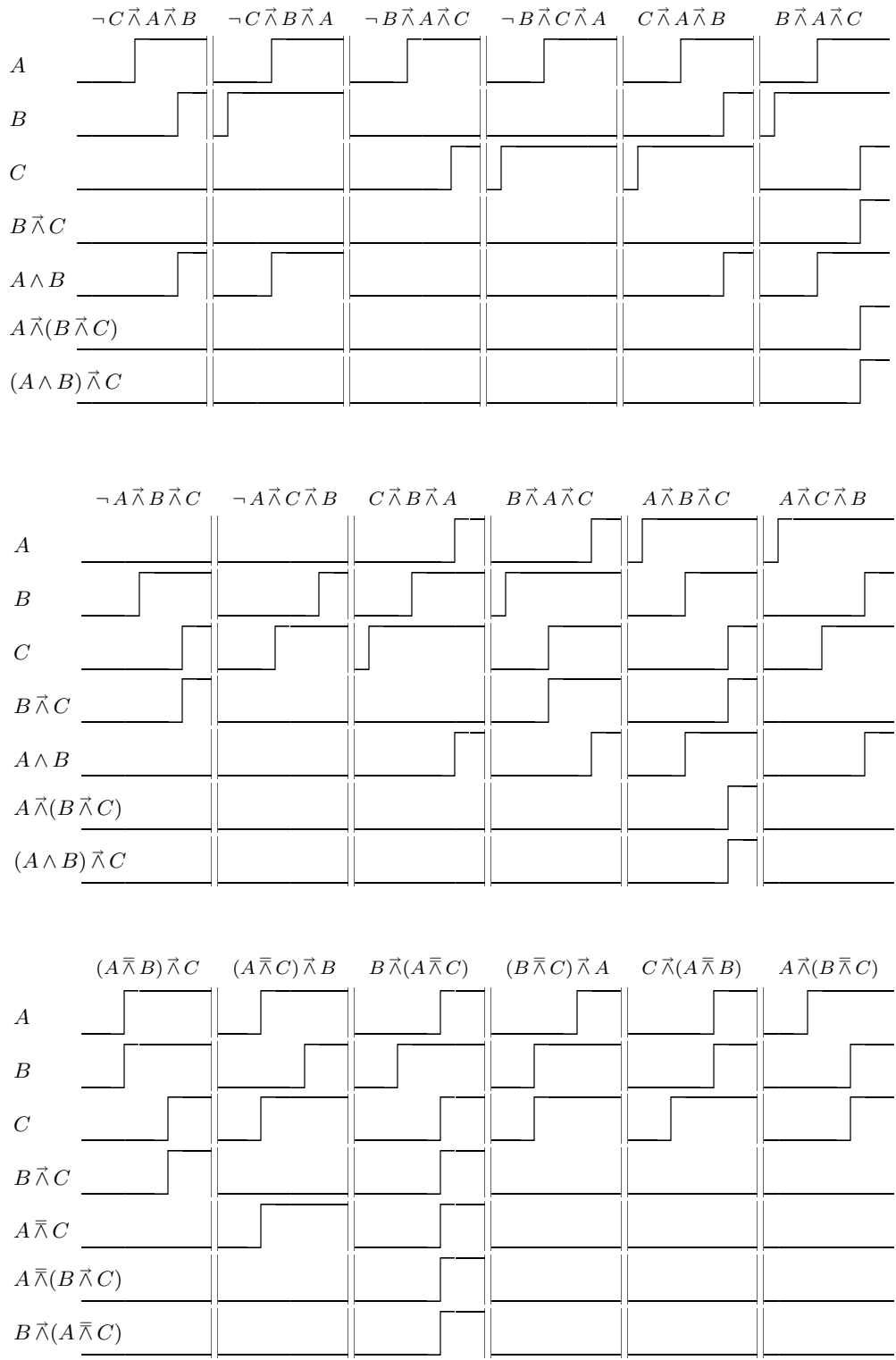

	\centering
		\input{pics/regeln_pand_pand_zu_and_pand}
~\\
 \input{pics/regeln_sand_pand_zu_pand_sand}
	\caption{Timing diagrams showing selected sequences from table \ref{tab:WahrtabelleWeitereGrundlegendeRegeln}, which demonstrate the correctness of  \eqref{080122-001} (upper two diagrams) and \eqref{080216-020} (lower diagram).}
	\label{fig:ablauf_grundlegende_ausdruecke1}
\end{figure}
%
\clearpage 
\subsection{Temporal Operations with Negated Events}
\label{080220-002}
\emph{Remark:}
The statements below exclusively relate to atomic negated events.
Specialities of non-atomic negated events are covered from page \pageref{090105001} on.
\subsubsection{How to Interpret Negated Events in {TFTA}}
In the {TFTA}, as well as in the conventional {FTA}, a non-negated event represents a failure of a real element, e.g.\ a component.
Therefore, a negated event represents the ``not-failing'' of a real element.

There are two possible interpretations for ``not-failing'':
\begin{enumerate}
 \item 
 An element, that has failed before, is repaired.
 The ``not-failing'' is an ``un-failing'', a transition from one state (failed) to another (repaired), and thus is an action.
 \item 
 An element has not yet failed and is still operational.
 The ``not-failing'' is a state.
\end{enumerate}
The temporal logic, as discussed in this theses and applied to the {TFTA}, relies on the assumtions of monotony of the temporal failure function as well as non-repairability of elements.

At first, at time $t=0$, all elements (components) are operational.
Failures occur at times $t>0$ and are represented in the temporal fault tree by (non-negated) failure events $X_{i}$.
The latter ``switch'' from $\False$ to $\True$ at times $t_{X_i}>0$.
Moreover, all elements are non-repairable.
Failure events that occurred (became $\True$) at $t_{X_i}$ stay $\True$.{}

Two things follow for negated events:
they are $\True$ until $t_{X_i}$ and then become $\False$;
and they cannot become $\True$ again after $t_{X_i}$.
Thus, a negated event in the {TFTA}
\begin{align}
\newcommand{\myLK}{[} 
 \boolnot X_i \ist{} \begin{cases}
      \True & \text{in } \myLK 0;t_{X_i}\myLK \qquad\text{and}\\
      \False & \text{in } \myLK t_{X_i};\infty\myLK ~, \end{cases} \label{090109017}
\end{align}
with $t_{X_i}>0$.

Therefore, the first interpretation of the meaning of negated events in the {TFTA} is to be rejected; in the {TFTA} negated failure events represent elements, that have not yet failed. 
%
\subsubsection{Using Negated Events in {TFTA}}
Negated events are used in two different ways within the {TFTA}; these are comparable to the two ways of using negated events in Boolean {FTA}.
\begin{enumerate}
\item{}
 Even if there are no NOT gates used explicitly in the fault tree, the temporal failure function may get negated events from logical transformations.
 For instance, the transformation of temporal expressions that are not mutually exclusive (not disjoint) into a disjoint form requires usage of negated events.
 \item{}
 NOT gates in the fault tree model allow explicit modelling of negated events.
 Such negations of basic events or non-atomic events (subtrees) are then input to other higher-level fault tree gates.
 Accordingly, the failure function then includes negated events.
\end{enumerate}
\paragraph{Negated Events Resulting From Logic Replacements}~\\
In the Boolean {FTA} non-disjoint expressions are transformed into a disjoint form using negated events \cite{Abraham1979,Heidtmann1989,Heidtmann1997}.
Thereby, negated events only occur within conjunctions (AND connected terms) in combination with at least one non-negated event.
The assumtion of monotony is not invalidated, because events are not substantially meshed by this transformation (the topic of substantial meshing is discussed in \cite{19811990DIN}).
Moreover, none of the transformation laws of the Boolean logic introduce new negated events -- de Morgan's theorems only discuss transformation of existing negated events.

The temporal logic of the {TFTA} also uses negated events for the transformation into a disjoint form, see chapter \ref{chap080809-001}.
But other than the Boolean logic, there are temporal transformation laws, specifically the temporal distributive laws in chapter \ref{080723-015}, that do introduce negated events.
These negated events only occur within conjunctions, though, and in combination with at least one non-negated event. 
In doing so, the assumption of monotony is not invalidated.
\paragraph{Using Negations Explicitely in Fault Trees}~\\
This kind of usage of negated events is restricted to cases where no substantially meshed negated events are used in order to not invalidate the assumption of monotony, see \cite{19811990DIN}.
Usually, this is limited to special use cases, e.g.\ if the results of one of the temporal laws of transformation (see above) shall explicitly be modelled with a temporal fault tree.

In general, {TFTA} statements like, e.g.,
\begin{itemize}
 \item{}
 ``A has not failed yet, before B has not failed yet'', i.e.\ $\boolnot A\pand{}\boolnot B$, or
 \item{}
 ``A and B have simultaneously not failed yet'', i.e.\ $\boolnot A\sand{}\boolnot B$, or
\item ``A has failed, because B has not failed yet, or C has failed'', i.e.\ $A\ist\boolnot B\boolor{}C$,
\end{itemize}
are neither logically meaningful nor allowed in {TFTA}.
Thus there is no necessity to use negated events explicitly as inputs to PAND or SAND gates, or to use them in combination with non-negated events as inputs to OR gates.

On the other hand, it is indeed permitted to model logical statements like $\boolnot A\booland{}B$ explicitly within the fault tree, if -- and only if -- the assumption of monotony still holds.

\subsubsection{Rules of Replacement for Negated Events in the Temporal Logic}\label{090123004}
The law of completion from \eqref{071217-017} must not be used on expressions where at least one of the operands of the conjunction (AND connection) is a negated event.

Therefore, the application of the other temporal laws of transformation also does not lead to negated events being input to PAND or SAND operators.
In case of the temporal distributiv laws all negated events are part of conjunction terms, see chapter \ref{080723-015}.
Furthermore, this leads to the conclusion that the Boolean logic rules may be used for handling of negated events, see chapter \ref{080129-001}.

Special considerations are necessary for ``mixed expressions'' where negated events and temporal expressions are both part of the same conjunction.
There are  
\begin{align}
 &\boolnot A \booland{} ( \ldots \pand{} A \pand{} \ldots) \ist{} \False \label{09010930} ~, \\
 &\boolnot A \booland{} ( \ldots \pand{} (A \sand{}\ldots) \pand{} \ldots) \ist{} \False \label{09010931} ~,
\end{align}
and
\begin{align}
 &(\boolnot A \booland B)\booland{}C \ist{} \bigl[\boolnot A\booland{}(B\booland{}C)\bigr] \boolor{}
         \bigl[(B\pand{}A)\booland{}C\bigr] \ist{} \nonumber\\
 &\hphantom{(\boolnot A \booland B)\booland{}C}\ist{}
  \bigl[\boolnot A\booland{}(B\booland{}C)\bigr] \boolor{}
         \bigl[B\pand{}A\pand{}C\bigr] \boolor{}
         \bigl[B\pand{}(A\sand{}C)\bigr] \label{09010932} ~, \\
 &(\boolnot A \booland B)\pand{}C \ist{} \bigl[\boolnot A\booland{}(B\pand{}C)\bigr] \boolor{}
         \bigl[B\pand{}A\pand{}C\bigr] \boolor{}
         \bigl[B\pand{}(A\sand{}C)\bigr]\label{09010933} ~, \\
 &(\boolnot A \booland B)\sand{}C \ist{} \boolnot A\booland{}(B\sand{}C) \label{09010934} ~, \\
 &C\pand{}(\boolnot A \booland B) \ist{} \boolnot A\booland{}(C\pand{}B) \label{09010935} ~.
\end{align}
Equation \eqref{09010932} shows the one main difference between temporal and Boolean logic with regards to usage of negated events.

In the Boolean logic the law of associativity from equation \eqref{080129-004} also applies to negated events.
But in the temporal logic negated events have a ``period of validity'', which is expressed by brackets.
For instance, $(\boolnot A \booland B)\booland{}C$ denotes two things: first, that at the point in time, at which event $B$ occurs, event $A$ has ``not yet'' occurred, and second, that $C$ has occurred; but there is no separate statement on the timing relationship between $C$ and the others.
On the other hand, $\boolnot A \booland{} ( B\booland{}C)$ expresses timing relationships between all three events; this expression denotes that at the point in time, at which ``$B$ and $C$'' occurs, the event $A$ has not yet occurred:
\begin{align}
 &
 \underbrace{\boolnot A}_{
  \text{\begin{minipage}{35mm}\input{pics/negierte_konzept_gueltigkeitsdauer1}
    \end{minipage}}
 }\booland{}
 \underbrace{B}_{
  \text{\begin{minipage}{35mm}\input{pics/negierte_konzept_gueltigkeitsdauer2}
    \end{minipage}}
 } \ist{}
 \underbrace{\boolnot A\booland{}B}_{
  \text{\begin{minipage}{35mm}\input{pics/negierte_konzept_gueltigkeitsdauer3}
    \end{minipage}}
 } \nonumber \\
 & ~ \nonumber\\
 &\underbrace{(\boolnot A\booland{}B)}_{
  \text{\begin{minipage}{35mm}\input{pics/negierte_konzept_gueltigkeitsdauer3}
    \end{minipage}}
 }\booland{}
 \underbrace{\vphantom{(B)}C}_{
  \text{\begin{minipage}{35mm}\input{pics/negierte_konzept_gueltigkeitsdauer4}
    \end{minipage}}
 } \ist{}
 \underbrace{(\boolnot A\booland{}B)\booland{}C}_{
  \text{\begin{minipage}{35mm}\input{pics/negierte_konzept_gueltigkeitsdauer5}
    \end{minipage}}
 }\nonumber \\
 & ~ \nonumber\\
 &\underbrace{B}_{
  \text{\begin{minipage}{35mm}\input{pics/negierte_konzept_gueltigkeitsdauer2}
    \end{minipage}}
 } \booland{}
 \underbrace{C}_{
  \text{\begin{minipage}{35mm}\input{pics/negierte_konzept_gueltigkeitsdauer4}
    \end{minipage}}
 } \ist{}
 \underbrace{B \booland{}C}_{
  \text{\begin{minipage}{35mm}\input{pics/negierte_konzept_gueltigkeitsdauer6}
    \end{minipage}}
 } \nonumber \\
 & ~ \nonumber\\
 &\underbrace{\vphantom{(B)}\boolnot A}_{
  \text{\begin{minipage}{35mm}\input{pics/negierte_konzept_gueltigkeitsdauer1}
    \end{minipage}}
 } \booland{}
 \underbrace{(B \booland{}C)}_{
  \text{\begin{minipage}{35mm}\input{pics/negierte_konzept_gueltigkeitsdauer6}
    \end{minipage}}
 } \ist{}
 \underbrace{\boolnot A \booland{} (B \booland{}C)}_{
  \text{\begin{minipage}{35mm}\input{pics/negierte_konzept_gueltigkeitsdauer7}
    \end{minipage}}
 } \nonumber
\end{align}
In particular, this also affects temporal expressions of the following type:
\begin{align}
 (\boolnot A \booland B)\sand{}A \ist{}& \boolnot A\booland{}(B\sand{}A) \ist{}\False\label{090123002} ~, \\
 A\pand{}(\boolnot A \booland B) \ist{} &\boolnot A\booland{}(A\pand{}B) \ist{}\False\label{090123003} ~, \\
 (\boolnot A \booland B)\pand{}A \ist{}& \bigl[\boolnot A\booland{}(B\pand{}A)\bigr] \boolor{}
         \bigl[B\pand{}A\pand{}A\bigr] \boolor{}
         \bigl[B\pand{}(A\sand{}A)\bigr]\ist{} \nonumber \\
 \ist{}& \False\boolor{}\False\boolor{} \bigl[B\pand{} A \bigr]\ist{} B\pand{}A ~. \label{090123001}
\end{align}
Chapter \ref{080909_005} discusses why and how these expressions are ``temporally (non-)minimal''.
%
\subsubsection{Conjunction of Negated Events}
The above discussion did not include conjunctions consisting of more than one negated event, as e.g.\ in \begin{align}
 \boolnot A \boolnot B & \ist{}\boolnot A \booland{} \boolnot B  ~. \label{090110003}
\end{align}
When applied to the {TFTA}, such conjunctions are interpreted as undividable entities; the rules for transformation and handling of negated events, as given above, apply to those entities analogously.  

From this follows that
\begin{align}
 \boolnot A \booland{} ( \boolnot B \booland{} C) & \ist{} (\boolnot A \boolnot B) \booland{} C ~.\label{090321011}
\end{align}
%
%
\subsubsection[Temporal Laws of Negation]{Temporal Laws of Negation, i.e. Negation of Non-Atomic Negated Events}\label{090105001}
So far, all statements regarding negated events have applied to atomic events (basic events) only.
Additional aspects have to be considered in case of negated non-atomic events, as e.g.\ in $\boolnot(A\pand{}B)$.

The Negation of Boolean non-atomic expressions like $\boolnot(A\booland{}B)$ or $\boolnot(A\boolor{}B)$ is done using de~Morgan's theoremes in \eqref{080129-008}.
The negation of SAND and PAND connected expressions can, for example, be deduced from figure \ref{fig:pandoramengen}; it yields:
\begin{align}\begin{split}
 \boolnot ( A \pand{} B )& \ist{}
  ( \minneg{A} \minneg{B} ) \boolor{}
  ( \minneg{B} \booland{} A ) \boolor{}
  ( \minneg{A} \booland{} B) \boolor{}
  ( B\pand{} A) \boolor{}
  ( A \sand{} B )
  \qquad \text{and}\end{split}\label{071219-01}\\
 \begin{split}
 \boolnot ( A \sand{} B )& \ist{}
  ( \minneg{A} \minneg{B} ) \boolor{}
  ( \minneg{B} \booland{} A ) \boolor{}
  ( \minneg{A} \booland{} B) \boolor{}
  ( A\pand{} B) \boolor{}
  ( B\pand{} A)
 ~. \end{split}
 \label{071219-011}
\end{align}
On the right hand side of the equations all terms are mutually exclusive (disjoint) and carry explicite (temporal) statements to all events involved, see chapter \ref{081018_002}.

In {TFTA} such non-atomic negated expressions can only exist as part of a conjunction expression together with non-negated events.
As such, they describe a system state where at a specific point in time a specific event sequence has ``not yet'' occurred.
The right hand sides of \eqref{080129-008} and \eqref{071219-01} and \eqref{071219-011} represent the different possibilities how this specific system state was reached. 

An example: the temporal expressions $\boolnot ( A \pand{} B ) \booland{} C$ represents a state in which at the time of occurrence of $C$ the event sequence $A \pand{} B$ has not occurred.
This implies either that at the time of occurrence of $C$
\begin{itemize}
\item neither $A$ nor $B$ have occurred~-- therefore $ (\boolnot A \boolnot B ) \booland{} C$~--
\item or $A$ has occurred, but $B$ has not~-- therefore $\minneg{B} \booland{} (A \booland{} C)$~--
\item or $B$ has occurred, but $A$ has not~-- therefore $\minneg{A} \booland{} (B \booland{} C)$~--
\item or $B$ has occurred before $A$ has occurred~-- therefore $(B \pand{} A ) \booland{} C$~--
\item or $A$ and $B$ have occurred simultaneously~-- therefore $(A \sand{} B ) \booland{} C$~.
\end{itemize}

The \emph{first temporal law of negation} is thus given as
\begin{align}\begin{split}
 \boolnot ( A \sand{} B ) \booland{} C& \ist{}
  \phantom{\boolor{}}
  \bigl[(\minneg{A} \minneg{B} ) \booland{} C \bigr]\boolor{}
  \bigl[ \minneg{B} \booland{} (A \booland{} C) \bigr]\boolor{}
  \bigl[\minneg{A} \booland{} (B \booland{} C) \bigr] \boolor \\
  & \phantom{\ist{}}\boolor{}
  \bigl[(A\pand{} B ) \booland{} C \bigr]\boolor{}
  \bigl[( B\pand{} A ) \booland{} C \bigr]
 ~. \end{split}
 \label{090110005}
\end{align}
Analogously, the \emph{second temporal law of negation} is given as
\begin{align}\begin{split}
 \boolnot ( A \pand{} B ) \booland{} C& \ist{}
  \phantom{\boolor{}}
  \bigl[(\minneg{A} \minneg{B} ) \booland{} C \bigr]\boolor{}
  \bigl[ \minneg{B} \booland{} (A \booland{} C) \bigr]\boolor{}
  \bigl[\minneg{A} \booland{} (B \booland{} C) \bigr] \boolor \\
  & \phantom{\ist{}}\boolor{}
  \bigl[( B\pand{} A ) \booland{} C \bigr] \boolor{}
   \bigl[(A\sand{} B ) \booland{} C \bigr]
 ~. \end{split}
 \label{090110004}
\end{align}
\subsection{\emph{True} and \emph{False} in Temporal Logics} \label{090111001}
Operations with the ``timeless'' expressions $\True$ and $\False$ should only be found in {TFTA} expressions, if a more complex temporal expression was reduced to $\True$ or $\False$ in a preceeding transformation step.

If $X$ and $X\neq\True$ themselves are not negated, then
\begin{align}
 && X \pand{} \True &\ist{} \False~, && X \sand{} \True &\ist{} \False ~, && \True &\pand{} X \ist{} X ~, \label{090110002} \\
 && X \pand{} \False &\ist{} \False ~, && X \sand{} \False &\ist{} \False ~, && \False &\pand{} X \ist{} \False ~.
\label{090110001}
\end{align}
Furthermore, 
\begin{align}
 &\True \pand{} \True \ist{} \False~,
 &&\True \sand{} \True \ist{} \True ~,
 &&\False \pand{} \True \ist{} \False~. \label{080909_004}
\end{align}
Given these rules, consistency to the Boolean logic rules, which are, of course, still valid, is obtained; thus,
\begin{align*}
 X \booland \mathrlap{\True}\hphantom{\False} \ist{}{}& ( X \pand{}\True ) \boolor{} ( X \sand{}\True ) \boolor{} (\True\pand{} X ) \ist{} \True\pand{} X \ist {} X ~, \\
 X \booland \False \ist{}{}& ( X \pand{}\False ) \boolor{} ( X \sand{}\False ) \boolor{} (\False\pand{} X ) \ist{} \False\pand{} X \ist {} \False ~.
\end{align*}
%
%
\subsection{Temporal Distributive Laws} \label{080723-015}
Boolean logic has the distributive law as given in \eqref{080129-005}.
Combined with the Boolean operators' property of associativity, see \eqref{080129-004}, this yields
\begin{align}
 (A\boolor{} B)\booland{} C & \ist C\booland{}(B\boolor{} A)\ist (A\booland{} C) \boolor{} (B\booland{} C) \ist (C\booland{} B) \boolor{} (C\booland{} A)~.
\end{align}
This distributive law is vital to the transformation of Boolean expressions into a disjunctive normal form ({DNF}).

Very similar, the SAND operator of the temporal logic also has the property of associativity; therefore, the temporal laws of associativity and commutativity apply, see \eqref{071218-001} and \eqref{080206-040}.

On the other hand, the PAND operator obviously lacks a law of commutativity, see \eqref{071218-002}; reason for that is that this operator ``transports'' a great part of its logic information in the sequence of events.

Therefore, at least the following has to be differentiated for something like a PAND's distributive law:
\begin{align}
 & A\pand{} (B\boolor{} C) && \text{, so-called \emph{type I},} & \text{and} & \label{080722-311}\\
 & (A\boolor{} B)\pand{} C && \text{, so-called \emph{type II}}. & &
\label{080722-011}
\end{align}
The following two sections discuss temporal distributive laws, first for PAND operators and expressions of type I and II, followed by the temporal dísributive law for SAND operatos; for the latter, no further discrimination of types is necessary.
\subsubsection{Distributive Law for PAND-OR Expressions of Type I}
The logic statment of expression $A\pand{} (B\boolor{} C)$ is:
``$A$ must occur, before the expression in brackets $(B\boolor{} C)$ occurs''.
This is \emph{not} equivalent to the logic statement ``$A$ must occur before $B$, or $A$ must occur before $C$'', as proven by table \ref{tab:WahrtabelleTypI} and figure \ref{fig:sequ_ausfallbaum_A_pand_B_or_A_pand_C}:
\begin{align}
 A\pand{}(B\boolor{} C) & \neq (A\pand{} B)\boolor{}(A\pand{} C)~, \label{080725-001}
\end{align}
and thus there is no simple temporal distributive law for expressions of type I.

In fact, the expression on the left hand side of \eqref{080725-001} makes no explicit statement on temporal dependencies between events $B$ and $C$; but is does include an implicit temporal dependency between $B$ and $C$. 
This temporal dependency not so much affects the occurrence of (further) events, but the non-occurence of $A\pand{}(B\boolor{} C)$ if one of the events $B$ or $C$ occurs 
before $A$.
This implicite dependency is lost in the right hand side of \eqref{080725-001}.

%
%
\begin{table}
\small
 \centering
 \begin{tabular}{ccc} \toprule
 \vphantom{\Large Ö} & $ A\!\pand{}\!(B\!\boolor{}\!C)$ & $(A\!\pand{}\! B)\!\boolor{}\!(A\!\pand{}\!C)$
 \\ \midrule 
\vphantom{\Large Ö} $\boolnot A\boolnot B\boolnot C   $ & $\False $ & $\False $\\
\vphantom{\Large Ö} $\boolnot B\boolnot C \booland{}A  $ & $\False $ & $\False $\\
\vphantom{\Large Ö} $\boolnot A\boolnot C \booland{}B  $ & $\False $ & $\False $\\
\vphantom{\Large Ö} $\boolnot A\boolnot B \booland{}C  $ & $\False $ & $\False $\\
\vphantom{\Large Ö} $\boolnot C\booland{}(A\pand{} B) $ & $\True $ & $\True $\\
\vphantom{\Large Ö} $\boolnot C\booland{}(B\pand{} A) $ & $\False $ & $\False $\\
\vphantom{\Large Ö} $\boolnot C\booland{}(A\sand{} B) $ & $\False $ & $\False $\\
\vphantom{\Large Ö} $\boolnot B\booland{}(A\pand{} C)  $ & $\True $ & $\True $\\
\vphantom{\Large Ö} $\boolnot B\booland{}(C\pand{} A)  $ & $\False $ & $\False $\\
\vphantom{\Large Ö} $\boolnot B\booland{}(A\sand{} C)  $ & $\False $ & $\False $\\
\vphantom{\Large Ö} $\boolnot A\booland{}(B\pand{} C)  $ & $\False $ & $\False $\\
\vphantom{\Large Ö} $\boolnot A\booland{}(C\pand{} B)  $ & $\False $ & $\False $\\
\vphantom{\Large Ö} $\boolnot A\booland{}(B\sand{} C)  $ & $\False $ & $\False $
 \\ \bottomrule
\end{tabular}
 \begin{tabular}{ccc} \toprule
 \vphantom{\Large Ö} & $ A\!\pand{}\!(B\!\boolor{}\!C)$ & $(A\!\pand{}\! B)\!\boolor{}\!(A\!\pand{}\!C)$
 \\ \midrule 
\vphantom{\Large Ö} $A\pand{} B\pand{} C       $ & $\True $ & $\True $\\
\vphantom{\Large Ö} $B\pand{} A\pand{} C$ & $\text{\textbf{\emph{False}}}$ & $\text{\textbf{\emph{True }}}$\\
\vphantom{\Large Ö} $A\pand{} C\pand{} B       $ & $\True $ & $\True $\\
\vphantom{\Large Ö} $C\pand{} A\pand{} B $ & $\text{\textbf{\emph{False}}}$ & $\text{\textbf{\emph{True }}}$\\
\vphantom{\Large Ö} $B\pand{} C\pand{} A       $ & $\False $ & $\False $\\
\vphantom{\Large Ö} $C\pand{} B\pand{} A       $ & $\False $ & $\False $\\
\vphantom{\Large Ö} $A\pand{} (B\sand{} C)       $ & $\True $ & $\True $\\
\vphantom{\Large Ö} $B\pand{} (A\sand{} C)       $ & $\False $ & $\False $\\
\vphantom{\Large Ö} $C\pand{} (A\sand{} B)       $ & $\False $ & $\False $\\
\vphantom{\Large Ö} $(A\sand{} B)\pand{} C $ & $\text{\textbf{\emph{False}}}$ & $\text{\textbf{\emph{True }}}$\\
\vphantom{\Large Ö} $(A\sand{} C)\pand{} B $ & $\text{\textbf{\emph{False}}}$ & $\text{\textbf{\emph{True }}}$\\
\vphantom{\Large Ö} $(B\sand{} C)\pand{} A       $ & $\False $ & $\False $\\
\vphantom{\Large Ö} $A\sand{} B\sand{} C        $ & $\False $ & $\False $
 \\ \bottomrule
\end{tabular}
 \caption{Truth table for expressions $A\pand{}(B\boolor{} C)$ and $(A\pand{} B)\boolor{}(A\pand{} C)$. 
 Including SANDs there are $26$ sequences, which are divided into two groups of $13$ each.
 As both expressions do not yield same results for all sequences (see deviations in bold), both expressions are not equivalent.}
 \label{tab:WahrtabelleTypI}
\end{table} %
\begin{figure}
\centering
 \hfill
 \input{pics/sequ_ausfallbaum_A_pand_B_or_C}
 \hfill
 \input{pics/sequ_ausfallbaum_A_pand_B_or_A_pand_C}
 \hfill ~

 \caption{Left side:
 Sequential failure tree for expression $A \pand{} (B \boolor{} C)$.
 Right side: Sequential failure tree for expression $(A \pand{} B)\boolor{}(A\pand{C})$.
 On the right side there are additional sequences, as each of the two sub-expressions $(A \pand{} B)$ and $(A\pand{}C)$ does not make any statements about the occurrence of the missing third event.}
 \label{fig:sequ_ausfallbaum_A_pand_B_or_A_pand_C}
\end{figure}
This problem is solved by explicitely stating the temporal dependencies which are only implied by the left side of \eqref{080725-001}.

The relevant expressions $(B\boolor{} C)$ splits into five possible sequences:
\begin{align}
 (B\boolor{} C)& =
   (\minneg{C} \pand{} B) \boolor{}
   (\minneg{B} \pand{} C) \boolor{}
   (B\pand{} C) \boolor{}
   (C\pand{} B) \boolor{}
   (B\sand{} C)
  ~. \nonumber
\end{align}
Only three of these sequences are minimal failure sequences, see figure \ref{fig:sequ_ausfallbaum_A_pand_B_or_A_pand_C} (left side):
\begin{align}
 (B\boolor{} C)& =
   (\boolnot C \booland{} B) \boolor{}
   (\boolnot B \booland{} C) \boolor{}
   (B\sand{} C)
  ~. \label{080728-006}
\end{align}
Inserting this into \eqref{080722-311} yields for temporal expressions of type I, that
\begin{align}
 & A\pand{} (B\boolor{} C) \ist
 A\pand{}\bigl[ (\boolnot C \booland{} B) \boolor{} (\boolnot B \booland{} C ) \boolor{} ( B\sand{} C) \bigr] ~.
\end{align}
At this point non-minimal sequences need not be considered.
The OR connected terms in brackets are on the right hand side of the PAND operator, and thus occur ``later'';
all non-minimal terms then occur ``later still''.
They are covered by the minimal sequences.

Now, with all temporal dependencies explicitly stated, a distribution of the expression is possible, thus
\begin{align}
 & A\pand{} (B\boolor{} C) \ist \bigl[A\pand{} (\boolnot C \booland{} B) \bigr]\boolor{}\bigl[ A\pand{} (\boolnot B \booland{} C) \bigr]\boolor{}\bigl[ A\pand{} (B \sand{} C) \bigr].
\end{align}
Further transformation of this according to chapter \ref{080220-002} then leads to the distributive law for temporal expression of type I:
\begin{align}
 & A\pand{} (B\boolor{} C) \ist \bigl[\boolnot C \booland{} (A \pand{} B) \bigr] \boolor{} \bigl[\boolnot B \booland{} (A \pand{} C)\bigr]
 \boolor{} \bigl[A\pand{} (B \sand{} C)\bigr] \label{080727-005}~.
\end{align}
The distributive law for temporal expression of type I therefore requires explicit statements on the (non-)occurrence of all of the relevant events, and requires such statements in every sub-expression which is OR connected.
Statements with that property are called \emph{temporal minterms} in analogy to Boolean \emph{minterms}.

If the temporal laws of negation are applied, \eqref{080727-005} holds for the case of non-atomic events $A$, $B$, $C$, too.

Terms on the right side of \eqref{080727-005} are mutually exclusive (disjoint).
This simplifies later probabilistic quantification, see chapter \ref{chap080401-033}.
\paragraph{Simplification if Terms are Disjoint}~\\
The relationship in \eqref{080727-005} also holds for the special case of disjoint events $B$ and $C$, i.e.\ $B\perp C$.
But $B\perp C$ implies that each of the events $B$ or $C$ occurs only if the other event does not occur and has not yet occurred.
Then, \eqref{080727-005} may be simplified to 
\begin{align}
 & A\pand{} (B\boolor{} C) \ist{} \bigl[A \pand{} B\bigr] \boolor{} \bigl[A\pand{C}\bigr]  ~,  \label{090317001}
\end{align}
if $B\perp C$.
%
%
\subsubsection{Distributive Law for PAND-OR Expressions of Type II}
\begin{figure}[H]
\centering
 \input{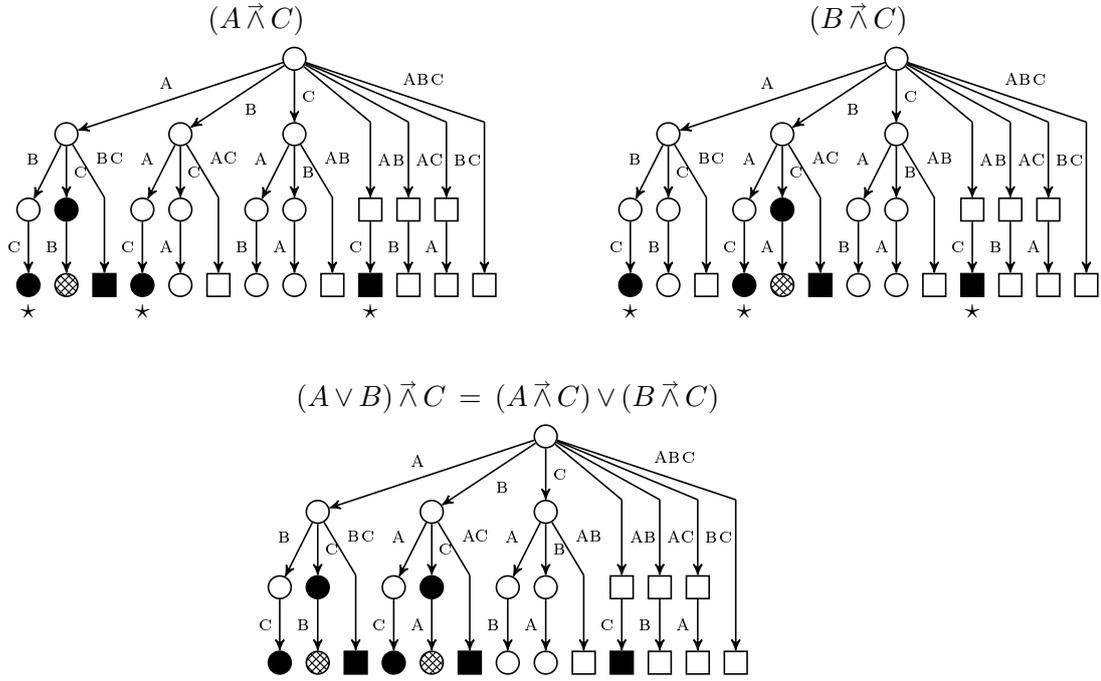}
 \caption{Distributive law for temporal expressions of type II:
 the sequential failure trees of $(A\boolor{} B)\pand{} C$, $(A\pand{} C)$, and $(B\pand{} C)$ show that $(A\pand{} C)$ and $(B\pand{} C)$ are minimal but not mutually exclusive (disjoint); joint sequences are marked with $\star$.}
 \label{080728-003}
\end{figure}

The logic statment of expression $(A\boolor{} B)\pand{} C$ is: ``the expression in brackets $(A\boolor{} B)$ must occur before $C$ occurs''.
This is equivalent to the logic statement ``$A$ must occur before $C$, or $B$ must occur before $C$'', as proven by the sequential failure trees in figure \ref{080728-003}, which correspond to the three expressions $(A\boolor{} B)\pand{} C$, $(A\pand{} C)$, and $(B\pand{} C)$.

Therefore, the distributive law for temporal expressions of type II is given as
\begin{align}
 (A\boolor{} B)\pand{} C & = (A\pand{} C)\boolor{}(B\pand{} C)~. \label{080728-002}
\end{align}
On the other hand, figure \ref{080728-003} also shows that $(A\pand{} C)$ and $(B\pand{} C)$ are not mutually exclusive.
The joint sequences, which are part of both expressions, are easily found by building the intersection, thus
\begin{align}
 (A\pand{} C)\booland{}(B\pand{} C) &\ist{} (A\booland{}B)\pand{}C\ist{} \bigl[A\pand{}B\pand{}C\bigr] \boolor{} \bigl[B\pand{}A\pand{}C\bigr] \boolor{} \bigl[(A\sand{}B)\pand{}C\bigr] ~.\nonumber
\end{align}
Figure \ref{080728-003} denotes these sequences with $\star$.
%
%
\subsubsection{Distributive Law for SAND-OR Expressions}
The logic statment of expression $A\sand{} (B\boolor{} C)$ is:
``$A$ must occur simultaneously with the expression in brackets $(B\boolor{} C)$''.
In analogy to the distributive law for temporal expressions of type I it is easily shown that this is \emph{not} equivalent to the logic statement ``$A$ occurs simultaneously with $B$, or $A$ occurs simultaneously with $C$'', as proven by figure \ref{090109016}.
In consequence, there is also no simple temporal distributive law for SAND-OR expressions.

Instead, the temporal distributive law for SAND-OR expressions looks similar to \eqref{080727-005} and is given as
\begin{align}
 & A\sand{} (B\boolor{} C) \ist \bigl[\boolnot C \booland{} (A \sand{} B) \bigr] \boolor{} \bigl[\boolnot B \booland{} (A \sand{} C)\bigr]
 \boolor{} \bigl[A\sand{} B \sand{} C\bigr] \label{090109015}~.
\end{align}
\begin{figure}[H]
\centering
 \hfill
 \input{pics/sequ_ausfallbaum_A_sand_B_or_C}
 \hfill
 \input{pics/sequ_ausfallbaum_A_sand_B_or_A_sand_C}
 \hfill ~

 \caption{Left side:
 Sequential failure tree for expression $A \sand{} (B \boolor{} C)$.
 Right side: Sequential failure tree for expression  $(A \sand{} B)\boolor{}(A\sand{C})$. 
 On the right side there are additional sequences, as each of the two sub-expressions $(A \sand{} B)$ and $(A\sand{}C)$ does not make any statements about the occurrence of the missing third event.}
 \label{090109016}
\end{figure}
\paragraph{Simplification if Terms are Disjunct}~\\
The relationship in \eqref{090109015} also holds for the special case of disjoint events $B$ and $C$, i.e.\ $B\perp C$.
But $B\perp C$ implies that each of the events $B$ or $C$ occurs only if the other event does not occur and has not yet occurred.
Then, \eqref{090109015} may be simplified to 
\begin{align}
 & A\sand{} (B\boolor{} C) \ist\bigl[A \sand{} B\bigr]\boolor{}\bigl[A\sand{C}\bigr]  \label{090317002}~.
\end{align}
if $B\perp C$.
\subsection{Temporal Laws of Absorption}\label{080902-100}
In analogy to the Boolean laws of absorption in \eqref{080129-007}, there are \emph{temporal laws of absorption}, as well.
Initially, it may seem that there are several temporal laws of absorption for different numbers of events involved; this intuition come mainly from the permutations that need to be taken into account when analysing event sequences.
On the other hand, it can be shown that the temporal laws of absorption really are specializations of the Boolean laws of absorption in \eqref{080129-007}:

Starting with the most simple case with only two events involved, the temporal laws of absorption may be derived from \eqref{080129-007} by using the law of completion in \eqref{071217-017}; this yields 
\begin{align}
 A \boolor{} (A\booland{}B) &\ist A \ist A \boolor{} \bigl[(A \pand{} B) \boolor{} (A \sand{} B) \boolor{} (B \pand{} A) \bigr] ~,
\end{align}
which may then be further transformed into
\begin{align}
 & A \boolor{} (A \pand{} B) \ist A \label{080901-013} ~,\\
 & A \boolor{} (B \pand{} A) \ist A \label{080901-014} ~, \\
 & A \boolor{} (A \sand{} B) \ist A \label{080901-101} ~.
\end{align}
The more ``general'' event $A$ absorbs the more ``concrete'' event, if the latter is a subset of $A$; this is the same for Boolean and temporal logic.
In general, if $\ES$ is an (extended) event sequence, then
\begin{align}
 & A \boolor{} \ES \ist A &&\text{, if } A\inplus \ES \text{, i.e.\ } \ES \subseteq A \label{090124011} ~.
\end{align}
This relation also holds for non-atomic events $A$.
Other than in the Boolean logic, with more complex temporal expressions it is increasingly difficult to spot subsets.
There are two major reasons for that:
the PAND operator has no law of commutativity;
and the invention of core events allows for nested events.

For instance, temporal law of absorption for three events are given as
\begin{align}
 & (A \pand{} B) \boolor{} (A \pand{} B \pand{} C )\ist A \pand{} B\label{080901-617} ~,\\
 & (A \pand{} B) \boolor{} (A \pand{} C \pand{} B )\ist A \pand{} B \label{080901-019} ~,\\
 & (A \pand{} B) \boolor{} (C \pand{} A \pand{} B )\ist A \pand{} B\label{080901-018} ~,\\
 & (A \pand{} B) \boolor{} ((A \sand{} C ) \pand{} B)\ist A \pand{} B\label{080901-102} ,\\
 & (A \pand{} B) \boolor{} ((A \pand{} B ) \sand{} C)\ist (A \pand{} B) \boolor{} (A\pand{}(B\sand{}C)) \ist{} A \pand{} B\label{080901-103} ~.
\end{align}
Indeed, \eqref{080901-617} to \eqref{080901-103} are simple reformulations of 
\begin{align}
 & (A \pand{} B) \boolor{} ((A \pand{} B) \booland{} C) \ist{} (A \pand{} B) ~,
\end{align}
as demonstrated by the following transformation:
\begin{align}
 &(A \pand{} B) \booland{} C \ist{}\bigl[(A \pand{} B) \pand{} C\bigr]
   \boolor{} \bigl[(A \pand{} B ) \sand{} C \bigr] \boolor{} \bigl[C \pand{}(A\pand{} B )\bigr] \ist{}\\
 &\qquad\ist{} \bigl[A \pand{} B \pand{} C\bigr]\boolor{} \bigl[A\pand{}(B\sand{}C)\bigr]
   \boolor{} \bigl[A \pand{} C \pand{} B\bigr] \boolor{} \bigl[C \pand{} A \pand{} B\bigr] \boolor{} \bigl[(A \sand{} C) \pand{} B\bigr] ~. \nonumber
\end{align}
Taking this concept one step further, the general temporal laws of absorption may then be given in complete analogy to its Boolean counterpart as
\begin{align}
 & \ES_i \boolor{} \ES_j \ist ES_i \label{090107418} \qquad\text{for }\ES_j \subseteq \ES_i ~.
\end{align}

The same holds true for the second Boolean law of absorption from \eqref{080129-007}; its temporal version reads as
\begin{align}
 & A \pand{} (A \boolor{} B) \ist{} \bigl[\boolnot B \booland{} (A\pand{} A)\bigr] \boolor{}
   \bigl[\boolnot A \booland{}(A\pand{} B) \bigr] \boolor{}
   \bigl[A\pand{} (A\sand{}B) \bigr] \ist{} \False ~, \label{080903_001}\\
 & (A \boolor{} B) \pand{} A \ist{} (A\pand{}A) \boolor{} (B\pand{}A) \ist{} B\pand{}A ~, \label{080903_002}\\
\begin{split}
& A \sand{} (A \boolor{} B) \ist{} \bigl[\boolnot B \booland{}(A\sand{} A)\bigr] \boolor{}
   \bigl[\boolnot A \booland{}(A\sand{} B) \bigr] \boolor{}
   \bigl[A\sand{} (A\sand{}B) \bigr] \ist{} \\
&\hphantom{A \sand{} (A \boolor{} B) }\ist{}
   (\boolnot B \booland{} A) \boolor{} (A\sand{}B) ~.
\end{split}\label{080903_003}
\end{align}
Allthough initially not very intuitive, these results are correct, as demonstrated by the following transformation:
On the one hand, 
\begin{align}
 \bigl[A \pand{} (A \boolor{} B)\bigr] \boolor{} \bigl[ (A \boolor{} B) \pand{} A\bigr] \boolor{} \bigl[A \sand{} (A \boolor{} B)\bigr] &\ist{}
   A\booland{}(A\boolor{}B) \ist{} A \nonumber~ .
\end{align}
And on the other hand, \eqref{080903_001} to \eqref{080903_003} yield
\begin{align}
 &\bigl[A \pand{} (A \boolor{} B)\bigr] \boolor{} \bigl[ (A \boolor{} B) \pand{} A\bigr] \boolor{} \bigl[A \sand{} (A \boolor{} B)\bigr] \ist{}
   (B\pand{}A) \boolor{} (\boolnot B \booland{} A) \boolor{} (A\sand{}B) ~.\nonumber
\end{align}
Furthermore, $\boolnot B \booland{} A$ covers the non-minimal sequence $A \pand{} B$, thus providing
\begin{align}
 &\bigl[A \pand{} (A \boolor{} B)\bigr] \boolor{} \bigl[ (A \boolor{} B) \pand{} A\bigr] \boolor{} \bigl[A \sand{} (A \boolor{} B)\bigr] \ist{}
   (B\pand{}A) \boolor{} (\boolnot B \booland{} A) \boolor{} (A\sand{}B) \ist{}\nonumber\\
 &\qquad \ist{} (\boolnot B \booland{} A) \boolor{} (A\pand{}B) \boolor{} (A\sand{}B) \boolor{} (B\pand{}A) \ist{} (\boolnot B \booland{} A) \boolor{} (A\booland{}B) \ist{} A~. \nonumber
\end{align}
These transformations illustrate that \eqref{080903_001} to \eqref{080903_003} really are only specializations of the Boolean laws of absorption.
\subsection{Temporal Law for Intersections}\label{090107010}
The introduction of PAND and SAND operators into the temporal {TFTA} logic leads to expressions like $A\booland{}(A\pand{}B)$, $B\booland{}(A\pand{}B)$, or $A\booland{}(A\sand{}B)$.
Such expressions are not easily covered by the temporal laws of absorption, as in their case, and other than in case of the laws of absorption, see above, the more ``general'' expression does not absorb the more ``concrete'' expression.
Therefore, a new \emph{temporal law for intersections} is proposed.

The temporal law for intersections describes conjunctions of two expression, one of which is an intersection of the other.
In the Boolean case, this can be solved by applying the laws of associativity and idempotency:
\begin{align}
 A \booland{} (A\booland{}B) &\ist{} A\booland{}A\booland{}B \ist{} A\booland{}B \label{090107011} ~.
\end{align}
In the temporal case, three different settings have to be considered:
\begin{align}
 A \booland{} (A\pand{}B) &\ist{} (A\pand{}B) \booland{} A \ist{} A\pand{}B \label{090107012} ~,\\
 B \booland{} (A\pand{}B) &\ist{} (A\pand{}B) \booland{} B \ist{} A\pand{}B \label{090107013} ~,\\
 A \booland{} (A\sand{}B) &\ist{} (A\sand{}B) \booland{} A \ist{} (B\sand{}A) \booland{} A \ist{} A \booland{} (B\sand{}A) \ist{} A\sand{}B \label{090107014} ~.
\end{align}
Correctness may be easily demonstrated using the temporal logic laws provided above.
For instance, 
\begin{align}
 A \booland{} (A\pand{}B) &\ist{} \bigl[A \pand{} (A\pand{}B)\bigr] \boolor{} \bigl[A \sand{} (A\pand{}B)\bigr] \boolor{} \bigl[(A\pand{}B) \pand{} A\bigr] \ist \nonumber\\
 &\ist{} (A \booland{} A)\pand{}B \boolor{} \False \boolor{} \False \ist{} A\pand{}B \nonumber ~.
\end{align}
The same holds true for more general cases with more complex expressions, as in
\begin{align}
 X_i \booland{} \ldots \booland{} X_j \booland{} (\ldots \pand{} X_i\pand{} \ldots) &\ist{}
 X_j \booland{} (\ldots \pand{} X_i\pand{} \ldots) \label{090107015} ~\text{and}\\
 X_i \booland{} \ldots \booland{} X_j \booland{} (\ldots \sand{} X_i\sand{} \ldots) &\ist{}
 X_j \booland{} (\ldots \sand{} X_i\sand{} \ldots) \label{090107016} ~,
\end{align}
as well as for expressions that include intersections with non-atomic core events, i.e.\
\begin{align}
 X_i \booland{} \ldots \booland{} X_j \booland{} (\ldots \pand{} ( X_i \sand{} \ldots) \pand{} \ldots) &\ist{}
 (\ldots \pand{} X_j \booland{} ( X_i \sand{} \ldots) \pand{} \ldots) \label{090107017} ~.
\end{align}
In general, the temporal law for intersections is therefore given as:
\begin{align}
 & \ES_i \booland{} \ES_j \ist ES_j \label{090107318} \qquad\text{for }\ES_j \subseteq \ES_i ~.
\end{align}

%
%
%
%
%
\section{Minimal and Disjoint Forms of TFTA Temporal Expressions}\label{chap080809-001}
%
%
%
\subsection{Minimal and Disjoint Forms of Boolean Expressions}\label{080901_003}
This chapter discusses two properties that {TFTA} temporal expressions may have.
Temporal expressions which are minimal or mutually exclusive (disjoint) have special meaning and importance within the TFTA's temporal logic; in this they are similar to the Boolean {FTA}.
In both cases, the Boolean as well as the temporal, any logic expression can be transformed into ``sum of product'' forms, i.e.\ DNF or TDNF, respecively, by using the laws of transformation given in chapter \ref{chap080401-031}.

In general, these cutsets (Boolean case) or event sequences (temporal logic) still include redundant information.
Therefore, further transformation into a minimal sum of products form, i.e.\ minimal cutsets and {MCSS}, respectively, is necessary and provides an even more useful representation of the (temporal) failure function. 

For further probabilistic calculation it is then helpful to transform this minimal form into a minterm form, where all minterms are mutually exclusive (disjoint), see chapter \ref{081018_002}.
%
%
%
\paragraph{Disjunctive Normal Form (Sum of Products)}~\\
Boolean expressions $\varphi$ are transformed into a DNF by applying the laws of Boolean algebra; in DNF
\begin{align}
  \varphi \ist{}{} \bigvee\limits_{\mathclap{j \ist{} 1}}^{\zeta} S_j \ist{}{} \bigvee\limits_{\mathclap{j\ist{}{} 1}}^{\zeta} \Bigl( \bigwedge\limits_{i\ist{}{} 1}^{n_j} X_{j, i} \Bigr)    ~,\label{080904_005}
\end{align}
where $\zeta$ denotes the number of \emph{cutsets} $S$ of $\varphi$, which are not necessarily already minimal, and $n_j$ denotes the number of events $X$ which constitute $S_{j}$.
%
%
\paragraph{Minimal DNF}~\\
In a next step, the cutsets $S$ of Boolean expressions $\varphi$ are minimal, if none of the cutsets ``includes'' another.
If so, they are called \emph{minimal cutsets} and are denoted with $MS$
for better discrimination.
Using the laws of Boolean algebra from chapter \ref{080129-001}, (monotone) Boolean expressions as in \eqref{080904_005} can be transformed into a minimal form, where
\begin{align}
  \varphi \ist{}{} \bigvee\limits_{\mathclap{j\ist{}{} 1}}^{\xi } \MS_j \ist{}{} \bigvee\limits_{\mathclap{j\ist{}{} 1}}^{\xi } \Bigl( \bigwedge\limits_{i\ist{}{} 1}^{n_j} X_{j, i} \Bigr)    ~,\label{080904_009}
\end{align}
where $\xi\leq\zeta$.{}

Each of these $\xi$ minimal cutsets $\MS_j$ and $\MS_{j'}$ with $j, j'\in\{1, 2, \ldots, \xi \}$ and $j'\neq j$ are pairwise mutually exclusive:
\begin{align}
  & \MS_j \booland{} \MS_{j'} \neq \MS_j \qquad \text{und} \qquad \MS_j \booland{} \MS_{j'} \neq \MS_{j'}~.\label{080904_008}
\end{align}
\paragraph{Simplifying Quantification By Using Disjoint Terms}~\\\label{080904_007}
In many cases it is helpful to transform logic functions into a equivalent form which is specifically well suited for a certain task.
For conventional fault trees the minimal cutset form of a system's failure function according to \eqref{080904_009} is, for example, especially illustrative and well suited for qualitative analyses; on the other hand, the form below is equivalent but much less easy to understand:
\begin{align}
  \varphi &\ist{}{} \bigvee_{\mathclap{j\ist{}{}1}}^{\xi} \Bigl( \MS_j \cdot \bigwedge_{\mathclap{i\ist{}{}1}}^{j-1} \boolnot \left(\MS_i\right)\Bigr)~. \label{080904_010}
\end{align}
This form aids probabilistic analyses because of its mutually exclusive (disjoint) OR connected terms; see chapter \ref{chap080401-033} for details. 

In general, two Boolean expressions $\varphi_1$ and $\varphi_2$ are mutually exclusive (disjoint), if their conjunction yields $\False$:
\begin{align}
  \varphi_1 \booland{} \varphi_2 &\ist{} \False && \Longleftrightarrow && \varphi_1 \perp \varphi_2 ~. \label{090109001}
\end{align}
\subsection{Minimal Temporal Expressions} \label{081018_001}
Minimalism of temporal logic expressions parallels the Boolean case.
Temporal logic expressions are minimal, if they ``do not include each other''.
In the temporal logic special care is necessary, though, because of three differences compared to the Boolean case: first, their are other and additional logic operators; second, negated events have special meaning; third, properties of commutativity and associativity are restricted. 
Moreover, temporal expressions can be structurally non-minimal as well as temporally non-minimal, see chapters \ref{080904_551} and \ref{080909_005}, respectively.
First some groundwork has to be laid, though.
%
%
%
\paragraph{Minimal Temporal Failure Function}~\\
Using the temporal transformation laws from above, temporal expressions $\varpi$ may be transformed into a TDNF, which is similar to the Boolean DNF.
For readability, \eqref{080828_001} is repeated here:
\begin{equation}
  \varpi \ist{}{} \bigvee\limits_{\mathclap{j=1}}^{\zeta} \ES_j \ist{}{} \ES_1 \vee{} \ES_2 \vee{} \ldots \vee{} \ES_{\zeta}~.
  \label{081019_001}
\end{equation}
$\zeta$ denotes the number of event sequences $\ES$ in $\varpi$, which need not to be minimal at this stage.

Then, the corresponding minimal form consists of $\xi$ \emph{minimal cutset sequences} (MCSS), which are OR connected:
\begin{equation}
  \varpi \ist{}{} \bigvee\limits_{\mathclap{j=1}}^{\xi} \MCSS_j \ist{}{} \MCSS_1 \vee{} \MCSS_2 \vee{} \ldots \vee{} \MCSS_{\xi} ~, with \xi\leq\zeta ~.
  \label{080901-007}
\end{equation}
%
\paragraph{Condition of Minimality}~\\
In the temporal logic ``minimal'' also means, that none of the $\MCSS_j$ ``covers'' or ``includes'' any other $\MCSS_{j'}$ (where $j, j'\in\{1, 2, \ldots, {\xi}\}$ and $j'\neq j$).

The sections below show that the criterion for temporal expressions being minimal is very similar to the Boolean criterion in \eqref{080904_008}.

Event sequences are minimal, if all pairs of $\MCSS_j$ and $\MCSS_{j'}$ with $j, j'\in\{1, 2, \ldots, \xi\}$ and $j'\neq j$ follow
\begin{align}
 & \MCSS_{j'} \nsubseteq \MCSS_j &&\Longleftrightarrow &&\MCSS_j \booland{} \MCSS_{j'} \neq \MCSS_{j'}  && \text{and} \label{090124010}\\
 & \mathrlap{\MCSS_{j}}\phantom{\MCSS_{j'}} \nsubseteq \MCSS_{j'}  &&\Longleftrightarrow &&\MCSS_j \booland{} \MCSS_{j'} \neq \MCSS_{j} ~. \label{080906_038}
\end{align}
For this relation a new operator is introduced:
\begin{align}
 & \MCSS_{j} \isMinimal{} \MCSS_{j'}\label{090215005}
\end{align}
implies that $\MCSS_j$ and $\MCSS_{j'}$ are minimal.

One difference to the Boolean case is that writing temporal expressions in their {TDNF} form usually requires the use of negated events; this comes from the temporal distributive laws, see chapter \ref{080723-015}, and requires a discussion on minimal temporal expressions with negated events.
%
\subsubsection{Structurally Non-Minimal Temporal Expressions}\label{080904_551}
Temporal expressions are \emph{structurally non-minimal}, if one of them is a special case of the other expression.
Structurally non-minimal expressions may be transformed into a minimal form by applying the temporal laws of absorption (chapter \ref{080902-100}) and the temporal law for intersections (chapter \ref{090107010}). 
%
\subsubsection{Temporally Non-Minimal Temporal Expressions}\label{080909_005}
Beyond the structural aspect of non-minimality there is the question of minimality in temporal expressions like
\begin{align}
  &(\boolnot B \booland{} A ) \boolor{} ( A\pand{} B ) ~.\label{080909_011}
\end{align}
Checking for minimality according to \eqref{080906_038} shows that these two terms are not minimal.

From
\begin{align}
  (\boolnot B \booland{} A ) \booland{} ( A\pand{} B )
\end{align}
follows with \eqref{09010932}, that
\begin{align}
   (\boolnot B \booland{}& A ) \booland{} ( A\pand{} B ) \ist{} \nonumber\\
      &\ist{}\bigl[\boolnot B \booland{} (A \booland{}( A\pand{} B ) )\bigr] \boolor{}
      \bigl[  ( A\pand{} B ) \pand{} ( A\pand{} B ) \bigr] \boolor{}
      \bigl[ A \pand{} ( B \sand{} ( A\pand{} B ) ) \bigr] ~.
\end{align}
The first sub-expression on the right side is then reduced by applying \eqref{09010930}, which yields
\begin{align}
   \boolnot B \booland{} (A \booland{}( A\pand{} B ) ) &\ist{}
      \boolnot B \booland{} ( A\pand{} B  ) \ist{} \False~.
\end{align}
The second sub-expression is then also reduced to $\False$ by applying the temporal law of contradiction, see \eqref{071217-011}.
Then, the remaining 
\begin{align}
   &(\boolnot B \booland{} A ) \booland{} ( A\pand{} B ) \ist{}
      A \pand{} ( B \sand{} ( A\pand{} B ) )  \ist{} A \pand{} ( A\pand{} B  )  \ist{} A\pand{} B \\
   &\underbrace{\boolnot B \booland{} A}_{
      \text{\begin{minipage}{35.2mm}\input{pics/minimaldisjunkt_sequ_afb_disjunkt_notB_pand_A}
                \end{minipage}}
  } \booland{}
  \underbrace{A\pand{} B}_{
      \text{\begin{minipage}{35.2mm}\input{pics/minimaldisjunkt_sequ_afb_disjunkt_A_pand_B}
                \end{minipage}}
  }\ist{}
  \underbrace{A\pand{} B}_{
      \text{\begin{minipage}{35.2mm}\input{pics/minimaldisjunkt_sequ_afb_disjunkt_A_pand_B}
                \end{minipage}}
  } \nonumber
\end{align}
does not satisfy the minimality condition from \eqref{080906_038}.
Therefore, \eqref{080909_011} is not minimal, which is also shown by the sequential failure trees, as the sub-expression $A\pand{} B$  consists only of such expressions that are non-minimal with regard to $\boolnot B \pand{} A$.
Thus, the minimal form ist given as $\boolnot B \pand{} A$, which ``covers'' the second term $A\pand{} B$.
%
%
\paragraph{Generalization}~\\
The example from above may be generalized with the laws of transformation for negated events from chapter \ref{090123004}.
From \eqref{090123001} follows $\boolnot X \booland{} \ES$ with $X\centernot{\inplus} \ES$ is temporally minimal to all temporal expressions with $\ES$ occuring before $X$, i.e.\ $\ES\pand{}X$.

As $(\boolnot X \booland{} \ES ) \boolor{} (\ES\pand{}X) $ with $X\centernot{\inplus} \ES$ is non-minimal because of the temporal sequence of the events, this effect is called \emph{temporal non-minimality}.
%
%
\paragraph{Two More Examples}~\\
$(\boolnot B \booland{} A ) \boolor{} ( C\pand{} A ) $
is already given in minimal form, as \eqref{09010932} and \eqref{080906_038} hold:
\begin{align}
  &\underbrace{\boolnot B\booland{} A}_{
      \text{\begin{minipage}{35.2mm}\input{pics/minimaldisjunkt_sequ_afb_disjunkt_notB_pand_A}
                \end{minipage}}
  }\booland{}
  \underbrace{C\pand{} A}_{
      \text{\begin{minipage}{35.2mm}\input{pics/minimaldisjunkt_sequ_afb_disjunkt_C_pand_A}
                \end{minipage}}
  } \ist{}
  \underbrace{\boolnot B \booland{} (C\pand{}  A) }_{
      \text{\begin{minipage}{35.2mm}\input{pics/minimaldisjunkt_sequ_afb_disjunkt_notB_and_C_pand_A}
                \end{minipage}}
  } \label{080910_003}~.
\end{align}
The sequential failure trees prove that each of the expressions includes failure nodes, which are unique to this expression and not part of the other. 

However, $\boolnot B \booland{} A$ is the minimal form of all such event sequences that include $A$ but not $B\pand{} A$, i.e.\ (without SAND)
$\boolnot B \booland{} (A \pand{} C)$,
$\boolnot C \booland{} (A \pand{} B)$,
$ A \pand{} B \pand{} C$,
$ A \pand{} C \pand{} B$, and
$ C \pand{} A \pand{} B$.
Exemplarily, this is shown with one of these expressions:
\begin{align}\begin{split}
  (\boolnot B \booland{} A ) \booland{}  ( A \pand{} C \pand{} B) &\ist{} \hphantom{\boolor{}}
    \bigl[\boolnot B (\booland{} A \booland{} ( A \pand{} C \pand{} B) ) \bigr]\boolor{} \\
    &\hphantom{\ist{}}\boolor{}\bigl[A \pand{} B \pand{} (( A \pand{} C \pand{} B) \pand{} A ) \bigr]\boolor{}\\
    &\hphantom{\ist{}}\boolor{}\bigl[A \pand{} (B \sand{} ( A \pand{} C \pand{} B) ) \bigr] \ist{}\\
    & \ist{} \hphantom{\boolor{}} \False \boolor{} \False \boolor{} A \pand{} (A \pand{} C \pand{} B) \ist{} A \pand{} C \pand{} B ~.
\end{split}\end{align}
As the sequential failure trees show, \eqref{080906_038} is not complied with; and $(\boolnot B \pand{} A ) \boolor{} ( A \pand{} C \pand{} B)$ is, thus, non-minimal. 
\subsection{Disjoint Temporal Expressions} \label{081018_002}
Minimal temporal expressions are not necessarily also mutually exclusive (disjoint).
For example, the failure function $\varpi\ist{}(\boolnot B \booland{} A ) \boolor{} ( C\pand{} A ) $ is given in minimal form.
But the two event sequences $\boolnot B \booland{} A $ and $C\pand{} A$ are not mutually exclusive; instead, $\boolnot B \booland{} (C\pand{}  A) $ is an intersection, see \eqref{080910_003}.

The sections below discuss mutually exclusive temporal expressions and a method for transforming them into mutually exclusive temporal expressions.
%
%
%
\subsubsection{Condition for Disjointness}\label{090122001}
In analogy to chapter \ref{080901_003}, two temporal expressions are mutually exclusive (disjoint), if their conjunction (AND connection) yields $\False$, i.e.\ if there is no intersection between them.
When illustrated by sequential failure trees, disjoint temporal expressions do not have any failure nodes in common.
In the following example, a temporal expression has three disjoint sub-expressions: 
\begin{align}
&
  \underbrace{(A\pand{} B\pand{} C)}_{
      \text{\begin{minipage}{35.2mm}\input{pics/minimaldisjunkt_sequ_afb_disjunkt_A_pand_B_pand_C}
                \end{minipage}}
  } \boolor
  \underbrace{(B \pand{} A\pand{} C)}_{
      \text{\begin{minipage}{35.2mm}\input{pics/minimaldisjunkt_sequ_afb_disjunkt_B_pand_A_pand_C}
                \end{minipage}}
  } \boolor
  \underbrace{(C \pand{} A)}_{
      \text{\begin{minipage}{35.2mm}\input{pics/minimaldisjunkt_sequ_afb_disjunkt_C_pand_A}
                \end{minipage}}
  } \nonumber~.
\end{align}
Thereby,
\begin{align}
 &( A\pand{} B\pand{} C ) \booland{} (B\pand{} A\pand{} C ) \ist{} \False~, \nonumber\\
 &\mathrlap{( A\pand{} B\pand{} C ) \booland{} (C\pand{} A) }
    \hphantom{( A\pand{} B\pand{} C ) \booland{} (B\pand{} A\pand{} C ) }\ist{} \False~, \nonumber\\
 &\mathrlap{(B\pand{} A\pand{} C )  \booland{} (C\pand{} A)}
     \hphantom{( A\pand{} B\pand{} C ) \booland{} (B\pand{} A\pand{} C ) }\ist{} \False~. \nonumber
\end{align}
On the other hand, there are intersections in the following example:
\begin{align} \underbrace{(A\pand{} B)}_{
      \text{\begin{minipage}{35.2mm}\input{pics/minimaldisjunkt_sequ_afb_disjunkt_A_pand_B}
                \end{minipage}}
  } \booland{}
  \underbrace{(A\pand{} C)}_{
      \text{\begin{minipage}{35.2mm}\input{pics/minimaldisjunkt_sequ_afb_disjunkt_A_pand_C}
                \end{minipage}}
  } &\ist{}
  \underbrace{\vphantom{(A\pand{}C)}\ldots }_{
      \text{\begin{minipage}{35.2mm}\input{pics/minimaldisjunkt_sequ_afb_disjunkt_A_pand_B_and_A_pand_C}
                \end{minipage}}
  } \nist \False ~, \nonumber
\end{align}
as
\begin{align}
    (A\pand{} B) \booland{} (A\pand{} C) &\ist{}
      \bigl[( A\pand{}B ) \pand{} (A\pand{}C )\bigr] \boolor{}
      \bigl[ ( A\pand{}B ) \sand{} (A\pand{}C )\bigr] \boolor{}
      \bigl[ ( A\pand{}C) \pand{} (A\pand{}B)\bigr]\ist{} \nonumber\\
    &\ist{}\bigl[A\pand{}B\pand{}C\bigr] \boolor{} \bigl[A\pand{}(B\sand{}C)\bigr]  \boolor{} \bigl[A\pand{}C\pand{}B\bigr] \nonumber ~.
\end{align}
%
%
\subsubsection{Structurally and Temporally Disjoint Temporal Expressions}\label{090125051}
In the TFTA's temporal logic there are two types of disjointness:
\begin{enumerate}
 \item An event can not be $\True$ and $\False$ at the same time.
 Therefore and in analogy to the Boolean logic, two expressions are disjoint, if one of them includes a non-negated event and the other expression includes the negation of the same event.
 For instance, $\boolnot A\booland{}B$ and $A\pand{} B$ are mutually exclusive (disjoint).
 In general, this type of disjointness is expressed in \eqref{09010930} and \eqref{09010931}.
  \item Other than Boolean expressions, temporal expressions can be mutually exclusive because of the possibility of temporal contradictions.
  Following from the temporal laws of completion and the temporal law of contradiction (see chapter \ref{080129-022} and \ref{080822-001}, respectively), two temporal expressions are disjoint, if the same events are included in both, but in different sequences.
  Therefore, $B\pand{}A$ und $A\pand{} B$ are, e.g., disjoint without any negated events.
\end{enumerate}
In both cases the lack of any intersections indicates that the expressions are mutually exclusive.
Therefore, the condition for disjointness from chapter \ref{090122001} is applicable for temporal as well as Boolean expressions, see \eqref{090109001}.
And in consequence, temporal and Boolean expressions do not differ significantly regarding being mutually exclusive.
%
%
\subsubsection{Disjoint Separation Using Temporal Minterms}\label{080914_001}
\emph{Temporal minterms} are event sequences, which consists of all $u$ parameters of a temporal logic function of size $u$, and each parameter is included exactly once. 

Temporal minterms are used in order to split a temporal expression into disjoint event sequences.
In this form they are especially well suited for later probabilistic quantification.
See chapter \ref{080901_003} for further background.

These expressions may be deduced using a method which is similar to Shannon's segmentation for Boolean expressions:
\begin{enumerate}\label{090122002}
 \item The relevant temporal function $\varpi$ with $u$ different parameters has to be given as TDNF.
 If not, $\varpi$ is transformed into a TDNF using the temporal logic laws from above.
 \item The first event sequence is chosen: $\ES\ist{}\ES_1$.{}
 \item If $\ES$ consists of all $u$ parameters, goto step seven.
 \item Choose the first parameter $X$ which is missing in $\ES$. 
 \item $\ES$ is then transformed into its disjoint form by using
      \begin{align}
          \ES & ~~\Longrightarrow~~ \ES\booland{}(\boolnot X \boolor{} X)\ist{}(\boolnot X \booland{} \ES ) \boolor{} ( X \booland{} \ES )~\label{080916_001}
      \end{align}
 \item Repeat step five for each of the other parameters that are missing in $\ES$.{}
 \item If the chosen $\ES$ is not the last event sequence in $\varpi$, choose the next event sequence $\ES$ and goto step three.
 \item Check whether the resulting expressions are minimal by applying the transformation laws of the temporal logic and specifically the temporal laws of absorption.
\end{enumerate}
This method and workflow are shown on two examples in appendix \ref{090111010}, see page \pageref{090111010}.
\section[Simplification Using Extended Event Sequences]{Simplification Using Extended Event Sequences and Extended {TDNF} and Extended {MCSS} }\label{080817-010}
Chapter \ref{090125001} discussed ``normal'' temporal expressions and the temporal logic, which allows to transform temporal expressions $\varpi$ into their -- possibly minimal and mutually exclusive (disjoint) -- TDNF.
The TDNF describes all the event sequences that lead to the occurrence of the TOP event; it is well suited for further qualitative cutset analyses, and it provides the basis for probabilistic quantification of the failure function.
%
%
\subsection{Motivation and Requirements}\label{090125002}
Allthough both of the TFTA's goals from chapter \ref{chap080401-021} are met with these ``normal'' temporal expressions, their practical useability is limited because of the high number of resulting event sequences.
For instance, the relatively simple temporal expression $A\pand{}(B\booland{}C)\pand{}(D\booland{E})$ already provides $32$ different temporal minterms (chapter \ref{080813-003}) -- and that is without even taking SANDs into account.
This combinatorial blow-up of the number of event sequences mainly stems from applying the temporal law of completion (see chapter \ref{080129-022}).

On the one hand, transformations according to the temporal logic are necessary for transforming complex expressions into manageable ones.
On the other hand, clarity and readability of the results depend very much on the (low) number of such sub-expressions.

It is, therefore, sensible to simplify a complex temporal expression only so far, as to obtain useable, and especially minimal, sub-expressions, while at the same time keep the number of such sub-expressions as small as possible.

Thus, there are certain requirements on such a simplified temporal form:
\begin{enumerate}
 \item The simplified form shall also allow qualitative as well as probabilistic analyses.
 \item The simplified form shall also be able to provide temporal expressions in a normal form.
 \item Each of the event sequences of this normal form shall be minimal.
 \item Each of the event sequences of this normal form shall be directly quantifiable.
 \item For probabilistic quantification, the event sequences shall be mutually exclusive.
\end{enumerate}
The extended {TDNF}, as introduced in chapter \ref{080817-002}, is one possibility to meet this requirements.

In extended {TDNF} temporal expressions consist of normal (atomic and non-atomic) core events as well as \emph{extended core events}, such as
\begin{align}
  \mathit{eK} & \ist{} X_1 \booland{} X_2 \booland{} \ldots ~~.
\end{align}
Event sequences with extended core events are called \emph{extended event sequences}, see the grammar of temporal logics in chapter \ref{090321009}.

Using this form is useful, if all sequences of specific events contribute equally to the TOP event.
The extended form combines these ``real'' events and reduces modelling effort, and allows concise presentation of temporal expressions.

Without the extended form, temporal expressions are transformed in order to generate their TDNF consisting of event sequences only, which themselves consist of core events.
Each core event stands for events which occur at a specific, though relative, point in time.
An expression $A\pand{}(B\sand{}C)$, for example, indicates, that an atomic core event $A$ occurred before later both events $B$ and $C$ happened simultaneously.
The event sequences indicates clearly, which event occurs when.

Now, with the extended form, temporal expressions are transformed in order to generate their extended {TDNF}.
The latter includes both, normal event sequences, consisting of normal core events, and extended event sequences, consisting of normal and extended core events.

Extended core events indicate, that at a given point in time certain events \emph{have happend}.
An expression $A\pand{}(B\booland{}C)$, for example, indicates, that an atomic event core event $A$ has occurred before later events $B$ and $C$ have occurred.
No statement is made on the real times at which the events $B$ and $C$ occurred that form the extended core event.
The extended form neither defines nor restricts the sequence between $B$ and $C$; it solely describes a ``latest possible'' time for occurrence.

Extended event sequences may contain more than one extended event sequence, as e.g.\ in $(A\booland{} B)\pand{}(C\booland{}D)$.
If events are included within the same extended event sequence more than once, then they need further transformation/simplification.

On the other hand, it disagrees with the extended {TDNF} to combine several (extended) event sequences with an AND.
Instead, further transformation/simplification is necessary first.
For example, only the simplification of $(A\pand{}B)\booland{}C$ according to the laws of temporal logic provides a correct extended {TDNF}:
\begin{align}
 (A\pand{}B)\booland{}C \ist{}& \bigl[  (A\booland{}C) \pand{} B \bigr] \boolor{}
                                                            \bigl[A\pand{}(B\sand{}C)\bigr] \boolor{}
                                                            \bigl[A\pand{}B\pand{}C\bigr] ~.
\end{align}
%
%
\subsection{Using Extended Temporal Expressions}\label{090124020}
The decision for using the extended form is taken during qualitative transformation of the temporal failure function:
\begin{itemize}
 \item The Boolean distributive law gets priority over the temporal law of completion.  
 \item AND connections are not broken up, if the AND connected events 
  \begin{itemize}
    \item are event sequences without negated events and
    \item are pairwise coprime as well as coprime to the rest of the (extended) event sequence which is currently looked at.
  \end{itemize}
\end{itemize}
In general, the temporal logic rules from chapter \ref{chap080401-031} and \ref{chap080809-001} apply to extended core events and extended event sequences, too.
Extended core events are handled as entities, i.e.\ they are handled in analogy to normal non-atomic core events like $X_1 \sand{} X_2 \sand{} \ldots$.

There are additional transformation laws specifically for the extended form.
These laws are discussed in the following sections.
%
%
%
\paragraph{Laws of Contradiction for Extended Event Sequences}~\\
The law of contradiction for normal temporal expressions (chapter \ref{080822-001}) does not directly apply to extended event sequences.
An example: the expression $(A\booland{}B)\pand{}(B\booland{}C)$ consists of two extended core events, which both include the same basic event $B$.
This does not yield $\False$, though.
Instead, it may be further transformed using \eqref{080122-001}, which yields
\begin{align}
 (A\booland{}B)\pand{}(B\booland{}C) \ist{}& (A\booland{}B\booland{}B)\pand{}C\ist{} (A\booland{}B)\pand{}C~.
\end{align}
On the other hand, extended event sequences may, of course, result in contradictions.
The following three cases differ from each other, and together they form the \emph{law of contradiction for extended event sequences}:

First and in analogy to \eqref{080216-105}, for extended event sequences with normal and extended core events $\mathit{eK}$ there is
\begin{align}
      &\mathit{eK}\!_1\pand{} \mathit{eK}\!_2\pand{} \ldots\pand{} \mathit{eK}\!_n \ist \mathit{False}~,
        \label{090317003}
\end{align}
if $\exists ~ \mathit{eK}\!_i\ist \mathit{eK}\!_j$ for $i, j\in\{1, 2, \ldots, n\}$ and $i\neq j$.
This may be shown by transforamtion of the extended form using \eqref{071217-017} and \eqref{090317001}.
For example, 
\begin{align}
 (A\booland{}B)\pand{}(A\booland{}B) \ist{}& (A\booland{}B)\pand{}\bigl[(A \pand{}B) \boolor{}  (B \pand{}A)\boolor{}(A\sand{}B)\bigr]\ist{}\nonumber\\
  \ist{} & \bigl[(A\booland{}B\booland{}A)\pand{}B\bigr] \boolor{}
                \bigl[(A\booland{}B\booland{}B)\pand{}A\bigr] \boolor{}
                \bigl[(A\booland{}B)\pand{}(A\sand{}B)\bigr]  \ist\nonumber\\
   \ist{} & \False ~.
\end{align}

Second, an extended event sequences yields $\False$ because of a contradiction if it has an extended core event $\mathit{eK}$ together with a normal core event $K$, which must occur \emph{later} in the event sequence, and if there is at least one event $X$ which apperas in $K$ as well as in $\mathit{eK}$:
\begin{align}
      &\mathit{eK}\!_1\pand{} \mathit{eK}\!_2\pand{} \ldots\pand{} K_j \pand{} \ldots \ist \mathit{False}~,
        \label{090321001}
\end{align}
if $\exists ~ (X \inplus \mathit{eK}\!_i ) \booland{} (X \inplus K_j) $ for  $i< j$.
$K$ may be an atomic or non-atomic core event.
For example, expression $(A\booland{}B)\pand{}B$ yields a contradiction, as it requires that $A$ as well as $B$ have occurred before $B$ occurs.
The expression $(A\booland{}B)\pand{}(A\sand{}C)$ also yields a contradiction, as it requires that $A$ as well as $B$ have occurred before $A$ and $C$ occur simultaneously.
In both cases, though, there is no contradiction, if the normal core event occurs \emph{before} the extended core event:
For instance, $A \pand{} (A\booland{}B)\ist{}A\pand{}B$ and $(A\sand{}C) \pand{} (A\booland{}B)\ist{}(A\sand{}C)\pand{}B$.

Third, an extended event sequences yields $\False$ because of a contradiction if it contains more than one normal core event, and the normal law of contradiction from \eqref{080216-005} applies to these core events.
%
%
\paragraph{Using Negated Events in Extended Event Sequences and Extended Core Events}~\\
Handling of negated events is also quite similar to the discussions from chapter \ref{080220-002}.
But there are certain additions for extended event sequences and extended core events.

Negation of extended event sequences is the same as in \eqref{090110004}, but extended core events are treated as entities.

Extended core events are negated by using de Morgan's theoremes:
\begin{align}
  \boolnot \mathit{eK} & \ist{} \boolnot \bigl( X_1 \booland{} X_2 \booland{} \ldots \bigr) \ist{}
                  \boolnot X_1 \boolor \boolnot  X_2 \boolor \ldots ~~.
\end{align}
Negated extended core events are negated events, and as such are included into (extended) event sequences with negated events; see chapter \ref{080220-002} for details. 
Additionally to \eqref{09010930} and \eqref{09010931}, 
\begin{align}
  &\boolnot A \booland{} ( \ldots \pand{} (A \booland{}\ldots) \pand{} \ldots) \ist{} \False \label{090321010} ~.
\end{align}
%
%
%
\paragraph{Temporal Laws for Intersections of Extended Event Sequences and Extended Core Events}~\\
There is a special law for intersections of extended event sequences and extended core events, which provides
\begin{align}
  & A \pand{} ( A \booland{} B \booland{} \ldots) \ist{} A \pand{} B \booland{} \ldots ~.
\end{align}
Its correctness is easily demonstrated by breaking up the extended core event.
\section{Summary}\label{090408}
The TFTA's temporal logic described in this chapter extends the conventional Boolean {FTA} for non-repairable components/failures; it allows to model and analyze event sequences.

The {TFTA} is an extension to Boolean algebra and logic and does not rely on state-based modelling techniques.
Apart from Boolean operators for the conventional conjunction, disjunction, and negation, the {TFTA} has two additional operators PAND and SAND; these are ``specialized conjunctions'' which differentiate between event sequences and simultaneous events.

Using conventional Boolean logic transformations and aditional laws of transformation for temporal expressions, it is possible to transform complex temporal expressions into a temporal disjunctive normal form ({TDNF}).
The {TDNF} consists of separated event sequences.
The latter may be reduced into their minimal form, so called {MCSS}.
The {TFTA} thus allows efficient and meaningful qualitative analyses, just as the conventional {FTA} does.

As an extension to the Boolean algebra, the TFTA's temporal logic is universally applicable and not at all restriced to certain failure rate distributions.  

In another step {MCSS} may be transformed into mutually exclusive expressions.
The latter are especially well suited for direct probabilistic quantification and thus allow probabilistic analyses of temporal expressions, see the next chapter \ref{chap080401-033}.

The {TFTA} follows the conventional {FTA} in notation, expressions, workflow-steps, and work products.
When compared to state based dynamic methods, the {TFTA}, therefore, has similar positive characteristics: its logic expressions and results are similarly intuitive in use, similarly readable and comprehensible, and it has good scalability.

Simplification of temporal expressions into a minimal form (and if necessary: mutually exclusive, disjoint form, too) requires heavy effort, which is an additional cost when compared to Boolean {FTA}.
This, on the other hand is no problem specific to the {TFTA}, and instead is, in principle, the same for all dynamic models. 

The {TFTA} allows for an efficient reduction of effort, though, by means of an ``extended logic form''.
If several sequences may be combined into a normal, i.e. Boolean, conjunction, then the extended form does not explicitely break them down.
This alone highly improves the calculatory effort, which otherwise grows exponentially. 
%
%
%

%
%
%
%
\clearpage
\clearpage
\ifx \printSprueche\undefined
\else
\renewcommand*{\dictumwidth}{.35\textwidth}

\setchapterpreamble[ur]{%
\dictum[Aristoteles]{Probable impossibilities are to be preferred to improbable possibilities.}
%
\vspace{3cm}
}
\fi

\chapter{Probabilistic Quantification of the TFTA Method}\label{chap080401-033}
~
%
%
%
%
%
The quantification of the {TFTA} method extends the qualitative analysis.
Allocation of failure rates and probabilities to basic events allows the calculation of the TOP event's failure parameters.
These are then used in order to assess system charateristics like its safety integrity or expected reliability. 

On the one hand, additional effort is necessary for the probabilistic quantification of the TOP event's parameters with consideration of event sequences.
On the other hand, the TFTA's quantification yields smaller values than the conventional Boolean FTA.

This chapter is structured in four sections:
\begin{itemize}
 \item Chapter \ref{090120001} starts with the basics of probabilistic quantification of the Boolean {FTA}.
 \item Chapter \ref{080207-001} describes the concept behind the quantification of the {TFTA}, which is based on failure densities.
 \item Chapter \ref{080522-001} discusses direct quantification of the PAND and SAND operations.
 \item Using these, chapter \ref{000020} then describes the quantification of entire temporal failure functions, i.e.\ the calculation of the TOP event's failure probability, failure density, and failure rate. 
 \item As these caluclations require exponentially increasing calculatory effort, chapter \ref{080724-010} introduces a simplification which provides approximated failure characteristics for temporal expressions. 
\end{itemize}
\emph{Note:}
In chapter \ref{_chap_080401-004} the qualitative TFTA was discussed for non-repairable components and their failures, only.
This restriction also applies to the concept of quantification including chapter \ref{080811-001}.
Chapter \ref{080222-001} then focusses on the special case where failure parameters are distributed exponentially.
\section{Quantification of the Boolean {FTA}}\label{090120001}
In the Boolean as well as the temporal {FTA} the probabilistic analysis of the TOP event is based on the system's TOP failure function as provided by a preceding qualitative analysis.
Usually, this logic expression is then transformed (using the transformation laws of Boolean or temporal logic) into a form, which is well suited for the task at hand (in this case: quantification).

For example, the minimal cutset form of the Boolean failure function of the system described in \eqref{080904_009} is given as
\begin{align}
  \varphi \ist{}{} \bigvee\limits_{\mathclap{j\ist{}{} 1}}^{\xi } \MS_j \ist{}{} \bigvee\limits_{\mathclap{j\ist{}{} 1}}^{\xi } \Bigl( \bigwedge\limits_{i\ist{}{} 1}^{n_j} X_{j, i} \Bigr)   ~.\nonumber
\end{align}
This form is very clear and well suited for qualitative analysis.
On the other hand, there is an equivalent but less clear form of the same failure function, as given in \eqref{080904_010}:
\begin{align}
  \varphi &\ist{}{} \bigvee_{\mathclap{j\ist{}{}1}}^{\xi} \Bigl( \MS_j \cdot \bigwedge_{\mathclap{i\ist{}{}1}}^{j-1} \boolnot \left(\MS_i\right)\Bigr)~. \nonumber
\end{align}
Here, the minimal cutsets are mutually exclusive (disjoint), which is less easily readable but simplifies probabilistic analyses.

The quantification of minimal cutsets of the conventional {FTA}, with Boolean AND and OR and NOT, is well known; it is mentioned here only for completeness. 

Assuming $n$ mutually independent events, there are
\begin{align}
  F_{\text{AND}}(t) \ist&  \prod_{\mathclap{i \ist 1}}^{n}F_{i}(t)\label{eq:07}~,\\
  F_{\text{OR}}(t) \ist&  1-\prod_{\mathclap{i \ist 1}}^{n}\left(1-F_{i}(t)\right)\label{eq:08}~,\\
  f_{\text{AND}}(t)\ist& \frac{\D}{\D t}F_{\text{AND}}(t)\ist  \sum_{i \ist 1}^{n}\left(f_{i}(t)\cdot\prod_{\mathclap{j \ist 1;j\neq
      i}}^{n}F_{j}(t)\right)\label{eq:09}~,\\
  f_{\text{OR}}(t) \ist&  \frac{\D}{\D t}F_{\text{OR}}(t)\ist
      \sum_{i \ist 1}^{n}\left(f_{i}(t)\cdot\prod_{\mathclap{j \ist 1;j\neq
      i}}^{n}  \left(1-F_{j}(t)\right)\right)\label{eq:10}\,.
\end{align}
Failure functions of fault trees are usually complex expressions with non-independent events and sub-expressions.
It is, thus, convenient to reduce such failure functions into their minimal cutset form before quantification, as well as to further transform the minimal cutsets into a mutually exclusive (disjoint) form.
This is, for example, described in \cite{Abraham1979,Heidtmann1989} (and for non-monotonous functions in \cite{Bertschy1996,Kohlas1995}).
Disjoint events simplify quantification; instead of the generic \eqref{eq:08} and \eqref{eq:10}, the much more simple  
\begin{align}
  F_{\text{OR}}(t) \ist&  \sum_{i \ist 1}^{n}F_{i}(t)\label{080110-002}~,\\
  f_{\text{OR}}(t) \ist&  \sum_{i \ist 1}^{n}f_{i}(t)\label{080110-003}
\end{align}
may be used.

In monotonous fault trees with non-repairable failure events, negated events are used exclusively as conditional events; and as such, there is no failure density of negated events.
This is also true in case of {TFTA}, as shown by the discussions in chapter \ref{080220-002}: negated events occur only prior to other (non-negated) events.

The probability of occurrence of a negated event $\boolnot X_i$  is then given by
\begin{align}
     &F_{\boolnot X_i}(t)\ist 1-F_{X_i}(t)\ist R_{X_i}(t) \label{080222-101}~.
\end{align}
\section{Quantification of the TFTA: Temporal Concept and Failure Frequencies}\label{080207-001}\label{080114-001}
Other than the Boolean {FTA}, the temporal logic of the {TFTA} permits restrictions on the sequence of event occurrence in conjunctions.
Any quantification of the {TFTA}, therefore, must also take only specific event sequences into account.
This chapter explains in general, how this may be accomplished.
Chapter \ref{080522-001} then uses these basics and derives specific rules for the quantification of the temporal operators PAND and SAND, respectively.

In general, failure probabilities, failure densities, and failure rates are given as \cite{Meyna2003Taschenbuch}
\begin{align}
	f_{X}(t)&\ist\frac{\D}{\D t}F_{X}(t)\label{080224-001}&\text{and}\\
 f_X(t)&\ist\lambda_X(t)\cdot\left(1-F_X(t)\right)\ist \lambda_X(t)\cdot R_X(t)~.&&\label{071213-002}
\end{align}
In case of constant failure rates the failure probabilities and failure densities are then given as
\begin{align}
  F_{X}(t)&\ist{} 1-\E^{-\lambda_X\,t}\qquad\text{and}\qquad
  f_X(t)\ist{} \lambda_X\cdot\E^{-\lambda_X\,t}~.\label{090213011}
\end{align}
\subsection{Sequences with Two Events}
In a concunction with independent inputs (basic events) $A$ and $B$ there is
\begin{align}
	F_{A\booland{} B}(t)&\ist F_A(t)\cdot F_B(t)~.\label{071213-001}
\end{align}
This is the probability, that at time $t$ both fault tree events $A$ and $B$ are $\True$.
This is also the probability, that the failures represented by $A$ and $B$ have both occurred at some time during interval $]0;t]$.
It is not possible, though, to make specific statements on either the sequence of these failures, nor on the absolute point in time at which the failures occurred.

Other than the failure probability $F(t)$, the failure density $f(t)$ does consider event sequences, as
\begin{align}
	f_{A\booland{} B}(t)&\ist\frac{\D}{\D t}F_{A\booland{} B}(t)\ist f_B(t)F_A (t)+f_A (t)F_B(t)\\
  \intertext{and thus, using \eqref{071213-002},}
	f_{A\booland{} B}(t)&\ist
	  F_A(t)R_B(t)\lambda_B(t)+F_B(t)R_A(t)\lambda_A(t)~.\label{071209-008}
\end{align}
Equation \eqref{071209-008} may be interpreted as the probability per time, that \cite{Alleman2000Fault-Tolerant}
\begin{itemize}
	\item either: $A$ has occurred at some time in interval $]0;t]$, i.e.\ $F_A(t)$, and $B$ has not occurred in interval $]0;t]$, i.e.\ $R_B(t)$, and $B$  will occur in the (infinitesimally) short period $]t;t+\Delta t]$ after $t$, i.e.\ $\lambda_B(t)$;
	\item or: $B$ has occurred at some time in interval $]0;t]$, i.e.\ $F_B(t)$, and $A$ has not occurred in interval $]0;t]$, i.e.\ $R_A(t)$, and $A$  will occur in the (infinitesimally) short period $]t;t+\Delta t]$ after $t$, i.e.\ $\lambda_A(t)$.
\end{itemize}
These two possibilites represent the two sequences ``$A$ first, and then $B$'' and ``$B$ first, and then $A$'', which are mutually exclusive.
Therefore, their probabilities may simply be added.

This makes it possible to quantify specific event sequences.
If, for example, only the event sequence ``$A$ first, and then $B$'' is relevant, then
\begin{align}
  \begin{split}
	f_{\text{``$A$ first, and then $B$''}}(t)&\ist F_A (t)\cdot\frac{\D}{\D t}F_B(t)\ist
      f_B(t)F_A (t)\ist\lambda_B(t)R_B(t)F_A(t)~.
	\end{split}\label{071208-025}
\end{align}
The corresponding failure probability is given by integration over the density:
\begin{align}
  \begin{split}
	F_{\text{``$A$ first, and then $B$''}}(t)\ist&\int\limits^t_0   f_{\text{``$A$ first, and then $B$''}}(\tau)\cdot\D\tau\ist 
      \int\limits^t_0 f_B(\tau)F_A (\tau)\cdot\D\tau~.
	 \end{split}\label{080218-020}
\end{align}
\subsection{Sequences with More Than Two Events}\label{090408002}
In case of more than two events, the sequence(s) of those events must also be considered that are not the ``last occurring'' events.
For an AND gate with three inputs $A$, $B$, and $C$, where event sequence ``$A$ first, and then $B$, and then $C$'' is relevant, it is thus not sufficient to simply take the derivative of $F_{A\booland{} B\booland{} C}(t)$, as 
\begin{align}\begin{split}
	f_{A\booland{} B\booland{} C}(t)&\ist
	f_A (t)F_B(t)F_C (t)+f_B(t)F_A (t)F_C (t)+
		f_C (t)F_A (t)F_B(t)~.
	\end{split}\label{071209-006}
\end{align}
None of the expressions on the right side of \eqref{071209-006} represents the relevant event sequence ``$A$ first, and then $B$, and then $C$''.
E.g., $f_C (t)F_A (t)F_B(t)$ is the density contribution of ``$A$ and $B$ first, and then $C$''; it thus represents both event sequences ``$A$ first, and then $B$, and then $C$'' and ``$B$ first, and then $A$, and then $C$''.

On the other hand, it is possible to correctly take the ``not-last-occurring'' events (here: $A$ and $B$) into account.
It is necessary to treat ``$A$ first, and then $B$'' as an entity by itself, thus
\begin{align*}
	&f_{\text{``$A$ first, and then $B$, and then $C$''}}(t)\ist\\
		&\qquad\ist{}
		f_{\text{``($A$ first, and then $B$) first, and then $C$''}}(t)\ist 
      f_C(t)F_{\text{``$A$ first, and then $B$''}}(t) ~.\nonumber
\end{align*}
Using \eqref{080218-020} the failure density is then given as
\begin{align}
	f_{\text{``$A$ first, and then $B$, and then $C$''}}(t)\ist&
			f_C (t)\int\limits^t_0 f_B(\tau)
			F_A (\tau)\cdot\D\tau~.\label{071208-024}
  \intertext{Finally, the failure probability is obtained by intergation:}
  \begin{split}
	F_{\text{``$A$ first, and then $B$, and then $C$''}}(t) \ist & 
        \int\limits^t_0  f_C (\tau)\int\limits^\tau_0  
		\vphantom{\int\limits^t_0}
			f_B(\tau')F_A (\tau')\cdot\D\tau'\cdot\D\tau~.
	\end{split}\label{080110-001}
\end{align}
This method allows quantification of arbitrarily complex sequences with more than two events.
%
\subsection{What Parameter to Use in Probabilistic Analyses?}\label{080114-002}
Safety standards, as e.g.\ {\IEC} or {\ISO}, require verification that systems meet specific failure rates $\lambda(t)$ \cite{Schilling2007c}; evidence to verify that may be provided using probabilistic {FTA}. 
If the failure probability $F(t)$ and failure frequency $f(t)$ are given, then the failure rate is derived from \eqref{071213-002}.

In most cases it is not necessary to provide the failure rate, though.
In the safety domain, the absolute probabilities of failure events occurring is usually so small that $F(t)\ll1$, and thus with \eqref{071213-002} 
\begin{align}
	f(t){}\approx{}\lambda(t)~.\label{071213-003}
\end{align}
In such cases, the failure frequency is a good approximation of the failure rate, and may be directly used as target value.
\section{Quantification of the PAND and SAND Operators}\label{080522-001}
Based on the generic method of quantification of event sequences in chapter \ref{080207-001}, the TFTA's temporal operations may now be quantified. 

But first it is helpful to grasp the temporal meaning of PAND and SAND operations probabilistically; this is accomplished in chapter \ref{080811-001}.
Chapter \ref{080222-001} compares the TFTA with a state-based model as reference, and thereby demonstrates the correctness of the TFTA's quantification.
\subsection{Quantification Using Logic Functions}\label{080811-001}
The failure probability is defined as the expectancy value for the occurrence of a failure \cite{Meyna2003Taschenbuch}, and thus
\begin{align}
	F_i(t)\ist& \EW\bigl[X_i(t)\ist \mathit{True}\bigr]\ist \EW\bigl[X_i(t)\bigr]~. \label{080522-002}
\end{align}
Accordingly, the failure frequency is defined as \cite{Schneeweiss1999a}
\begin{align}
	f_i(t)
	\ist&
		\lim_{\Delta t\rightarrow0}\frac{1}{\Delta t}\EW\bigl[
			\left(X_i(t)\ist \mathit{False}\right)\booland
	  \left(X_i(t+\Delta t)\ist \mathit{True}\right)\bigr]\ist\nonumber\\
	\ist& \lim_{\Delta t\rightarrow0}\frac{1}{\Delta t}\EW\bigl[
			\boolnot X_i(t)\booland{}X_i(t+\Delta t)\bigr]~. \label{080522-003}
\end{align}
By simple transformation an equivalent form is provided, which is specifically helpful for the further discussion:
\begin{align}
	f_i(t)\Delta t+\oInfinitOperat
	\ist&
		\EW\bigl[
			\boolnot X_i(t)\booland{}X_i(t+\Delta t)\bigr]
      \qquad \text{where} \qquad
	\lim_{\Delta t\rightarrow0}\frac{\oInfinitOperat }{\Delta t}
	\ist 0 ~. \label{080522-005}
\end{align}
\paragraph{PAND Operation}~\\
The PAND operator in $A\pand{} B$ describes the occurrence of $B$ at time $t$ after $A$ has already occurred.
Non-infinitesimally, this implies that
\begin{itemize}
 \item at time $t$ event $A$ has already occurred, and event $B$ has not yet occurred, and
 \item at $t+\Delta t$ both, event $A$ as well as event $B$, have occurred.
\end{itemize}
Therefore,
\begin{align}
	&	A(t) ~\booland~ \boolnot  B(t) ~~\booland{}~~ A(t+\Delta t)~\booland{}~ B(t+\Delta t)~, \label{080522-006}
\end{align}
from which with \eqref{080522-005} follows (assuming independent events $A$ and $B$), that
\begin{align}
	f_{A\pand{} B}(t)\Delta t+\oInfinitOperat
	\ist 	&
		\EW\bigl[A(t)\booland\boolnot B(t)\booland{} A(t+\Delta t)\booland{} B(t+\Delta t)\bigr]\ist \nonumber\\
	\ist	 & \EW\bigl[A(t)\booland{}A(t+\Delta t)\bigr]\cdot \EW\bigl[\boolnot B(t)\booland{}B(t+\Delta t)\bigr]~.
\end{align}
The expectancy value $\EW\bigl[\boolnot B(t)\booland{}B(t+\Delta t)\bigr]$ may be directly replaced by \eqref{080522-005}.
The expectancy value $\EW\bigl[A(t)\booland{}A(t+\Delta t)\bigr]$, on the other hand, is not equal to the simple product of the expectancy values of events $A(t)$ und $A(t+\Delta t)$, as they are not independent from each other. 
Instead, 
\begin{align}
  \EW\bigl[A(t)\booland{}A(t+\Delta t)\bigr]&\ist\EW\bigl[A(t+\Delta t)\mid A(t)\bigr]\cdot\EW\bigl[A(t)\bigr]\ist \EW\bigl[A(t)\bigr]~,
\end{align}
as a failure, that has occurred at time $t$, ``is still occurred'' at $t+\Delta t$.
Thus, 
\begin{align}
  &f_{A\pand{} B}(t)\Delta t+\oInfinitOperat
  \ist F_A(t)\cdot \bigl[ f_B(t)\Delta t+\oInfinitOperat \bigr]~.
\end{align}
Division by $\Delta t$, and $\Delta t\rightarrow0$, yields
\begin{align}
	f_{A\pand{} B}(t)&\ist
		\lim_{\Delta t\rightarrow0} \left(F_A(t)\cdot \left[ f_B(t)+\frac{\oInfinitOperat }{\Delta t}\right]-\frac{\oInfinitOperat }{\Delta t}\right)~,
\end{align}
and finally
\begin{align}
	f_{A\pand{} B}(t)	&\ist F_A(t)\cdot f_B(t)~. 	\label{080522-007}
\end{align}
Obviously, $A\pand{} B$ from \eqref{080522-007} is therefore equal to the sequence ``$A$ first, and then $B$'' from \eqref{071208-025}.
This allows to state the failure probability function of the PAND operator:
\begin{align}
  \begin{split}
  F_{A\pand{} B}(t)\ist&\int\limits^t_0 F_A(\tau)f_B (\tau)\cdot\D\tau~.
   \end{split}\label{080218-320}
\end{align}
\paragraph{SAND Operation}\label{080822-020} \label{080726-011}~\\%
The SAND operator in $A\sand{} B$ describes the exact simultaneous occurrence of $A$ and $B$ at time $t$.
Non-infinitesimally, this implies that
\begin{itemize}
 \item at time $t$ neither event $A$ nor event $B$ has already occurred, and
 \item at $t+\Delta t$ both, event $A$ as well as event $B$, have occurred.
\end{itemize}
Therefore,
\begin{align}
  & \boolnot A(t)~\booland~ \boolnot  B(t)~~\booland{}~~ A(t+\Delta t)~\booland{} ~B(t+\Delta t)~, \label{080524-006}
\end{align}
from which follows (assuming independent events $A$ and $B$), that
\begin{align}
  f_{A\sand B}(t)\Delta t+\oInfinitOperat
  \ist
  &
    \EW\bigl[\boolnot A(t)\booland{}A(t+\Delta t)\bigr]\cdot \EW\bigl[\boolnot B(t)\booland{}B(t+\Delta t)\bigr]\ist\nonumber\\
  \ist &  \bigl[ f_A(t)\Delta t+\oInfinitOperat \bigr]\cdot \bigl[ f_B(t)\Delta t+\oInfinitOperat \bigr]\ist \nonumber\\
  \ist & f_A(t)\cdot\Delta t\cdot f_B(t)\Delta t+\oInfinitOperat \cdot \bigl[\ldots\bigr]~.\nonumber
\end{align}
Division by $\Delta t$, and $\Delta t\rightarrow0$, yields
\begin{align}
  f_{A\sand B}(t)&\ist
    \lim_{\Delta t\rightarrow0} \left(f_A(t)f_B(t)\Delta t+\frac{\oInfinitOperat }{\Delta t}\bigl[\ldots\bigr]-\frac{\oInfinitOperat }{\Delta t}\right)\,,
\end{align}
and finally
\begin{align}
  f_{A\sand B}(t) &\ist 0~.  \label{080524-007}
\end{align}
This implies that the probability of exact simultaneous occurrence of two independent events is always $0$; every small deviation from simultaneousness is already covered -- probabilistically -- by the two PAND sequences of these events.
Therefore, 
\begin{align}
  F_{A\sand B}(t) &\ist 0~, \label{080524-008}\\
  \lambda_{A\sand B}(t) &\ist 0~. \label{080524-009}
\end{align}
Allthough the SAND operator may seem unnecessary from this probabilistic point of view, it is essential for the qualitative transformation of temporal expressions, as well as for qualitative analyses.
Specifically, it provides the temporal law of idempotency in \eqref{090107019}, which serves as an important filter for the simplification of temporal expressions.
\subsection{Quantification Using Comparison with State Diagrams}\label{080222-001}
\emph{Note:}
The statements up to \eqref{080524-009} apply universally.
After that, the further statements discuss exponentially distributed parameters, only.

Looking back, chapter \ref{080207-001} approaches the question of quantification of the PAND operation from the definitions of the relevant parameters.
Chapter \ref{080522-001} then demonstrates, that the logical meaning of PAND and SAND operations yields identical results, respectively.

In this chapter these results are compared to a reference model in order to confirm them absolutely.

\begin{figure}[h]
  \centering
  \input{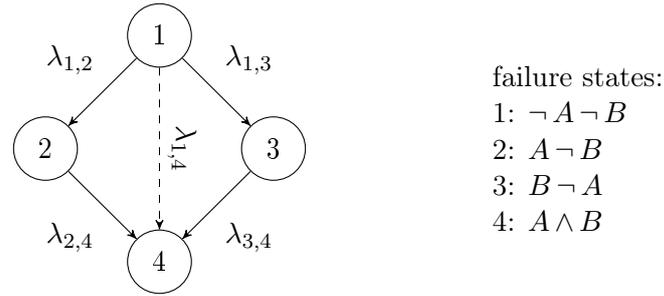}
  \caption{State diagram of a system consisting of two non-repairable components $A$ and $B$ which have constant transition- and failure rates $\lambda_{i,j}$.}
  \label{fig:bsp_markov_nicht_rep}
\end{figure}

This comparison is split into two parts.
First, the Boolean AND and OR operations are quantified, then the quantification is extended to the temporal PAND and SAND operations using the law of completion from chapter \ref{080129-022}.
\paragraph{Boolean Operations}\label{080526-003}~\\%
Figure \ref{fig:bsp_markov_nicht_rep} shows the state diagram of an example system consisting of two non-repairable components $A$ and $B$ which have constant transition- and failure rates $\lambda_{i,j}$;
this diagram is the same as in figure \ref{fig:bsp_zweiausfaelle_zustaende}.

The state probabilities $P_i(t)$ are given by the following system of differential equations:
\begin{align}
			\begin{bmatrix}
				\dot{P_1}(t)\\
				\dot{P_2}(t)\\
				\dot{P_3}(t)\\
				\dot{P_4}(t)\\
			\end{bmatrix}
			& \ist
			\begin{bmatrix}
        -(\lambda_{1,2}+\lambda_{1,3}+\lambda_{1,4})&0&0&0\\
				\lambda_{1,2}&-\lambda_{2,4}&0&0\\
				\lambda_{1,3}&0&-\lambda_{3,4}&0\\
				\lambda_{1,4}&\lambda_{2,4}&\lambda_{3,4}&0\\
			\end{bmatrix} \cdot
			\begin{bmatrix}
				P_1(t)\\
				P_2(t)\\
				P_3(t)\\
				P_4(t)\\
			\end{bmatrix}
\label{eq:123001} ~.
\end{align}
Assuming markov conditions are valid, event $A$ and $B$ have constant failure rates, and thus
\begin{align}
				\lambda_A& \ist \lambda_{1,2} \ist \lambda_{3,4}\qquad\text{und}\qquad
				\lambda_B \ist \lambda_{1,3} \ist \lambda_{2,4}~.\label{eq:0103}
\end{align}
Solving the system of differential equations \eqref{eq:123001} provides four state probabilities $P_1(t)$ to $P_4(t)$.
Looking from a reliability and safety point of view, these probabilities may be interpreted, depending on how components $A$ and $B$ interact:
\begin{itemize}
  \item
  In case of parallel connection (redundant components) the system fails, if both components, $A$ and $B$, fail.
  The system's failure function is $\varphi \ist A\booland{} B$, and thus state $4$ represents the system failure.
  As a consequence, 
  $F_{\varphi}(t) \ist P_4(t)$ und $R_{\varphi}(t) \ist 1-P_4(t) \ist P_1(t)+P_2(t)+P_3(t)$.
  \item
  In case of series connection the system fails, if either $A$ or $B$ or both, $A$ and $B$, fail.
  The system's failure function is $\varphi \ist A\boolor B$, and thus states $2$ and $3$ and $4$ represent the system failure.
  As a consequence, 
  $F_{\varphi}(t) \ist P_2(t)+P_3(t)+P_4(t)$ und $R_{\varphi}(t) \ist 1-P_4(t) \ist P_1(t)$.            
\end{itemize}
\paragraph{Simplification}~\\
As a first step and using chapter \ref{080811-001}, the transition representing the SAND is discarded, i.e.\ $\lambda_{1,4} \ist 0$.
This is done assuming structural independence between $A$ and $B$. 

For the example system in \eqref{eq:123001} and with \eqref{eq:0103} this yields
\begin{align}
    \begin{split}
      F_{A\booland{} B}(t)& \ist P_4(t) \ist (1-\E^{-\lambda_At})(1-\E^{-\lambda_Bt}) \ist F_A(t)F_B(t)
    \end{split}\label{eq-0104}
    \intertext{and}
    \begin{split}
      F_{A\boolor B}(t)& \ist P_2(t)+P_3(t)+P_4(t) \ist 1-\E^{-(\lambda_A+\lambda_B)t}
             \ist 1-\bigl[1-F_A(t)\bigr]\bigl[1-F_B(t)\bigr]~.
    \end{split}\label{eq-0105}
\end{align}
Generalization of these ideas again leads to the rules for quantification of Boolean operators, as already mentioned in \eqref{eq:07} to \eqref{080110-003}. 
\paragraph{PAND and SAND Operations}~\\
Temporal fault trees are quantified using their MCSS the same way as conventional fault trees are quantified using their minimal cutsets.
State-transition diagrams show the correctness of the laws of completition, and they allow to derive an approach to quantification of temporal operations.
\begin{figure}[h]
  \centering
  \input{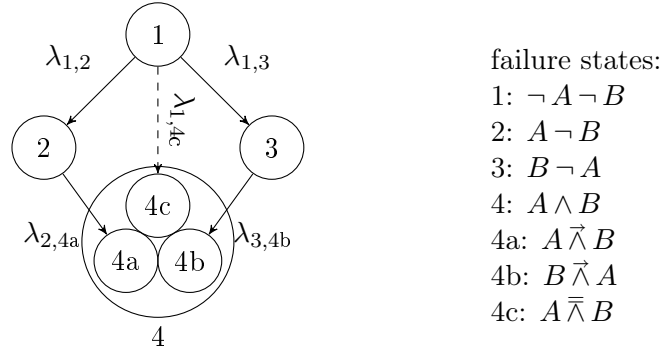}
  \caption{Markov model of the example system from figure \ref{fig:bsp_markov_nicht_rep} with division of state $4$ ``$A$ and $B$ have occurred'' into three substates $4\text{a}$, $4\text{b}$, and $4\text{c}$.}
  \label{fig:bsp_markov_nicht_rep_temp}
\end{figure}

Figure \ref{fig:bsp_markov_nicht_rep_temp} shows the example system from chapter \ref{080526-003} with its different event sequences.
Other than figure \ref{fig:bsp_markov_nicht_rep}, state $4$ (``$A$ and $B$ failed'') is now divided into three substates.
State $4\text{a}$ describes the system, where $A$ has occurred first, and then $B$ has occurred.
State $4\text{b}$ describes the system, where $B$ has occurred first, and then $A$ has occurred.
State $4\text{c}$ describes the system, where $A$ and $B$ have occurred simultaneously.
These state diagrams are really sequential failure trees, see page \pageref{chap080726-021}.

These three possibilites are mutually exclusive (disjoint) and they are complete, i.e.\ there are no more possible ways for ``$A$ and $B$ have occurred''.
The probability for ``superstate'' $4$ is then given as
\begin{align}
		& P_{4}(t) \ist P_{4\text{a}}(t)+P_{4\text{b}}(t)+P_{4\text{c}}(t)~.  \label{eq-0102}
\end{align}

The corresponding differential equation system of the states' probabilites may be given as the following matrix:
\begin{align}
			&\begin{bmatrix}
				\dot{P_1}(t)\\
				\dot{P_2}(t)\\
				\dot{P_3}(t)\\
				\dot{P_4}(t)\\
				\dot{P_{4\text{a}}}(t)\\
				\dot{P_{4\text{b}}}(t)\\
				\dot{P_{4\text{c}}}(t)
			\end{bmatrix}
			 \ist
			\begin{bmatrix}
        -(\lambda_{1,2}+\lambda_{1,3}+\lambda_{1,{4\text{c}}})&0&0&0&0&0&0\\
				\lambda_{1,2}&-\lambda_{2,{4\text{a}}}&0&0&0&0&0\\
				\lambda_{1,3}&0&-\lambda_{3,{4\text{b}}}&0&0&0&0\\
				\lambda_{1,{4\text{c}}}&\lambda_{2,{4\text{a}}}&\lambda_{3,{4\text{b}}}&0&0&0&0\\
				0&\lambda_{2,{4\text{a}}}&0&0&0&0&0\\
				0&0&\lambda_{3,{4\text{b}}}&0&0&0&0\\
				\lambda_{1,{4\text{c}}}&0&0&0&0&0&0
			\end{bmatrix}
			\begin{bmatrix}
				P_1(t)\\
				P_2(t)\\
				P_3(t)\\
				P_4(t)\\
				P_{4\text{a}}(t)\\
				P_{4\text{b}}(t)\\
				P_{4\text{c}}(t)
			\end{bmatrix}.\label{080508-006}
\end{align}

\begin{figure}[h]
 \centering
   \input{pics/fig_markov_zwei_bool_and_or}
  \caption{Markov model of the example system corresponding to failure functions $A\booland{} B$ and $A\boolor B$ and the set of differential equations in \eqref{080508-006}.
  System failure states are marked in bold.}
  \label{fig:bsp_markov_nicht_rep_oder}
\end{figure}
%
%
%
%
%
%
%
%
%
%
%
%
%
%
%
%
\begin{figure}[h]
 \centering \vskip1em \input{pics/fig_markov_zwei_temp_pand_pand_por}
  \caption{Markov model of the example system corresponding to failure functions 
  $A\pand{} B$,
  $B\pand{} A$, and
  $A\sand{} B$ and the set of differential equations in \eqref{080508-006}.
  System failure states are marked in bold.}
  \label{fig:bsp_markov_nicht_rep_sand}
\end{figure}

Figure \ref{fig:bsp_markov_nicht_rep_oder} shows the markov modells corresponding to the example system's $A\booland{} B$ and $A\boolor B$.
Assuming markovian conditions yields
\begin{align}
				\lambda_A& \ist \lambda_{1,2} \ist \lambda_{3,{4\text{b}}}\qquad\text{and}\qquad
				\lambda_B \ist \lambda_{1,3} \ist \lambda_{2,{4\text{a}}}~.\label{eq:0108}
	\end{align}%
The solution of the set of differential equations in \eqref{080508-006} for $\lambda_{1,{4\text{c}}} \ist 0$ yields the two equations known from \eqref{eq-0104} and \eqref{eq-0105}:
\begin{align}
    \begin{split}
      F_{A\booland{} B}(t)& \ist P_4(t) \ist (1-\E^{-\lambda_At})(1-\E^{-\lambda_Bt}) \ist F_A(t)F_B(t)
    \end{split}\label{eq-0194}
    \intertext{and}
    \begin{split}
      F_{A\boolor B}(t)& \ist P_2(t)+P_3(t)+P_4(t) \ist
                1-\E^{-(\lambda_A+\lambda_B)t} \ist 1-\bigl[1-F_A(t)\bigr]\bigl[1-F_B(t)\bigr]~.
    \end{split}\label{eq-0195}
\end{align}
The law of completeness from chapter \ref{080129-022} allows representing an AND operation by PAND and SAND operations.
Figure \ref{fig:bsp_markov_nicht_rep_sand} shows the relevant state diagrams and failure functions.

Solving the set of differential equations in \eqref{080508-006} then yields
\begin{align}
		\begin{split}
			F_{A\pand{} B}(t)&\ist P_{4\text{a}}(t) \ist \int\limits^t_0 f_B(\tau)F_A(\tau)\cdot\D\tau~,
		\end{split}\label{080513-009}\\
		\begin{split}
			F_{B\pand{} A}(t)&\ist P_{4\text{b}}(t) \ist \int\limits^t_0 f_A(\tau)F_B(\tau)\cdot\D\tau~,
		\end{split}\label{080513-010}		\\
		F_{A\sand B}(t)& \ist P_{4\text{c}}(t) \ist 0~.\label{080513-011}
\end{align}
Insertion of \eqref{080513-009}, \eqref{080513-010}, and \eqref{080513-011} in \eqref{eq-0102} provides
\begin{align}\begin{split}
    F_{A\booland{} B}(t)& \ist F_{A\pand{} B}(t)+F_{B\pand{} A}(t)+F_{A\sand B}(t)\ist
         \\
      & \ist  \int\limits^t_0 \bigl(f_B(\tau)F_A(\tau) + f_A(\tau)F_B(\tau) \bigr)
      		\cdot\D\tau + 0 \ist F_A(t)\cdot F_B(t)~.
    \end{split}
      \label{080513-129}
\end{align}
This demonstrates that the law of completition holds also from a probabilistic point of view, and shows the correctness of the calculations in chapter \ref{080522-001}.

The corresponding failure frequencies and failure rates are then given by \eqref{080224-001} and \eqref{071213-002}, respectively.

\section{Quantification of the Temporal Failure Function}\label{000020}
Chapter \ref{080207-001} shows the basic concept of quantifing event sequences.
Applying this concept to arbitrary temporal expressions in {TDNF} allows the quantification of temporal fault trees, i.e.\ calculation of their events' -- and especially their TOP event's -- failure probabilities and failure rates.
\subsection{Quantification of Event Sequences and MCSS}\label{080222-004}
The probabilistic quantification of a fault tree requires, firstly, to determin its {MCSS}, i.e.\ all the critical event combinations (including their sequences) in minimal form.
This is done using the rules for qualitative transformations from chapter \ref{chap080401-031} and \ref{chap080809-001}.
In a next step, the probabilistic parameters are determined for each of the MCSS; these parameters are then used to calculate the TOP event's parameters.
\paragraph{Simplification for Independent Failure Events}\label{090121001}~\\%
In case of independent failure events an essential simplification is possible:
According to chapter \ref{080522-001} all {MCSS} may be omitted that include at least one SAND. They are omitted after transforming the temporal expression into its {MCSS} but before the {MCSS} are quantified.
{MCSS} including SANDs are only relevant for the qualitative analysis and provide no probabilistic contribution to the failure rates, failure frequencies, and failure probabilities of the temporal failure function.
Only {MCSS} without SAND are then quantified.
Thus, the quantification is carried out for MCSS of the following type:
\begin{align}
			& X_1\pand{} X_2\pand{} \ldots\pand{} X_n ~,\label{080216-985}
			\intertext{possibly also in conjunction with negated events}
 			& (\boolnot X_{I} \booland \boolnot  X_{II} \cdots ) \booland{} (X_1\pand{} X_2\pand{} \ldots\pand{} X_n)
				 ~. \label{080222-198}
\end{align}
{MCSS} according to \eqref{080216-985} may be directly quantified using convolutions of the failure frequencies, see \eqref{080522-007} and \eqref{080218-320}, thus
\begin{align}
  f_{\mathit{MCSS}}(t)\ist & f_{X_1\pand{} X_2\pand{} \ldots\pand{} X_n}(t) \ist \label{080525-001}\\
 \begin{split}\ist &
      f_{X_n}(t) \int\limits_0^t f_{X_{n-1}}(\tau^{\{1\}})\int\limits_0^{\mathclap{\tau^{\{1\}}}} f_{X_{n-2}}(\tau^{\{2\}})\cdots
      \int\limits_0^{\mathclap{\tau^{\{n-2\}}}}
      f_{X_{2}}(\tau^{\{n-1\}})
      \int\limits_0^{\mathclap{\tau^{\{n-1\}}}}
       f_{X_1}(\tau^{\{n\}})\cdot\\
& \qquad\qquad\qquad\qquad\qquad\qquad\qquad\qquad\cdot \D \tau^{\{n\}} \cdot \D \tau^{\{n-1\}}\cdots\D \tau^{\{2\}}\cdot\D \tau^{\{1\}} ~.
             \end{split}
  \nonumber
\end{align}
\paragraph{MCSS with Negated Events}~\\
The probabilities of negated events, which are part of MCSS, are multiplied to these results.
Therefore, the MCSS' quantification according to \eqref{080222-198} is given as
\begin{align}
  \begin{split}
  f_{\mathit{MCSS}}(t)\ist& f_{ (\boolnot X_{I} \booland \boolnot  X_{II} \cdots ) \booland{} (X_1\pand{} X_2\pand{} \ldots\pand{} X_n)}(t) \ist \\
    \ist &  f_{X_1\pand{} X_2\pand{} \ldots\pand{} X_n}(t)\cdot R_{X_I}(t)\cdot R_{X_{II}}(t)\cdots~,
  \end{split}\label{080525-002}
\end{align}
where $f_{X_1\pand{} X_2\pand{} \ldots\pand{} X_n}(t)$ comes from \eqref{080525-001}.
%
%
\paragraph{Failure Probability and Failure Rate}~\\
The failure probabilities and failure rates corresponding to \eqref{080525-001} and \eqref{080525-002} are then given by using the generic equations \eqref{080224-001} and \eqref{071213-002}.
%
\subsection{Quantification of Extended Event Sequences}\label{080817-001}
Extended event sequences and extended {MCSS} include at least one extended core event.
They are, therefore, a mixture of a Boolean and a temporal logic expression.
In their logical statement extended {MCSS} combine several real {MCSS} and thus cover several event sequences, see chapter \ref{080817-010}.

All extended {MCSS} may be omitted that include at least one SAND connection; for independent events, these do not contribute probabilistically to the event probabilities.
\paragraph{Extended MCSS with One Extended Core Event}~\\
Let
\begin{align}\begin{split}
    & X_1 \pand{} \ldots \pand{} X_{k-1} \pand{} X_k \pand{} X_{k+1} \pand{} \ldots \pand{} X_{n-1} \pand{} X_n \qquad \text{with}
        \\
    & X_k  \ist  X_{k,1} \booland{} X_{k,2} \booland{} \ldots X_{k,r}
  \end{split}  \label{080822_001}
\end{align}
be an extended {MCSS} with one extended core event ($w=1$) at position $k$ within the PAND chain, and let the extended core event consist of $r$ basic events that are AND connected.

Using \eqref{eq:09}, the failure frequency for $X_k(t)$ is then given as
\begin{align}
    f_{X_k}(t) & \ist
          \sum_{i \ist 1}^{r}\left(f_{k,i}(t)\cdot\prod_{{j \ist 1;j\neq
      i}}^{r}  \left(F_{k,j}(t)\right)\right)
              ~. \label{080822_002}
\end{align}
Event sequences (and thus {MCSS}, too) must not include the same basic event more than once, as stated by the laws of contradiction in \eqref{080216-005} for normal and \eqref{090317003} for extended temporal expressions.

All events in an (extended) MCSS are thus mutually independent; therefore, the failure frequency of an extended core event may be calculated independently from the rest of the expression and using \eqref{080822_002}.
It is then inserted into the overall failure frequency of the extended {MCSS}:
\begin{align} \begin{split}
    f_{\mathit{MCSS}}(t)  \ist  & f_{X_1 \pand{} \ldots \pand{} X_{k-1} \pand{} X_k \pand{} X_{k+1} \pand{} \ldots \pand{} X_{n-1} \pand{} X_n}(t) {}=  \\
         \ist  &
        f_{X_n}(t)
                   \int\limits_0^t f_{X_{n-1}}(\tau^{\{1\}})  \cdots
                    \int\limits_0^{\mathclap{\tau^{\{n-(k+1)\}}}}
                                f_{X_{k+1}}(\tau^{\{n-k\}}) \cdot \\
             &    \cdot   \int\limits_0^{\mathclap{\tau^{\{n-k\}}}}  \underbrace{ f_{X_{k}}(\tau^{\{n-(k-1)\}})}_
                              {\text{from \eqref{080822_002}}}
                    \int\limits_0^{\mathclap{\tau^{\{n-({k-1})\}}}}  f_{X_(k-1)}(\tau^{\{n-(k-2)\}}) \cdots
                    \int\limits_0^{\mathclap{\tau^{\{n-1\}}}}  f_{X_{1}}(\tau^{\{n\}}) \cdot \\
        & \qquad   \cdot   \D\tau^{\{n\}} \cdot \D\tau^{\{n-1)\}} \cdots \D\tau^{\{n-({k-1})\}} \cdots \D\tau^{\{n-({k+1}))\}}
              \cdots \D\tau^{\{1\}} ~.
\end{split}
\label{080822_005}
\end{align}
\paragraph{Extended MCSS with Several Extended Core Events}~\\
In case of extended {MCSS} with more than one extended core event, thus $w>1$, 
\begin{enumerate}
 \item the $f_{X_{k}}(t)$ are calculated according to \eqref{080822_002} for each $k\in\{1,2,\ldots,w\}$, and
  \item the resulting $w$ failure frequencies are then inserted into the overall failure frequency of the extended {MCSS}; this is the same as in case of $w=1$ from \eqref{080822_005}.
\end{enumerate}
\paragraph{MCSS with Negated Events}~\\
The probabilities of negated events that are part of extended {MCSS} are considered in analogy to \eqref{080525-002}.
%
\paragraph{Failure Probability and Failure Rate}~\\
An extended MCSS' failure probability is given using \eqref{080224-001} by integrating over \eqref{080822_005}; the corresponding failure rate is then given by \eqref{071213-002}.
%
\subsection{Quantification of the Temporal Failure Function on TOP Level}\label{080810-001}
{MCSS} resulting from the method in chapter \ref{chap080809-001} are mutually exclusive (disjoint).

Therefore, the TOP event's failure probability and failure frequency is given by \eqref{080110-002} and \eqref{080110-003} and is the simple sum of the probabilistic contributions of the disjoint {MCSS}:
\begin{align}
	F_{\mathit{TOP}}(t)\ist & \sum\limits_{i \ist 1}^\xi F_{\mathit{MCSS}_i}(t)\label{080224-002}~,\\
	f_{\mathit{TOP}}(t)\ist & \sum\limits_{i \ist 1}^\xi f_{\mathit{MCSS}_i}(t)\label{080224-003}~.
\end{align}

The parameters of the disjoint {MCSS} come
\begin{itemize}
 \item from chapter \ref{080222-004} in case of normal {MCSS} and
 \item from chapter \ref{080817-001} in case of extended {MCSS}.
\end{itemize}
%
%
\section{Reducing the Computing Time}\label{080724-010}
The calculatory effort necessary for the multiple integrals in \eqref{080525-001}, \eqref{080525-002}, and \eqref{080822_005} is high; this is especially true for complex temporal fault trees and their complex failure functions.
This is not helpful to the TFTA's declared goal to faciliate modelling of event sequences for large and complex systems 

The following chapter therefore presents an approximatory approach to the calculaion of failure probabilities, failure frequencies, and MCSS in order to significantly reduce the calculatory effort.
Essential prerequesites to this approximation are
\begin{itemize}
 \item constant failure rates of all basic events, i.e.\ exponentially distributed failure probabilities, and 
  \item ``small enough'' failure probabilities and failure rates, i.e.\ 
  the ``small value assumption'' from \eqref{071213-003} must be valid that $\lambda\,t\ll 1$ and thus $f(t){}\approx{}\lambda(t)$; in a safety context this is usually a given. 
\end{itemize}
%
\subsection{Temporal Terms in MCSS Format}\label{080813-002}
First, temporal expressions in {MCSS} form are discussed; they result e.g.\ from qualitative transformations of a {TFTA} according to chapter \ref{chap080809-001}.
%
%
%
\paragraph{MCSS Without Negated Events}~\\
The failure probability and failure rate of {MCSS} without negated events, which include at least one SAND, is always zero according to the discussion following page \pageref{090121001}.

Therefore, the quantification is again based on {MCSS} without negated events as shown in \eqref{080216-985}.
The corresponding failure probability is given by integration over \eqref{080525-001} which yields
\begin{align}
  F_{\mathit{MCSS}}(t)   \ist  & F_{X_1\pand{} X_2\pand{} \ldots\pand{} X_n}(t) \ist
        \int\limits_0^t f_{X_1\pand{} X_2\pand{} \ldots\pand{} X_n}(\tau) \cdot \D \tau \nonumber  {}=\\
   \ist  &
      \int\limits_0^t  f_{X_n}(\tau)\cdot
            \int\limits_0^\tau f_{X_{n-1}}(\tau^{\{1\}})\cdots
                  \int\limits_0^{\mathclap{\tau^{\{n-2\}}}} f_{X_{2}}(\tau^{\{n-1\}})\cdot
                  \int\limits_0^{\mathclap{\tau^{\{n-1\}}}}
                              f_{X_1}(\tau^{\{n\}})\cdot  \label{080813-001} \\
      &\qquad\qquad\qquad\qquad\qquad\qquad\qquad\qquad\cdot \D \tau^{\{n\}} \cdot
                  \D \tau^{\{n-1\}}\cdots\D \tau^{\{1\}} \cdot \D \tau ~.
\nonumber
\end{align}
With a total of $n$ basic events that constitute an {MCSS}, each {MCSS} represents exactly one event sequence of the $n!$ possible permutations.
The probability that all $n$ events included in an {MCSS} have occurred at time $t$ is given by \eqref{eq:07} for the case that no event sequences are distinguished; this yields
\begin{align}
  &F_{X_1 \booland{} X_2 \booland{} \ldots \booland{} X_n}(t) \ist
        F_{X_1}(t)\cdot F_{X_2}(t)\cdots F_{X_n}(t) \ist
        \prod_{\mathclap{i \ist 1}}^n F_{X_i}(t)  ~.
  \label{080813-005}
\end{align}
For exponentially distributed and very small failure rates equation \eqref{071213-003} then allows the approximation that
\begin{align}
  f(t) {}\approx{} & \lambda(t)  \ist  \lambda &&\text{and therefore}
        \label{080819-001}\\
  F(t) {}\approx{} & \lambda\cdot t &&\text{for }\lambda\cdot t\ll 1 ~.  \label{080819-002}
\end{align}
Then, 
\begin{align}
  &F_{X_1 \booland{} X_2 \booland{} \ldots \booland{} X_n}(t) {}\approx{}
        \lambda_{X_1}\,t \cdot \lambda_{X_2}\,t \cdots \lambda_{X_n}\,t  \ist
          \prod_{\mathclap{i \ist 1}}^n \left(\lambda_{X_i}\, t\right)  ~.
  \label{080813-006}
\end{align}
If all $n$ failure rates $\lambda_{X}\ist{}\lambda_{X_1}\ist{}\ldots\ist{}\lambda_{X_n}$ are equal, all $n!$ possible permutations of the event sequences occur with the same probability; thus, for each {MCSS}
\begin{align}
  &F_{\mathit{MCSS}}(t) \ist
        \frac{1}{n!}\prod_{\mathclap{i \ist 1}}^n F_{X_i}(t) {}\approx{}
        \frac{1}{n!}\prod_{\mathclap{i \ist 1}}^n \left(\lambda_{X_i}\, t\right)  ~.
  \label{080813-007}
\end{align}
Equation \eqref{080813-007} is also a generic approximation in case of different failure rates, if the highest of the $n$ failure rates satisfies the condition that
\begin{align}
    & \max \bigl(\lambda_{X_1};\lambda_{X_2};\cdots;\lambda_{X_n}\bigr) \cdot t \ll 1 ~.
\end{align}
Thus, 
\begin{align}
  &F_{\mathit{MCSS}}(t){}\approx{} \frac{1}{n!}\prod_{\mathclap{i \ist 1}}^n \left(\lambda_{X_i}\, t\right)  ~.
  \label{080813-008}
\end{align}

The approximation for an MCSS' failure frequency is provided accordingly.
Let, without restriction to the general case, be $X_n$ the last occurring event in a {MCSS} ẃith $n$ involved events; then
\begin{align}
  f_{\mathit{MCSS}}(t) \ist  & f_{X_1\pand{} X_2\pand{} \ldots\pand{} X_n}(t)  \ist  f_{X_n}(t) \cdot F_{X_1\pand{} X_2\pand{} \ldots\pand{} X_{n-1}}(t) ~;
  \intertext{from this follows with \eqref{080819-001} and \eqref{080819-002} that}
  f_{\mathit{MCSS}}(t){}\approx{} & \frac{1}{(n-1)!} \cdot \lambda_{X_n} \cdot \prod_{\mathclap{i \ist 1}}^{\mathclap{n-1}} \left(\lambda_{X_i}\, t\right) ~.
  \label{080813-009}
\end{align}
%
%
%
\paragraph{{MCSS} with Negated Events}~\\
An approximation for {MCSS} with negated events combines the procedure of chapter \ref{080222-004} -- with the probability of negated events from \eqref{080222-101} -- and the quantification approach to {MCSS} without negated events, as discussed above.
Using \eqref{080813-008} and \eqref{080813-009} this yields
\begin{align}
  \begin{split}
  F_{\mathit{MCSS}}(t)  \ist &
      F_{ (\boolnot X_{I} \booland{} \boolnot X_{II} \cdots ) \booland{} (X_1\pand{} X_2\pand{} \ldots\pand{} X_n )}(t)\cdots {}\approx{}\\
      {}\approx{}& \frac{1}{n!}\cdot\prod_{\mathclap{i \ist 1}}^n \left(\lambda_{X_i}\, t\right) \cdot R_{X_I}(t)\cdot R_{X_{II}}(t)\cdots~,
  \end{split}\label{080813-010}\\
  \begin{split}
  f_{\mathit{MCSS}}(t)  \ist &
      f_{X_1\pand{} X_2\pand{} \ldots\pand{} X_n}(t)\cdot R_{X_I}(t)\cdot R_{X_{II}}(t)\cdots {}\approx{}\\
      {}\approx{}& \frac{1}{(n-1)!} \cdot  \lambda_{X_n}\cdot  \prod_{\mathclap{i \ist 1}}^{\mathclap{n-1}} \left(\lambda_{X_i}\, t\right) \cdot R_{X_I}(t)\cdot R_{X_{II}}(t)\cdots~.
  \end{split}\label{080813-011}
\end{align}
%
\subsection{Temporal Terms in an Extended {MCSS} Format}\label{080813-003}
The assumptions from chapter \ref{080813-002} still hold; specifically, no SAND connections are considered, as they do not contribute probabilistically.

Extending the method with minimized computational effort to extended {MCSS} requires discussing how many normal {MCSS} are covered by an extended {MCSS}.

In a very simple example, the extended {MCSS} $(X_1 \booland{} X_2) \pand{} X_3$ covers two normal {MCSS}, $X_1 \pand{} X_2 \pand{} X_3$ and $X_2 \pand{} X_1 \pand{} X_3$, which are disjoint.
Using \eqref{080813-007}, each of these two normal {MCSS} has a probability of
\begin{align}
  F_{X_1\pand{} X_2\pand{} X_3}(t) {}{}{} \ist {}{}{} & F_{X_2\pand{} X_1\pand{} X_3}(t) {}\approx{}
        \frac{1}{6} F_{X_1} F_{X_2} F_{X_3}~.
  \label{080818-001}
\end{align}
Accordingly,
\begin{align}
  F_{(X_1\booland{} X_2) \pand{} X_3}(t) {{}\approx{}} &
        2 \cdot \frac{1}{6} F_{X_1} F_{X_2} F_{X_3}  \ist  \frac{1}{3} F_{X_1} F_{X_2} F_{X_3} ~.
  \label{080818-002}
\end{align}
All normal {MCSS} that are covered by an extended {MCSS} are mutually exclusive (disjoint) because of the temporal law of completition.
An extended MCSS' failure probability and failure frequency is therefore given as simple sum of the failure probabilities and failure frequencies of the normal {MCSS} that are covered by the extended {MCSS}.

In general and withouth SAND connections, the number $\varUpsilon$ of normal {MCSS} that are covered by an extended {MCSS} dependes 
\begin{itemize}
 \item on $w$, which is the number of extended core events in the extended {MCSS}, and
 \item on $r_{i}$ for each extended core event $i\in \{1,\ldots,w\}$, which is the number of its AND connected basic events, and 
 \item on $k_i$, which is the corresponding extended core event's position in the {MCSS}.
\end{itemize}
Some examples:
\begin{align}
  & (X_1 \booland{} X_2) \pand{} X_3 && \rightarrow \quad w \ist 1 ~;~ r \ist 2 ~;~ k \ist 1 ~,\nonumber\\
  & X_1 \pand{} ( X_2 \booland{} X_3 ) &&\rightarrow \quad w \ist 1 ~;~ r \ist 2 ~;~ k \ist 2 ~,\nonumber\\
  & (X_1 \booland{} X_2) \pand{} (X_3 \booland{} X_4) && \rightarrow \quad w \ist 2 ~;~ r_1 \ist r_2 \ist 2 ~;~ k_1 \ist 1 ~;~ k_2 \ist 3 ~, \nonumber\\
  & X_1 \pand{} ( X_2 \booland{} X_3 \booland{} X_4) && \rightarrow \quad w \ist 1 ~;~ r \ist 3 ~;~ k \ist 2 ~. \nonumber
\end{align}
In the third example it is noteworthy, that $k_2 \ist 3$.
The position of the $i\in\{2,\ldots,w\}$-th core event is calculated including all events; even those events in ``preceding'' core events are considered, i.e.\ events on the left side of the $i$-th extended core event in the {MCSS}.
SAND connections are omitted, though:
\begin{align}
  (X_1 \booland{} X_2) \pand{} (X_3 \booland{} X_4) &  \ist
     \Bigl[ X_1 \pand{} X_2 \pand{} (X_3 \booland{} X_4)  \Bigr] \boolor{} \Bigl[ X_2 \pand{} X_1 \pand{} (X_3 \booland{} X_4)  \Bigr] ~.
\end{align}
The position of the second extended core event is therefore $k_2=3$.

In general, each extended core event $i$ with $r_{i}$ basic events and standing at position $k_{i}$ covers
\begin{align}
  \varUpsilon_i & \ist  \binom{(k_i-1)+(r_i-1)}{(k_i-1)}\cdot r_i!
\end{align}
normal {MCSS}.
This follows from $r_i!$ possible permutations within the extended core event.
For each permutation $(k_i-1)$  preceding events (left of the extended core event) may then hold $(k_i-1)+(r_i-1)$ possible positions, as described in \eqref{080122-001}.

Some examples:
\begin{itemize}
 \item
$  (X_1 \booland{} X_2) \pand{} X_3 \hphantom{{}\pand{} X_4} \qquad\qquad\rightarrow \quad w \ist 1 ~;~ r \ist 2 ~;~ k \ist 1
      \quad \rightarrow \quad \varUpsilon \ist 2~:$ \\
$\rightarrow\qquad X_1 \pand{} X_2 \pand{} X_3 ~,~ X_2 \pand{} X_1 \pand{} X_3 ~. $
\item
$  X_1 \pand{}{}( X_2 \booland{} X_3 ) \hphantom{{}\pand{} X_4} \qquad\qquad \rightarrow \quad w \ist 1 ~;~ r \ist 2 ~;~ k \ist 2
      \quad \rightarrow \quad \varUpsilon \ist 4~:$\\
$ \rightarrow\qquad X_1 \pand{} X_2 \pand{} X_3 ~,~ X_1 \pand{} X_3 \pand{} X_2 ~,~ X_2 \pand{} X_1 \pand{} X_3  ~,~ X_3 \pand{} X_1 \pand{} X_2 ~. $
\item
$ X_1 \pand{} X_2 \pand{} (X_3 \booland{} X_4) \qquad\qquad\rightarrow \quad w \ist 1 ~;~ r \ist 2 ~;~ k \ist 3  \quad \rightarrow \quad \varUpsilon \ist 6~:$\\
$\rightarrow\qquad X_1 \pand{} X_2 \pand{} X_3 \pand{} X_4 ~,~ X_1 \pand{} X_2 \pand{} X_4 \pand{} X_3 ~,~
     X_1 \pand{} X_3 \pand{} X_2 \pand{} X_4 ~,~$\\
$ \hphantom{\rightarrow{}}\qquad   X_1 \pand{} X_4 \pand{} X_2 \pand{} X_3 ~,~
      X_3 \pand{} X_1 \pand{} X_2 \pand{} X_4 ~,~ X_4 \pand{} X_1 \pand{} X_2 \pand{} X_3 ~.$
\end{itemize}
With $w>1$ extended core events the total number of covered permutations is then given as
\begin{align}
  \varUpsilon & \ist  \prod_{\mathclap{i \ist 1}}^w \varUpsilon_i ~.
\end{align}
For example, the extended {MCSS} $(X_1 \booland{} X_2) \pand{} (X_3 \booland{} X_4)$ with $w\!\ist\!2$, $r_1\!\ist\!r_2\!\ist\!2$, $k_1\!\ist\!1$, $k_2\ist3$ covers a total of  $\varUpsilon  \ist  \varUpsilon_1 \cdot \varUpsilon_2  \ist  2\cdot 6  \ist 12{}$ permutations.
\begin{align}
  &\qquad X_1 \pand{} X_2 \pand{} X_3 \pand{} X_4 ~,~ X_1 \pand{} X_2 \pand{} X_4 \pand{} X_3 ~,~
          X_1 \pand{} X_3 \pand{} X_2 \pand{} X_4 ~,~\nonumber\\
  &\qquad  X_1 \pand{} X_4 \pand{} X_2 \pand{} X_3 ~,~
          X_3 \pand{} X_1 \pand{} X_2 \pand{} X_4 ~,~ X_4 \pand{} X_1 \pand{} X_2 \pand{} X_3 ~,~
     \nonumber \\
  &\qquad X_2 \pand{} X_1 \pand{} X_3 \pand{} X_4 ~,~ X_2 \pand{} X_1 \pand{} X_4 \pand{} X_3 ~,~
          X_2 \pand{} X_3 \pand{} X_1 \pand{} X_4 ~,~\nonumber\\
  &\qquad  X_2 \pand{} X_4 \pand{} X_1 \pand{} X_3 ~,~
          X_3 \pand{} X_2 \pand{} X_1 \pand{} X_4 ~,~ X_4 \pand{} X_2 \pand{} X_1 \pand{} X_3 ~.
     \nonumber
\end{align}
In analogy to \eqref{080813-010}, the failure probability of an extended {MCSS} is approximated as 
\begin{align}
  \begin{split}
  F_{\mathit{MCSS}}(t) {}\approx{}&
      \varUpsilon \cdot \frac{1}{n!} \cdot \prod_{\mathclap{i \ist 1}}^n \left(\lambda_{X_i}\, t\right) \cdot R_{X_I}(t)\cdot R_{X_{II}}(t)\cdots~.
  \end{split}\label{080818-991}
\end{align}
In analogy to \eqref{080813-011}, the approximated failure frequency is then given by
\begin{align}
  \begin{split}
   f_{\mathit{MCSS}}(t)  {}\approx{}&
      \varUpsilon \cdot  \frac{1}{(n-1)!} \cdot  \lambda_{X_n}\cdot  \prod_{\mathclap{i \ist 1}}^{\mathclap{n-1}} \left(\lambda_{X_i}\, t\right) \cdot R_{X_I}(t)\cdot R_{X_{II}}(t)\cdots~.
  \end{split}\label{080819-003}
\end{align}
\paragraph{Summary of Chapter \ref{080724-010} \nameref{080724-010} }~\\For constant failure rates and ``small enough'' failure probabilities the probabilities and rates of occurrence of each possible permutation of the events in an {MCSS} do not significantly differ among each other.
The calculation of $F_{\mathit{MCSS}}(t)$ and $f_{\mathit{MCSS}}(t)$ is therefore almost independent of the exact event sequence information.
This is beneficial, as the quantification with exact sequence information requires calculation of multiply nested integrals (see chapter \ref{080522-001}) which is very costly.
On the other hand, the approximation method provided in this chapter allows an estimation of $F_{\mathit{MCSS}}(t)$ and $f_{\mathit{MCSS}}(t)$ solely based on the number of events in an {MCSS} and their respective failure rates, see \eqref{080813-010} and \eqref{080813-011}.
It is not necessary to explicitely take the exact sequence information into consideration.
Extended {MCSS} may also be quantified using this approximation, as shown in \eqref{080818-991} and \eqref{080819-003}.

%
%
%

%
%
%
%
\clearpage
\clearpage
\ifx \printSprueche\undefined
\else
\renewcommand*{\dictumwidth}{.3333\textwidth}

\setchapterpreamble[ur]{%
\dictum[Henry Fielding]{Much may be said on both sides.}
\vspace{3cm}
}
\fi

\chapter{Comparing {TFTA} to Other Dynamic Modelling Approaches}\label{chap080104_005}

%
%
%
%
%
In this chapter the advantages of using the {TFTA} method are demonstrated and discussed; in order to do so, an example system (see chapter \ref{090212001}) is modelled and analyzed  
\begin{itemize}
 \item as conventional Boolean {FTA} in chapter \ref{090212002},
 \item as dynamic fault tree ({DFT}) in chapter \ref{chap080401-011}, and
 \item as markov model in chapter \ref{090212004},
\end{itemize}
and these are then compared with the new {TFTA} approach, see chapter \ref{090212005}.
The comparison models are created and analyzed using the Isograph FaultTree+ tool \cite{Isograph2005FaultTree+}.
%
\section{An Example System}\label{090212001}
\begin{figure}
\centering
\input{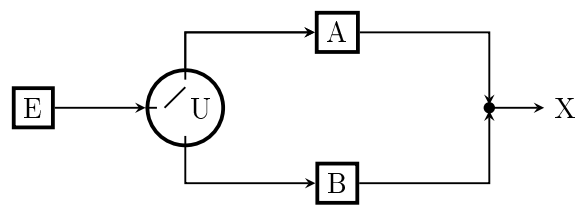}
\caption{An example system used for comparing the Boolean {FTA}, the {DFT}, the markov model, and the {TFTA}.}
\label{090213001}
\end{figure}

An example system from \cite{Reinschke1988} is shown in figure \ref{090213001}.
%
\paragraph{System Description}~\\
The relevant system function of the system under consideration is to supply point $X$ with power.
The power supply $E$ delivers energy via switch $U$ and two redundant paths $A$ and $B$.
First, $U$ is switched to allow energy flow via path $A$.
In case of a fault in $A$, switch $U$ will redirect the energy flow via path $B$ in order to sustain the system function.

The following component faults are considered here:
\begin{itemize}
 \item[E:] $E$ fails to supply energy; the corresponding failure rate is $\lambda_E\ist{}1\cdot10^{-9}\frac{1}{\sjshour}$.
 \item[U:] $U$ fails to switch from $A$ to $B$; the corresponding failure rate is $\lambda_U\ist{}5\cdot10^{-6}\frac{1}{\sjshour}$.
 \item[A:] Internal fault of $A$ inhibiting energy flow; the corresponding failure rate is $\lambda_A\ist{}1\cdot10^{-6}\frac{1}{\sjshour}$.
 \item[B:] Internal fault of $B$ inhibiting energy flow; the corresponding failure rate is $\lambda_B\ist{}1\cdot10^{-6}\frac{1}{\sjshour}$.
\end{itemize}
All components are non-repairable; all failure rates are constant; the mission time is $T_{M}\ist{}400\sjshour$.{}
The failure sequence is relevant because the failure of $U$ before failure of $A$ leads to a system failure, but the failure of $U$ after switching from $A$, i.e.\ after failure of $A$, does not lead to a system failure.
The qualitative and probabilistic results of modelling this example system using the different modelling techniques are listed in tables \ref{090214011} and \ref{090214010} on page \pageref{090214011}.
%
\section{Comparison with the Boolean {FTA}}\label{090212002}
\begin{figure}
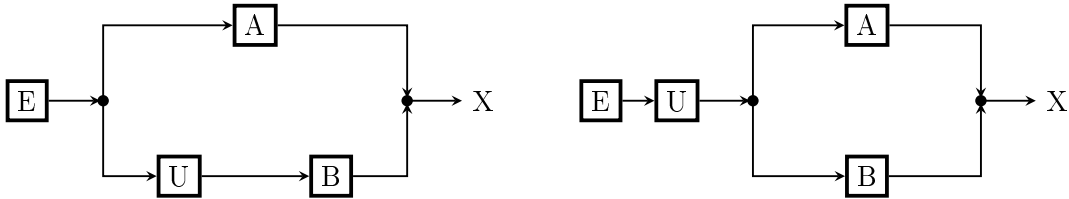

\hfill
\input{pics/anwendung_BSP2_system_bool1}
\hfill
\input{pics/anwendung_BSP2_system_bool2}
\hfill\hfill
\caption{Two possible versions of Boolean approximations of the example system from figure \ref{090213001} as basis for a conventional Boolean {FTA}. The left side is called ``Bool 1'', and the right side is called ``Bool 2''.}
\label{090213002}
\end{figure}

\begin{figure}
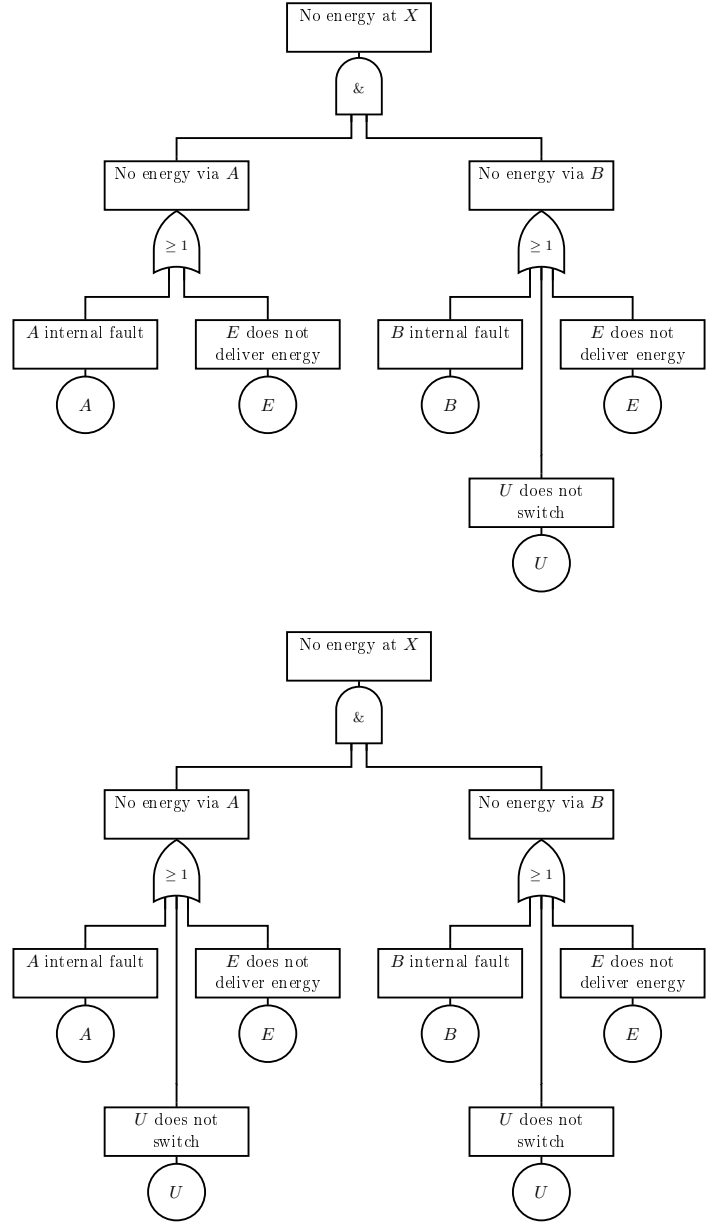

\hfill
\input{pics/anwendung_BSP2_fta_bool1}
\hfill\hfill\\

\hfill
\input{pics/anwendung_BSP2_fta_bool2}
\hfill
\hfill
\caption{Boolean fault trees corresponding to ``Bool 1'' (top) and ``Bool 2'' (bottom).}
\label{090213008}
\end{figure}

The Boolean model is not able to take sequence information into account as relevant for this example system's failure behaviour.
As an approximation to the real system diagram from figure \ref{090213001}, one of the versions from figure \ref{090213002} must be chosen as basis for the Boolean fault tree model \cite{Reinschke1988}.
Figure \ref{090213008} shows the Boolean fault trees corresponding to these two versions, which are called ``Bool 1'' and ``Bool 2''.
%
%
\paragraph{Qualitative and Probabilistic Calculation}~\\
The components' failure probabilites and failure frequencies at the end of the mission time are calculated using \eqref{090213011}; this yields
\begin{align}
  F_A(T_M) & \ist{} 3,9992\cdot10^{-4}~, &
          f_A(T_M) & \ist{} 9,9960 \cdot10^{-7}\tfrac{1}{\sjshour} ~, \\
  F_B(T_M) & \ist{} 3,9992\cdot10^{-4}~,  &
          f_B(T_M) & \ist{} 9,9960 \cdot10^{-7}\tfrac{1}{\sjshour} ~, \\
  F_U(T_M) & \ist{} 1,9960\cdot10^{-3}~,  &
          f_U(T_M) & \ist{} 4,9900 \cdot10^{-6}\tfrac{1}{\sjshour} ~, \\
  F_E(T_M) & \ist{} 4,0000\cdot10^{-7}~,  &
          f_E(T_M) & \ist{} 1,0000 \cdot10^{-9}\tfrac{1}{\sjshour}~.\label{090213030}
\end{align}
The failure function $\varphi$ is 
\begin{align}
  \varphi_{\text{Bool 1}} &\ist{} (A\boolor{}E) \booland{} (B\boolor{}U\boolor{}E) \ist{}  \bigl[A\booland{} B \bigr] \boolor{} \bigl[A\booland{} U \bigr] \boolor{} \bigl[E \bigr]  \qquad \text{and} \label{090213020}\\
  \varphi_{\text{Bool 2}} &\ist{} (A\boolor{}U\boolor{}E) \booland{} (B\boolor{}U\boolor{}E) \ist{} \bigl[A\booland{} B\bigr]  \boolor{} \bigl[U \bigr] \boolor{} \bigl[E  \bigr] \label{090213021}~.
\end{align}
It may be transformed into a disjunctive normal form of mutually exclusive expressions:
\begin{align}
  \varphi_{\text{Bool 1}} &\ist{} \bigl[ A \booland{} B \booland{} \boolnot E \booland{} \boolnot U \bigr] \boolor{} \bigl[A\booland{} U \booland{} \boolnot E \bigr] \boolor{}\bigl[ E \bigr]  \qquad \text{and} \\
  \varphi_{\text{Bool 2}} &\ist{}  \bigl[A\booland{} B \booland{} \boolnot E \booland{} \boolnot U \bigr] \boolor{} \bigl[U \booland{} \boolnot E  \bigr] \boolor{} \bigl[E  \bigr] ~.
\end{align}
Using the failure data from above for quantification, the TOP event provides
\begin{align}
  F_{\text{Bool 1}}(T_M) &\ist{} 1,3587 \cdot10^{-6}~, &  f_{\text{Bool 1}}(T_M) &\ist{} 5,7899 \cdot10^{-9}  \tfrac{1}{\sjshour} \qquad \text{and} \\
 F_{\text{Bool 2}}(T_M)&\ist{} 1,9986 \cdot10^{-3}~, &  f_{\text{Bool 2}}(T_M) &\ist{} 4,9918 \cdot10^{-6}\tfrac{1}{\sjshour} ~.
\end{align}
These results were verified using the FaultTree+ tool.
\paragraph{Discussion on Creating the Fault Trees}~\\
In both cases the fault tree is derived systematically from the system diagrams by following the energy flow backwards through the system, i.e.\ from output $X$ to input $E$.
The modeller needs not think about possible event duplications, as the Boolean logic correctly eliminates those.
\paragraph{Discussion on Results}~\\
Qualitative analysis of the minimal cutsets shows that both cases provide system failures where no real system failure are occurring.
In case of ``Bool 1'' the inaccuracy lies in minimal cutset $\bigl[ A\booland{}U \bigr]$, and in case of ``Bool 2'' the inaccuracy lies in minimal cutset $\bigl[ U \bigr]$.
Therefore, ``Bool 2'' is an especially conservative approximation:
qualitatively, the fault tree has one additional and unnecessary single point failure;
probabilistically, the fault tree yields much higher values for the TOP level failure paramters.
Comparing both Boolean versions it appears clear that ``Bool 1'' is the more realistic model.   
%
\section{Comparison with Dynamic {FTA} ({DFT} Method)}\label{090212003}
Other than the Boolean modell, the {DFT} fault tree uses PAND gates to consider event sequences, that are relevant to the system failure behaviour. 

\begin{figure}
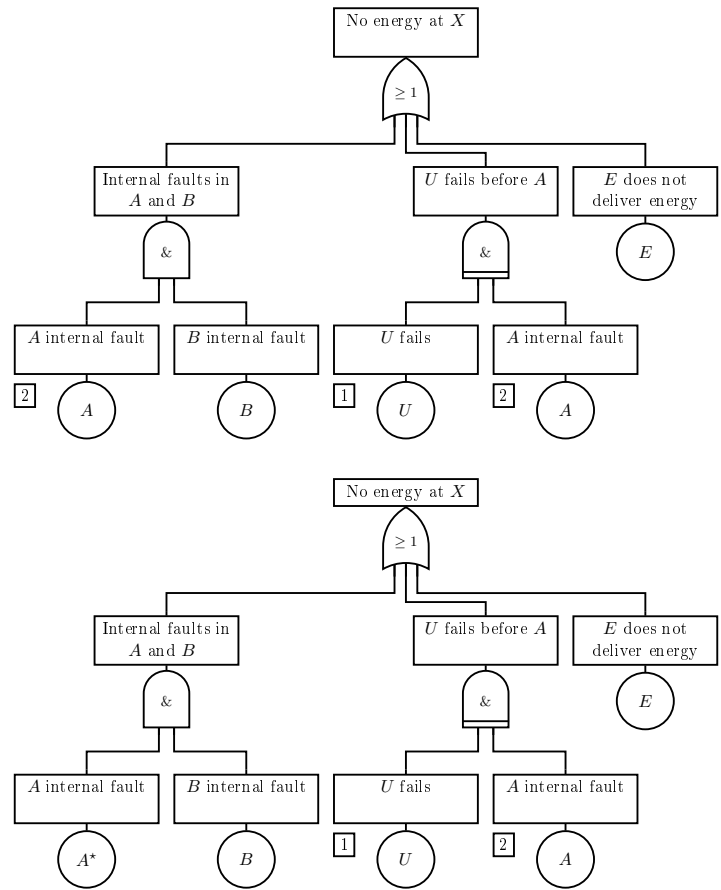

\hfill
\input{pics/anwendung_BSP2_fta_dft2}
\hfill\hfill\\

\hfill
\input{pics/anwendung_BSP2_fta_dft1}
\hfill
\hfill
\caption{{DFT} fault trees in two versions, ``DFT 1'' (top) and ``DFT 2'' (bottom).
In ``DFT 1'' event $A$ is illegally meshed between the Boolean part of the fault tree and the dynamic module.
In ``DFT 2'' the same real world failure of component $A$ is represented by two different basic events $A$ and $A^\star$, which breaks the meshing.
Both versions provide only approximative probabilistic results, though, and do not provide a qualitative analysis that also includes sequence information.}
\label{090213009}
\end{figure}

Figure \ref{090213009} shows two versions ``DFT 1'' and ``DFT 2'' which include a dynamic module, i.e.\ the gate ``$U$ fails before $A$''; this module represents a markov model, see figure \ref{fig:stand_der_technik_markov_petri}. 
For better understanding, in these figures the PAND gate is shown with its original {DFT} symbol from the {DFT} \cite{Dugan1992}, i.e.\ an AND gate with double bars, instead of the {TFTA} PAND gate symbol (an AND gate with horizontal left-to-right arrow). 

In ``DFT 1'' basic event $A$ is meshed between the dynamic module and the Boolean part of the fault tree.
Basic event $A$ has a set sequence flag, and because of the meshing this flag is also set where $A$ is input to the Boolean AND gate ``Internal failure of $A$ and $B$''.
But this sequence information is errornous with regard to event $B$; it provides prababilistically optimistic results, i.e.\ to small failure values.

In ``DFT 2'' this meshing is broken up.
In order to do so, the identical failure of the one component $A$ has to be represented by two different basic events $A$ and $A^\star$.{}
In complex fault trees this method is not feasible, is costly, and complicates clear analysis.
Moreover, the probabilistic results are conservativ as possible intersections between these events are not taken into account.
%
%
\paragraph{Qualitative and Probabilistic Calculation}~\\
At the end of the mission time each component's failure probability and failure frequency equals those of the Boolean model from page \pageref{090213030}.

One feature of the {DFT} approach is that the qualitative calculation of the failure function interprets the PAND gate as conventional AND gate.
This certainly is a sensible conservative approach; as a consequence, though, the event sequence information is not present in the qualitative results.
The failure function $\varphi$ yields
\begin{align}
  \varphi_{\text{DFT 1}} &\ist{} \bigl[A\booland{} B \bigr] \boolor{} \bigl[U\booland{} A \bigr] \boolor{} \bigl[E \bigr]  \qquad \text{and} \label{090213040}\\
  \varphi_{\text{DFT 2}} &\ist{} \bigl[A^\star \booland{} B \bigr] \boolor{} \bigl[U\booland{} A \bigr] \boolor{} \bigl[E \bigr]  \label{090213041}~.
\end{align}
Isograph FaultTree+ provides the following results:
\begin{align}
  F_{\text{DFT 1}}(T_M) &\ist{} 8,7933 \cdot10^{-7}~, &  f_{\text{DFT 1}}(T_M) &\ist{} 3,3946 \cdot10^{-9} \tfrac{1}{\sjshour}  \qquad \text{and} \\
  F_{\text{DFT 2}}(T_M) &\ist{} 9,5962 \cdot10^{-7}~, &  f_{\text{DFT 2}}(T_M) &\ist{} 3,7967 \cdot10^{-9}\tfrac{1}{\sjshour} ~.
\end{align}

\section{Comparison with Markov Diagrams}\label{090212004}
\begin{figure}[H]
\hfill
\input{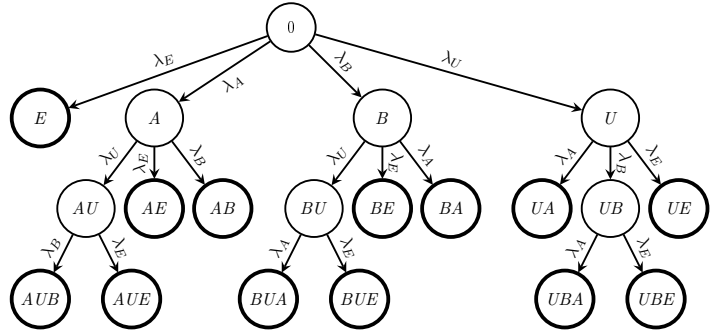}
\hfill
\hfill
\caption{Markov Diagram (and also sequential failure tree) of the example system.
System failure states ``no energy at $X$'' are denoted in bold.}
\label{090214005}
\end{figure}

The example system's markov model in this chapter is used as a reference for probabilistic calculations.
Figure \ref{090214005} shows the corresponding markov diagram, where all system failure states ``no energy at $X$'' are denoted in bold.
Event sequence information between $U$ and $A$ is taken into account. 

Using $T_M\ist{}400\sjshour$, Isograph FaultTree+ provides the following results:
\begin{align}
  F_{\text{MAR}}(T_M) &\ist{} 9,5940 \cdot10^{-7}~, &  f_{\text{MAR}}(T_M) &\ist{} 3,7955 \cdot10^{-9} \tfrac{1}{\sjshour} ~.
\end{align}
This modelling method does not allow for qualitative analysis like the analysis of minimal cutsets.

In comparison to the fault tree modelling methods from above the higher complexity of the markov method is apparent, which in real life inhibits the use of markov methods for analysis of many systems. 
%
%
%
%
%
%
\section{Dynamic {FTA} According to the {TFTA} Method}\label{090212005}
\begin{figure}
\hfill
\input{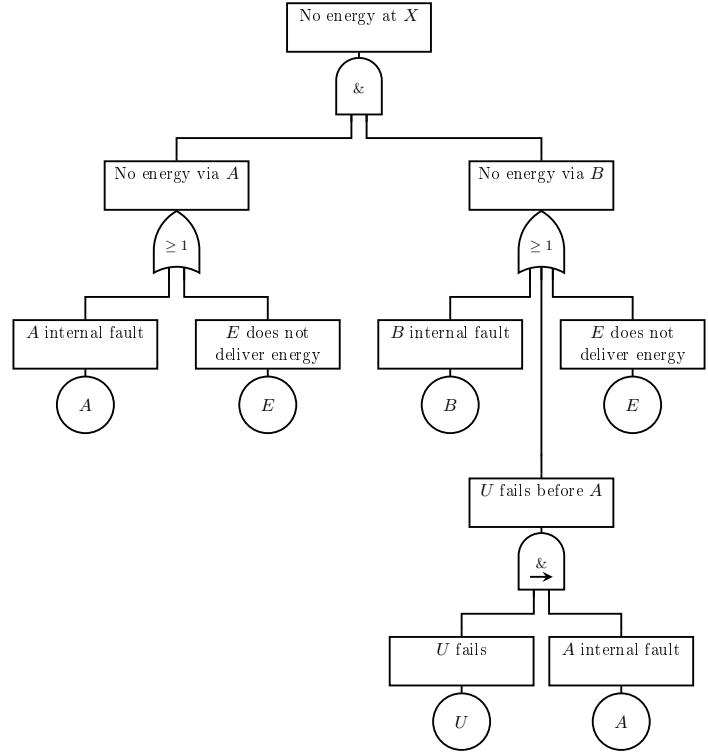}
\hfill
\hfill
\caption{{TFTA} fault tree of the example system.
It correctly takes the meshing of event $A$ into account, as well as the sequence information between events $U$ and $A$, and allows for a ``schematic-driven built-process''.
Probabilistically, the correct results are calculated, too.}
\label{090214001}
\end{figure}

Figure \ref{090214001} shows the temporal TFTA fault tree corresponding to the example system.  
One main benefit of the {TFTA} over the {DFT} approach is the way in which the fault tree structure is built.
Just like the conventional Boolean {FTA}, it is possible to apply a ``schematic-driven built-process'';
i.e.\ to proceed backwards through the system, from its outputs to its inputs, and following the signal paths.
This method is very intuitive as well as very systematic, thus reducing modelling errors.
If there are meshings in the {TFTA} fault tree, they are broken up and resolved by the temporal logic.
The same approach is generally not possible with the {DFT} because of its separated modules.
%
%
\paragraph{Qualitative and Probabilistic Calculation}~\\
The temporal system function of the temporal fault tree shown in figure \ref{090214001} is given as
\begin{align}
 \varpi & \ist{} (A \boolor{} E ) \booland{} \bigl( B \boolor{} E \boolor{} (U \pand{} A ) \bigr) \ist{} \nonumber \\
            & \ist{} \bigl[ A \booland{} B \bigr] \boolor{}
                           \bigl[ A \booland{} E \bigr] \boolor{}
                           \bigl[ A \booland{} (U \pand{} A ) \bigr] \boolor{}
                           \bigl[ E \booland{} B \bigr] \boolor{}
                           \bigl[ E \bigr] \boolor{}
                           \bigl[ E \booland{} (U \pand{} A ) \bigr] \ist{} \nonumber \\
            & \ist{} \bigl[ A \booland{} B \bigr] \boolor{}
                           \bigl[ U \pand{} A \bigr] \boolor{}
                           \bigl[ E \bigr] ~. \label{090214050}
\end{align}
Its three event sequences are already minimal according to chapter \ref{081018_001}, as
\begin{align*}
   \bigl[ A \booland{} B \bigr] & \isMinimal{} \bigl[ U \pand{} A \bigr] &\text{arnd} & &
   \bigl[ A \booland{} B \bigr] &\isMinimal{}  \bigl[ E \bigr] &\text{and} & &
   \bigl[ E \bigr]                         &\isMinimal{}  \bigl[ U \pand{} A \bigr]  ~.
\end{align*}
These event sequences are also {MCSS} and thus starting point for further qualitative evaluation.
Qualitative analysis of the {MCSS} shows that the {MCSS} are indeed correctly calculated and do include the sequence information between events $U$ and $A$.
Further qualitative analysis then requires the transformation of the {MCSS} into a mutually exclusive (disjoint) form.
The transformation according to chapter \ref{081018_002} yields an extended {TDNF} with mutually exclusive expressions:
\begin{align}
  \varpi &\ist{} \bigl[ \boolnot E\booland{} (A \booland{} B) \bigr] \boolor{} \bigl[\boolnot B \boolnot E \booland{} (U \pand{} A) \bigr] \boolor{}\bigl[ E \bigr]  \qquad ~. \label{090214015}
\end{align}
Using the components' failure data from page \pageref{090213030}, direct quantification is then possible:
\begin{align}
 F_{{TFTA}}(t) &\ist{} \bigl(1-F_E(t) \bigr)\cdot F_A(t)  \cdot F_B(t) ~ + \nonumber\\
                              &\phantom{\ist{}} + \bigl(1-F_E(t) \bigr) \bigl(1-F_B(t) \bigr) \cdot \int^t_0 F_U(\tau) \cdot f_A(\tau) \cdot \D\,\tau + F_E(t)~,\nonumber\\
   F_{{TFTA}}                       (T_M)& \ist{} 9,5940\cdot 10^{-7} ~, \\
f_{{TFTA}}(t) &\ist{} \bigl(1-F_E(t) \bigr) \cdot f_A(t)  \cdot F_B(t)  + \bigl(1-F_E(t) \bigr)  \cdot F_A(t)  \cdot f_B(t)  ~+ \nonumber\\
                         &\phantom{\ist{}} +  \bigl(1-F_E(t) \bigr)\bigl(1-F_B(t) \bigr) \cdot F_U(t) \cdot f_A(t) + f_E(t) \nonumber\\
  f_{{TFTA}} (T_M)& \ist{} 3,7955\cdot 10^{-9} \tfrac{1}{\sjshour}~.
\end{align}
Comparison with the reference results from the markov model (see chapter \ref{090212004}) shows that the {TFTA} provides exact probabilistic results, too. 
%
%
\paragraph{Approximation}~\\
Instead of using this exact calculation method, the TOP event's failure parameters may also be approximated using the approach with reduced calculatory effort from chapter \ref{080724-010}.

First, this approach is used on the extended {TDNF} of the temporal failure function from \eqref{090214015}; $T_M\ist{}400\sjshour$ then yields
\begin{align}
   F_{{TFTA}}(t) &\approx{} (1-\lambda_E\,t )\cdot \lambda_A\,t \cdot \lambda_B\,t + \frac{1}{2} (1-\lambda_E\,t )(1-\lambda_B\,t )\cdot \lambda_U\,t \cdot \lambda_A\,t + \lambda_E\,t ~, \nonumber\\
   F_{{TFTA}} (T_M)  & \ist{} 9,5984 \cdot 10^{-7} ~, \label{090214070}\\
   f_{{TFTA}}(t) &\approx{} 2(1-\lambda_E\,t )\cdot \lambda_A \cdot \lambda_B\,t + \frac{1}{2} (1-\lambda_E\,t )(1-\lambda_B\,t )\cdot \lambda_U \cdot \lambda_A\,t + \lambda_E  ~, \nonumber\\
   f_{{TFTA}} (T_M) & \ist{} 3,7988 \cdot 10^{-7}  \tfrac{1}{\sjshour}~.\label{090214071}
\end{align}
Further significant simplification is possible using \eqref{090214050} instead of the temporal failure function from \eqref{090214015}.
The quantification of \eqref{090214050} yields
\begin{align}
   F_{{TFTA}}(t) &\approx{} \lambda_A\,t \cdot \lambda_B\,t + \frac{1}{2} \,\lambda_U\,t \cdot \lambda_A\,t + \lambda_E\,t~, \nonumber\\
F_{{TFTA}} (T_M) &\approx{} 9,6000 \cdot 10^{-7} ~, \label{090214072}\\
   f_{{TFTA}}(t) &\approx{} 2 \, \lambda_A \cdot \lambda_B\,t + \frac{1}{2} \, \lambda_U \cdot \lambda_A\,t + \lambda_E  ~, \nonumber\\
f_{{TFTA}} (T_M) &\approx{}3,8000 \cdot 10^{-7}  \tfrac{1}{\sjshour}~. \label{090214073}
\end{align}
On the one hand, it is no longer necessary to carry out the - possibly very costly - transformation into a disjoint form.
On the other hand, the results are conservative approximations, usually good enough for at least a first assessment during a multi-step analysis.
\begin{table}[h]
\centering
\begin{tabular}{llllllll} \toprule
Cutsets/Sequ. & Bool 1  & Bool 2  & {DFT} 1  & {DFT} 2  & Markov  & {TFTA} \\
\midrule \addlinespace[7pt]
%
%
%
%
1.	&	$E$	&	$E$	&	$E$	&	$E$	&		--	&	$E$	\\
2.	&	$A\booland{}U$	&	$U$	&	$U\booland{}A$	&	$U\booland{}A$	&		--	&	$U\pand{}A$	\\
3.	&	$A\booland{}B$	&	$A\booland{}B$	&	$A\booland{}B$	&	$A^\star\booland{}B$	&		--	&	$A\booland{}B$	\\

\bottomrule
\end{tabular}
\caption{Comparison of the qualitative results of the different modelling methods for the example system from chapter \ref{090212001}.
Minimal cutsets of the ``Bool \ldots'' and the ``DFT \ldots'' methods do not include event sequence information.
As a consequence, there are failure combinations, that do not lead to a real life system failure, but are taken for system failures.
The results of ``Bool 2'' and ``DFT 2'' deviate the most from the correct results represented by the {MCSS} of the {TFTA}.
The markov model does not provide comparable qualitative results at all.}
\label{090214011}
\end{table}
%
%
\begin{table}[h]
\centering
\begin{tabular}{llll} \toprule
Method & $F(T_M)~~[.]=1$ & $f(T_M)~~[.]\ist{}\frac{1}{\sjshour}$ & $\lambda(T_M)~~[.]\ist{}\frac{1}{\sjshour}$  \\
\midrule \addlinespace[7pt]
%
%
%
%
Bool 1	&	$1,3587\cdot 10^{-6}$	&	$5,7899\cdot 10^{-9}$	&	$5,7899\cdot 10^{-9}$			\\
Bool 2	&	$1,9986\cdot 10^{-3}$	&	$4,9918\cdot 10^{-6}$	&	$5,0019\cdot 10^{-6}$		\\
\ac{DFT} 1	&	$8,7933\cdot 10^{-7}$	&	$3,3946\cdot 10^{-9}$	&	$3,3946\cdot 10^{-9}$			\\
\ac{DFT} 2 	&	$9,5962\cdot 10^{-7}$	&	$3,7967\cdot 10^{-9}$	&	$3,7967\cdot 10^{-9}$			\\
Markov	&	$9,5940\cdot 10^{-7}$	&	$3,7955\cdot 10^{-9}$	&	$3,7955\cdot 10^{-9}$			\\
\ac{TFTA}	&	$9,5940\cdot 10^{-7}$	&	$3,7955\cdot 10^{-9}$	&	$3,7955\cdot 10^{-9}$			\\
\ac{TFTA} (Approx. 1)	&	$9,5984\cdot 10^{-7}$	&	$3,7988\cdot 10^{-9}$	&	$3,7988\cdot 10^{-9}$			\\
\ac{TFTA} (Approx. 2)	&	$9,6000\cdot 10^{-7}$	&	$3,8000\cdot 10^{-9}$	&	$3,8000\cdot 10^{-9}$		\\

\bottomrule
\end{tabular}
\caption{Comparison of the probabilistic results of the different modelling methods from chapters \ref{090212002} to \ref{090212005}; the mission time is set to $T_M\ist{}400\sjshour$.{}
Obviously, the Boolean results are comparably conservative.
The markov model is used as reference.
The {TFTA} provides identical and therefore correct results, too.
The last two rows show the results of the approximations of the probabilistic {TFTA}.
``Approx 1'' corresponds to the temporal failure function after it is transformed into a mutually exclusive (disjoint) form;
``Approx 2'' corresponds to the temporal failure function in a {TDNF} before being transformed into disjoint minterms, see \eqref{090214070} to \eqref{090214073}.\vspace{5mm}}
\label{090214010}
\end{table}

\section{Summarizing the Results}\label{090212007}
The side-by-side comparision of Boolean {FTA}, {DFT} approach, markov model, and the new {TFTA} approach shows that the {TFTA} combines and surpasses the benefits of the other more conventional methods.

The {TFTA} adopts the basic steps of creating fault trees from the Boolean {FTA}.
Most notably, it allows for a ``schematic-driven built-process''; this assures a very systematic design and few modelling errors.
The basic steps of the fault tree's qualitative and probabilistic evaluation are also very similar between both methods.
The failure function is qualitatively simplified into a minimal {DNF}; in a next step, this is then further qualitatively analysed, as well as transformed into mutually exclusive (disjoint) sub-expressions; these are then quantified.
Other than the Boolean {FTA}, the {TFTA} takes relevant event sequence information into account qualitatively as well as probabilistically.

Looking at the qualitative results, only the {TFTA} provides minimal combinations of component failures that lead to a system failure, which include event sequence information, see table \ref{090214011}.
The {DFT} and the conventional {FTA} provide minimal cutsets without event sequence information instead.
Furthermore, the necessity of modules in the {DFT} is noteworthy:
Meshing of events between Boolean and dynamic modules may lead to modelling errors which are difficult to discern and thus distort the qualitative results.
It is possible to break such meshing up by using several ``copied'' events for one real world failure event; this provides good probabilistic approximations, but it reduces the significance and reliability of the qualitative results, as they contain nonsensical or even impossible event combinations.

The {TFTA} also provides correct probabilistic failure parameters at TOP event level; this is shown by comparison with the morkov referrence, see table \ref{090214010}. 
The Boolean models are comparatively conservative.
The {DFT} provides correct results only for those fault trees that do not have events meshed between Boolean and dynamic modules.
If such meshings are necessary, then the {DFT} usually provides optimistic (i.e.\ too small) probabilistic results. 

The {TFTA} is also well suited for a multi-step approach of modelling, where the results' accuracy is improved step by step. The TFTA's approach with reduced calculatory effort provides conservative probabilistic approximations as well as, qualitatively, the minimal failure sequences. 
\renewcommand*{\dictumwidth}{.3333\textwidth}
\clearpage
\ifx \printSprueche\undefined
\else
\setchapterpreamble[ur]{%
\dictum[Erich Fromm]{Insight separated from practice remains ineffective.}
\vspace{3cm}
}
\fi

\chapter{{TFTA} Analysis of an Automotive ECU Architecture}
\label{chap080401_006}
This chapter uses the TFTA method on a more complex example and shows how TFTA may be applied to more than academic minimal examples.
%
%
%
%
%
%
%
%
%
\section{The Example System}\label{chap080401-050}
\begin{figure}[H]
 \centering
\input{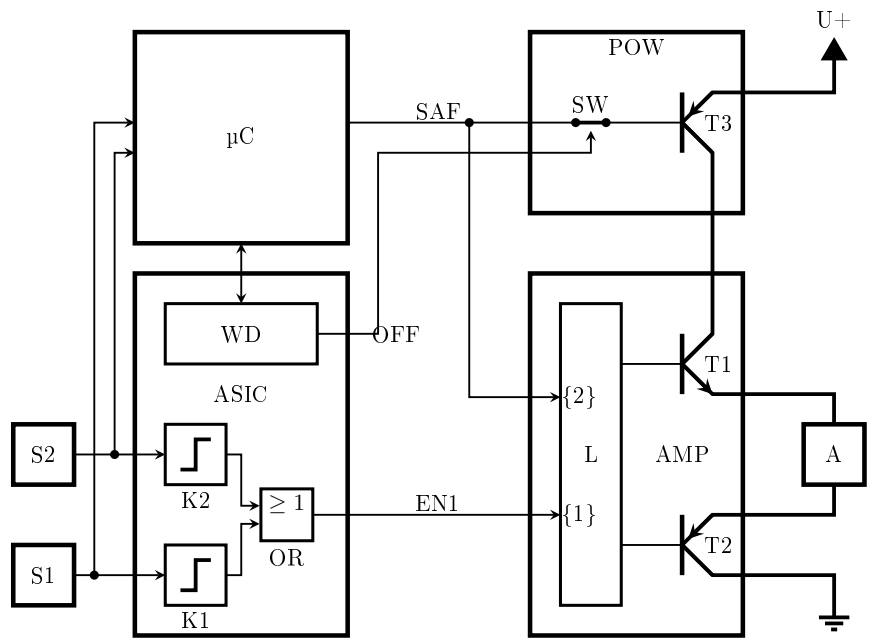}
\caption{A real world example systems which is analyzed using the {TFTA} method.}
\label{090203001}
\end{figure}

The example system in figure \ref{090203001} is an abstraction of a system architecture typically used in the automotive domain for safety critical systems up to SIL~3 according to {\IEC} or ASIL~D according to {\ISO}.

The structure of this chapter:
In chapter \ref{090426001} the temporal fault tree corresponding to the example system is shown.
The qualitative analysis in chapter \ref{090215100} and the probabilistic evaluation in chapter \ref{090215101} are followed by a discussion of the results in chapter \ref{090215102}.
%
%
\subsection{System Description, Safety Goal and Safe State}
\paragraph{Scope}~\\
The example system consists of the components and signals wlisted in table \ref{090207001}.
\begin{table}[H]
\centering
 %
%
%
%
{\small
\begin{tabular}{lll} \toprule
Component  & Subcomponent & Description  \\
\midrule
S1	&		&	sensor 1	\\
S2	&		&	sensor 2	\\
\textmu C	&		&	microcontroller	\\
ASIC	&		&	system-ASIC	\\
	&	WD	&	watchdog for \textmu C	\\
	&	K1	&	comparator 1	\\
 	&	K2	&	comparator 2	\\
	&	OR	&	OR gate	\\
POW	&		&	power switch	\\
	&	SW	&	emergency switch	\\
	&	T3	&	power transistor	\\
AMP	&		&	driver IC	\\
 	&	L	&	logic 	\\
	&	T1	&	high side power stage	\\
	&	T2	&	low side power stage	\\
A	&		&	actuator	\\
\toprule
Signal	& & 	Description			\\
\midrule
EN	& &	enable signal for the logic in the driver IC\\
SAF& & enable signal for power transistor and driver IC\\
OFF	& &	disable/cutoff signal from watchdog		\\
\bottomrule
\end{tabular}
}

\caption{Components and signals of the example system in figure \ref{090203001}}
\label{090207001}
\end{table}
%
\paragraph{Functional Description and Safety Concept}~\\
The example system is used to safely activate actuator A based on some sensor information.
The actuator shall be activated, if (and only if) the sensor input shows that some threshold level is exceeded.
If the sensor input is below this threshold, the actuator shall be deactivated.
The system includes several redundancy measures in order to increase its functional safety.

Both sensors S1 and S2 record some physical parameters from the surrounding.
Each sensor sends its data over a separate serial port to microcontroller \textmu C as well as the system ASIC.
The transmission is protected using CRC and alive counters.

Microcontroller \textmu C evaluates the sensor data of both sensors S1 and S2.
If at least one of the sensors' data is below the threshold, output SAF of the \textmu C is deactivated.
If both of the sensors' data are above the threshold, \textmu C activates the power transistor T3 via the SAF signal.
At the same time, \textmu C activates the power stages T1 and T2 in the AMP driver via AMP's enabler input $\{2\}$.
Meanwhile, the microcontroller serves the intelligent watchdog in the system ASIC via an additional bidirectional port.

The system ASIC evaluates the same sensor data as the microcontroller.
It has two hardware comparators K1 and K2.
Comparator K1 evaluates data from sensor S1.
Comparator K2 evaluates data from sensor S2.
If at least one of the hardware comparators detects that the corresponding threshold is exceeded, it activates its output EN.
Additionally, the system ASIC includes an intelligent watchdog WD.
Using several mechanisms, the watchdog monitors that the \textmu C hardware is operable and the operating system and the application software on \textmu C run correctly. 
This is accomplished, first, using a window watchdog triggered by special waypoints within the program software;
second, WD queries \textmu C and monitors the provided answers.
If \textmu C answers too early or too late or provides a wrong answer, WD activates (opens) a separate emergency switch SW via the OFF signal.
If SW is open, T3 is deactivated independently of SAF; the power supply to the power stages and thus to the actuator is interrupted.	

Driver AMP consists of the two power stages T1 and T2 as well as an internal logic L.
L activates the power stages, if (and only if) enable input $\{1\}$ is activated first, and then enable input $\{2\}$ is activated second.
Every other sequence does not activate the power stages. 

Normally, the activation abides the sequence $\{1\}, \{2\}$:
on the one hand, data from S1 and S2 do not occur at exactly the same time, e.g.\ because S1 and S2 are spatially separated.
Then, signal EN will always be activated first, when the first sensor data indicates an exceeding of the threshold.
On the other hand, the software in \textmu C also carries some latency to EN, which leads to an internally delayed activation of SAF.

\paragraph{Safety Goal, Safe State, and Fault Tolerance Time Span}~\\
The system's hazard and risk analysis yields the following safety goal: ``prevent errornous current feed through the actuator''.
The corresponding safety state is ``no current feed through actuator''.
The fault tolerant time span is $0$ seconds, i.e.\ current feeds are considered immediately dangerous and are thus not allowed even for very short times. 
%
%
\subsection{Failures} \label{090429005}
Using the simplification that all connections between components S1, S2, \textmu C, K1, K2, WD, SW, T1, T2, T3, L, and A are ideal and have no faults, the components' failures listed in table \ref{090207002} remain.
The failures' dangerousness depends on their potential to contribute to an infraction of the safety goal.
The listed safety measures prevent a direct infraction of the safety goal by the failures.
\begin{sidewaystable}
\centering
{\tiny
  \begin{tabular}{llllll} \toprule
  Comp. &Nr.&Failure &Failure consequence&Dangerous&Prevention against direct infraction of the safety goal\\
  \midrule \addlinespace[7pt]
  %
%
%
%
S1	
&	1	
&	wrongly provide value above threshold	
&	\textmu C and ASIC recognize activation criterion 
&	yes  
&	A is activated only if second fault in S2 \\
	
&	2	
&	wrongly provide value below threshold
&	\textmu C and ASIC don't recognize activation criterion 
&	no \\
	
&	3	
&	no communication with \textmu C 	
&	\textmu C doesn't recognize activation criterion 
&	no \\
	
&	4	
&	no communication with ASIC	
&	ASIC may enable EN with only S2 
&	no \\

S2	
&	5	
&	wrongly provide value above threshold	
&	\textmu C and ASIC recognize activation criterion  
&	yes  
&	A is activated only if second fault in S1 \\

&	6	
&	wrongly provide value below threshold
&	\textmu C and ASIC don't recognize activation criterion 
&	no \\
	
&	7	
&	no communication with \textmu C 	
&	\textmu C doesn't recognize activation criterion 
&	no \\
	
&	8	
&	no communication with ASIC	
&	ASIC may enable EN with only S1 
&	no \\

\textmu C	
&	9	
&	\textmu C stuck-at failure	
&	\textmu C can't change output SAF 
&	no 
& 	WD detects \textmu C failure and activates cutoff\\

& 
& 
&	and can't serve WD\\
	
&	10	
&	address-, program counter-, or IO-failure 	
&	\textmu C arbitrarily changes SAF output;  
&	yes 
&	WD detects \textmu C failure and activates cutoff\\

& 
& 
&	WD is not correctly served	
&	
&	sequence logic in AMP-L inhibts activation of A\\
	
&	11	
&	input S1 stuck-at	
&	no evaluation of S1 	
&	no	\\

&	12  
&	input S2 stuck-at	
&	no evaluation of S2 	
&	no	\\
	
&	13	
&	data from S1 is wrongly interpreted 
&	\textmu C recognizes activation criterion  
&	yes  
&	A is activated only if second fault in S2 \\
  
& 
&	to be above threshold 
&	\\
	
&	14	
&	data from S1 is wrongly interpreted
&	\textmu C doesn't recognize activation criterion 
&	no \\  
  
& 
& 	to be below threshold
&	\\
	
&	15	
&	data from S2 is wrongly interpreted 
&	\textmu C recognizes activation criterion  
&	yes  
&	A is activated only if second fault in S1 \\
  
& 
&	to be above threshold 
&	\\
	
&	16	
&	data from S2 is wrongly interpreted
&	\textmu C doesn't recognize activation criterion 
&	no \\  
  
& 
& 	to be below threshold
&	\\
	
&	17	
&	no communication with sensors 
&	\textmu C doesn't recognize activation criterion 
&	no \\

WD	
&	18	
&	wrongly not detect \textmu C failure	
&	WD does not activated Signal OFF and POW-SW
&	yes	
&	A is activated only if second fault in AMP, or\\

& 
& 
&	
& 
&	EN occurs before SAF activated\\
	
&	19	
&	wrongly detect \textmu C failure	
&	WD activates signal OFF and cutoff signal, 
&	no	
&	\\

& 
& 
&	no supply of AMP (= safe state) \\

K1	
&	20	
&	wrongly interpret data from S1 as above threshold
&	ASIC activates EN  
&	yes  
&	A is activated only if second fault in \textmu C or AMP\\
	
&	21	
&	wrongly interpret data from S1 as below threshold
& 	ASIC does not activate EN 
&	no  
&   \\
  
K2	
&	22	
&	wrongly interpret data from S2 as above threshold
&	ASIC activates EN  
&	yes  
&	A is activated only if second fault in \textmu C or AMP\\
	
&	23	
&	wrongly interpret data from S2 as below threshold
& 	ASIC does not activate EN
&	no  
&   \\
  
OR	
&	24	
&	wrongly decide to activate without request by ASIC-K1 or K2
& 	ASIC activates EN  
&	yes  
&	A is activated only if second fault in \textmu C or AMP\\
  
&	25  
&	wrongly ignore activation request by ASIC-K1 or K2	
&	ASIC does not activate EN  
&	no  
&	\\

SW	
&	26	
&	open cutoff wrongly without request 
&	no activation of T3 when SAF is activated  
&	no  
&   \\
	
&	27	
&	not open cutoff despite OFF signal	
&	activation of T3	
&	yes	
&	A is activated only if E2 is also wrongly activated, or\\

& 
& 
& 
& 
&	with another multiple point fault in ASIC or AMP	\\

T3	
&	28	
&	swich on without request by SAF	
&	high side power stage T1 is supplied with energy	
&	yes	
&	A is activated only if SAF is also wrongly activated, or\\

& 
& 
& 
& 
& 	with another multiple point fault in ASIC or AMP \\
	
&	29	
&	not switch on despite request by SAF	
&	no energy supply to T1 
&	no	
&	\\

T1	
&	30	
&	switch on without request by AMP-L	
&	T3 is connected to actuator A
&	yes	
&	A is activated only if T2 is also wrongly switched on and\\

& 
& 
& 
& 
&	another multiple point fault in \textmu C or POW  \\
	
&	31	
&	not switch on despite request by AMP-L	
&	A is not supplied with energy
&	no	
&	\\

T2	
&	32	
&	switch on without request by AMP-L	
&	actuator A is connected to ground
&	yes	
&	A is activated only if T1 is also wrongly switched on and\\

& 
& 
& 
& 
&	another multiple point fault in \textmu C or POW  \\
	
&	33	
&	not switch on despite request by AMP-L
&	A has no connection to ground
&	no	
&	\\

L	
&	34	
&	activate AMP-T1 without request
&	see 30 
&	yes 
&	see 30 \\
	
&	35	
&	not activate AMP-T1 despite request 
&	see 31 
&	no \\
	
&	36	
&	activate AMP-T2 without request
&	see 32  
&	yes  
&	see 32  \\
	
&	37	
&	not activate AMP-T2 despite request 
&	see 33 
&	no \\
	
&	38	
&	activate AMP-T1 and T2 without request
&	T3 is connected to actuator A and
&	yes	
&	A is activated only in combination with a\\

& 
& 
&	actuator A is connected to ground 
& 
&	second fault in \textmu C or POW  \\

A	
&	39	
&	no action despite correct energy supply 
& 	safe state 
&	no\\

  \bottomrule\end{tabular}
}
\vspace{-3mm}\caption{Overview over the possible failures of the example system from figure \ref{090203001}.}
\label{090207002}
\end{sidewaystable}
For a dynamic failure analysis two areas of the system are specifically interesting.
First, there is a sequence logic in L, and second the are dangerous failures of WD and SW (numbers 18 and 27 in table \ref{090207002}, respectively) in combination with a failure of the microcontroller.
These failures of the watchdog or switch SW are relevant, if (and only if) at least one of them occurs before failures of \textmu C.
But if \textmu C fails first, while WD as well as SW are operational, i.e.\ have not failed, or have failed, but ``in a safe direction'', it is assumed, that this was detected and thus the system is disabled.
Further dangerous consequences are then ruled out.
Furthermore, dependent failures, and especially common cause failures (CCF), are not considered in this example.

Failures of \textmu C may not be easily attributed to specific hardware faults, as \textmu C's functionality is largely realised in software.
It is assumed, that the different failures of \textmu C -- numbers $9$ to $17$ in table \ref{090207002} -- occur independent from each other.
%
%
\section{Temporal Fault Tree} \label{090426001}
A temporal fault tree for the example system is to be created.
It shall provide evidence that no dangerous single failure leads to a direct infraction of the safety goal; this is called ``single failure resistance''.
Furthermore, an {MCSS} analysis shall provide the most relevant combinations of dangerous failures.
A probabilistic quantification shall then provide evidence that the system's failure rate stays below the threshold as defined for ASIL D in {\ISO}.

\begin{figure}[H]
 \centering
\input{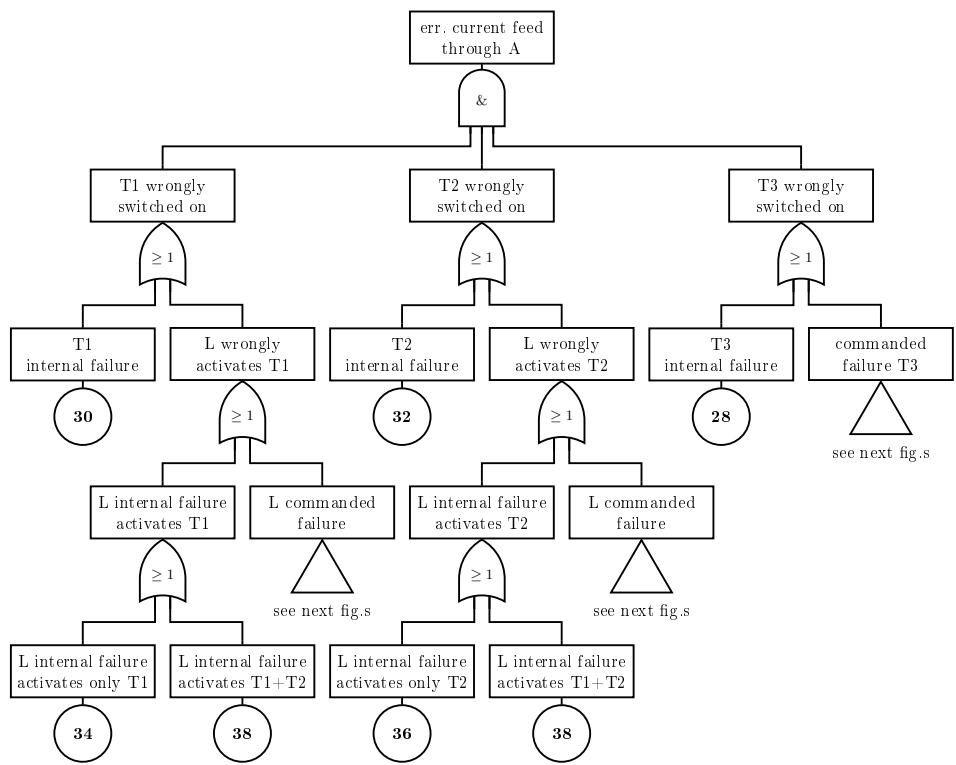}
\caption{Fault tree of the example system from figure \ref{090203001}, part 1.
The basic events' numbers correspond to those in table \ref{090207002}.}
\label{090208001}
\vfill
\end{figure}

The TOP event of the fault tree is the ``infraction of the safety goal'', i.e.\ the ``errornous current feed through the actuator''.
As the system has time-dependencies between its components' failures, it is necessary to use temporal fault tree gates.
Figures \ref{090208001} bis \ref{090208003} show the temporal fault tree for the example system, split into three parts.
The basic events' numbers correspond to those in table \ref{090207002}.
 
In total the temporal fault tree consists of $32$ gates and $34$ basic events.
There are $16$ meshed gates and $18$ meshed basic events.
Two of the gates are PAND gates, which appear three times because of meshings.
These temporal gates represent sub fault trees with ten different basic events and ten different gates.  
\begin{figure}[H]
\centering
\input{pics/anwendung_bsp_system_ft2}
\caption{Fault tree of the example system from figure \ref{090203001}, part 2.
}
\label{090208002}
\end{figure}
\begin{figure}[H]
\centering
\input{pics/anwendung_bsp_system_ft3}
\caption{Fault tree of the example system from figure \ref{090203001}, part 3.
}
\label{090208003}
\end{figure}
%
%
\section{Qualitative Analysis of the Temporal Fault Tree}\label{090215100}
%
%
\subsection{Temporal Failure Function}\label{090502005}
The failure function for the TOP event is directly read from the fault tree in figures \ref{090208001} to \ref{090208003}:
\begin{align}
   \varpi&\ist{}  \phantom{\booland{}\vphantom{.}}
        \Biggl(
          X_{30} \boolor{} X_{34} \boolor{} X_{38} \boolor{}
          \biggl[
            \Bigl(
              \underbrace{
                X_{24} \boolor{} X_{20} \boolor{} X_{1} \boolor{} X_{22} \boolor{} X_{5}
              }_A
            \Bigr) \pand{} \nonumber\\
           &\hphantom{\ist{} \booland{}\vphantom{.}\Biggl(\Biggr.} \pand{}
            \Bigl(
               \underbrace{
                     X_{10} \boolor{} [
                       (X_{1} \boolor{} X_{13} ) \booland{} (X_{5} \boolor{} X_{15} ) ]
               }_B
             \Bigr)
          \biggr]
        \Biggr) \booland{} \nonumber\\
       &\hphantom{\ist{}} \booland{}
        \Biggl(
          X_{32} \boolor{} X_{36} \boolor{} X_{38} \boolor{}
          \biggl[ A \pand{} B \biggr]
        \Biggr)\booland{}\nonumber\\
       &\hphantom{\ist{}} \booland{}
        \Biggl( \underbrace{
                 X_{28} \boolor{} [
                       (X_{1} \boolor{} X_{13} ) \booland{} (X_{5} \boolor{} X_{15} ) ] \boolor{} \Bigl[
                   (X_{27} \boolor{} X_{18} ) \pand{} X_{10}
                   \Bigr]
               }_C \Biggr) ~. \label{090502010}
\end{align}
Substitutions $A$, $B$, and $C$ facilitate further simplification:
\begin{align}
\varpi&\ist{}
        \Bigl(
          X_{30} \boolor{} X_{34} \boolor{} X_{38} \boolor{}
          \bigl[ A \pand{} B \bigr]
        \Bigr) \booland{}
        \Bigl(
          X_{32} \boolor{} X_{36} \boolor{} X_{38} \boolor{}
          \bigl[ A \pand{} B \bigr]
        \Bigr) \booland{}
        \Bigl( C \Bigr) \ist{} \nonumber \\
%
&\ist{} \phantom{\boolor{}\vphantom{.}}
        \bigl[ X_{30} \booland{} X_{32} \booland{} C \bigr] \boolor{}
        \bigl[ X_{30} \booland{} X_{36} \booland{} C \bigr] \boolor{}
        \bigl[ X_{30} \booland{} X_{38} \booland{} C \bigr] \boolor{}
        \bigl[ X_{30} \booland{} (A \pand{} B) \booland{} C \bigr] \boolor{}
        \nonumber\\&\hphantom{\ist{}}  \boolor{}
        \bigl[ X_{34} \booland{} X_{32} \booland{} C \bigr] \boolor{}
        \bigl[ X_{34} \booland{} X_{36} \booland{} C \bigr] \boolor{}
        \bigl[ X_{34} \booland{} X_{38} \booland{} C \bigr] \boolor{}
        \bigl[ X_{34} \booland{} (A \pand{} B) \booland{} C \bigr] \boolor{}
        \nonumber\\&\hphantom{\ist{}}  \boolor{}
        \bigl[ X_{38} \booland{} X_{32} \booland{} C \bigr] \boolor{}
        \bigl[ X_{38} \booland{} X_{36} \booland{} C \bigr] \boolor{}
        \bigl[ X_{38} \booland{} X_{38} \booland{} C \bigr] \boolor{}
        \bigl[ X_{38} \booland{} (A \pand{} B) \booland{} C \bigr] \boolor{}
        \nonumber\\&\hphantom{\ist{}}  \boolor{}
        \bigl[ (A \pand{} B) \booland{} X_{32} \booland{} C \bigr] \boolor{}
        \bigl[ (A \pand{} B) \booland{} X_{36} \booland{} C \bigr] \boolor{}
        \bigl[ (A \pand{} B) \booland{} X_{38} \booland{} C \bigr] \boolor{}
        \nonumber\\&\hphantom{\ist{}}  \boolor{}
        \bigl[ (A  \pand{} B) \booland{} (A \pand{} B) \booland{} C \bigr] ~.
\end{align}
Applying the laws of absorption and idempotency yields
\begin{align}
\varpi&\ist{} \phantom{\booland{}\vphantom{.}}
        \bigl[ X_{30} \booland{} X_{32} \booland{} C \bigr] \boolor{}
        \bigl[ X_{30} \booland{} X_{36} \booland{} C \bigr] \boolor{}
        \bigl[ X_{34} \booland{} X_{32} \booland{} C \bigr] \boolor{}
        \nonumber\\
       &\hphantom{\ist{}}  \boolor{}
        \bigl[ X_{34} \booland{} X_{36} \booland{} C \bigr] \boolor{}
        \bigl[ X_{38}  \booland{} C \bigr] \boolor{}
        \bigl[ (A  \pand{} B) \booland{} C \bigr] ~. \label{090420001}
\end{align}

The next chapter transforms the temporal failure function from \eqref{090420001} according to the laws of temporal logic.
The analysis of the resulting {MCSS} of $\varpi$ follows in chapter \ref{090502007}.
%
%
\subsection{Transformation According to the Temporal Logic Rules}\label{090502006}
%
%
%
%
\paragraph{{MCSS} of the First Five Terms in \textbf{\eqref{090420001}}:}~\\
The temporal failure function in \eqref{090420001} has five parts
\begin{align}
        & \bigl[ X_{30} \booland{} X_{32} \booland{} C \bigr] ,
        \bigl[ X_{30} \booland{} X_{36} \booland{} C \bigr] ,
        \bigl[ X_{34} \booland{} X_{32} \booland{} C \bigr],
        \bigl[ X_{34} \booland{} X_{36} \booland{} C \bigr],
        \bigl[ X_{38}  \booland{} C \bigr] \label{090501002}
\end{align}
that have no reference to event $A$.
Basic events $X_{30}, X_{32}, X_{34}, X_{36}, X_{38}$ are not also included in $C$.
If each of these five expressions is combined with the {TDNF} of $C$, i.e.\ 
\begin{align}
  C \ist{} X_{28} \boolor{} (X_{1} X_{5}) \boolor{} (X_{1} X_{15})  \boolor{} (X_{5}X_{13} ) \boolor{} (X_{13} X_{15}) \boolor{} (X_{27} \pand{} X_{10}) \boolor{} (X_{18} \pand{} X_{10})~,
\end{align}
They provide nine different event sequences each, as shown here for the one example with $X_{38} \booland{} C$:
\begin{align}
   X_{38}  \booland{}C  \ist{}
        X_{38}  \booland{} \bigl[ X_{28} & \boolor{} (X_{1} X_{5}) \boolor{} (X_{1} X_{15})  \boolor{} (X_{5}X_{13} ) \boolor{} (X_{13} X_{15}) \boolor{}&\nonumber\\
         & \boolor{} (X_{27} \pand{} X_{10}) \boolor{} (X_{18} \pand{} X_{10}) \bigr ]~.&
\end{align}
Next, this provides five event sequences each, like in 
\begin{align}
        &
        \bigl[ X_{38}   X_{1}   X_{5} \bigr] ,
        \bigl[ X_{38}   X_{1}   X_{15}\bigr] ,
        \bigl[ X_{38}   X_{5}   X_{13} \bigr] ,
        \bigl[ X_{38}   X_{13}   X_{15} \bigr] ,
        \bigl[ X_{28}   X_{38} \bigr] ~. \label{090501901}
\end{align}
Furthermore, there are four additional event sequences (without SAND) from $X_{38} \booland{} (X_{18} \pand{} X_{10}) $ und $X_{38} \booland{} (X_{27} \pand{} X_{10}) $:
\begin{align}
        &
        \bigl[ (X_{18}   X_{38})   \pand{}  X_{10}\bigr] ,
        \bigl[ X_{18}    \pand{}  X_{10} \pand{} X_{38} \bigr] ,
        \bigl[ (X_{27}   X_{38})   \pand{}  X_{10}\bigr] ,
        \bigl[ X_{27}    \pand{}  X_{10} \pand{} X_{38} \bigr] ~. \label{090501501}
\end{align}
In total there are $45$ event sequences, as shown in table \ref{090501034}.
%
\begin{table}[H]
\centering
{\small
\bgroup
\tabcolsep=1pt
  \begin{tabular}{clccl}
\toprule
\multicolumn{5}{c}{(extended) {MCSS} of rank two:}\\
1: & $ X_{28}   X_{38} $\\
\midrule
\multicolumn{5}{c}{ (extended) {MCSS} of rank three:}
\\
1:&
$   X_{28}   X_{30}   X_{32}   $ &$\qquad\qquad\qquad$& 7: &  $  X_{28}   X_{30}   X_{36}  $ \\
2:&
$   X_{28}   X_{32}   X_{34}  $ &~~&  8: &  $  X_{28}   X_{34}   X_{36}   $\\
3:&
$   X_{18} \pand{} X_{10} \pand{} X_{38}  $ &~~& 9: &
$  X_{27} \pand{} X_{10} \pand{} X_{38}   $ \\
4:&
$   (X_{18} X_{38} )\pand{} X_{10}    $ &~~& 10: &
$  (X_{27} X_{38} )\pand{} X_{10}   $ \\
5:&
$   X_{38}   X_{1}   X_{5}  $ &~~& 11: &
$  X_{38}   X_{1}   X_{15}  $ \\
6:&
$   X_{38}   X_{5}   X_{13}  $ &~~& 12: &
$  X_{38}   X_{13}   X_{15}   $\\
\midrule
\multicolumn{5}{c}{(extended) {MCSS} of rank four:}\\
1:&
$   X_{1}   X_{5}    X_{30}   X_{32}   $ &~~& 17: &  $  X_{1}   X_{5}    X_{30}   X_{36}  $ \\
2:&
$   X_{1}   X_{15}   X_{30}   X_{32}   $ &~~& 18: &  $  X_{1}   X_{15}    X_{30}   X_{36}  $ \\
3:&
$   X_{13}   X_{5}    X_{30}   X_{32}   $ &~~& 19: &  $  X_{13}   X_{5}    X_{30}   X_{36}  $ \\
4:&
$   X_{13}   X_{15}    X_{30}   X_{32}   $ &~~& 20: &  $  X_{13}   X_{15}   X_{30}   X_{36}  $ \\
5:&
$   X_{1}   X_{5}    X_{32}   X_{34}  $ &~~&  21: &  $  X_{1}   X_{5}   X_{34}   X_{36}   $\\
6:&
$   X_{1}   X_{15}    X_{32}   X_{34}  $ &~~&  22: &  $  X_{1}   X_{15}    X_{34}   X_{36}   $\\
7:&
$   X_{13}   X_{5}    X_{32}   X_{34}  $ &~~&  23: &  $  X_{13}   X_{5}    X_{34}   X_{36}   $\\
8:&
$   X_{13}   X_{15}    X_{32}   X_{34}  $ &~~&  24: &  $ X_{13}   X_{15} X_{34}   X_{36}   $\\
9:&
$  X_{18}    \pand{}  X_{10} \pand{} (X_{30}   X_{32})    $ &~~& 25: &
$  X_{27}  \pand{}  X_{10} \pand{} (X_{30}   X_{32})   $ \\
10:&
$  X_{18}    \pand{}  X_{10} \pand{} (X_{30}   X_{36})    $ &~~& 26: &
$  X_{27}  \pand{}  X_{10} \pand{} (X_{30}   X_{36})   $ \\
11:&
$  X_{18}    \pand{}  X_{10} \pand{} (X_{32}   X_{34})    $ &~~& 27: &
$  X_{27}  \pand{}  X_{10} \pand{} (X_{32}   X_{34})   $ \\
12:&
$  X_{18}    \pand{}  X_{10} \pand{} (X_{34}   X_{36})    $ &~~& 28: &
$  X_{27}  \pand{}  X_{10} \pand{} (X_{34}   X_{36})   $ \\
13:&
$  (X_{18} X_{30}   X_{32} )\pand{} X_{10}    $ &~~& 29: &  $  (X_{27} X_{30}   X_{32} )\pand{} X_{10}    $ \\
14:&
$   (X_{18} X_{30}   X_{32} )\pand{} X_{10}   $ &~~& 30: &  $  (X_{27} X_{30}   X_{32} )\pand{} X_{10}    $ \\
15:&
$   (X_{18} X_{32}   X_{34} )\pand{} X_{10}    $ &~~& 31: &  $  ( X_{27} X_{32}   X_{34})\pand{} X_{10}    $ \\
16:&
$   (X_{18} X_{34}   X_{36} )\pand{} X_{10}   $ &~~& 32: &  $  (X_{27} X_{34}   X_{36})\pand{} X_{10}    $ \\
\bottomrule
\end{tabular}
\egroup
}
\caption{{MCSS} of ranks two, three, and four, resulting from the first five expressions in \eqref{090420001}. }
\label{090501034}
\end{table}%
%
%
\paragraph{Simplification of $\mathbf{A \pand{} B }$:}~\\
First, $A \pand{} B $ has to be broken apart.
Because of limited space in this thesis, only the first transformational steps are shown, as relevant for understanding the basic concept.
$B$
may be transformed into the following {DNF}:
\begin{align}
  B  & \ist{}
              X_{10} \boolor{}
              (X_{1}X_{5} )\boolor{} (X_{1}X_{15} )\boolor{}
              (X_{5}X_{13}) \boolor{} (X_{13}X_{15})
          \ist{} X_{10} \boolor{} \eta ~.
\end{align}
According to the temporal distributive law for temporal expressions of type I -- see \eqref{080727-005} --, 
\begin{align}
  A \pand{} B & \ist{}
              \bigl[
              \boolnot\eta \booland{}
              (A \pand{} X_{10})
              \bigr]\boolor{}
              \bigl[
              \boolnot X_{10} \booland{}
              (A \pand{} \eta )
              \bigr]\boolor{}
              \bigl[
              A \pand{} \bigl(
              X_{10} \sand{}\eta
              \bigr)
              \bigr] \ist{}
              \nonumber\\
          & \ist{}\hphantom{\boolor{}}
              \bigl[
              \boolnot (
              X_{1}X_{5} \boolor{} X_{1}X_{15} \boolor{}
              X_{5}X_{13} \boolor{} X_{13}X_{15}) \booland{}
              (A \pand{} X_{10})
              \bigr]\boolor{} \nonumber\\
          & \hphantom{\ist{}\vphantom{.}} \boolor{}
              \bigl[
              \boolnot X_{10} \booland{}
              (A \pand{} (
              X_{1}X_{5} \boolor{} X_{1}X_{15} \boolor{}
              X_{5}X_{13} \boolor{} X_{13}X_{15}))
              \bigr]\boolor{}\nonumber\\
          & \hphantom{\ist{}\vphantom{.}} \boolor{}
              \bigl[
              A \pand{} \bigl(
              X_{10} \sand{} (
              X_{1}X_{5} \boolor{} X_{1}X_{15} \boolor{}
              X_{5}X_{13} \boolor{} X_{13}X_{15})
              \bigr)
              \bigr] \ist{} \nonumber\\
              & \ist{}
              \eta_1 \boolor{} \eta_2 \boolor{} \eta_3 \label{090418001}
              ~.
\end{align}
Expression $\eta_1 $ may then easily be transformed into a {TDNF}:
\begin{align}
  \eta_1
          & \ist{} \hphantom{\boolor{}}
              \boolnot (
              X_{1}X_{5} \boolor{} X_{1}X_{15} \boolor{}
              X_{5}X_{13} \boolor{} X_{13}X_{15}) \booland{}
              (A \pand{} X_{10})
              \ist{}
             \nonumber\\
          & \ist{} \hphantom{\boolor{}}
              \bigl[
              (\boolnot X_{1} \boolnot X_{13} ) \booland{}
              (A \pand{} X_{10})
              \bigr]
              \boolor{}
              \bigl[
              (\boolnot X_{5} \boolnot X_{15} ) \booland{}
              (A \pand{} X_{10})
              \bigr] ~.
             \label{090418002}
\end{align}
Expression $\eta_2 $ is more complex and thus is transformed step by step:
\begin{align}
  \eta_2
          & \ist{}\hphantom{\boolor{}}
              \boolnot X_{10} \booland{}
              \Bigl(
              \boolnot (
              X_{1}X_{15} \boolor{}
              X_{5}X_{13} \boolor{} X_{13}X_{15}) \booland{} (A \pand{} (X_{1}X_{5}))
              \Bigr)
              \boolor{}\nonumber\\
          & \hphantom{\ist{}\vphantom{.}} \boolor{}
              \boolnot X_{10} \booland{}
              \Bigl(
              \boolnot (X_{1}X_{5}) \booland{} (A \pand{}(
              X_{1}X_{15} \boolor{}
              X_{5}X_{13} \boolor{} X_{13}X_{15}) )
              \Bigr)
              \boolor{}\nonumber\\
          & \hphantom{\ist{}\vphantom{.}} \boolor{}
              \boolnot X_{10} \booland{}
              \Bigl(
              A \pand{} \bigl(
              (X_{1}X_{5}) \sand{} (
              X_{1}X_{15} \boolor{}
              X_{5}X_{13} \boolor{} X_{13}X_{15})
              \bigr)
              \Bigr) \ist{} \nonumber\\
           & \ist{}
              \eta_{2\text{a}} \boolor{} \eta_{2\text{b}} \boolor{}\eta_{2\text{c}}   \label{090418003}
             ~.
\end{align}
The first expression in \eqref{090418003} provides three event sequences --
\begin{align}
  \eta_{2\text{a}}
          & \ist{}\hphantom{\boolor{}}
              \boolnot X_{10} \booland{}
              \bigl[
              ( [ \boolnot X_{1}\boolnot X_{13} ] \boolor{} [ \boolnot X_{5} \boolnot X_{15} ] \boolor{} [ \boolnot X_{13} \boolnot X_{15} ]) \booland{} (A \pand{} (X_{1}X_{5}))
              \bigr] \ist{} \nonumber\\
          & \ist{}\hphantom{\boolor{}}
              \bigl[
              ( \boolnot X_{1}\boolnot X_{10} \boolnot X_{13} ) \booland{} (A \pand{} (X_{1}X_{5}))
              \bigr] \boolor{}
              \nonumber\\
          & \hphantom{\ist{}\vphantom{.}} \boolor{}
              \bigl[
              ( \boolnot X_{5} \boolnot X_{10} \boolnot X_{15} ) \booland{} (A \pand{} (X_{1}X_{5}))
              \bigr] \boolor{}
              \nonumber\\
          & \hphantom{\ist{}\vphantom{.}} \boolor{}
              \bigl[
              ( \boolnot X_{10} \boolnot X_{13} \boolnot X_{15} ) \booland{} (A \pand{} (X_{1}X_{5}))
              \bigr]
             ~\text{--}
\end{align}
but only the third of these does not yield $\False$, if rules \eqref{09010930} and \eqref{09010931} are applied.

Therefore, 
\begin{align}
  \eta_{2\text{a}}
          & \ist{}
              \bigl[
              ( \boolnot X_{10} \boolnot X_{13} \boolnot X_{15} ) \booland{} (A \pand{} (X_{1}X_{5}))
              \bigr]
             ~.\label{090418005}
\end{align}
The second expression in \eqref{090418003} itself provides three expressions:
\begin{align}
  \eta_{2\text{b}}
          & \ist{}\hphantom{\boolor{}}
              (\boolnot X_{10} \boolnot (X_{1}X_{5})) \booland{}
              \Bigl(
              \boolnot (
              X_{5}X_{13} \boolor{} X_{13}X_{15}) \booland{}
              (A \pand{}(X_{1}X_{15}))
              \Bigr) \boolor{} \nonumber\\
          & \hphantom{\ist{}\vphantom{.}} \boolor{}
              (\boolnot X_{10} \boolnot  (X_{1}X_{5})) \booland{}
              \Bigl(
              \boolnot (X_{1}X_{15}) \booland{}
              (A \pand{} (
              X_{5}X_{13} \boolor{} X_{13}X_{15}) )
              \Bigr) \boolor{} \nonumber\\
          & \hphantom{\ist{}\vphantom{.}} \boolor{}
              (\boolnot X_{10} \boolnot  (X_{1}X_{5})) \booland{}
              \Bigl(
              A \pand{} ((X_{1}X_{15}) \sand{}
              (
              X_{5}X_{13} \boolor{} X_{13}X_{15}) )
              \Bigr) \ist{} \nonumber\\
          & \ist{}
              \eta_{2\text{b}1} \boolor{} \eta_{2\text{b}2} \boolor{}\eta_{2\text{b}3}  \label{090418006}~.
\end{align}
Using the rules in \eqref{09010930} and \eqref{09010931} on
\begin{align}
  \eta_{2\text{b}1}
          & \ist{}\hphantom{\boolor{}}
              (\boolnot X_{10} \boolnot (X_{1}X_{5})) \booland{}
              \bigl[
              ( \boolnot X_{13} \boolor{} [ \boolnot X_{5}\boolnot X_{15} ] ) \booland{} (A \pand{} (X_{1}X_{15}))
              \bigr] \ist{} \nonumber\\
          & \ist{}\hphantom{\boolor{}}
              \bigl[
              ( \boolnot X_{1}\boolnot X_{10} \boolnot X_{13} ) \booland{} (A \pand{}(X_{1}X_{15}))
              \bigr] \boolor{}
              \nonumber\\
          & \hphantom{\ist{}\vphantom{.}} \boolor{}
              \bigl[
              ( \boolnot X_{5} \boolnot X_{10} \boolnot X_{13} ) \booland{} (A \pand{} (X_{1}X_{15}))
              \bigr] \boolor{}
              \nonumber\\
          & \hphantom{\ist{}\vphantom{.}} \boolor{}
              \bigl[
              ( \boolnot X_{1} \boolnot X_{5} \boolnot X_{10} \boolnot X_{15} ) \booland{} (A \pand{} (X_{1}X_{15}))
              \bigr] \boolor{}
              \nonumber\\
          & \hphantom{\ist{}\vphantom{.}} \boolor{}
              \bigl[
              ( \boolnot X_{5} \boolnot X_{10} \boolnot X_{15} ) \booland{} (A \pand{} (X_{1}X_{15}))
              \bigr]
             \label{090418078}
\end{align}
leaves only 
\begin{align}
  \eta_{2\text{b}1}
          & \ist{}
              \bigl[
              ( \boolnot X_{5} \boolnot X_{10} \boolnot X_{13} ) \booland{} (A \pand{} (X_{1}X_{15}))
              \bigr]~.
             \label{090418008}
\end{align}

The second part of \eqref{090418006} again provides three expressions, i.e.\
\begin{align}
  \eta_{2\text{b}2}
          & \ist{}\hphantom{\boolor{}}
              (\boolnot X_{10} \booland{} (\boolnot X_{1} \boolor{} [ \boolnot X_{5}\boolnot X_{15} ] ) ) \booland{}
              \Bigl(
              \boolnot (X_{13}X_{15}) \booland{}
              (A \pand{} (
              X_{5}X_{13}) )
              \Bigr) \boolor{} \nonumber\\
          & \hphantom{\ist{}\vphantom{.}} \boolor{}
              (\boolnot X_{10} \booland{} (\boolnot X_{1} \boolor{} [ \boolnot X_{5}\boolnot X_{15} ]) ) \booland{}
              \Bigl(
              \boolnot( X_{5}X_{13} ) \booland{} (A \pand{} (X_{13}X_{15})   )
              \Bigr) \boolor{} \nonumber\\
          & \hphantom{\ist{}\vphantom{.}} \boolor{}
              (\boolnot X_{10} \booland{} (\boolnot X_{1} \boolor{}  [ \boolnot X_{5}\boolnot X_{15} ] ) ) \booland{}
              \Bigl(
              A \pand{} ( ( X_{5}X_{13} ) \sand{} (X_{13}X_{15})   )
              \Bigr)  \ist{} \nonumber\\
           & \ist{}
              \eta_{2\text{b}2\text{a}} \boolor{} \eta_{2\text{b}2\text{b}} \boolor{}\eta_{2\text{b}2\text{c}} \label{090418009}
              ~.
\end{align}
Because of rules \eqref{09010930} and \eqref{09010931}, the first of these expressions may be simplified to 
\begin{align}
  \eta_{2\text{b}2\text{a}}
          & \ist{}\hphantom{\boolor{}}
              \bigl[
              \boolnot X_{10} \booland{} (\boolnot X_{1} \boolor{} [ \boolnot X_{5}\boolnot X_{15} ] ) \booland{} \boolnot (X_{13}X_{15})
              \bigr]\booland{}
              (A \pand{} ( X_{5}X_{13}) )
              \ist{} \nonumber\\
          & \ist{}\hphantom{\boolor{}}
              (\boolnot X_{1} \boolnot X_{10} \boolnot X_{13 } ) \booland{}
              (A \pand{} ( X_{5}X_{13}) )
              \boolor{} \nonumber\\
          & \hphantom{\ist{}\vphantom{.}} \boolor{}
              (\boolnot X_{1} \boolnot X_{10} \boolnot X_{15 } ) \booland{}
              (A \pand{} ( X_{5}X_{13}) )
              \boolor{} \nonumber\\
          & \hphantom{\ist{}\vphantom{.}} \boolor{}
              (\boolnot X_{5} \boolnot X_{10} \boolnot X_{15} ) \booland{}
              (A \pand{} ( X_{5}X_{13}) )
              \boolor{} \ist{} \nonumber\\
          & \ist{}
              \bigl[ (\boolnot X_{1} \boolnot X_{10} \boolnot X_{15 } ) \booland{}
              (A \pand{} ( X_{5}X_{13}) ) \bigr]
             ~.\label{090418010}
\end{align}

The same steps repeated for the second expression yield 
\begin{align}
  \eta_{2\text{b}2\text{b}}
          & \ist{}
              \bigl[
              \boolnot X_{10} \booland{} (\boolnot X_{1} \boolor{}  [ \boolnot X_{5}\boolnot X_{15} ] ) \booland{} \boolnot (X_{5}X_{13})
              \bigr] \booland{}
              (A \pand{} ( X_{13}X_{15} ) )
              \ist{} \nonumber\\
          & \ist{}
              \bigl[(\boolnot X_{1} \boolnot X_{5} \boolnot X_{10 } ) \booland{}
              (A \pand{} (X_{13}X_{15} ) ) \bigr]
             ~.\label{090429002}
\end{align}

Because of
\begin{align}
    ( X_{5}X_{13} ) \sand{} (X_{13}X_{15})  & \ist{} \hphantom{\boolor{}}
                \bigl[ ( X_{5}X_{15} ) \pand{} X_{13} \bigr] \boolor{} \bigl[ X_{13} \pand{} ( X_{5} \sand{} X_{15} )  \bigr] \boolor{}
                \bigl[ X_{15} \pand{} ( X_{5} \sand{} X_{13} )  \bigr] \boolor{} \nonumber \\
          & \hphantom{\ist{}\vphantom{.}} \boolor{}
                \bigl[ X_{5} \pand{} ( X_{13} \sand{} X_{15} )  \bigr] \boolor{} \bigl[ X_{5} \sand{} X_{13} \sand{} X_{15} \bigr]
\end{align}
the third expression in \eqref{090418009} provides
\begin{align}
  \eta_{2\text{b}2\text{c}}
          & \ist{}\hphantom{\boolor{}}
              ( \boolnot X_{1}  \boolnot X_{10} ) \booland{}
              \Bigl(
              A \pand{} \bigl[ ( X_{5}X_{15} ) \pand{} X_{13} \bigr] \boolor{} A \pand{} \bigl[ X_{13} \pand{} ( X_{5} \sand{} X_{15} )  \bigr] \boolor{} \Bigr. \label{090429003} \\
          & \hphantom{\ist{}\vphantom{.}} \boolor{} \Bigl.
                 A \pand{} \bigl[ X_{15} \pand{} ( X_{5} \sand{} X_{13} )  \bigr] \boolor{}
                 A \pand{} \bigl[ X_{5} \pand{} ( X_{13} \sand{} X_{15} )  \bigr] \boolor{} A \pand{} \bigl[ X_{5} \sand{} X_{13} \sand{} X_{15} \bigr]
              \Bigr)
             ~,\nonumber
\end{align}
but only event sequence $( \boolnot X_{1}  \boolnot X_{10} ) \booland{}\bigl[(A \booland{} X_{5} \booland{} X_{15} ) \pand{} X_{13} \bigr]$ is free of SANDs.
Therefore, only this one event sequence is taken into account, as in this example dependent failures are not considered, see chapter \ref{090429005}.

Inserting \eqref{090429003} and \eqref{090429002} and \eqref{090418010} into \eqref{090418009} provieds three event sequences
\begin{align}
  \eta_{2\text{b}2}
          & \ist\hphantom{\boolor{}}
              \bigl[ (\boolnot X_{1} \boolnot X_{10} \boolnot X_{15 } ) \booland{}
              (A \pand{} ( X_{5}X_{13}) ) \bigr] \boolor{}
              \nonumber\\
          & \hphantom{\ist{}\vphantom{.}} \boolor{}
              \bigl[(\boolnot X_{1} \boolnot X_{5} \boolnot X_{10 } ) \booland{}
              (A \pand{} (X_{13}X_{15} ) ) \bigr] \boolor{}
              \nonumber\\
          & \hphantom{\ist{}\vphantom{.}} \boolor{}
              \bigl[
              ( \boolnot X_{1}  \boolnot X_{10} ) \booland{} ( (A \booland{} X_{5} \booland{} X_{15} ) \pand{} X_{13} )\bigr]
             ~.\label{090418020}
\end{align}
The third expression from \eqref{090418006} is still open.
Using the same steps, it may be simplified to
\begin{align}
  \eta_{2\text{b}3}
          & \ist{}\hphantom{\boolor{}}
              (\boolnot X_{10} \boolnot  (X_{1}X_{5})) \booland{}
              \Bigl(
              A \pand{} ((X_{1}X_{15}) \sand{}
              ( X_{5}X_{13} \boolor{} X_{13}X_{15}) )
              \Bigr) \ist{}
               \nonumber\\
          & \ist{}\hphantom{\boolor{}}
              (\boolnot X_{10} \boolnot  (X_{1}X_{5})) \booland{}
              \Bigl( \boolnot( X_{13}X_{15} ) \booland{} (
              A \pand{} ((X_{1}X_{15}) \sand{}
              ( X_{5}X_{13} )) )
              \Bigr)  \boolor{} \nonumber\\
          & \hphantom{\ist{}\vphantom{.}} \boolor{}
              (\boolnot X_{10} \boolnot  (X_{1}X_{5})) \booland{}
              \Bigl( \boolnot( X_{5}X_{13} ) \booland{} (
              A \pand{} ((X_{1}X_{15}) \sand{}
              ( X_{13}X_{15} )) )
              \Bigr)  \boolor{} \nonumber\\
          & \hphantom{\ist{}\vphantom{.}} \boolor{}
              (\boolnot X_{10} \boolnot  (X_{1}X_{5})) \booland{}
              \Bigl(
              A \pand{} ((X_{1}X_{15}) \sand{}
              ( X_{5}X_{13} ) \sand{} ( X_{13}X_{15} ) )
              \Bigr) .\nonumber
\end{align}
Applying rules \eqref{09010930} and \eqref{09010931} provides a simplified $\eta_{2\text{b}3}$:
\begin{align}
          \eta_{2\text{b}3}
          & \ist{}\hphantom{\boolor{}}
              \False  \boolor{}
              \bigl[
              (\boolnot X_{10} \boolnot  (X_{1}X_{5}) \boolnot( X_{5}X_{13} ) ) \booland{} (
              (A \booland{} X_{1} \booland{} X_{13} ) \pand{} X_{15}
              ) \bigr] \boolor{} \False
              \ist{} \nonumber\\
          & \ist{}\hphantom{\boolor{}}
              \bigl[
              (\boolnot X_{5} \boolnot  X_{10}) \booland{} (
              (A \booland{} X_{1} \booland{} X_{13} ) \pand{} X_{15}
              ) \bigr]
          ~.
        \label{090429020}
\end{align}

The results in \eqref{090429020} and \eqref{090418020} and \eqref{090418008} are inserted into \eqref{090418006}, which provides the five event sequences (again withouth SANDs) of 
$\eta_{2\text{b}} $.

Transformation of expressions $\eta_{2\text{c}}$ and $\eta_{2}$ and $\eta_{3}$ is carried out analogously to the detailled steps from above.
This is not described explicitely.

Expression $\eta_{2\text{c}}$ from \eqref{090418003} provides two expressions (again without SAND):
\begin{align}
  \eta_{2\text{c}}
          & \ist{}\hphantom{\boolor{}}
              \bigl[
              (\boolnot X_{10} \boolnot  X_{15}) \booland{} (
              (A \booland{} X_{1} \booland{} X_{13} ) \pand{} X_{5}
              ) \bigr]
              \boolor{} \nonumber\\
          & \hphantom{\ist{}\vphantom{.}} \boolor{}
              \bigl[
              (\boolnot X_{10} \boolnot  X_{13}) \booland{} (
              (A \booland{} X_{5} \booland{} X_{15} ) \pand{} X_{1}
              ) \bigr]
          ~.
        \label{0904306601}
\end{align}
Together with \eqref{090418005} and \eqref{090418006} $\eta_{2}$ therefore yields eigth event sequences (without SAND).

Then, expression $\eta_{3}$ provides only event sequences with at least one SAND and is therefore not considered further.

In total, $A \pand{} B$ therefore yields two event sequences without SAND from $\eta_{1}$, see \eqref{090418002}, and eigth event sequences from $\eta_{2}$:
\begin{align}
  A \pand{} B
          & \ist{}\hphantom{\boolor{}}
              \bigl[
              (\boolnot X_{1} \boolnot X_{13} ) \booland{}
              (A \pand{} X_{10})
              \bigr]
              \boolor{}
              && \langle\text{ES1}\rangle
              \nonumber\\
          & \hphantom{\ist{}\vphantom{.}} \boolor{}
              \bigl[
              (\boolnot X_{5} \boolnot X_{15} ) \booland{}
              (A \pand{} X_{10})
              \bigr]
              \boolor{}
              && \langle\text{ES2}\rangle
              \nonumber\\
          & \hphantom{\ist{}\vphantom{.}} \boolor{}
              \bigl[
              ( \boolnot X_{10} \boolnot X_{13} \boolnot X_{15} ) \booland{} (A \pand{} (X_{1}X_{5}))
              \bigr]
              \boolor{}
              && \langle\text{ES3}\rangle
              \nonumber\\
          & \hphantom{\ist{}\vphantom{.}} \boolor{}
               \bigl[
              ( \boolnot X_{5} \boolnot X_{10} \boolnot X_{13} ) \booland{} (A \pand{} (X_{1}X_{15}))
              \bigr]
              \boolor{}
              && \langle\text{ES4}\rangle
              \nonumber\\
          & \hphantom{\ist{}\vphantom{.}} \boolor{}
              \bigl[ (\boolnot X_{1} \boolnot X_{10} \boolnot X_{15 } ) \booland{}
              (A \pand{} ( X_{5}X_{13}) ) \bigr] \boolor{}
              && \langle\text{ES5}\rangle
              \nonumber\\
          & \hphantom{\ist{}\vphantom{.}} \boolor{}
              \bigl[(\boolnot X_{1} \boolnot X_{5} \boolnot X_{10 } ) \booland{}
              (A \pand{} (X_{13}X_{15} ) ) \bigr] \boolor{}
              && \langle\text{ES6}\rangle
              \nonumber\\
          & \hphantom{\ist{}\vphantom{.}} \boolor{}
              \bigl[
              ( \boolnot X_{1}  \boolnot X_{10} ) \booland{} ( (A \booland{} X_{5} \booland{} X_{15} )  \pand{} X_{13} )\bigr]
              \boolor{}
              && \langle\text{ES7}\rangle
              \nonumber\\
          & \hphantom{\ist{}\vphantom{.}} \boolor{}
              \bigl[
              (\boolnot X_{5} \boolnot  X_{10}) \booland{} (
              (A \booland{} X_{1} \booland{} X_{13} )  \pand{} X_{15}
              ) \bigr]
              \boolor
              && \langle\text{ES8}\rangle
              \nonumber\\
          & \hphantom{\ist{}\vphantom{.}} \boolor{}
              \bigl[
              (\boolnot X_{10} \boolnot  X_{15}) \booland{} (
              (A \booland{} X_{1} \booland{} X_{13} )  \pand{} X_{5}
              ) \bigr]
              \boolor{}
              && \langle\text{ES9}\rangle
              \nonumber\\
          & \hphantom{\ist{}\vphantom{.}} \boolor{}
              \bigl[
              (\boolnot X_{10} \boolnot  X_{13}) \booland{} (
              (A \booland{} X_{5} \booland{} X_{15} )  \pand{} X_{1}
              ) \bigr]           ~.
              && \langle\text{ES10}\rangle
        \label{090430001}
\end{align}
Below, identifiers $\langle\text{ES1}\rangle$ to $\langle\text{ES10}\rangle$ are used as a reference to the respective event sequence.
The transformation of $A$ is done using the temporal distributive law for temporal expressions of type II according to \eqref{080728-002}.
Applying \eqref{090430001} and further simplification then yields $28$ different event sequences for $A\pand{}B$.
\begin{align}
  ( X_{1} \boolor{} &X_{5} \boolor{} X_{20} \boolor{} X_{22} \boolor{}  X_{24} )  \pand{} B \ist{} \ldots \ist{} \nonumber\\
          & \ist{}\hphantom{\boolor{}}
              \bigl[
              (\boolnot X_{1} \boolnot X_{13} ) \booland{}
              (
              ( X_{5} \boolor{} X_{20} \boolor{} X_{22} \boolor{}  X_{24} )
              \pand{} X_{10})
              \bigr]
              \boolor{}&& \langle\text{from ES1}\rangle\nonumber\\
          & \hphantom{\ist{}\vphantom{.}} \boolor{}
              \bigl[
              (\boolnot X_{5} \boolnot X_{15} ) \booland{}
              (
              ( X_{1} \boolor{} X_{20} \boolor{} X_{22} \boolor{}  X_{24} )
              \pand{} X_{10})
              \bigr]
              \boolor{}&& \langle\text{from ES2}\rangle\nonumber\\
          & \hphantom{\ist{}\vphantom{.}} \boolor{}
              \bigl[
              ( \boolnot X_{10} \boolnot X_{13} \boolnot X_{15} ) \booland{} (
              (X_{20} \boolor{} X_{22} \boolor{}  X_{24} )
              \pand{} (X_{1}X_{5}))
              \bigr]
              \boolor{}&& \langle\text{from ES3}\rangle\nonumber\\
          & \hphantom{\ist{}\vphantom{.}} \boolor{}
              \bigl[
              ( \boolnot X_{10} \boolnot X_{13} \boolnot X_{15} ) \booland{}
              ( X_{1} \pand{} X_{5} )
              \bigr]
              \boolor{}&& \langle\text{from ES3}\rangle\nonumber\\
          & \hphantom{\ist{}\vphantom{.}} \boolor{}
              \bigl[
              ( \boolnot X_{10} \boolnot X_{13} \boolnot X_{15} ) \booland{}
              ( X_{5} \pand{} X_{1} )
              \bigr]
              \boolor{}&& \langle\text{from ES3}\rangle\nonumber\\
          & \hphantom{\ist{}\vphantom{.}} \boolor{}
               \bigl[
              ( \boolnot X_{5} \boolnot X_{10} \boolnot X_{13} ) \booland{} (
              ( X_{20} \boolor{} X_{22} \boolor{}  X_{24} )
              \pand{} (X_{1}X_{15}))
              \bigr]
              \boolor{}&& \langle\text{from ES4}\rangle\nonumber\\
          & \hphantom{\ist{}\vphantom{.}} \boolor{}
               \bigl[
              ( \boolnot X_{5} \boolnot X_{10} \boolnot X_{13} ) \booland{}
              ( X_{1} \pand{} X_{15})
              \bigr]
              \boolor{}&& \langle\text{from ES4}\rangle\nonumber\\
          & \hphantom{\ist{}\vphantom{.}} \boolor{}
              \bigl[ (\boolnot X_{1} \boolnot X_{10} \boolnot X_{15 } ) \booland{}
              (
              ( X_{20} \boolor{} X_{22} \boolor{}  X_{24} )
               \pand{} ( X_{5}X_{13}) ) \bigr] \boolor{} && \langle\text{from ES5}\rangle
              \nonumber\\
          & \hphantom{\ist{}\vphantom{.}} \boolor{}
              \bigl[ (\boolnot X_{1} \boolnot X_{10} \boolnot X_{15 } ) \booland{}
              (
              X_{5}\pand{} X_{13}) \bigr] \boolor{} && \langle\text{from ES5}\rangle
              \nonumber\\
          & \hphantom{\ist{}\vphantom{.}} \boolor{}
              \bigl[(\boolnot X_{1} \boolnot X_{5} \boolnot X_{10 } ) \booland{}
              (
              ( X_{20} \boolor{} X_{22} \boolor{}  X_{24} )
               \pand{} (X_{13}X_{15} ) ) \bigr] \boolor{} && \langle\text{from ES6}\rangle
              \nonumber\\
          & \hphantom{\ist{}\vphantom{.}} \boolor{}
              \bigl[
              ( \boolnot X_{1}  \boolnot X_{10} ) \booland{} ( (
              X_{5} \booland{} X_{15} )  \pand{} X_{13} )\bigr]
              \boolor{}&& \langle\text{from ES7}\rangle\nonumber\\
          & \hphantom{\ist{}\vphantom{.}} \boolor{}
              \bigl[
              (\boolnot X_{5} \boolnot  X_{10}) \booland{}
              (
              ( X_{1}\booland{} X_{13} )  \pand{} X_{15}
              ) \bigr]
              \boolor&& \langle\text{from ES8}\rangle\nonumber\\
          & \hphantom{\ist{}\vphantom{.}} \boolor{}
              \bigl[
              (\boolnot X_{10} \boolnot  X_{15}) \booland{}
              (
              ( X_{1} \booland{} X_{13} )  \pand{} X_{5}
              ) \bigr]
              \boolor{}&& \langle\text{from ES9}\rangle\nonumber\\
          & \hphantom{\ist{}\vphantom{.}} \boolor{}
              \bigl[
              (\boolnot X_{10} \boolnot  X_{13}) \booland{}
              (
              ( X_{5} \booland{} X_{15} )  \pand{} X_{1}
              ) \bigr]
          ~. && \langle\text{from ES10}\rangle
        \label{090430005}
\end{align}
Thus, $A \pand{} B$ alone provides $12$ event sequences of rank two and $16$ event sequences of rank three.
\paragraph{Simplification of $\mathbf{(A  \pand{} B) \booland{} C}$:}~\\
Using the TFTA's temporal logic, the meshing between event $B$ and $C$ in the sixth and last sub-expression of \eqref{090420001} may be solved.

According to \eqref{090502010} $B$ and $C$ are given as
\begin{align}
  B &\ist{}  X_{10} \boolor{} [
                       (X_{1} \boolor{} X_{13} ) \booland{} (X_{5} \boolor{} X_{15} ) ]
        \quad\text{and} \label{090502011} \\
  C &\ist{} X_{28} \boolor{} [
                       (X_{1} \boolor{} X_{13} ) \booland{} (X_{5} \boolor{} X_{15} ) ] \boolor{} [
                   (X_{27} \boolor{} X_{18} ) \pand{}  X_{10} ]\, .\label{090502012}
\end{align}
Further substitution with
\begin{align}
    D  &\ist{}  (X_{1} \boolor{} X_{13} ) \booland{} (X_{5} \boolor{} X_{15} ) \ist{}
                        X_{1} X_{5} \boolor{} X_{1} X_{15}  \boolor{} X_{5}X_{13}  \boolor{} X_{13} X_{15} \label{090502013}
\end{align}
uncovers the relationship between $B$ and $C$:
\begin{align}
    B &\ist{}  X_{10} \boolor{} D   \quad\text{and} \label{090502015} \\
    C &\ist{}  X_{28} \boolor{} D \boolor{} ( (X_{27} \boolor{} X_{18} ) \pand{} X_{10} ) ~. \label{090502014}
\end{align}
Applying \eqref{090502015} and \eqref{090502014} provides
\begin{align}
(A  \pand{} & B) \booland{} C   \ist{}
                              (A  \pand{} B) \booland{} \bigl (
                              X_{28} \boolor{} D \boolor{} ( (X_{27} \boolor{} X_{18} ) \pand{} X_{10} )
                   \bigr) \ist\nonumber\\
                & \ist{}
                              \bigl[ (A  \pand{} B) \booland{} X_{28} \bigr]  \boolor{}
                              \bigl[ (A  \pand{} B) \booland{} D \bigr]  \boolor{}
                              \bigl[ (A  \pand{} B) \booland{} \bigl( (X_{27} \boolor{} X_{18} ) \pand{} X_{10}  \bigr) \bigr] \label{090503001} ~.
\end{align}
The first expression yields (without SAND)
\begin{align}
   (A  \pand{} B) \booland{} X_{28} & \ist{}
                  \big[ A  \pand{} B \pand{} X_{28} \bigr] \boolor{}
                  \bigl[ (A  \booland{} X_{28} ) \pand{} B \bigr]
            \label{0905030021} ~.
\end{align}
The {TDNF} of $(A  \pand{} B) \booland{} X_{28} $ consists of $56$ {MCSS} in total.
$ \big[ A  \pand{} B \pand{} X_{28} \bigr] $ provides $28$ {MCSS}, each similar to those in \eqref{090430005} but extended by an additional $X_{28}$.
$\bigl[ (A  \booland{} X_{28} ) \pand{} B \bigr]$ also provides $28$ {MCSS} similar to those in \eqref{090430005}.
Instead of $A$ the expression $A  \booland{} X_{28} $ is used, respectively.
$24$ of the {MCSS} are of rank three and $32$ of the {MCSS} are of rank four.

The second expression in \eqref{090503001} provides (without SAND)
\begin{align} 
   (A  \pand{} B) \booland{} D & \ist{}
            (A  \pand{} (X_{10} \boolor{} D ) ) \booland{} D \ist{} \ldots \ist{} \nonumber\\
      & \ist{}
                  \big[ \boolnot X_{10} \booland{} (A  \pand{} D ) \bigr] \boolor{}
                  \bigl[ A  \pand{} X_{10} \pand{} D \bigr]
            \label{0905030026} ~.
\end{align}
$ \big[ \boolnot X_{10} \booland{} (A  \pand{} D ) \bigr] $ provides $20$ {MCSS} similar to those in \eqref{090430005}.
As $D$ does not include event $X_{10}$ (other than $B$), the first eight event sequences may be dropped, i.e.\ the first two rows in \eqref{090430005}.
In the other rows the $\boolnot X_{10}$ are also dropped. 
Therefore, 
\begin{align}
  A  \pand{} D \ist{} A  \pand{} B \vphantom{\Bigl)\Bigr(} \text{\Large $\arrowvert$}_{X_{10}\ist{}\False} ~.
\end{align}
For expression $ \big[ \boolnot X_{10} \booland{} (A  \pand{} D ) \bigr] $ only four {MCSS} of rank two and $16$ {MCSS} of rank three remain, see \eqref{0988001}.
\begin{align}
   \boolnot X_{10} \booland{} (A  \pand{} D ) & \ist{}\hphantom{\boolor{}}
              \bigl[
              ( \boolnot X_{10} \boolnot X_{13} \boolnot X_{15} ) \booland{} (
              (X_{20} \boolor{} X_{22} \boolor{}  X_{24} )
              \pand{} (X_{1}X_{5}))
              \bigr]
              \boolor{}\nonumber\\
          & \hphantom{\ist{}\vphantom{.}} \boolor{}
              \bigl[
              ( \boolnot X_{10} \boolnot X_{13} \boolnot X_{15} ) \booland{}
              ( X_{1} \pand{} X_{5} )
              \bigr]
              \boolor{}\nonumber\\
          & \hphantom{\ist{}\vphantom{.}} \boolor{}
              \bigl[
              ( \boolnot X_{10} \boolnot X_{13} \boolnot X_{15} ) \booland{}
              ( X_{5} \pand{} X_{1} )
              \bigr]
              \boolor{}\nonumber\\
          & \hphantom{\ist{}\vphantom{.}} \boolor{}
               \bigl[
              ( \boolnot X_{5} \boolnot X_{10} \boolnot X_{13} ) \booland{} (
              ( X_{20} \boolor{} X_{22} \boolor{}  X_{24} )
              \pand{} (X_{1}X_{15}))
              \bigr]
              \boolor{}\nonumber\\
          & \hphantom{\ist{}\vphantom{.}} \boolor{}
               \bigl[
              ( \boolnot X_{5} \boolnot X_{10} \boolnot X_{13} ) \booland{}
              ( X_{1} \pand{} X_{15})
              \bigr]
              \boolor{}\nonumber\\
          & \hphantom{\ist{}\vphantom{.}} \boolor{}
              \bigl[ (\boolnot X_{1} \boolnot X_{10} \boolnot X_{15 } ) \booland{}
              (
              ( X_{20} \boolor{} X_{22} \boolor{}  X_{24} )
               \pand{} ( X_{5}X_{13}) ) \bigr] \boolor{}
              \nonumber\\
          & \hphantom{\ist{}\vphantom{.}} \boolor{}
              \bigl[ (\boolnot X_{1} \boolnot X_{10} \boolnot X_{15 } ) \booland{}
              (
              X_{5}\pand{} X_{13}) \bigr] \boolor{}
              \nonumber\\
          & \hphantom{\ist{}\vphantom{.}} \boolor{}
              \bigl[(\boolnot X_{1} \boolnot X_{5} \boolnot X_{10 } ) \booland{}
              (
              ( X_{20} \boolor{} X_{22} \boolor{}  X_{24} )
               \pand{} (X_{13}X_{15} ) ) \bigr] \boolor{}
              \nonumber\\
          & \hphantom{\ist{}\vphantom{.}} \boolor{}
              \bigl[
              ( \boolnot X_{1}  \boolnot X_{10} ) \booland{} ( (
              X_{5} \booland{} X_{15} )  \pand{} X_{13} )\bigr]
              \boolor{}\nonumber\\
          & \hphantom{\ist{}\vphantom{.}} \boolor{}
              \bigl[
              (\boolnot X_{5} \boolnot  X_{10}) \booland{}
              (
              ( X_{1}\booland{} X_{13} )  \pand{} X_{15}
              ) \bigr]
              \boolor\nonumber\\
          & \hphantom{\ist{}\vphantom{.}} \boolor{}
              \bigl[
              (\boolnot X_{10} \boolnot  X_{15}) \booland{}
              (
              ( X_{1} \booland{} X_{13} )  \pand{} X_{5}
              ) \bigr]
              \boolor{}\nonumber\\
          & \hphantom{\ist{}\vphantom{.}} \boolor{}
              \bigl[
              (\boolnot X_{10} \boolnot  X_{13}) \booland{}
              (
              ( X_{5} \booland{} X_{15} )  \pand{} X_{1}
              ) \bigr]
          ~.
        \label{0988001}
\end{align}
The expression in \eqref{0988001} provides $20$ {MCSS}.
Because of the additional $X_{10}$, four of those {MCSS} are of rank three and $16$ are of rank four, see \eqref{0988002}.
\begin{align}
   \boolnot X_{10} \booland{} (A  \pand{} D ) & \ist{}\hphantom{\boolor{}}
              \bigl[
              (  \boolnot X_{13} \boolnot X_{15} ) \booland{} (
              (X_{20} \boolor{} X_{22} \boolor{}  X_{24} ) \pand{} X_{10}
              \pand{} (X_{1}X_{5}))
              \bigr]
              \boolor{}\nonumber\\
          & \hphantom{\ist{}\vphantom{.}} \boolor{}
              \bigl[
              ( \boolnot X_{13} \boolnot X_{15} ) \booland{}
              ( X_{1} \pand{} X_{10} \pand{} X_{5} )
              \bigr]
              \boolor{}\nonumber\\
          & \hphantom{\ist{}\vphantom{.}} \boolor{}
              \bigl[
              ( \boolnot X_{10} \boolnot X_{13} \boolnot X_{15} ) \booland{}
              ( X_{5} \pand{} X_{10} \pand{} X_{1} )
              \bigr]
              \boolor{}\nonumber\\
          & \hphantom{\ist{}\vphantom{.}} \boolor{}
               \bigl[
              ( \boolnot X_{5} \boolnot X_{13} ) \booland{} (
              ( X_{20} \boolor{} X_{22} \boolor{}  X_{24} ) \pand{} X_{10}
              \pand{} (X_{1}X_{15}))
              \bigr]
              \boolor{}\nonumber\\
          & \hphantom{\ist{}\vphantom{.}} \boolor{}
               \bigl[
              ( \boolnot X_{5} \boolnot X_{10} \boolnot X_{13} ) \booland{}
              ( X_{1}\pand{} X_{10} \pand{} X_{15})
              \bigr]
              \boolor{}\nonumber\\
          & \hphantom{\ist{}\vphantom{.}} \boolor{}
              \bigl[ (\boolnot X_{1}\boolnot X_{15 } ) \booland{}
              (
              ( X_{20} \boolor{} X_{22} \boolor{}  X_{24} ) \pand{} X_{10}
               \pand{} ( X_{5}X_{13}) ) \bigr] \boolor{}
              \nonumber\\
          & \hphantom{\ist{}\vphantom{.}} \boolor{}
              \bigl[ (\boolnot X_{1} \boolnot X_{10} \boolnot X_{15 } ) \booland{}
              (
              X_{5} \pand{} X_{10} \pand{} X_{13}) \bigr] \boolor{}
              \nonumber\\
          & \hphantom{\ist{}\vphantom{.}} \boolor{}
              \bigl[(\boolnot X_{1} \boolnot X_{5} ) \booland{}
              (
              ( X_{20} \boolor{} X_{22} \boolor{}  X_{24} ) \pand{} X_{10}
               \pand{} (X_{13}X_{15} ) ) \bigr] \boolor{}
              \nonumber\\
          & \hphantom{\ist{}\vphantom{.}} \boolor{}
              \bigl[
              \boolnot X_{1}  \booland{} ( (
              X_{5} \booland{} X_{15} )  \pand{} X_{10} \pand{} X_{13} )\bigr]
              \boolor{}\nonumber\\
          & \hphantom{\ist{}\vphantom{.}} \boolor{}
              \bigl[
              \boolnot X_{5} \booland{}
              (
              ( X_{1}\booland{} X_{13} )  \pand{} X_{10} \pand{} X_{15}
              ) \bigr]
              \boolor\nonumber\\
          & \hphantom{\ist{}\vphantom{.}} \boolor{}
              \bigl[
              \boolnot  X_{15}\booland{}
              (
              ( X_{1} \booland{} X_{13} ) \pand{} X_{10}  \pand{} X_{5}
              ) \bigr]
              \boolor{}\nonumber\\
          & \hphantom{\ist{}\vphantom{.}} \boolor{}
              \bigl[
              \boolnot  X_{13}\booland{}
              (
              ( X_{5} \booland{} X_{15} )   \pand{} X_{10} \pand{} X_{1}
              ) \bigr]
          ~.
        \label{0988002}
\end{align}
The transformation of the third expression $\bigl[ (A  \pand{} B) \booland{} \bigl( (X_{27} \boolor{} X_{18} ) \pand{} X_{10}  \bigr) \bigr]$, see \eqref{090503001}, is best demonstrated separately for each of the event sequences $\langle\text{ES1}\rangle$ to $\langle\text{ES10}\rangle$ in \eqref{090430001}.

$\langle\text{ES1}\rangle$ and $\langle\text{ES2}\rangle$ differ in the relevant events; therefore
\begin{align}
  \langle\text{ES1}\rangle  : & \phantom{\ist{}\vphantom{.}}
              \bigl[ (\boolnot X_{1} \boolnot X_{13} ) \booland{}
              (
              ( X_{5} \boolor{} X_{20} \boolor{} X_{22} \boolor{}  X_{24} )
              \pand{} X_{10}) \bigr]
              \booland{} \bigl( (X_{27} \boolor{} X_{18} ) \pand{} X_{10}  \bigr) \ist{} \nonumber\\
          & \ist (\boolnot X_{1} \boolnot X_{13} ) \booland{}
                    \bigl( ( X_{5} X_{18} ) \boolor{} ( X_{20} X_{18} ) \boolor{} ( X_{22} X_{18} ) \boolor{} ( X_{24} X_{18} ) \boolor{} \nonumber\\
          & \qquad\qquad \boolor{}
                      ( X_{5} X_{27} ) \boolor{} ( X_{20} X_{27} ) \boolor{} ( X_{22} X_{27} ) \boolor{} ( X_{24} X_{27} ) \bigr) \pand{} X_{10}
                                   \qquad\text{and}\nonumber\\
  \langle\text{ES2}\rangle  : & \phantom{\ist{}\vphantom{.}}
               (\boolnot X_{5} \boolnot X_{15} ) \booland{}
                    \bigl( ( X_{1}  X_{18} ) \boolor{} ( X_{20} X_{18} ) \boolor{} ( X_{22} X_{18} ) \boolor{} ( X_{24} X_{18} ) \boolor{} \nonumber\\
          & \qquad\qquad \boolor{}
                      ( X_{1}  X_{27} ) \boolor{} ( X_{20} X_{27} ) \boolor{} ( X_{22} X_{27} ) \boolor{} ( X_{24} X_{27} ) \bigr) \pand{} X_{10}
                                    ~. \label{090345671}
\end{align}
The first part of $\langle\text{ES3}\rangle$ provides
\begin{align}
 \langle\text{ES3}\rangle  : & \phantom{\ist{}\vphantom{.}}
              \bigl[( \boolnot X_{10} \boolnot X_{13} \boolnot X_{15} ) \booland{} (
              (X_{20} \boolor{} X_{22} \boolor{}  X_{24} )
              \pand{} (X_{1}X_{5})) \bigr] \booland{} \bigl( (X_{27} \boolor{} X_{18} ) \pand{} X_{10}  \bigr)~.
\end{align}
Further simplification yields only event sequences of rank five and higher.
These are not further considered, as they are far more improbable than the other {MCSS}, which contribute significantly more.
Such a reduction of the necessary effort is state of the art in conventional {FTA}, too.
The same is true for the simplification of the first part of $\langle\text{ES4}\rangle$ and $\langle\text{ES5}\rangle$, as well as for all of $\langle\text{ES6}\rangle$ to $\langle\text{ES10}\rangle$.

The second part of $\langle\text{ES3}\rangle$ provides four {MCSS} of rank four:
\begin{align}
 \langle\text{ES3}\rangle  :
              \bigl[ ( \boolnot X_{10} &\boolnot X_{13} \boolnot X_{15} ) \booland{}
              ( X_{1} \pand{} X_{5} ) \bigr]
               \booland{} \bigl( (X_{27} \boolor{} X_{18} ) \pand{} X_{10}  \bigr) \ist{} \nonumber\\
           & \ist{}\phantom{\boolor{}\vphantom{.}}
              \bigl[ ( \boolnot X_{13} \boolnot X_{15}  \boolnot X_{27} ) \booland{}
              ( X_{1} \pand{} X_{5} \pand{} X_{18}  \pand{} X_{10} ) \bigr] \boolor{} \nonumber\\
          & \phantom{\ist{}\vphantom{.}}\boolor{}
              \bigl[ ( \boolnot X_{13} \boolnot X_{15}  \boolnot X_{18} ) \booland{}
              ( X_{1} \pand{} X_{5} \pand{} X_{27}  \pand{} X_{10} ) \bigr] \boolor{} \nonumber\\
          & \phantom{\ist{}\vphantom{.}}\boolor{}
              \bigl[ ( \boolnot X_{13} \boolnot X_{15}  \boolnot X_{27} ) \booland{}
              ( (X_{1}X_{18}  ) \pand{} X_{5} \pand{} X_{10} ) \bigr] \boolor{} \nonumber\\
         & \phantom{\ist{}\vphantom{.}}\boolor{}
              \bigl[ ( \boolnot X_{13} \boolnot X_{15}  \boolnot X_{18} ) \booland{}
              ( (X_{1}X_{27}  ) \pand{} X_{5} \pand{} X_{10} ) \bigr] ~. \label{090345672}
\end{align}

Analogously, the third part of $\langle\text{ES3}\rangle$ and the second parts of $\langle\text{ES4}\rangle$ and $\langle\text{ES5}\rangle$ also provide four {MCSS} of rank four, respectively: 
\begin{align}
 \langle\text{ES3}\rangle  :
              \bigl[ ( \boolnot X_{10} &\boolnot X_{13} \boolnot X_{15} ) \booland{}
              ( X_{5} \pand{} X_{1} ) \bigr]
               \booland{} \bigl( (X_{27} \boolor{} X_{18} ) \pand{} X_{10}  \bigr) \ist{} \nonumber\\
           & \ist{}\phantom{\boolor{}\vphantom{.}}
              \bigl[ ( \boolnot X_{13} \boolnot X_{15}  \boolnot X_{27} ) \booland{}
              ( X_{5} \pand{} X_{1} \pand{} X_{18}  \pand{} X_{10} ) \bigr] \boolor{} \nonumber\\
          & \phantom{\ist{}\vphantom{.}}\boolor{}
              \bigl[ ( \boolnot X_{13} \boolnot X_{15}  \boolnot X_{18} ) \booland{}
              ( X_{5} \pand{} X_{1} \pand{} X_{27}  \pand{} X_{10} ) \bigr] \boolor{} \nonumber\\
          & \phantom{\ist{}\vphantom{.}}\boolor{}
              \bigl[ ( \boolnot X_{13} \boolnot X_{15}  \boolnot X_{27} ) \booland{}
              ( (X_{5}X_{18}  ) \pand{} X_{1} \pand{} X_{10} ) \bigr] \boolor{} \nonumber\\
         & \phantom{\ist{}\vphantom{.}}\boolor{}
              \bigl[ ( \boolnot X_{13} \boolnot X_{15}  \boolnot X_{18} ) \booland{}
              ( (X_{5}X_{27}  ) \pand{} X_{1} \pand{} X_{10} ) \bigr] ~.  \label{090345673} \\
 \langle\text{ES4}\rangle  :
              \bigl[ (  \boolnot X_{5} & \boolnot X_{10} \boolnot X_{13} ) \booland{}
              ( X_{1} \pand{} X_{15} ) \bigr]
               \booland{} \bigl( (X_{27} \boolor{} X_{18} ) \pand{} X_{10}  \bigr) \ist{} \nonumber\\
           & \ist{}\phantom{\boolor{}\vphantom{.}}
              \bigl[ ( \boolnot X_{5} \boolnot X_{13}   \boolnot X_{27} ) \booland{}
              ( X_{1} \pand{} X_{15} \pand{} X_{18}  \pand{} X_{10} ) \bigr] \boolor{} \nonumber\\
          & \phantom{\ist{}\vphantom{.}}\boolor{}
              \bigl[ ( \boolnot X_{5} \boolnot X_{13}   \boolnot X_{18} ) \booland{}
              ( X_{1} \pand{} X_{15} \pand{} X_{27}  \pand{} X_{10} ) \bigr] \boolor{} \nonumber\\
          & \phantom{\ist{}\vphantom{.}}\boolor{}
              \bigl[ ( \boolnot X_{5} \boolnot X_{13}  \boolnot X_{27} ) \booland{}
              ( (X_{1}X_{18}  ) \pand{} X_{15} \pand{} X_{10} ) \bigr] \boolor{} \nonumber\\
         & \phantom{\ist{}\vphantom{.}}\boolor{}
              \bigl[ ( \boolnot X_{5} \boolnot X_{13}  \boolnot X_{18} ) \booland{}
              ( (X_{1}X_{27}  ) \pand{} X_{15} \pand{} X_{10} ) \bigr] ~. \label{090345674} \\
 \langle\text{ES5}\rangle  :
              \bigl[ ( \boolnot X_{1} &\boolnot X_{10} \boolnot X_{15} ) \booland{}
              ( X_{5} \pand{} X_{13} ) \bigr]
               \booland{} \bigl( (X_{27} \boolor{} X_{18} ) \pand{} X_{10}  \bigr) \ist{} \nonumber\\
           & \ist{}\phantom{\boolor{}\vphantom{.}}
              \bigl[ ( \boolnot X_{1} \boolnot X_{15}  \boolnot X_{27} ) \booland{}
              ( X_{5} \pand{} X_{13} \pand{} X_{18}  \pand{} X_{10} ) \bigr] \boolor{} \nonumber\\
          & \phantom{\ist{}\vphantom{.}}\boolor{}
              \bigl[ ( \boolnot X_{1} \boolnot X_{15}  \boolnot X_{18} ) \booland{}
              ( X_{5} \pand{} X_{13} \pand{} X_{27}  \pand{} X_{10} ) \bigr] \boolor{} \nonumber\\
          & \phantom{\ist{}\vphantom{.}}\boolor{}
              \bigl[ ( \boolnot X_{1} \boolnot X_{15}  \boolnot X_{27} ) \booland{}
              ( (X_{5}X_{18}  ) \pand{} X_{13} \pand{} X_{10} ) \bigr] \boolor{} \nonumber\\
         & \phantom{\ist{}\vphantom{.}}\boolor{}
              \bigl[ ( \boolnot X_{1} \boolnot X_{15}  \boolnot X_{18} ) \booland{}
              ( (X_{5}X_{27}  ) \pand{} X_{13} \pand{} X_{10} ) \bigr] ~. \label{090345675}
\end{align}


%
%
\subsection{Analyis of the {MCSS} }\label{090502007}
The {MCSS} of the temporal failure function $\varpi$ are derived from the event sequences of the sub-expressions in \eqref{090420001}, which are not necessarily already {MCSS}, i.e.\ there could be intersections and overlaps between these individual expressions.
In general, {MCSS} of smaller rank are those with higher importance.
Therefore, the following discussion focusses on {MCSS} of rank two and three.
\paragraph{Event Sequences of the Resulting Expressions}~\\
The expresions' event sequences of rank two and three are listed in table \ref{090502001}.
They are derived from table \ref{090501034} on page \pageref{090501034} as well as the equations \eqref{0905030021}, \eqref{0988001}, \eqref{0988002}, and \eqref{090345671}.
\begin{table}[H]
\centering
{\small
  \begin{tabular}{clcclccl}
  \toprule
\multicolumn{8}{c}{(extended) event sequences of rank two:}
\\
1:&$  X_{1} \pand{} X_{5} $ &\qquad\qquad&
3:&$    X_{1} \pand{} X_{15}  $  &\qquad\qquad&
5:&$   X_{28} \booland{} X_{38}  $
\\
2:&$    X_{5} \pand{} X_{1}  $ &\qquad\qquad&
4:&$    X_{5} \pand{} X_{13}  $  &\qquad\qquad&
\\
\midrule
\multicolumn{8}{c}{(extended) event sequences of rank three:}
\\
1:&$  X_{28} \booland{} X_{30} \booland{} X_{32}    $ &~~~&
25:&$    (X_{5} \booland{} X_{28}) \pand{} X_{10}         $  &~~~&
49:&\sout{$       X_{1} \pand{} X_{10} \pand{} X_{5}      $}
\\
2:&$  X_{28} \booland{} X_{30} \booland{} X_{36}    $ &~~~&
26:&$     (X_{20} \booland{} X_{28}) \pand{} X_{10}     $  &~~~&
50:&\sout{$       X_{5} \pand{} X_{10} \pand{} X_{1}      $}
\\
3:&$    X_{28} \booland{} X_{32} \booland{} X_{34}  $ &~~~&
27:&$     (X_{22} \booland{} X_{28}) \pand{} X_{10}      $  &~~~&
51:&\sout{$        X_{1} \pand{} X_{10} \pand{} X_{15}    $}
\\
4:&$     X_{28} \booland{} X_{34} \booland{} X_{36}    $ &~~~&
28:&$     (X_{24} \booland{} X_{28}) \pand{} X_{10}     $  &~~~&
52:&\sout{$       X_{5} \pand{} X_{10} \pand{} X_{13}    $}
\\
5:&$        X_{18} \pand{} X_{10} \pand{} X_{38}      $ &~~~&
29:&$     (X_{1} \booland{} X_{28}) \pand{} X_{10}     $  &~~~&
53:&$     (X_{5} \booland{} X_{18}) \pand{} X_{10}        $
\\
6:&$      (X_{18} \booland{} X_{38}) \pand{} X_{10}    $ &~~~&
30:&\uwave{$     (X_{20} \booland{} X_{28}) \pand{} X_{10}     $} &~~~&
54:&$      (X_{18} \booland{} X_{20}) \pand{} X_{10}     $
\\
7:&\sout{$     X_{1} \booland{} X_{5} \booland{} X_{38}      $} &~~~&
31:&\uwave{$     (X_{22} \booland{} X_{28}) \pand{} X_{10}     $}&~~~&
55:&$     (X_{18} \booland{} X_{22}) \pand{} X_{10}    $
\\
8:&$     X_{1} \booland{} X_{15} \booland{} X_{38}     ~\lozenge$ &~~~&
32:&\uwave{$     (X_{24} \booland{} X_{28}) \pand{} X_{10}       $} &~~~&
56:&$      (X_{18} \booland{} X_{24}) \pand{} X_{10}    $
\\
9:&$     X_{5} \booland{} X_{13} \booland{} X_{38}      ~\lozenge$ &~~~&
33:&\sout{$      (X_{1} \booland{} X_{28}) \pand{} X_{5}     $} &~~~&
57:&$       (X_{1} \booland{} X_{18}) \pand{} X_{10}   $
\\
10:&$     X_{13} \booland{} X_{15} \booland{} X_{38}    $ &~~~&
34:&\sout{$     (X_{5} \booland{} X_{28}) \pand{} X_{1}    $} &~~~&
58:&\uwave{$       (X_{18} \booland{} X_{20}) \pand{} X_{10}   $}
\\
11:&$   X_{27} \pand{} X_{10} \pand{} X_{38}        $ &~~~&
35:&\sout{$     (X_{1} \booland{} X_{28}) \pand{} X_{15}    $}  &~~~&
59:&\uwave{$      (X_{18} \booland{} X_{22}) \pand{} X_{10}    $}
\\
12:&$     (X_{27} \booland{} X_{38}) \pand{} X_{10}      $ &~~~&
36:&\sout{$    (X_{5} \booland{} X_{28}) \pand{} X_{13}     $}&~~~&
60:&\uwave{$      (X_{18} \booland{} X_{24}) \pand{} X_{10}    $}
\\
13:&$    X_{5} \pand{} X_{10} \pand{} X_{28}      $ &~~~&
37:&\sout{$   X_{20}  \pand{}    (X_{1} \booland{} X_{5})       $}  &~~~&
61:&$      (X_{5} \booland{} X_{27}) \pand{} X_{10}    $
\\
14:&$    X_{20} \pand{} X_{10} \pand{} X_{28}      $ &~~~&
38:&\sout{$    X_{22}  \pand{}    (X_{1} \booland{} X_{5})     $}  &~~~&
62:&$     (X_{20} \booland{} X_{27}) \pand{} X_{10}    $
\\
15:&$   X_{22} \pand{} X_{10} \pand{} X_{28}     $ &~~~&
39:&\sout{$   X_{24}  \pand{}    (X_{1} \booland{} X_{5})      $}  &~~~&
63:&$     (X_{22} \booland{} X_{27}) \pand{} X_{10}    $
\\
16:&$    X_{24} \pand{} X_{10} \pand{} X_{28}    $ &~~~&
40:&$    X_{20}  \pand{}    (X_{1} \booland{} X_{15})    ~\lozenge$  &~~~&
64:&$      (X_{24} \booland{} X_{27}) \pand{} X_{10}   $
\\
17:&$    X_{1} \pand{} X_{10} \pand{} X_{28}    $ &~~~&
41:&$   X_{22}  \pand{}    (X_{1} \booland{} X_{15})      ~\lozenge$  &~~~&
65:&$      (X_{1} \booland{} X_{27}) \pand{} X_{10}   $
\\
18:&\uwave{$    X_{20} \pand{} X_{10} \pand{} X_{28}    $}&~~~&
42:&$    X_{24}  \pand{}    (X_{1} \booland{} X_{15})   ~\lozenge$  &~~~&
66: &\uwave{$     (X_{20} \booland{} X_{27}) \pand{} X_{10}    $}
\\
19:&\uwave{$    X_{22} \pand{} X_{10} \pand{} X_{28}    $}&~~~&
43:&$    X_{20}  \pand{}    (X_{5} \booland{} X_{13})    ~\lozenge$  &~~~&
67: &\uwave{$    (X_{22} \booland{} X_{27}) \pand{} X_{10}     $}
\\
20:&\uwave{$     X_{24} \pand{} X_{10} \pand{} X_{28}   $}&~~~&
44:&$   X_{22}  \pand{}    (X_{5} \booland{} X_{13})    ~\lozenge$  &~~~&
68:&\uwave{$     (X_{24} \booland{} X_{27}) \pand{} X_{10}    $}
\\
21:&\sout{$   X_{1} \pand{} X_{5} \pand{} X_{28}     $}&~~~&
45:&$    X_{24}  \pand{}    (X_{5} \booland{} X_{13})   ~\lozenge$  &~~~&
69:&\sout{$     (X_{5} \booland{} X_{15}) \pand{} X_{13}    $}
\\
22:&\sout{$     X_{5} \pand{} X_{1} \pand{} X_{28}     $} &~~~&
46:&$    X_{20}  \pand{}    (X_{13} \booland{} X_{15})    $  &~~~&
70:&\sout{$       (X_{1} \booland{} X_{13}) \pand{} X_{15}    $}
\\
23:&\sout{$     X_{1} \pand{} X_{15} \pand{} X_{28}     $}&~~~&
47:&$   X_{22}  \pand{}    (X_{13} \booland{} X_{15})      $  &~~~&
71:&\sout{$       (X_{1} \booland{} X_{13}) \pand{} X_{5}    $}
\\
24:&\sout{$      X_{5} \pand{} X_{13} \pand{} X_{28}    $} &~~~&
48:&$    X_{24}  \pand{}    (X_{13} \booland{} X_{15})     $  &~~~&
72:&\sout{$      (X_{5} \booland{} X_{15}) \pand{} X_{1}     $}
\\
\bottomrule
\end{tabular}
}
\caption{Event sequences of rank two and three. Event sequences which are included more than once have an \uwave{``underwave''}, non-minimal event sequences are \sout{striked through}, ``partly'' non-minimal event sequences are marked with $\lozenge$.}
\label{090502001}
\end{table}%

\paragraph{Minimal Form and {MCSS} of the Failure Funktion}~\\
A total of $12$ of the $77$ event sequences in table \ref{090502001} are included at least twice and may be omitted using the law of idempotency.
A further $20$ event sequences are non-minimal and also omitted.
The extended event sequences number $8$ and $9$ and $40$ to $45$, i.e.\
\begin{align}
    &
    \bigl[    X_{1} \booland{} X_{15} \booland{} X_{38}    \bigr],
    \bigl[    X_{5} \booland{} X_{13} \booland{} X_{38}    \bigr], \nonumber\\
     &
     \bigl[  X_{20}  \pand{}    (X_{1} \booland{} X_{15}) \bigr],
     \bigl[  X_{22}  \pand{}    (X_{1} \booland{} X_{15}) \bigr],
     \bigl[  X_{24}  \pand{}    (X_{1} \booland{} X_{15}) \bigr] , \nonumber\\
     &
     \bigl[X_{20}  \pand{}    (X_{5} \booland{} X_{13})\bigr]  ,
     \bigl[X_{22}  \pand{}    (X_{5} \booland{} X_{13}) \bigr],
     \bigl[X_{24}  \pand{}    (X_{5} \booland{} X_{13})\bigr],
\end{align}
are ``partly'' non-minimal with respect to the {MCSS} of rank two, i.e.\
\begin{align}
  &\bigl[ X_{1} \pand{} X_{15}\bigr] \qquad\text{and} \qquad \bigl[ X_{5} \pand{} X_{13}\bigr]~.
\end{align}
Therefore, it is necessary to break up the extended event sequences in order to separate their minimal and non-minimal parts.

For example, the event sequence $X_{20}  \pand{}    (X_{1} \booland{} X_{15}) $ provides (without SAND) two non-extended (normal) event sequences, i.e.\
\begin{align}
     X_{20}  \pand{}    (X_{1} \booland{} X_{15}) &\ist{}
            \bigl[(X_{1}X_{20} ) \pand{} X_{15}\bigr]   \boolor{} \bigl[  (X_{15}X_{20} ) \pand{} X_{1}\bigr]~,
\end{align}
where the first is non-minimal with respect to $ X_{1} \pand{} X_{15}$.

In analogy to that, 
\begin{align}
     X_{1} \booland{} X_{15} \booland{} X_{38}&\ist{}
           \bigl[ (X_{1} \pand{} X_{15}) \booland{} X_{38}\bigr]   \boolor{} \bigl[ (X_{15} \pand{} X_{1}) \booland{} X_{38}\bigr] ~.
\end{align}
Only the second event sequence is minimal.
It is first transformed into a {TDNF}, thus
\begin{align}
     (X_{15} \pand{} X_{1}) \booland{} X_{38}&\ist{}
            \bigl[ X_{15} \pand{} X_{1} \pand{} X_{38}\bigr]   \boolor{} \bigl[ (X_{15}\booland{} X_{38}) \pand{} X_{1}  \bigr] ~.
\end{align}
Therefore, the two partly minimal event sequences number $8$ and $9$ provide four minimal {MCSS}.

Table \ref{0905878787} shows a cleaned up list, in which only {MCSS} of rank two and three of the failure function $\varpi$ are shown.
%

\begin{table}[H]
\centering
{\small
  \begin{tabular}{clcclccl}
  \toprule
\multicolumn{8}{c}{(extended) {MCSS} of rank two:}
\\
1:&$  X_{1} \pand{} X_{5} $ &\qquad\qquad&
3:&$    X_{1} \pand{} X_{15}  $  &\qquad\qquad&
5:&$   X_{28} \booland{} X_{38}  $
\\
2:&$    X_{5} \pand{} X_{1}  $ &\qquad\qquad&
4:&$    X_{5} \pand{} X_{13}  $  &\qquad\qquad&
\\
\midrule
\multicolumn{8}{c}{(extended) {MCSS} of rank three:}
\\
1:&$  X_{28} \booland{} X_{30} \booland{} X_{32}    $ &~~~&
15:&$    (X_{5} \booland{} X_{28}) \pand{} X_{10}         $  &~~~&
29:&$    X_{20} \pand{} X_{10} \pand{} X_{28}      $
\\
2:&$  X_{28} \booland{} X_{30} \booland{} X_{36}    $ &~~~&
16:&$     (X_{20} \booland{} X_{28}) \pand{} X_{10}     $  &~~~&
30:&$   X_{22} \pand{} X_{10} \pand{} X_{28}     $
\\
3:&$    X_{28} \booland{} X_{32} \booland{} X_{34}  $ &~~~&
17:&$     (X_{22} \booland{} X_{28}) \pand{} X_{10}      $  &~~~&
31:&$    X_{24} \pand{} X_{10} \pand{} X_{28}    $
\\
4:&$     X_{28} \booland{} X_{34} \booland{} X_{36}    $ &~~~&
18:&$     (X_{24} \booland{} X_{28}) \pand{} X_{10}     $  &~~~&
32:&$    X_{1} \pand{} X_{10} \pand{} X_{28}    $
\\
5:&$        X_{18} \pand{} X_{10} \pand{} X_{38}      $ &~~~&
19:&$     (X_{1} \booland{} X_{28}) \pand{} X_{10}     $  &~~~&
33:&$     (X_{5} \booland{} X_{18}) \pand{} X_{10}        $
\\
6:&$      (X_{18} \booland{} X_{38}) \pand{} X_{10}    $ &~~~&
20:&$   (X_{15} \booland{} X_{20} ) \pand{} X_{1}   $   &~~~&
34:&$      (X_{18} \booland{} X_{20}) \pand{} X_{10}     $
\\
7:&$   X_{15} \pand{} X_{1} \pand{} X_{38}  $&~~~&
21:&$   (X_{15} \booland{} X_{22} ) \pand{} X_{1}  $     &~~~&
35:&$     (X_{18} \booland{} X_{22}) \pand{} X_{10}    $
\\
8:&$    (X_{15}\booland{} X_{38}) \pand{} X_{1}     $ &~~~&
22:&$    (X_{15} \booland{} X_{24} ) \pand{} X_{1}  $    &~~~&
36:&$      (X_{18} \booland{} X_{24}) \pand{} X_{10}    $
\\
9:&$    X_{13} \pand{} X_{5} \pand{} X_{38}   $ &~~~&
23:&$     (X_{13} \booland{} X_{20} ) \pand{} X_{5}   $    &~~~&
37:&$       (X_{1} \booland{} X_{18}) \pand{} X_{10}   $
\\
10:&$    (X_{13}\booland{} X_{38}) \pand{} X_{5}     $ &~~~&
24:&$    (X_{13} \booland{} X_{22} ) \pand{} X_{5}   $    &~~~&
38:&$      (X_{5} \booland{} X_{27}) \pand{} X_{10}    $
\\
11:&$     X_{13} \booland{} X_{15} \booland{} X_{38}    $  &~~~&
25:&$    (X_{13} \booland{} X_{24} ) \pand{} X_{5}  $     &~~~&
39:&$     (X_{20} \booland{} X_{27}) \pand{} X_{10}    $
\\
12:&$   X_{27} \pand{} X_{10} \pand{} X_{38}        $ &~~~&
26:&$    X_{20}  \pand{}    (X_{13} \booland{} X_{15})    $     &~~~&
40:&$     (X_{22} \booland{} X_{27}) \pand{} X_{10}    $
\\
13:&$     (X_{27} \booland{} X_{38}) \pand{} X_{10}      $ &~~~&
27:&$   X_{22}  \pand{}    (X_{13} \booland{} X_{15})      $    &~~~&
41:&$      (X_{24} \booland{} X_{27}) \pand{} X_{10}   $
\\
14:&$    X_{5} \pand{} X_{10} \pand{} X_{28}      $ &~~~&
28:&$    X_{24}  \pand{}    (X_{13} \booland{} X_{15})     $      &~~~&
42:&$      (X_{1} \booland{} X_{27}) \pand{} X_{10}   $
\\
\bottomrule
\end{tabular}
}
\caption{{MCSS} of rank two and three.
This table is a version of table \ref{090502001}, but stripped of non-minimal event sequences and duplicates.}
\label{0905878787}
\end{table}%

\paragraph{Results}~\\
The {MCSS} of the failure function are all of ranks two and higher.
Therefore, no single failure within the system as modelled leads directly to an infraction of the safety goal.
The example system thus satisfies the requirement of single-failure-resistance, as described in chapter \ref{090426001}.

The most important combinations of dangerous failures, that lead to an infraction of the safety goal, are {MCSS} of rank two and three.
The five {MCSS} of rank two are
\begin{enumerate}
 \item either failures of the two sensors following each other.
 In this case EN1 would be activated by the first sensor failure, and SAF would be activated by the second sensor failure.
 These two failures may occur in arbitrary sequence.
 \item or one sensor failure in combination with a failure of \textmu C.
 The sensor failure needs to occur before the failure of the \textmu C, otherwise the sequence logic in L would not be activated.
 \item or am failure of T3 in combination with a failure of L, which activates both power stages.
 These two failures may occur in arbitrary sequence.
\end{enumerate}
{MCSS} of rank three are e.g.\
\begin{enumerate}
 \item a double failure of the high side and the low side of the driver in combination with an internal failure in T3.
 No sequence logic has to be respected here.
 Specifically, numbers $1$ to $4$ in table \ref{0905878787} are combinations of this type.
 \item failures of the system ASIC in combination with failures of the \textmu C and/or sensor failures.
 Specifically, numbers $20$ to $28$ in table \ref{0905878787} are combinations of this type.
\item a failure of the watchdog or of the emergency switch in combination with an ASIC failure, where both occur before an additional failure of the \textmu C, see, for instance, numbers $34$ to $36$ and $39$ to $41$ in table \ref{0905878787}. 
\item a failure in one of the sensors in combination with a failure of the watchdog or the emergency switch, where both occur before an additional failure of the \textmu C, see, for instance, numbers $33$, $37$, $38$, and $42$ in table \ref{0905878787}. 
\end{enumerate}
%
%
\section{Probabilistic Analysis of the TOP Failure Parameters}\label{090215101}
The qualitative analysis of the temporal fault tree is used as evidence that the system stays below the threshold for failure rates as required by {\ISO}  for ASIL D systems.
This threshold is given as $\leq 1\cdot 10^{-8}\,\tfrac{1}{\sjshour}$ for any operating hour during the whole mission time.

In order to do so, it has to be demonstrated, that the failure rate of the TOP event $\lambda_{\mathit{TOP}}$ stays below this threshold.

Because of $f_{\mathit{TOP}}(T_M)\approx{}\lambda_{\mathit{TOP}}$, see \eqref{080819-001}, it is sufficient to use the TOP event's failure frequency as a good approximation.

Furthermore, an iterative multi-step approach is chosen, that reduces effort and is used in similar fashion in many real world {FTA} analyses.
First, an approximation with conservative estimations of the failure rates is used that allows for a first overview.

The evidence is sufficiently produced if, using this approach, the thresholds, as required by the safety standard, are not exceeded.
If this can not be shown, the next step is to determine the failure rates more exactly and/or use exact calculations instead of approximations -- and to possibly restrict the further analysis to the most important contributors as identified in the first step's overview.
The termination condition for these steps is that the thresholds, as required by the safety standard, are no longer exceeded.

Because of this, in the following discussion the {MCSS} are not transformed into a mutually exclusive (disjoint) form.
Instead, the approximation approach from chapter \ref{080724-010} is used.
This corresponds to the bottom most path in figure \ref{fig02} on page \pageref{fig02}.

Quantification of the failure function $\varpi$ is carried out using its {MCSS} from table \ref{0905878787}.
All basic events are allocated the same failure rate of $\lambda \ist{}  10^{-6} \,\tfrac{1}{\sjshour}$.

Table \ref{09058787345} shows failure probabilities and failure frequencies according to \eqref{080818-991} and \eqref{080819-003} for each {MCSS} from table \ref{0905878787}.
The mission time is given as $T_M\ist{}1000\sjshour$.
\begin{table}[H]
\centering
{\small
  \begin{tabular}{clcclccl}
  \toprule
\multicolumn{8}{c}{{MCSS} of rank two:}
\\
1:&{\footnotesize $F\!=\!5\!\cdot\!10^{-7};f\!=\!1\!\cdot\!10^{-9}\,\tfrac{1}{\sjshour}$} &\quad&
3:&{\footnotesize $F\!=\!5\!\cdot\!10^{-7};f\!=\!1\!\cdot\!10^{-9}\,\tfrac{1}{\sjshour}$}  &\quad&
5:&{\footnotesize $F\!=\!1\!\cdot\!10^{-6};f\!=\!2\!\cdot\!10^{-9}\,\tfrac{1}{\sjshour}$}
\\
2:&{\footnotesize $F\!=\!5\!\cdot\!10^{-7};f\!=\!1\!\cdot\!10^{-9}\,\tfrac{1}{\sjshour}$} &\quad&
4:&{\footnotesize $F\!=\!5\!\cdot\!10^{-7};f\!=\!1\!\cdot\!10^{-9}\,\tfrac{1}{\sjshour}$}  &\quad&
\\
\midrule
\multicolumn{8}{c}{{MCSS} of rank three:}
\\
1:&{\footnotesize $F\!=\!1\!\cdot\!10^{-9};f\!=\!3\!\cdot\!10^{-12}\,\tfrac{1}{\sjshour}$}  &&
15:&{\footnotesize $F\!=\!\tfrac{1}{3}\!\cdot\!10^{-9};f\!=\!1\!\cdot\!10^{-12}\,\tfrac{1}{\sjshour}$}  &&
29:&{\footnotesize $F\!=\!\tfrac{1}{6}\!\cdot\!10^{-9};f\!=\!1\!\cdot\!10^{-12}\,\tfrac{1}{\sjshour}$}
\\
2:&{\footnotesize $F\!=\!1\!\cdot\!10^{-9};f\!=\!3\!\cdot\!10^{-12}\,\tfrac{1}{\sjshour}$}  &&
16:&{\footnotesize $F\!=\!\tfrac{1}{3}\!\cdot\!10^{-9};f\!=\!1\!\cdot\!10^{-12}\,\tfrac{1}{\sjshour}$}    &&
30:&{\footnotesize $F\!=\!\tfrac{1}{6}\!\cdot\!10^{-9};f\!=\!1\!\cdot\!10^{-12}\,\tfrac{1}{\sjshour}$}
\\
3:&{\footnotesize $F\!=\!1\!\cdot\!10^{-9};f\!=\!3\!\cdot\!10^{-12}\,\tfrac{1}{\sjshour}$}  &&
17:&{\footnotesize $F\!=\!\tfrac{1}{3}\!\cdot\!10^{-9};f\!=\!1\!\cdot\!10^{-12}\,\tfrac{1}{\sjshour}$}    &&
31:&{\footnotesize $F\!=\!\tfrac{1}{6}\!\cdot\!10^{-9};f\!=\!1\!\cdot\!10^{-12}\,\tfrac{1}{\sjshour}$}
\\
4:&{\footnotesize $F\!=\!1\!\cdot\!10^{-9};f\!=\!3\!\cdot\!10^{-12}\,\tfrac{1}{\sjshour}$}  &&
18:&{\footnotesize $F\!=\!\tfrac{1}{3}\!\cdot\!10^{-9};f\!=\!1\!\cdot\!10^{-12}\,\tfrac{1}{\sjshour}$}   &&
32:&{\footnotesize $F\!=\!\tfrac{1}{6}\!\cdot\!10^{-9};f\!=\!1\!\cdot\!10^{-12}\,\tfrac{1}{\sjshour}$}
\\
5:&{\footnotesize $F\!=\!\tfrac{1}{6}\!\cdot\!10^{-9};f\!=\!1\!\cdot\!10^{-12}\,\tfrac{1}{\sjshour}$} &&
19:&{\footnotesize $F\!=\!\tfrac{1}{3}\!\cdot\!10^{-9};f\!=\!1\!\cdot\!10^{-12}\,\tfrac{1}{\sjshour}$}   &&
33:&{\footnotesize $F\!=\!\tfrac{1}{3}\!\cdot\!10^{-9};f\!=\!1\!\cdot\!10^{-12}\,\tfrac{1}{\sjshour}$}
\\
6:&{\footnotesize $F\!=\!\tfrac{1}{3}\!\cdot\!10^{-9};f\!=\!1\!\cdot\!10^{-12}\,\tfrac{1}{\sjshour}$}   &&
20:&{\footnotesize $F\!=\!\tfrac{1}{3}\!\cdot\!10^{-9};f\!=\!1\!\cdot\!10^{-12}\,\tfrac{1}{\sjshour}$}     &&
34:&{\footnotesize $F\!=\!\tfrac{1}{3}\!\cdot\!10^{-9};f\!=\!1\!\cdot\!10^{-12}\,\tfrac{1}{\sjshour}$}
\\
7:&{\footnotesize $F\!=\!\tfrac{1}{6}\!\cdot\!10^{-9};f\!=\!1\!\cdot\!10^{-12}\,\tfrac{1}{\sjshour}$}&&
21:&{\footnotesize $F\!=\!\tfrac{1}{3}\!\cdot\!10^{-9};f\!=\!1\!\cdot\!10^{-12}\,\tfrac{1}{\sjshour}$}      &&
35:&{\footnotesize $F\!=\!\tfrac{1}{3}\!\cdot\!10^{-9};f\!=\!1\!\cdot\!10^{-12}\,\tfrac{1}{\sjshour}$}
\\
8:&{\footnotesize $F\!=\!\tfrac{1}{3}\!\cdot\!10^{-9};f\!=\!1\!\cdot\!10^{-12}\,\tfrac{1}{\sjshour}$}  &&
22:&{\footnotesize $F\!=\!\tfrac{1}{3}\!\cdot\!10^{-9};f\!=\!1\!\cdot\!10^{-12}\,\tfrac{1}{\sjshour}$}   &&
36:&{\footnotesize $F\!=\!\tfrac{1}{3}\!\cdot\!10^{-9};f\!=\!1\!\cdot\!10^{-12}\,\tfrac{1}{\sjshour}$}
\\
9:&{\footnotesize $F\!=\!\tfrac{1}{6}\!\cdot\!10^{-9};f\!=\!1\!\cdot\!10^{-12}\,\tfrac{1}{\sjshour}$} &&
23:&{\footnotesize $F\!=\!\tfrac{1}{3}\!\cdot\!10^{-9};f\!=\!1\!\cdot\!10^{-12}\,\tfrac{1}{\sjshour}$}      &&
37:&{\footnotesize $F\!=\!\tfrac{1}{3}\!\cdot\!10^{-9};f\!=\!1\!\cdot\!10^{-12}\,\tfrac{1}{\sjshour}$}
\\
10:&{\footnotesize $F\!=\!\tfrac{1}{3}\!\cdot\!10^{-9};f\!=\!1\!\cdot\!10^{-12}\,\tfrac{1}{\sjshour}$}  &&
24:&{\footnotesize $F\!=\!\tfrac{1}{3}\!\cdot\!10^{-9};f\!=\!1\!\cdot\!10^{-12}\,\tfrac{1}{\sjshour}$}    &&
38:&{\footnotesize $F\!=\!\tfrac{1}{3}\!\cdot\!10^{-9};f\!=\!1\!\cdot\!10^{-12}\,\tfrac{1}{\sjshour}$}
\\
11:&{\footnotesize $F\!=\!1\!\cdot\!10^{-9};f\!=\!3\!\cdot\!10^{-12}\,\tfrac{1}{\sjshour}$}  &&
25:&{\footnotesize $F\!=\!\tfrac{1}{3}\!\cdot\!10^{-9};f\!=\!1\!\cdot\!10^{-12}\,\tfrac{1}{\sjshour}$}     &&
39:&{\footnotesize $F\!=\!\tfrac{1}{3}\!\cdot\!10^{-9};f\!=\!1\!\cdot\!10^{-12}\,\tfrac{1}{\sjshour}$}
\\
12:&{\footnotesize $F\!=\!\tfrac{1}{6}\!\cdot\!10^{-9};f\!=\!1\!\cdot\!10^{-12}\,\tfrac{1}{\sjshour}$} &&
26:&{\footnotesize $F\!=\!\tfrac{2}{3}\!\cdot\!10^{-9};f\!=\!2\!\cdot\!10^{-12}\,\tfrac{1}{\sjshour}$}         &&
40:&{\footnotesize $F\!=\!\tfrac{1}{3}\!\cdot\!10^{-9};f\!=\!1\!\cdot\!10^{-12}\,\tfrac{1}{\sjshour}$}
\\
13:&{\footnotesize $F\!=\!\tfrac{1}{3}\!\cdot\!10^{-9};f\!=\!1\!\cdot\!10^{-12}\,\tfrac{1}{\sjshour}$}  &&
27:&{\footnotesize $F\!=\!\tfrac{2}{3}\!\cdot\!10^{-9};f\!=\!2\!\cdot\!10^{-12}\,\tfrac{1}{\sjshour}$}    &&
41:&{\footnotesize $F\!=\!\tfrac{1}{3}\!\cdot\!10^{-9};f\!=\!1\!\cdot\!10^{-12}\,\tfrac{1}{\sjshour}$}
\\
14:&{\footnotesize $F\!=\!\tfrac{1}{6}\!\cdot\!10^{-9};f\!=\!1\!\cdot\!10^{-12}\,\tfrac{1}{\sjshour}$} &&
28:&{\footnotesize $F\!=\!\tfrac{2}{3}\!\cdot\!10^{-9};f\!=\!2\!\cdot\!10^{-12}\,\tfrac{1}{\sjshour}$}        &&
42:&{\footnotesize $F\!=\!\tfrac{1}{3}\!\cdot\!10^{-9};f\!=\!1\!\cdot\!10^{-12}\,\tfrac{1}{\sjshour}$}
\\
\bottomrule
\end{tabular}
}
\caption{Failure probabilites and failure frequencies according to \eqref{080818-991} and \eqref{080819-003} for each {MCSS} from table \ref{0905878787}.}
\label{09058787345}
\end{table}%

Failure characteristics at TOP event level are then calculated using \eqref{080224-002} and \eqref{080224-003}, respectively, as sum of the individual MCSS' contributions.
Using the values from table \ref{09058787345} yields
\begin{align}
  F_{\mathit{TOP}}(T_M) {}\approx{}& 3,017 \cdot 10^{-6} \qquad\text{and} \label{999123}\\
  f_{\mathit{TOP}}(T_M)  {}\approx{} &6,055 \cdot 10^{-9}\,\tfrac{1}{\sjshour} \label{999124} ~.
\end{align}

This first approximation already provides the evidence for meeting the {\ISO} standard's requirements for ASIL D; the TOP event's failure frequency in \eqref{999124} stays well below the threshold of $1\cdot 10^{-8}\,\tfrac{1}{\sjshour}$.

\emph{Remark:}~
With conventional {FTA} the PAND gate would have to be replaced by normal AND gates.
This would affect the failure frequencies of minimal cutsets of rank two the most.
These minimal cutsets would be the same as the {MCSS} of rank two, only using AND operators instead of the PANDs.
Accordingly, in an Boolean {FTA} the TOP event's failure frequency would nearly double compared to the TFTA's result, yielding $>1\cdot10^{-8}\,\tfrac{1}{\sjshour} $ and, thus, exceeding the threshold limit.
%
%
\section{Discussion}\label{090215102}
The analysis of this real world example system in chapter \ref{chap080401-050} demonstrates that the {TFTA} method is not limited to modelling only very small examples.
Chapter \ref{chap080401_006} thereby extends the theoretical discussions on the {TFTA} approach in chapters \ref{_chap_080401-004} and \ref{chap080401-033}, as well as the statements on basic application of the {TFTA} in chapter \ref{chap080104_005}.

The analogy to the conventional {FTA} is shown during the creation of the temporal fault tree in figures \ref{090208001} to \ref{090208003}.
In this process no additional effort is necessary in comparison to the Boolean {FTA} apart from choosing temporal fault tree gates.

In this temporal fault tree there are several meshings of basic events as well as of whole sub trees.
For instance, events beneath ``\textmu C signal failure'' are found beneath a temporal gate (``L commanded failure'').
The same events are also found in the purely Boolean part of the fault tree below of ``commanded failure T3''.
Additionally, the basic event ``10~-- \textmu C generic failure activates SAF'' is found in different and otherwise separated subtrees beneath different PAND gates.

Using such meshing in e.g.\ an {DFT} approach would dramatically increase the effort;
the necessary separation into different dynamic and non-dynamic modules would require that almost the whole fault tree had to be modelled as a dynamic module, i.e.\ in case of the {DFT} it had to be modelled using markov methods.

The detailed qualitative analysis of the temporal fault tree in chapter \ref{090215100} demonstrates that the {TFTA} is able to solve these meshings by use of its temporal transformation laws.

On the one hand it is true that the calculatory effort for these transformations increases rapidly, specifically because of the temporal distributive laws.
On the other hand, the required calculations are mostly limited to string-manipulations.
As a general rule, these are less costly than solving exponentially growing markov models or simulating big petri nets, as necessary for the other methods.

The analysis of the {MCSS} in chapter \ref{090502007} is, then, very similar to the Boolean {FTA}.
Among others, it is demonstrated that the {TFTA} is well suited for real qualitative analysis.
As described in chapter \ref{chap080104_005}, this is one of the main advantages of the {TFTA}. 

The probabilistic quantification, as demonstrated in chapter \ref{090215101}, is based on a step-by-step approach, as is best praxis.
This allows adjusting modelling precision to the issue at hand -- which implies adjustable effort --, as well as concentrating all ressources on the most important contributors.
Both is not possible to the same extend when using the {DFT}.   
\clearpage
\clearpage
\ifx \printSprueche\undefined
\else
  \renewcommand*{\dictumwidth}{.35\textwidth}

  \setchapterpreamble[ur]{%
  \dictum[Thomas Alva Edison]{I want electricity to become so cheap that only the rich can afford candles.}\vspace{3cm}
}
\fi

\chapter{Summary and Outlook}\label{chap080401-008}
The new approach to temporal fault tree analysis presented in this thesis is called {TFTA}; it extends the Boolean {FTA} in order to include event sequences.
In comparison to the conventional {FTA} this allows a more realistic model of the failure behaviour of complex and dynamic systems.

The new {TFTA} uses a new temporal logic described in this thesis.
With this logic it differs significantly from most existing approaches with similar aims.
These transform the {FTA} model completely or partially into a state based model;
temporal effects are then handled in the state space, and the results are then transfered back into the fault tree.
{TFTA} contrasts with such state based methods in that
\begin{itemize}
 \item it uses an extension to Boolean algebra and logic, 
 \item its notation, terms, and its workflow and work products are taken from the conventional {FTA}, 
 \item it allows qualitative as well as probabilistic analyses and calculations including event sequence information.
\end{itemize}

In comparison to other known approaches that also use a ``temporal logic'' to include temporal information into the fault tree the {TFTA} is significantly leaner.

Specifically, {TFTA} is not another attempt to create a formal {FTA} logic for modelling of software systems.
Instead, {TFTA} emphasises practise-oriented characteristics like intuitive  applicability, readability, comprehensible logic expressions and results, transferability of real world failure effects into the model, and scalability.

The temporal logic of the {TFTA} uses the Boolean operations of conjunction, disjunction, and negation.
Additionally, two new temporal operations (PAND and SAND) represent two ``special conjunctions'' that describe event sequences and simultaneous events, respectively.
 
Using the well known Boolean algebra and a set of new temporal transformation laws, it is possible to transform complex temporal expressions into their temporal disjunctive normal form ({TDNF}) which consists of separate event sequences.
In analogy to the Boolean fault tree cutsets these event sequences are reduced to a minimal form, the so-called minimal cutset sequences ({MCSS}).

Then, {MCSS} are made mutually exclusive (i.e.\ disjoint).
This disjoint form is especially well suited for direct quantification and makes probabilistic analysis possible.

Other than conventional {FTA}, probabilistic {TFTA} allows to calculate reliability characteristics like failure probability, failure frequency, and failure rate of a fault tree TOP event with consideration of event sequence information, and without the need to change into the state space.

\paragraph{Evaluation of this Thesis}~\\
Originally, the development of an own temporal logic aimed primarily at solving some of the problems that arise with the known dynamic extensions of the {FTA} which are based on markov methods.
The {DFT} method \cite{Dugan1992} is a well known representative of such dynamic extensions, and thus it is an obvious choice to compare what this thesis achieved with the {DFT} method.

With regard to the calculatory effort, the consideration of event sequences always implies additional cost when compared to the Boolean {FTA}.
This is true for state based extensions, as well as for extended logics covering temporal effects.
This additional cost is a concern, even more so, as the determination of disjoint minimal cutsets in Boolean {FTA} already carries exponentially growing complexity.
On the other hand, the {TFTA} method does not aim at solving this.

Some of the TFTA's problems are fundamentally connected to the kind of temporal logic that is used.
Event sequence statements only cover the points in time at which events occur.
Therefore, ``time-limited'' failure events, i.e.\ events with a defined time span of being $\True$, can not be represented by PAND and SAND.
Instead, such effects need to be represented by conventional AND gates.
This, however, is no deterioration in comparison to the {DFT} method.
The markov chains that the {DFT} uses are also only able of capturing state transitions resulting from ``initiating'' failure events; it is not able to capture ``time-limited'' failure events.
The {DFT} only hides this shortcoming better, because of the necessary modularization and because meshing is impossible.

One major shortcoming of the {DFT} is modularization.
In some cases, it makes it impossible to mesh events beyond single dynamic fault tree gates logically correctly.
Compared to that, {TFTA} allows for such meshing.
It, thus, is possible to consider more event sequence effects.

Another major shortcoming of the {DFT} concerns qualitative evaluation of minimal cutsets.
The transformation into the state space either forces the use of ``meta events'' in addition to basic events; these meta event represent complete markov models.
As an alternative, qualitative analysis is restricted to not include event sequence information.
Compared to that, the (extended) event sequences in {TFTA} show exact event sequence information of all basic events that contribute to the TOP failure.
As such, the {TFTA} permits more meaningful and efficient qualitative analyses than the {DFT}.

Both, the {TFTA} as well as the {DFT} allow for probabilistic evaluation of the TOP event's failure rate and failure probability.
On the one hand, with this quantification it is possible to determine the precise TOP event's failure characteristics at comparably high calculatory costs. 
On the other hand, an approximation for the TFTA is provided, which reduces the necessary effort significantly. 

Three more arguments support the {TFTA} with regard to calculatory costs:
first, the size of the differential equations system, necessary for solving the {DFT}, grows exponentially with the number of component failures that are within a dynamic module.
Therefore, the overhead of {TFTA} (compared to Boolean {FTA}) is at least comparable with the DFT's overhead -- and the {TFTA} provides more meaningful results, as discussed above.
Second, calculations in the {TFTA} are mainly string-manipulations.
These usually require less effort than solving exponentially growing state models. 
Third, the {TFTA} offers approximation methods, which provide a real possibility to reduce overhead effectively, while accepting a certain degree of impreciseness; this may be used e.g.\ as a first step within a multi-step analysis.

Therefore, the {TFTA} is a capable replacement for the DFT's PAND gates, and furthermore provides some advantages methodology-wise, as well as for its useability.
\paragraph{Possible Further Research}~\\
During this theses several additional topics were discovered that could not be completely covered and solved within this work.
For instance, SAND connections are defined as (structural) dependencies between failure events, and they are considered qualitatively, but they are not taken into account probabilistically.
Because of the significance of dependent failures, which are sometime just called common cause failures ({CCF}), it seems promising to extend the {TFTA} method, as described in this thesis, by such dependencies.
Furthermore, this thesis restricts itself to non-repairable failures.
It seems possible that the TFTA's temporal logic, as well as the probabilistic aspects of the {TFTA}, may be extended to repairable failures.
It could also be interesting to develop advanced methods to determine mutually exclusive (disjoint) expressions from a given {TDNF}.
One possible way could be to follow segmentation-methods, like Abraham \cite{Abraham1979} or Heidtmann \cite{Heidtmann1989} proposed for Boolean algebra.
Furthermore, it seems promising to investigate possible synergies between the {TFTA} logic and the {BDD} method in \cite{Sinnamon1996}.
In general, there certainly is a demand for improved algorithms for using the {TFTA} in practise. 
In this regard, contributing to open source fault tree tools (like e.g.\ OpenFTA \cite{FormalSoftwareConstructionLtd2005}) could be an interesting possibility.

%
%
\nocite{Meyna2003Taschenbuch}
\bibliographystyle{unsrtnat}
\bibliography{sjsliteratur_jabref_generated_only_used_entries_and_translated_to_english}
%
%
%
%
%
%
%
\appendix 
\renewcommand{\thechapter}{\Alph{chapter}} 
%
%
%
\addtocontents{toc}{\vspace{1em}}
\addcontentsline{toc}{chapter}{Appendix}
\part*{Appendix}\label{_chap_00999}
%
%
%
%
%
\chapter{Further Explanations on Selected Topics}\label{chap080404-005}
\section{Reliability Characteristics}\label{chap080412-003}
The probabilistic description of the failure behaviour of systems is done using \emph{characteristics}, see table
\ref{tab:KenngrossenQuantitativerZuverlassigkeitsBzwSicherheitsanalysen}.
These are stochastic or probabilistic values, as the deterministic failure behaviour of an individual component or an individual system is usually not known in advance.
Taking the probability distributions into account that result from such values is difficult in many real applications, in particular because of the effort necessary to assemble knowlegde on the kind of distribution.
In many cases constant or mean values are thus used instead of distributed values.
\begin{table}
	{\small
  \begin{center}
		\begin{tabular}{lclc} \toprule
 \multicolumn{4}{c}{\textbf{non repairable systems}}
 		\\ \addlinespace[7pt]
 \multicolumn{2}{c}{\textbf{reliability}} & \multicolumn{2}{c}{\textbf{safety}}
 		\\ \addlinespace[7pt]
 \emph{charact.} & \emph{symbol} & \emph{charact.} & \emph{symbol}
 		\\ \midrule \addlinespace[7pt]
 failure probability & $F(t)$ & hazard-probability & $G(t)$ \\
 reliability & $R(t)$ & safety-probability & $S(t)$ \\
 failure frequency & $f(t)$ & hazard-density & $g(t)$ \\
 failure rate & $h(t)$ & hazard-rate & $\delta(t)$ \\
 ~~~if constant: & $\lambda$ & \\
 \toprule
 \multicolumn{4}{c}{\textbf{repairable systems}}
 		\\ \addlinespace[7pt]
 \multicolumn{2}{c}{\textbf{reliability}} & \multicolumn{2}{c}{\textbf{safety}}
 		\\ \addlinespace[7pt]
 \emph{charact.} & \emph{symbol} & \emph{charact.} & \emph{symbol}
 		\\ \midrule \addlinespace[7pt]
 repair rate & $\mu(t)$ & safety-restoration rate & $\nu(t)$ \\
 probability of restoration & $M(t)$ & probability of safety- & $W(t)$ \\
  && restoration \\
 repair frequency & $m(t)$ & frequency of safety- & $w(t)$ \\
  && restoration \\
 availability & $V(t)$ & safety-availability & $V_S(t)$ \\
 unavailability & $U(t)$ & ``safety-unavailability`` & $U_S(t)$ \\
 \bottomrule
\end{tabular}
\end{center}
}
	\caption{Characteristics of reliability and safety analysis according to \cite{Meyna2003Taschenbuch}}
	\label{tab:KenngrossenQuantitativerZuverlassigkeitsBzwSicherheitsanalysen}
\end{table}

This thesis uses the terms failure probability, failure frequency and failure rate, even if it originates in a safety backgound, as
\begin{itemize}
	\item the essential statements apply to the field of general reliability analogously and
	\item the use of such terms, that originally come from general reliability, is very common in the context of safety; see e.g.\ the relevant safety standards {\ISO} \cite{ISO2009} and {\IEC} \cite{2002IECa}.
\end{itemize}
\section{Creating and Using Sequential Failure Trees in the TFTA}\label{091009020}
Sequential failure trees allow visualization of temporal-logical expressions, as well as manual verification of transformations according to the laws of TFTA's temporal logic.
Creating a sequential failure tree corresponding to a complex temporal expression requires some effort, but it is based on only a few basic steps.
\paragraph{Choosing the Right Failure Tree}~\\
The number of basic events within a temporal expression determines what kind of sequential failure tree needs to be chosen.
The failure tree must at least support the number of basic events, but it may be bigger, too.
Depending on the particular application, the simplified sequential failure tree without SAND may be sufficient. 

An example: the following figure shows two sequential failure trees, that are both suited for the expression $\varpi\ist A \booland{} B$ and are not yet filled in.

\begin{center}
\input{pics/anhang_sequ_ausfallbaum1.tex}
\end{center}

\paragraph{Transforming the Temporal Expression}~\\
If the temporal expression is too complex, then, in a first step, simple sub-expressions need to be identified, and for these sequential fault trees are then created.
As an extreme example, the basic events of the temporal expression are chosen.
The following steps are then repeated for all these sub-expressions.

For instance, the two sub-expressions $A$ and $B$ are chosen for the expression $\varpi\ist A \booland{} B$.
\paragraph{Minimal Failure Nodes}~\\
Starting with the top-node all branches of the sequential failure tree are walked along, until in each branch the currently chosen sub-expression has occurred (or the branch has ended), and the minimal failure nodes are tagged.

An example is presented in the next step.
\paragraph{Non-Minimal Failure Nodes}~\\
All nodes beneath a minimal failure node are tagged as successor nodes.

An example: the following figures show the minimal (on the left side) as well as minimal and successor failure nodes (right side) corresponding to the temporal expression $\varpi\ist{}A$.
\begin{center}
\input{pics/anhang_sequ_ausfallbaum2.tex}
\end{center}

\paragraph{Negated Events}~\\
Starting with the sequential failure tree corresponding to an event, all original non-failure nodes are marked as new minimal failure nodes; and all original failure nodes (minimal as well as successor) are marked as non-failure nodes.
\emph{No new} non-minimal failure nodes are added.

The following figure shows the sequential failure tree for the example of $\boolnot A$.
\begin{center}
\input{pics/anhang_sequ_ausfallbaum3.tex}
\end{center}

\paragraph{Conjunction/AND Relationship}~\\
The sequential failure tree of the conjunction of two temporal expressions is the ``intersection'' of the individual expressions' sequential failure trees.
Minimal failure nodes thereby absorb non-minimal failure nodes.
In a next step, non-minimal failure nodes are added as necessary; this is especially necessary in case of negated events.

An example is presented in the next step.

\paragraph{Disjunction/OR Relationship}~\\
The sequential failure tree of the disjunction of two temporal expressions is the ``union`` of the individual expressions' sequential failure trees.
Non-minimal failure nodes thereby absorb minimal failure nodes.
In a next step, non-minimal failure nodes are added as necessary; this is especially necessary in case of negated events.

An example: the following figure shows (from left to right) two simplified sequential failure trees, as well as their ``intersection'' and ``union``, respectively.
\begin{center}
\input{pics/anhang_sequ_ausfallbaum4.tex}
\end{center}

\paragraph{PAND Relationship}~\\
The sequential failure tree of the PAND connection of two temporal expressions, i.e.\ $\varpi_1\pand{}\varpi_2$, is generated as follows:
All those nodes are marked as minimal failure nodes that are minimal failure nodes of $\varpi_2$ together with being non-minimal failure nodes of $\varpi_1$.{}
In a next step, non-minimal failure nodes are added as necessary.

An example: the following figure shows (from left to right) two simplified sequential failure trees and their PAND connection.
\begin{center}
\input{pics/anhang_sequ_ausfallbaum5.tex}
\end{center}

\paragraph{SAND Relationship}~\\
The sequential failure tree of the SAND connection of two temporal expressions, i.e.\ $\varpi_1\sand{}\varpi_2$, is generated as follows:
All those nodes are marked as minimal failure nodes that are minimal failure nodes of $\varpi_2$ together with being minimal failure nodes of $\varpi_1$.{}
In a next step, non-minimal failure nodes are added as necessary.

An example: the following figure shows (from left to right) two simplified sequential failure trees and their SAND connection.
\begin{center}
\input{pics/anhang_sequ_ausfallbaum6.tex}
\end{center}

\section{Examples: Mutually Exclusive (Disjoint) Temporal Expressions}\label{090111010}
%
%
The following assumes $n=3$ and failure events $A$, $B$, and $C$.
\paragraph{First Example}~\\
The failure function $\varpi\ist{}B$ is already given as a {TDNF} with only one sub-expression; it is not a minterm, though, as not all possible failure events are included in this expression.
Using the method provided on page \pageref{090122002} yields a {TDNF} of mutually exclusive (disjoint) and minimal event sequences, that are temporal minterms, too:
\begin{align*}
  B & \ist{} B\booland{}(\boolnot A\boolor{}A)\booland{}(\boolnot C\boolor{}C)\ist{}\\
 & \ist{}
              \bigl[A \booland{} B \booland{} C\bigr] \boolor{}\bigl[\boolnot C \booland{}  ( A \booland{} B )\bigr]\boolor{}
          \bigl[\boolnot A \booland{} ( B \booland{} C )\bigr] \boolor{}\bigl[(\boolnot A \boolnot C )\booland{} B\bigr] ~.
\end{align*}
For better readability, the four resulting sub-expressions are inspected separately.
\begin{align*}
  \eta_1 & \ist{} A \booland{} B \booland{} C ~.
\end{align*}
Using the law of completion twice yields
\begin{align*}
 \eta_1 & \ist{} \phantom{\boolor{}}
                        \bigl[ ( A \booland{} B ) \pand{} C \bigr] \boolor{}
                        \bigl[ ( A \booland{} B ) \sand{} C \bigr] \boolor{}
                        \bigl[ C \pand{} ( A \booland{} B ) \bigr] \ist \\
    & \ist{}
        \phantom{\boolor{}}
                        \bigl[ ( A \pand{} B \boolor{} A \sand{} B \boolor{} B \pand{} A ) \pand{} C \bigr] \boolor{}
                        \bigl[ ( A \pand{} B \boolor{} A \sand{} B \boolor{} B \pand{} A ) \sand{} C \bigr] \boolor{}\\
    & \phantom{\ist{}}\boolor{}
                        \bigl[ C \pand{} ( A \pand{} B \boolor{} A \sand{} B \boolor{} B \pand{} A ) \bigr]~.
\end{align*}
As the expressions in round brackets are mutually exclusive (disjoint),
\begin{align*}
 \eta_1 & \ist{}
        \phantom{\boolor{}}\bigl[A \pand{} B \pand{} C\bigr] \boolor{} \bigl[(A \sand{} B) \pand{} C\bigr] \boolor{} \bigl[B \pand{} A \pand{} C \bigr]\boolor{}
        \bigl[(A \pand{} B) \sand{} C\bigr] \boolor{}\bigl[ (A \sand{} B) \sand{} C\bigr] \boolor{}\\
    & \phantom{\ist{}}\boolor{}
\bigl[(B \pand{} A) \sand{} C\bigr]\boolor{}
         \bigl[C \pand{} ( A \pand{} B)\bigr]\boolor{} \bigl[C \pand{} (A \sand{} B) \bigr]\boolor{} \bigl[C \pand{} (B \pand{} A ) \bigr]~.
\end{align*}
Applying the transformation laws of the temporal logic then yields
\begin{align*}
 \eta_1 & \ist{}
\phantom{\boolor{}}\bigl[A \pand{} B \pand{} C\bigr] \boolor{}  \bigl[(A \sand{} B) \pand{} C \bigr]\boolor{}  \bigl[B \pand{} A \pand{} C\bigr]\boolor{}
         \bigl[A \pand{} (B \sand{} C) \bigr]\boolor{}  \bigl[A \sand{} B \sand{} C \bigr]\boolor{} \\
    & \phantom{\ist{}}\boolor{}
\bigl[B \pand{} (A \sand{} C ) \bigr]\boolor{}
         \bigl[A \pand{} C \pand{} B \bigr]  \boolor{}   \bigl[(A \sand{} C) \pand{} B\bigr] \boolor{}   \bigl[C \pand{} A \pand{} B \bigr]\boolor{}\\
    & \phantom{\ist{}}\boolor{}
         \bigl[B \pand{} C \pand{} A \bigr]\boolor{}  \bigl[ (B \sand{} C) \pand{} A \bigr]\boolor{}  \bigl[ C \pand{} B \pand{} A\bigr] \boolor{} \bigl[C \pand{} (A \sand{} B) \bigr] ~.
\end{align*}
With this the transformation of the first sub-expression is completed.

Now, applying the law of completion on the second sub-expression, i.e.\
\begin{align*}
  \eta_2 & \ist \boolnot C \booland{}  (A \booland{} B)
\end{align*}
yields already disjoint expressions, thus
\begin{align*}
  \eta_2 & \ist \boolnot C \booland{} \bigl((A \pand{} B \boolor{} (A \sand{} B) \boolor{} (B \pand{} A )\bigr) \ist{} \\
    & \ist
    \bigl[\boolnot C\booland{}  (A \pand{} B) \bigr]\boolor{}
              \bigl[\boolnot  C \booland{}  (A \sand{} B) \bigr]\boolor{}\bigl[\boolnot C \booland{} (B \pand{} A ) \bigr]~.
\end{align*}
The third sub-expression is transformed analogously, thus
\begin{align*}
  \eta_3 & \ist \boolnot A \booland{} (B \booland{} C) \ist{}
\bigl[\boolnot A\booland{}  (B \pand{} C) \bigr]\boolor{}
              \bigl[\boolnot  A \booland{}  (B\sand{} C) \bigr]\boolor{}\bigl[\boolnot A \booland{} (C \pand{} B ) \bigr]~.
\end{align*}
The fourth sub-expression consists of one event sequence, that cannot be further simplified:
\begin{align*}
  \eta_4 & \ist (\boolnot A \boolnot C)\booland{} B ~.
\end{align*}
Combining these results, the three-variables minterm form of expression $\varpi\ist{}B$ is given as (meaning of underlines, see below):
\begin{align*}
  \varpi & \ist{} B \ist{}
          \eta_1 \boolor{} \eta_2 \boolor{} \eta_3 \boolor{} \eta_4 \ist{} \\
     &\ist{}   \phantom{\boolor{}}
          \bigl[\underline{\vphantom{(\pand{})}A \pand{} B \pand{} C}\bigr] \boolor{}
          \bigl[\underline{(A \sand{} B) \pand{} C} \bigr]\boolor{}
          \bigl[\underline{\vphantom{(\pand{})}B \pand{} A \pand{} C}\bigr] \boolor{}
          \bigl[A \pand{} (B \sand{} C) \bigr]\boolor{}
          \bigl[A \sand{} B \sand{} C \bigr]\boolor{} \\
    & \phantom{\ist{}}\boolor{}
          \bigl[\underline{B \pand{} (A \sand{} C ) }\bigr]\boolor{}
          \bigl[A \pand{} C \pand{} B \bigr]  \boolor{}
          \bigl[(A \sand{} C) \pand{} B\bigr] \boolor{}
          \bigl[C \pand{} A \pand{} B \bigr]\boolor{}
          \bigl[\underline{\vphantom{(\pand{})}B \pand{} C \pand{} A} \bigr]\boolor{} \\
    & \phantom{\ist{}}\boolor{}
          \bigl[\underline{ (B \sand{} C) \pand{} A }\bigr]\boolor{}
          \bigl[ \underline{\vphantom{(\pand{})}C \pand{} B \pand{} A}\bigr] \boolor{}
          \bigl[ \boolnot C \booland{}  (A \pand{} B) \bigr]\boolor{}
          \bigl[ \boolnot  C \booland{}  (A \sand{} B) \bigr]\boolor{}\\
& \phantom{\ist{}}\boolor{}
          \bigl[\underline{\boolnot C \booland{} (B \pand{} A ) } \bigr]\boolor{}
          \bigl[\underline{\boolnot A\booland{}  (B \pand{} C) }\bigr]\boolor{}
          \bigl[\boolnot  A \booland{}  (B\sand{} C)  \bigr]\boolor {}
          \bigl[\boolnot A \booland{} (C \pand{} B ) \bigr]\boolor{}\\
& \phantom{\ist{}}\boolor{}
          \bigl[C \pand{} (A \sand{} B) \bigr]\boolor{}
          \bigl[(\boolnot A \boolnot C)\booland{} B \bigr] ~.
\end{align*}
In this form $\varpi$ is not yet minimal.
As shown in figure \ref{080914_002}, only eleven of the $20$ nodes, in which $B\ist{} \True$, are really minimal.
The minterms corresponding to these non-minimal nodes are underlined in the figure above.
Applying the temporal laws of absorption provides the following minimal form, where
\begin{align*}
  \varpi & \ist{} B \ist{} \eta_1 \boolor{} \eta_2 \boolor{} \eta_3 \boolor{} \eta_4 \ist{} \\
    &\ist{}   \phantom{\boolor{}}
         \bigl[A \pand{} (B \sand{} C) \bigr]\boolor{}
         \bigl[A \sand{} B \sand{} C \bigr]\boolor{}
         \bigl[A \pand{} C \pand{} B \bigr]  \boolor{}
         \bigl[(A \sand{} C) \pand{} B\bigr] \boolor{}
         \bigl[C \pand{} A \pand{} B \bigr]\boolor{}\\
& \phantom{\ist{}}\boolor{}
         \bigl[\boolnot C\booland{} (A \pand{} B) \bigr]\boolor{}
          \bigl[\boolnot C\booland{} (A \sand{} B) \bigr]\boolor {}
          \bigl[\boolnot A\booland{} (B \sand{} C) \bigr]\boolor {}\\
& \phantom{\ist{}}\boolor{}
          \bigl[\boolnot A\booland{}( C \pand{} B )\bigr]\boolor{}
          \bigl[C \pand{} (A \sand{} B) \bigr]\boolor{}
          \bigl[(\boolnot A \boolnot C) \booland{} B \bigr] ~.
\end{align*}
Specifically, the structurally and temporally non-minimal temporal expressions (see chapter \ref{081018_001}) demonstrate that
\begin{align*}
  & (\boolnot A \boolnot C) \booland{}B  &&\text{covers}~
          \bigl[B \pand{} (A \sand{} C ) \bigr]~,~
           \bigl[\boolnot A\booland{} (B \pand{} C )\bigr]~, ~
          \bigl[\boolnot C \booland{} (B \pand{} A ) \bigr] ~, \\
  & \boolnot A\booland{} (B \pand{} C ) && \text{covers} ~ B \pand{} C \pand{} A ~, \\
  & \boolnot C\booland{} (B \pand{} A) && \text{covers} ~ B \pand{} A \pand{} C ~, \\
  & \boolnot C\booland{} (A \pand{} B) && \text{covers} ~ A \pand{} B \pand{} C ~, \\
  & \boolnot A\booland{} (C \pand{} B )&& \text{covers} ~ C \pand{} B \pand{} A ~, \\
&  \boolnot C\booland{}( A \sand{} B) && \text{covers} ~ ( A \sand{} B) \pand{} C ~, \\
&  \boolnot A\booland{}( B \sand{} C) && \text{covers} ~ ( B \sand{} C) \pand{} A ~. \\
\end{align*}
%
\begin{figure}[H]
  \centering
\input{pics/minimaldisjunkt_sequ_ausfallbaum_B_sand_gross}
\caption{Sequential failure tree corresponding to $\varpi\ist{} B$ with eleven minimal failure nodes and nine non-minimal failure nodes.
Nine failure nodes also include at least one SAND connection.}
  \label{080914_002}
\end{figure}
%
\paragraph{Second Example}~\\
The failure function $\varpi\ist{} (A \boolor{} B ) \pand C$ is not presented in a {TDNF}.
First, the transformation laws of temporal logic are used in order to create a {TDNF}:
\begin{align*}
  \varpi \ist{}& (A \boolor{} B ) \pand C \ist{} (A \pand{} C ) \boolor ( B \pand{} C )~.
\end{align*}
Both sub-expressions on the right side do not include all three relevant variables.
Each sub-expression is therefore transformed according to \eqref{080916_001} as to include the missing variables.
\begin{align*}
  \varpi &\ist{}   \phantom{\boolor{}}
      \bigl[\boolnot B \booland{} (A \pand{} C ) \bigr]\boolor{}
      \bigl[B \booland{} (A \pand{} C ) \bigr]\boolor{}
      \bigl[\boolnot A\booland{}  (B \pand{} C ) \bigr]\boolor{}
      \bigl[A \booland{}  (B \pand{} C )\bigr] \ist\\
  & \ist{}\phantom{\boolor{}}
      \bigl[\boolnot B\booland{} (A \pand{} C) \bigr]\boolor{}
      \bigl[B \pand {} A \pand{} C \bigr] \boolor{}
      \bigl[A \pand {} B \pand{} C \bigr] \boolor{}
      \bigl[ (A \sand {} B) \pand{} C  \bigr]\boolor{}\\
  & \phantom{\ist{}}\boolor{}
      \bigl[A \pand {} (B \sand{} C ) \bigr]\boolor{}
      \bigl[A \pand {} C \pand{} B\bigr]  \boolor{}
      \bigl[\boolnot A\booland{} ( B \pand{} C )\bigr] \boolor{}
      \bigl[A \pand{} B \pand{} C \bigr]\boolor{}\\
  & \phantom{\ist{}}\boolor{}
      \bigl[B \pand {} A \pand{} C \bigr] \boolor{}
      \bigl[(A \sand {} B) \pand{} C \bigr] \boolor{}
      \bigl[B \pand {} (A \sand{} C ) \bigr]\boolor{}
      \bigl[B \pand {} C \pand{} A\bigr] ~.
\end{align*}
The expressions $A \pand{} B \pand{} C $ and $B \pand {} A \pand{} C  $ and $ (A \sand {} B) \pand{} C $ are listed twice each.
Moreover, $\boolnot A\booland{} ( B \pand{} C )$ and $\boolnot B\booland{} (A \pand{} C)$ cover the non-minimal expressions $B \pand {} C \pand{} A$ und $A \pand {} C \pand{} B$.
Thus, the minterm-form of the failure function is given as
\begin{align*}
  \varpi &\ist{}   \phantom{\boolor{}}
      \bigl[\boolnot B\booland{} (A \pand{} C)\bigr]\boolor{}
      \bigl[B \pand {} A \pand{} C \bigr] \boolor{}
      \bigl[A \pand {} B \pand{} C \bigr] \boolor{}
      \bigl[(A \sand {} B) \pand{} C \bigr] \boolor{}\\
  & \phantom{\ist{}}\boolor{}
      \bigl[A \pand {} (B \sand{} C ) \bigr]\boolor{}
      \bigl[\boolnot A\booland{} ( B \pand{} C ) \bigr]\boolor{}
      \bigl[B \pand {} (A \sand{} C )\bigr] ~.
\end{align*}
Figure \ref{080915_001} shows the sequential failure tree of this second example, including its seven minimal and two non-minimal failure nodes.
%
\begin{figure}[H]
  \centering
\input{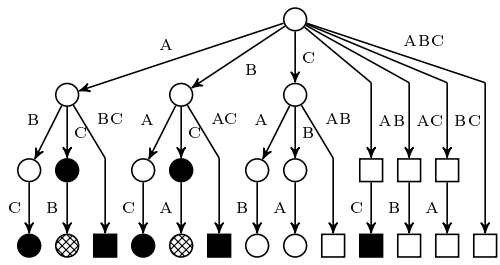}
\caption{Sequential failure tree corresponding to $\varpi\ist{} (A \boolor{} B ) \pand C$ seven minimal and two non-minimal failure nodes.
Three failure nodes include at least one SAND connection.
}
  \label{080915_001}
\end{figure}
%

%
%
%
%
%
\chapter{Abbreviations/Acronyms}
\acused{BDMP}
\acused{FAA}
\acused{SAND}
\acused{PAND}
\acused{POR}
\acused{DGL}
\acused{CCF}
\acused{BDD}
\acused{BDMP}
\acused{CCF}
\acused{DFT}
\acused{DGL}
\acused{DNF}
\acused{DRBD}
\acused{EE}
\acused{FAA}
\acused{FMEA}
\acused{FT}
\acused{FTA}
\acused{HRA}
\acused{MoCaS}
\acused{MCSS}
\acused{PAND}
\acused{POR}
\acused{RBD}
\acused{SAND}
\acused{TDNF}
\acused{TFTA}
\acused{ZSA}
{\raggedright
\begin{multicols}{2}
\begin{acronym}[MoCaS]
 \acro{BDD}{binary decision diagram}
 \acro{BDMP}{Boolean logic driven markov processes}
 \acro{CCF}{common cause failure}
 \acro{DFT}{dynamic fault tree}
 \acro{DGL}{differential equation}
 \acro{DNF}{disjunctive normal form}
 \acro{DRBD}{dynamic reliability block diagram}
 \acro{EE}[E/E]{electric/electronic}
 \acro{FAA}{federal aviation administration}
 \acro{FMEA}{failure modes and effects analysis}
 \acro{FT}{fault tree}
 \acro{FTA}{fault tree analysis}
 \acro{HRA}{human reliability analysis}
 \acro{MoCaS}{monte-carlo-simulation}
 \acro{MCSS}{minimal cutset sequences}
 \acro{PAND}{priority AND}
 \acro{POR}{priority OR}
 \acro{RBD}{reliability block diagram}
 \acro{SAND}{simultaneous AND}
 \acro{TDNF}{temporal disjunctive normal form}
 \acro{TFTA}{temporal fault tree analysis}
 \acro{ZSA}{reliability and safety analyses}
\end{acronym}
\end{multicols}
}
%
%
%
%
%

\chapter{Notation}\label{ch20071202-001}
\newdimen\sjsnotationenspalteeins
\newdimen\sjsnotationenspaltezwei
\sjsnotationenspalteeins=1.5cm      
\sjsnotationenspaltezwei=\linewidth     
\advance \sjsnotationenspaltezwei by -1.0\sjsnotationenspalteeins
\advance \sjsnotationenspaltezwei by -7mm
\begin{longtable}{p{\sjsnotationenspalteeins}p{\sjsnotationenspaltezwei}} 
 Symbol  & Meaning \\
\addlinespace[7pt]
\endfirsthead
\multicolumn{2}{l}{\sl \small continued} \\\addlinespace[7pt]
 Symbol  & Meaning \\
\addlinespace[7pt]
\endhead
 \\
\addlinespace[7pt]
 \multicolumn{2}{r}{\sl \small continued on next page}\\
\endfoot
~
\endlastfoot
$.(t)$						&time dependend parameter $.$\\
$._{i\vphantom{j}}$ & parameter $.$ for element $i$\\
$\oInfinitOperat$ & function with $\lim_{\Delta t\rightarrow0}\tfrac{\oInfinitOperat }{\Delta t}\ist 0$\\
$\booland{}$ & Boolean AND\\
$\boolor{}$ & Boolean OR\\
$\boolnot{}$ & Boolean NOT\\
$\pand{}$ & temporal PAND\\
$\sand{}$ & temporal SAND\\
$\subset; \subseteq$ & proper subset; subset\\
$\perp$ & are disjoint (for events, e.g.\ $A \pand{} B ~ \perp ~ B \pand{} A$)\\
$\in$ & is element of (for sets, e.g.\ $1 \in \{1, 2, \ldots, n\}$)\\
$\inplus$ & is part of (for events, e.g.\ $A \inplus A \pand{} B$)\\
$ \exists$ & there is \\
$ \isMinimal $ & is minimal\\
$A,B,C,D$ & failure events (within examples), see $X$\\
$\AEtoken$      & token for atomic events\\
$\CEtoken$& token for core events\\
$\EW$ & expectancy value\\
$ \mathit{eK} $ & extended core event\\
$\eCEtoken$& token for extended core events\\
$\ES$ & event sequence\\
$\EStoken$& token for event sequences\\
$\eES$& extended event sequence\\
$\eEStoken$& token for extended event sequences\\
$\eTTtoken$& token for extended temporal expressions in {TDNF}\\
$ \eta $ & temporal (sub)expression (in chapter \ref{chap080401_006} and appendix \ref{chap080404-005})\\
$f$ &failure density (density function of the failure probability)\\
$F$ &failure probability/unavailability\\
$ i $&index\\
$ j $&index\\
$ k $&index\\
$ k $ & position of an extended core event within an extended {MCSS}\\
$K$ & core event\\
$\vec{K}$ & system-state-vector/-node (sequential failure tree)\\
$\vec{K}'$ & predecessor node (sequential failure tree)\\
$\vec{K}''$ & successor node (sequential failure tree)\\
$ l $&index\\
$\lambda$ &failure rate\\
$\lambda_{i,j}$ & transition rate between states $i$ and $j$\\
$\text{max}(.)$ & maximum\\
$\MS$ & minimal cutset\\
$ \MCSS $ & minimal cutset sequence\\
$n$&index\\
$\NAEtoken$& token for negated atomic events\\
$\NCEtoken$& token for negated core events\\
$\NEStoken$& token for event sequences with negated events\\
$\NeEStoken$& token for extended event sequences with negated events\\
$\text{O}\{x\}$& order of complexity $x$\\
$P$ & state probability\\
$\dot{P}$ & derivative of the state probability\\
$\varphi$ & Boolean failure function\\
$\varpi$ & temporal failure function\\
$r$ & system state (sequential failure tree)\\
$ r $& number of AND-connected basic events within an extended core event\\
$R$ &reliability\\
$S$ & cutset (as in minimal cutset)\\
$t$ &time\\
$t_{X}$ & time of occurence of event $X$ (at this time the failure represented by $X$ occurs)\\
$T$&life expectancy\\
$T_M$&mission time\\
$\tau$ & time (parameter in integrations)\\
$\tau^{\{i\}}$ & $i$-th parameter in integrations in multiple integrals\\
$\Delta t$ & (infinitesimally) small time span\\
$\TTtoken$& token for temporal expressions in {TDNF}\\
$u$&index\\
$U$ &unavailability\\
$ w $ & number of extended core events within an extended {MCSS}\\
$X$ &Boolean event (failure logic: $X=1$ $\rightarrow$ failed, $X=0$ $\rightarrow$ not failed)\\
$\varUpsilon $& number of MCSS covered by an extended MCSS\\
$\zeta$ & number of cutsets\\
$ \xi $ & number of minimal cutsets\\
\end{longtable}

\vfill{}
\vfill{}

\centering{{\tiny (END OF DOCUMENT)}}

\end{document}